%% file: HGFv2.0.tex
\documentclass[usenatbib]{mn2e}
\pdfoutput=1
\usepackage{color}
\usepackage{ifthen}
\usepackage[figuresleft]{rotating}
\usepackage{caption}
\usepackage{colortbl}

\def\gsim{~\rlap{$>$}{\lower 1.0ex\hbox{$\sim$}}}
\def\lsim{~\rlap{$<$}{\lower 1.0ex\hbox{$\sim$}}}

\def\G{{\rm G}}
\def\clight{{\rm c}}
\def\deriv{{\rm d}}
\def\e{{\rm e}}

\def\gf{{\sc Galform}}
\def\grasil{{\sc Grasil}}

\newboolean{ColourTables}
\setboolean{ColourTables}{true}

\newcounter{AGNDone}
\def\AGN{\ifthenelse{\equal{\arabic{AGNDone}}{0}}{active galactic nuclei (AGN) \setcounter{AGNDone}{1}}{AGN}}
\newcounter{BTDone}
\def\BT{\ifthenelse{\equal{\arabic{BTDone}}{0}}{bulge-to-total ratio (B/T) \setcounter{BTDone}{1}}{B/T}}
\newcounter{CDMDone}
\def\CDM{\ifthenelse{\equal{\arabic{CDMDone}}{0}}{cold dark matter (CDM) \setcounter{CDMDone}{1}}{CDM}}
\newcounter{CMBDone}
\def\CMB{\ifthenelse{\equal{\arabic{CMBDone}}{0}}{cosmic microwave background (CMB) \setcounter{CMBDone}{1}}{CMB}}
\newcounter{IGMDone}
\def\IGM{\ifthenelse{\equal{\arabic{IGMDone}}{0}}{intergalactic medium (IGM) \setcounter{IGMDone}{1}}{IGM}}
\newcounter{IMFDone}
\def\IMF{\ifthenelse{\equal{\arabic{IMFDone}}{0}}{initial mass function (IMF) \setcounter{IMFDone}{1}}{IMF}}
\newcounter{IRDone}
\def\IR{\ifthenelse{\equal{\arabic{IRDone}}{0}}{infrared (IR) \setcounter{IRDone}{1}}{IR}}
\newcounter{ISMDone}
\def\ISM{\ifthenelse{\equal{\arabic{ISMDone}}{0}}{interstellar medium (ISM) 
\setcounter{ISMDone}{1}}{ISM}}
\newcounter{NFWDone}
\def\NFW{\ifthenelse{\equal{\arabic{NFWDone}}{0}}{Navarro-Frenk-White (NFW) \setcounter{NFWDone}{1}}{NFW}}
\newcounter{PAHDone}
\def\PAH{\ifthenelse{\equal{\arabic{PAHDone}}{0}}{polycyclic aromatic hydrocarbon (PAH) \setcounter{PAHDone}{1}}{PAH}}
\newcounter{PCADone}
\def\PCA{\ifthenelse{\equal{\arabic{PCADone}}{0}}{principal components analysis (PCA) \setcounter{PCADone}{1}}{PCA}}
\newcounter{SDSSDone}
\def\SDSS{\ifthenelse{\equal{\arabic{SDSSDone}}{0}}{Sloan Digital Sky Survey (SDSS) \setcounter{SDSSDone}{1}}{SDSS}}
\newcounter{SEDDone}
\def\SED{\ifthenelse{\equal{\arabic{SEDDone}}{0}}{spectral energy distribution (SED) \setcounter{SEDDone}{1}}{SED}}
\newcounter{SPHDone}
\def\SPH{\ifthenelse{\equal{\arabic{SPHDone}}{0}}{spectral energy distribution (SPH) \setcounter{SPHDone}{1}}{SPH}}
\newcounter{SNeDone}
\def\SNe{\ifthenelse{\equal{\arabic{SNeDone}}{0}}{supernovae (SNe) \setcounter{SNeDone}{1}}{SNe}}
\newcounter{TdFDone}
\def\TdF{\ifthenelse{\equal{\arabic{TdFDone}}{0}}{Two-degree Field Galaxy Redshift Survey (2dFGRS) \setcounter{TdFDone}{1}}{2dFGRS}}
\newcounter{TMASSDone}
\def\TMASS{\ifthenelse{\equal{\arabic{TMASSDone}}{0}}{Two-Micron All Sky Survey (2MASS) \setcounter{TMASSDone}{1}}{2MASS}}
\newcounter{UVDone}
\def\UV{\ifthenelse{\equal{\arabic{UVDone}}{0}}{ultraviolet (UV) \setcounter{UVDone}{1}}{UV}}
\newcounter{WMAPDone}
\def\WMAP{\ifthenelse{\equal{\arabic{WMAPDone}}{0}}{\emph{Wilkinson Microwave Anisotropy Probe} (WMAP) \setcounter{WMAPDone}{1}}{WMAP}}

\newcommand{\pcite}[1]{\citep{#1}}
\newcommand{\citenote}[2]{\citeauthor{#1} (\citeyear{#1}; #2)}

\input{Data/data}

\title{Galaxy Formation Spanning Cosmic History}
\author[Andrew J. Benson \& Richard Bower]{Andrew J. Benson$^1$ and Richard Bower$^2$\\
$^1$Mail Code 350-17, California Institute of Technology, Pasadena, CA~91125, U.S.A. (e-mail: {\tt abenson@caltech.edu})\\
$^2$Institute for Computational Cosmology, University of Durham, Durham, U.K.}

\begin{document}

\maketitle

\begin{abstract}
Over the past several decades, galaxy formation theory has met with significant successes. In order to test current theories thoroughly we require predictions for as yet unprobed regimes. To this end, we describe a new implementation of the \gf\ semi-analytic model of galaxy formation. Our motivation is the success of the model described by Bower et al. in explaining many aspects of galaxy formation. Despite this success, the Bower et al. model fails to match some observational constraints and certain aspects of its physical implementation are not as realistic as we would like. The model described in this work includes substantially updated physics, taking into account developments in our understanding over the past decade, and removes certain limiting assumptions made by this (and most other) semi-analytic models. This allows it to be exploited reliably in high-redshift and low mass regimes. Furthermore, we have performed an exhaustive search of model parameter space to find a particular set of model parameters which produce results in good agreement with a wide range of observational data (luminosity functions, galaxy sizes and dynamics, clustering, colours, metal content) over a wide range of redshifts. This model represents a solid basis on which to perform calculations of galaxy formation in as yet unprobed regimes.
\end{abstract}


\section{Introduction}\label{sec:Intro}

Understanding the physics of galaxy formation has been an active field of study ever since it was demonstrated that galaxies are stellar systems external to our own Milky Way. Modern galaxy formation theory grew out of early studies of cosmology and structure formation and is set within the cold dark matter cosmological model and therefore proceeds via a fundamentally hierarchical paradigm. Observational evidence and theoretical expectations indicate that galaxy formation is an ongoing process which has been occurring over the vast majority of the Universe's history. The goal of galaxy formation theory then is to describe how underlying physical principles give rise to the complicated set of phenomena which galaxies encompass.

Approaches to modelling the complex and non-linear processes of galaxy formation fall into two broad categories: direct hydrodynamical simulation and semi-analytic modelling. The division is of a somewhat fuzzy nature: semi-analytic models frequently make use of N-body simulation merger trees and calibrations from simulations, while simulations themselves are forced to include semi-analytical prescriptions for sub-resolution physics. The direct simulation approach has the advantage of, in principle, providing precise solutions (in the limit of large number of particles and assuming that numerical artifacts are kept under control), but require substantial investments of computing resources and are, at present (and for the foreseeable future), more fundamentally limited by our incomplete understanding of the various sub-resolution physical processes incorporated into them. The semi-analytical approach is less precise, but allows for rapid exploration of a wide range of galaxy properties for large, statistically useful samples. A primary goal of the semi-analytic approach is to develop insights into the process of galaxy formation that are comprehensible in terms of fundamental physical processes or emergent phenomena\footnote{A good example of an emergent phenomenon here is dynamical friction. Gravity (in the non-relativistic limit) is described entirely by $1/r^2$ forces and at this level makes no mention of frictional effects. The phenomenon of dynamical friction emerges from the interaction of large numbers of gravitating particles.}.

The problem is therefore one of complexity: can we tease out the underlying mechanisms that drive different aspects of galaxy formation and evolution from the numerous and complicated physical mechanisms at work. The key here is then ``understanding''. One can easily comprehend how a $1/r^2$ force works and can, by extrapolation, understand how this force applies to the billions of particles of dark matter in an N-body simulation. However, it is not directly obvious (at least not to these authors) how a $1/r^2$ force leads to the formation of complex filamentary structures and collapsed virialized objects. Instead, we have developed simplified analytic models (e.g. the Zel'dovich approximation, spherical top-hat collapse models etc.) which explain these phenomena in terms more accessible to the human intellect. It seems that this is what we must strive for in galaxy formation theory---a set of analytic models that we can comprehend and which allow us to understand the physics and a complementary set of precision numerical tools to allow us to determine the quantitative outcomes of that physics (in order to make precision tests of our understanding). As such, it is our opinion that no set of numerical simulations of galaxy formation, no matter how precise, will directly result in understanding. Instead, analytic methods, perhaps of an approximate nature, must always be developed (and, of course, checked against those numerical simulations) to allow us to understand galaxy formation.

Modern semi-analytic models of galaxy formation began with \cite{white_galaxy_1991}, drawing on earlier work by \cite{rees_cooling_1977} and \cite{white_core_1978}. Since then numerous studies \citep{kauffmann_formation_1993,cole_recipe_1994,baugh_semianalytic_1999,baugh_epoch_1998,somerville_semi-analytic_1999,cole_hierarchical_2000,benson_effects_2002,hatton_galics-_2003,monaco_morgana_2007} have extended and improved this original framework. Current semi-analytic models have been used to investigate many aspects of galaxy formation including:
\begin{itemize}
\item galaxy counts \pcite{kauffmann_faint_1994,devriendt_galaxy_2000};

\item galaxy clustering \pcite{diaferio_clustering_1999,kauffmann_clustering_1999,kauffmann_clustering_1999-1,baugh_modellingevolution_1999,benson_dependence_2000,benson_nature_2000,wechsler_galaxy_2001,blaizot_galics_2006};

\item galaxy colours and metallicities \pcite{kauffmann_chemical_1998,springel_populatingcluster_2001,lanzoni_galics-_2005,font_colours_2008,nagashima_metal_2005-1};

\item sub-mm and \IR\ galaxies \pcite{guiderdoni_semi-analytic_1998,granato_infrared_2000,baugh_canfaint_2005,lacey_galaxy_2008};

\item abundance and properties of Local Group galaxies \pcite{benson_effects_2002-1,somerville_can_2002};

\item the reionization of the Universe \pcite{devriendt_contribution_1998,benson_non-uniform_2001,somerville_star_2003,benson_epoch_2006};

\item the heating of galactic disks \pcite{benson_heating_2004};

\item the properties of Lyman-break galaxies \pcite{governato_seeds_1998,blaizot_predicting_2003,blaizot_galics-_2004};

\item supermassive black hole formation and \AGN feedback \pcite{kauffmann_unified_2000,croton_many_2006,bower_breakinghierarchy_2006,malbon_black_2007,somerville_semi-analytic_2008,fontanot_many_2009};

\item damped Lyman-$\alpha$ systems \pcite{maller_damped_2001,maller_damped_2003};

\item the X-ray properties of galaxy clusters \pcite{bower_impact_2001,bower_flip_2008};

\item chemical enrichment of the ICM and \IGM\ \pcite{de_lucia_chemical_2004,nagashima_metal_2005};

\item the formation histories and morphological evolution of galaxies \pcite{kauffmann_age_1996,de_lucia_formation_2006,fontanot_reproducingassembly_2007,somerville_explanation_2008}.
\end{itemize}
The goal of this approach is to provide a coherent framework within which the complex process of galaxy formation can be studied. Recognizing that our understanding of galaxy formation is far from complete these models should not be thought of as attempting to provide a ``final theory'' of galaxy formation (although that, of course, remains the ultimate goal), but instead to provide a means by which new ideas and insights may be tested and by which quantitative and observationally comparable predictions may be extracted in order to test current theories.

In order for these goals to be met we must endeavour to improve the accuracy and precision of such models and to include all of the physics thought to be relevant to galaxy formation. The complementary approach of direct numerical (N-body and/or hydrodynamic) simulation has the advantage that it provides high precision, but is significantly limited by computing power, resulting in the need for inclusion of semi-analytic recipes in such simulations. In any case, while a simulation of the entire Universe with infinite resolution would be impressive, the goal of the physicist is to understand Nature through relatively simple arguments\footnote{For example, while it is clear from N-body simulations that the action of $1/r^2$ gravitational forces in a \CDM\ universe lead to dark matter halos with approximately \NFW\ density profiles, there is a clear drive to provide simple, analytic models to demonstrate that we understand the underlying physics of these profiles \protect\pcite{taylor_phase-space_2001,barnes_density_2007,barnes_velocity_2007}.}

The most recent incarnation of the \gf\ model was described by \cite{bower_breakinghierarchy_2006}. The major innovation of that work was the inclusion of feedback from \AGN\ which allowed it to produce a very good match to the observed local luminosity functions of galaxies. In particular, the \cite{bower_breakinghierarchy_2006} model was designed to explain the phenomenon of ``down sizing''. While the \cite{bower_breakinghierarchy_2006} model turned out to also give a good match to several other datasets---including stellar mass functions at higher redshifts, the luminosity function at $z=3$ \pcite{marchesini_assessingpredictive_2007}, the abundance of $5<z<6$ galaxies \pcite{mclure_luminosity_2009},  overall colour bimodality \pcite{bower_breakinghierarchy_2006}, morphology \pcite{parry_galaxy_2009}, the global star formation rate and the black hole mass vs. bulge mass relation \pcite{bower_breakinghierarchy_2006}---it fails in several other areas, such as the mass-metallicity relation for galaxies, the sizes of galactic disks \pcite{gonzalez_testing_2009}, the small-scale clustering amplitude \pcite{seek_kim_modelling_2009}, the normalization and environmental dependence of galaxy colours \pcite{font_colours_2008} and the X-ray properties of groups and clusters \pcite{bower_parameter_2010}. Additionally, while the implementation of physics in semi-analytic models must always involve approximations, there are several aspects of the \cite{bower_breakinghierarchy_2006} model which call out for improvement and updating. Chief amongst these is the cooling model---crucial to the implementation of \AGN\ feedback---which retained assumptions about dark matter halo ``formation'' events which make implementing feedback physics difficult. Our motivation for this work is therefore to attempt to rectify these shortcomings of the \cite{bower_breakinghierarchy_2006} model, by updating the physics of \gf, removing unnecessary assumptions and approximations, and adding in new physics that is thought to be important for galaxy formation but which has previously been neglected in \gf. In addition, we will systematically explore the available model parameter space to locate a model which agrees as well as possible with a wide range of observational constraints.

In this current work, we describe the advances made in the \gf\ semi-analytic model over the past nine years. Our goal is to present a comprehensive model for galaxy formation that agrees as well as possible with current experimental constraints. In future papers we will utilize this model to explore and explain features of the galaxy population through cosmic history.

The remainder of this paper is structured as follows. In \S\ref{sec:Model} we describe the details of our revised \gf\ model. In \S\ref{sec:Selection} we describe how we select a suitable set of model parameters. In \S\ref{sec:Results} we present some basic results from our model, while in \S\ref{sec:Effects} we explore the effects of certain physical processes on the properties of model galaxies. Finally, in \S\ref{sec:Discussion} we discuss their implications and in \S\ref{sec:Conclusions} we give our conclusions. Readers less interested in the technicalities of semi-analytic models and how they are constrained may wish to skip \S\ref{sec:Model}, \S\ref{sec:Selection} and most of \S\ref{sec:Results}, and jump directly to \S\ref{sec:Predictions} where we present two interesting predictions from our model and \S\ref{sec:Effects} in which we explore the effects of varying key physical processes.

\section{Model}\label{sec:Model}

In this section we provide a detailed description of our model. 

\subsection{Starting Point}

The starting point for this discussion is \cite{cole_hierarchical_2000} and we will refer to that work for details which have not changed in the current implementation. We choose \cite{cole_hierarchical_2000} as a starting point for the technical description of our model as it represents the last point at which the details of the \gf\ model were presented as a coherent whole in a single document. As noted in \S\ref{sec:Intro} however, the scientific predecessor of this work is that of \cite{bower_breakinghierarchy_2006}. That paper, and several others, introduced many improvements relative to \cite{cole_hierarchical_2000}, many of which are described in more detail here. A brief chronology of the development of \gf\ from \cite{cole_hierarchical_2000} to the present is as follows:\\
\begin{itemize}
\item \cite{cole_hierarchical_2000}: Previous full description of the \gf\ model.

\item \cite{granato_infrared_2000}: Detailed dust modelling utilizing \grasil\ (see \S\ref{sec:DustModel}).

\item \cite{benson_non-uniform_2001}: Treatment of reionization and the evolution of the \IGM\ (see \S\ref{sec:IGM}).

\item \cite{bower_impact_2001}: Treatment of heating and ejection of hot material from halos due to energy input (see \S\ref{sec:AGNFeedback}).

\item \cite{benson_effects_2002-1}: Back reaction of reionization and photoionizing background on galaxy formation (see \S\ref{sec:IGM}) and detailed treatment of satellite galaxy dynamics (a somewhat different approach to this is described in \S\ref{sec:Merging} and \S\ref{sec:Stripping}).

\item \cite{benson_what_2003}: Effects of thermal conduction on cluster cooling rates and ``superwind'' feedback from supernovae (described in further detail by \citealt{baugh_canfaint_2005}).

\item \cite{benson_heating_2004}: Heating of galactic disks by orbiting dark matter halos.

\item \cite{nagashima_metal_2005}: Detailed chemical enrichment models (incorporating delays and tracking of individual elements; see \S\ref{sec:NonInstGasEq}).

\item \cite{bower_breakinghierarchy_2006}: Feedback from \AGN\ (see \S\ref{sec:AGNFeedback}).

\item \cite{malbon_black_2007}: Black hole growth (see \S\ref{sec:AGNFeedback}) as applied to the \cite{baugh_canfaint_2005}---see \cite{fanidakis_agn_2009} for a similar (and more advanced) treatment of black holes in the \cite{bower_breakinghierarchy_2006} model.

\item \cite{stringer_formation_2007}: Radially resolved structure of galactic disks.

\item \cite{font_colours_2008}: Ram pressure stripping of cold gas from galactic disks (see \S\ref{sec:Stripping}).
\end{itemize}

\subsection{Executive Summary}

Having developed these treatments of various physical processes one-by-one, our intention is to integrate them into a single baseline model. In addition to the accumulation of many of these improvements (many of which have not previously been utilized simultaneously), the two major modifications to the \gf\ model introduced in this work are:\\
\begin{itemize}
\item The removal of discrete ``formation'' events for dark matter halos (which previously occurred each time a halo doubled in mass and caused calculations of cooling and merging times to be reset). This has facilitated a major change in the \gf\ cooling model which previously made fundamental reference to these formation events.

\item The inclusion of arbitrarily deep levels of subhalos within subhalos and, as a consequence, the possibility of mergers between satellite galaxies.
\end{itemize}

Aspects of the model that are essentially unchanged from \cite{cole_hierarchical_2000} are listed in \S\ref{sec:Unchanged}. Before launching into the detailed discussion of the model, \S\ref{sec:Changes} provides a quick overview of what has changed between \cite{cole_hierarchical_2000} and the current implementation. In addition to changes to the physics of the model, the \gf\ code has been extensively optimized and made OpenMP parallel to permit rapid calculation of self-consistent galaxy/\IGM\ evolution (see \S\ref{sec:IGM}).

\subsection{Unchanged Aspects}\label{sec:Unchanged}

Below we list aspects of the current implementation of \gf\ that are unchanged relative to that published in \cite{cole_hierarchical_2000}.

\begin{itemize}
 \item \emph{Virial Overdensities:} Virial overdensities of dark matter halos are computed as described by \cite{cole_hierarchical_2000}, i.e. using the spherical top-hat collapse model for the appropriate cosmology and redshift. Given the mass and virial overdensity of each halo the corresponding virial radii and velocities are easily computed.
 \item \emph{Star Formation Rate:} The star formation rate in disk galaxies is given by
 \begin{equation}
  \dot{\phi} = M_{\rm cold}/\tau_\star \hbox{ where } \tau_\star = \epsilon_\star^{-1} \tau_{\rm disk} (V_{\rm disk}/200\hbox{km s}^{-1})^{\alpha_\star},
 \end{equation}
 where $M_{\rm cold}$ is the mass of cold gas in the disk, $\tau_{\rm disk}=r_{\rm disk}/V_{\rm disk}$ is the dynamical time of the disk at the half mass radius $r_{\rm disk}$ and $V_{\rm disk}$ is the circular velocity of the disk at that radius. The two parameters $\epsilon_\star$ and $\alpha_\star$ control the normalization of the star formation rate and its scaling with galaxy circular velocity respectively.
 \item \emph{Mergers/Morphological Transformation:} The classification of merger events as minor or major follows the logic of \citenote{cole_hierarchical_2000}{\S4.3.2}. However, the rules which determine when a burst of star formation occurs are altered to become:
 \begin{itemize}
  \item Major merger?
  \begin{itemize}
   \item Requires $M_{\rm sat}/M_{\rm cen}>f_{\rm burst}$.
  \end{itemize}
  \item Minor merger?
  \begin{itemize}
   \item Requires $\left\{\begin{array}{c}M_{\rm cen(bulge)}/M_{\rm cen} < {\rm B/T}_{\rm burst} \\ {\it and} \\ M_{\rm cen(cold)}/M_{\rm cen} \ge f_{\rm gas,burst}.\end{array}\right.$
  \end{itemize}
 \end{itemize}
 where $M_{\rm cen}$ and $M_{\rm sat}$ are the baryonic masses of the central and satellite galaxies involved in the merger respectively, $M_{\rm cen(bulge)}$ is the mass of the bulge component in the central galaxy and $f_{\rm burst}$, $f_{\rm gas,burst}$ and B/T$_{\rm burst}$ are parameters of the model. The parameter B/T$_{\rm burst}$ is intended to inhibit minor merger-triggered bursts in systems that are primarily spheroid dominated (since we may expect that in such systems the minor merger cannot trigger the same instabilities as it would in a disk dominated system and therefore be unable to drive inflows of gas to the central regions to fuel a burst). We would expect that the value of this parameter should be of order unity (i.e. the system should be spheroid dominated in order thatthe burst triggering be inhibited).

 \item \emph{Spheroid Sizes:} The sizes of spheroids formed through mergers are computed using the approach described by \citenote{cole_hierarchical_2000}{\S4.4.2}.
 \item \emph{Calculation of Luminosities:} The luminosities and magnitudes of galaxy are computed from their known stellar populations as described by \citenote{cole_hierarchical_2000}{\S5.1}. (Although note that the treatment of dust extinction has changed; see \S\ref{sec:DustModel}.)
\end{itemize}

\subsection{Overview of Changes}\label{sec:Changes}

We list below the changes in the current implementation of \gf\ relative to that published in \cite{cole_hierarchical_2000}. These are divided into ``minor changes'', which are typically simple updates of fitting formulas, and ``major changes'', which are significant additions to or modifications of the physics and structure of the model.

\subsubsection{Minor changes}\label{sec:MinorChanges}

\begin{itemize}
 \item \emph{Dark matter halo mass function:} [See \S\ref{sec:HaloMassFunction}] \cite{cole_hierarchical_2000} use the \cite{press_formation_1974} mass function for dark matter halos. In this work, we use the more recent determination of \cite{reed_halo_2007} which is calibrated against N-body simulations over a wide range of masses and redshifts.

 \item \emph{Dark matter merger trees:} [See \S\ref{sec:MergerTrees}] \cite{cole_hierarchical_2000} use a binary split algorithm utilizing halo merger rates inferred from the extended Press-Schechter formalism \pcite{lacey_merger_1993}. We use an empirical modification of this algorithm proposed by \cite{parkinson_generating_2008} and which provides a much more accurate match to progenitor halo mass functions as measured in N-body simulations.

 \item \emph{Density profile of dark matter halos:} [See \S\ref{sec:HaloProfiles}] \cite{cole_hierarchical_2000} employed \NFW\ \pcite{navarro_universal_1997} density profiles. We instead use Einasto density profiles \pcite{einasto__1965} consistent with recent findings (\citealt{navarro_inner_2004}; \citealt{merritt_universal_2005}; \citealt{prada_far_2006}).

 \item \emph{Density and angular momentum of halo gas:} [See \S\ref{sec:HotGasDist}] \cite{cole_hierarchical_2000} adopted a cored isothermal profile for the hot gas in dark matter halos and furthermore assumed a solid body rotation, normalizing the rotation speed to the total angular momentum of the gas (which was assumed to have the same average specific angular momentum as the dark matter). We choose to adopt the density and angular momentum distributions measured in hydrodynamical simulations by \cite{sharma_angular_2005}.

 \item \emph{Dynamical friction timescales:} [See \S\ref{sec:DynFric}] \cite{cole_hierarchical_2000} estimated dynamical friction timescales using the expression derived by \cite{lacey_merger_1993} for isothermal dark matter halos and the distribution of orbital parameters found by \cite{tormen_rise_1997}. In this work, we adopt the fitting formula of \cite{jiang_fitting_2008} to compute dynamical friction timescales and the orbital parameter distribution of \cite{benson_orbital_2005}.

 \item \emph{Disk stability:} We retain the same test of disk stability as did \citet{cole_hierarchical_2000} and similarly assume that unstable disks undergo bursts of star formation resulting in the formation of a spheroid\footnote{While the implementation of this physical process is unchanged, \protect\cite{cole_hierarchical_2000} actually ignored this process in their fiducial model, while we include it in our work.}. One slight difference is that we assume that the instability occurs at the largest radius for which the disk is deemed to be unstable rather than at the rotational support radius as \citet{cole_hierarchical_2000} assumed. This prevents galaxies with very low angular momenta from contracting to extremely small sizes (and thereby becoming very highly self-gravitating and unstable) before the stability criterion is tested. Additionally, we allow for different stability thresholds for gaseous and stellar disks. We employ the stability criterion of \cite{efstathiou_stability_1982} such that disks require
\begin{equation}
{V_{\rm d} \over (\G M_{\rm d}/R_{\rm s})^{1/2}} > \epsilon_{\rm d},
\end{equation}
to be stable, where $V_{\rm d}$ is the disk rotation speed at the half-mass radius,$M_{\rm d}$ is the disk mass and $R_{\rm s}$ is the disk radial scale length. \cite{efstathiou_stability_1982} found a value of $\epsilon_{\rm d,\star}=1.1$ was applicable for purely stellar disks. \cite{christodoulou_new_1995} demonstrate that an equivalent result for gaseous disks gives $\epsilon_{\rm d,gas}=0.9$. We choose to make $\epsilon_{\rm d,gas}$ a free parameter of the model and enforce $\epsilon_{\rm d,\star}=\epsilon_{\rm d,gas}+0.2$. For disks containing a mixture of stars and gas we linearly interpolate between $\epsilon_{\rm d,\star}$ and $\epsilon_{\rm d,gas}$ using the gas fraction as the interpolating variable. As has been recently pointed out by \cite{athanassoula_disc_2008}, this treatement of the process of disk destabilization, similar to that in other semi-analytic models, is dramatically oversimplified. As \cite{athanassoula_disc_2008} also describes, a more realistic model would need both a much more careful assessment of the disk stability and a consideration of the process of bar formation. This currently remains beyond the ability of our model to address, although it should clearly be a priority area in which semi-analytic models should strive to improve. In \gf\, we can consider an alternative disk instability treatment in which during an instability event only just enough mass is transferred from the disk to the spheroid component to restabilize the disk. While this does not explore the full range of uncertainties arising from the treatment of this process, it gives at least some idea of how significant they may be. We find that the net result of switching to the alternative treatment of instabilities is to slightly increase the number of bulgeless galaxies at all luminosities, with a corresponding decrease in the numbers of intermediate and pure spheroid galaxies. The changes, however, do not alter the qualitative trends of morphological mix with luminosity nor global properties of galaxies such as sizes and luminosity functions at $z=0$. At higher redshifts (e.g. $z\ge5$), the change is more significant, with a reduction in star formation rate by a factor of 2--3 resulting from the lowered frequency of bursts of star formation. This change could be offset by adjustments in other parameters, but demonstrates the need for a refined understanding and modelling of the disk instability process in semi-analytic models.

 \item \emph{Sizes of galaxies:} [See \S\ref{sec:Sizes}]. Sizes of disks and spheroids are determined as described by \citet{cole_hierarchical_2000}, although the equilibrium is solved for in the potential corresponding to an Einasto density profiles (used throughout this work) rather than the \NFW\ profiles assumed by \citet{cole_hierarchical_2000} and adiabatic contraction is computed using the methods of \cite{gnedin_response_2004} rather than that of \cite{blumenthal_contraction_1986}.

\end{itemize}

While we class the above as minor changes, the effects of some of these changes can be significant in the sense that reverting to the previous implementation would change some model predictions by an amount comparable to or greater than the uncertainties in the relevant observational data. However, none of these modifications lead to fundamental changes in the behaviour of the model and their effects could all be counteracted by small adjustments in model parameters. This is why we classify them as ``minor'' and do not explore their consequences in any greater detail.

\subsubsection{Major changes}

\begin{itemize}
 \item \emph{Spins of dark matter halo:} [See \S\ref{sec:HaloSpins}] In \cite{cole_hierarchical_2000} spins of dark matter halos were assigned randomly by drawing from the distribution of \cite{cole_structure_1996}. In this work, we implement an updated version of the approach described by \cite{vitvitska_origin_2002} to produce spins correlated with the merging history of the halo and consistent with the distribution measured by \cite{bett_spin_2007}.
 \item \emph{Removal of discrete formation events:} [See \S\ref{sec:FormEvents}] The discrete ``formation'' events (associated with mass doublings) in merger trees which \cite{cole_hierarchical_2000} utilized to reset cooling and merging calculations are no more. Instead, cooling, merging and other processes related to the merger tree evolve smoothly as the tree grows.
 \item \emph{Cooling model:} [See \S\ref{sec:Cooling}] The cooling model has been revised to remove the dependence on halo formation events, allow for gradual recooling of gas ejected by feedback and accounts for cooling due to molecular hydrogen and Compton cooling and for heating from a photon background.
 \item \emph{Ram pressure and tidal stripping} [See \S\ref{sec:Stripping}] Ram pressure and tidal stripping of both hot halo gas and stars and \ISM\ gas in galaxies are now accounted for.
 \item \emph{\IGM\ interaction} [See \S\ref{sec:IGM}] Galaxy formation is solved simultaneously with the evolution of the intergalactic medium in a self-consistent way: emission from galaxies and \AGN\ ionize and heat the \IGM\ which in turn suppresses the formation of future generations of galaxies.
 \item \emph{Full hierarchy of subhalos} [See \S\ref{sec:Merging}] All levels of the substructure hierarchy (i.e. subhalos, sub-subhalos, sub-sub-subhalos\ldots) are included in calculations of merging. This allows for satellite-satellite mergers.
 \item \emph{Non-instantaneous recycling} [See \S\ref{sec:NonInstGasEq}] The instantaneous recycling approximation for mass loss, chemical enrichment and feedback has been dropped and the full time and metallicity-dependences included. All models presented in this work utilize fully non-instantaneous recycling, metal production and supernovae feedback.
\end{itemize}

\subsection{Dark Matter Halos}

\subsubsection{Mass Function}\label{sec:HaloMassFunction}

We assume that the masses of dark matter halos at any given redshift are distributed according to the mass function found by \cite{reed_halo_2007}. Specifically, the mass function is given by
\begin{eqnarray}
{\deriv n \over\deriv\ln M_{\rm v}} &=& \sqrt{2\over\pi} {\Omega_0\rho_{\rm crit}\over M_{\rm v}} \left|{\deriv\ln\sigma\over\deriv\ln M}\right| \nonumber \\
 & & \times [1+1.047(\omega^{-2p})+0.6G_1+0.4G_2] A^\prime \omega \nonumber \\
 & & \times \exp\left(-{1\over 2} \omega^2-0.0325 {\omega^{2p}\over (n_{\rm eff}+3)^2}\right),
\label{eq:HaloMF}
\end{eqnarray}
where $\deriv n /\deriv\ln M_{\rm v}$ is the number of halos with virial mass $M_{\rm v}$ per unit volume per unit logarithmic interval in $M_{\rm v}$, $\sigma(M)$ is the fractional mass root-variance in the linear density field in top-hat spheres containing, on average, mass $M$, $\delta_{\rm c}(z)$ is the critical overdensity for spherical top-hat collapse at redshift $z$ \pcite{eke_cluster_1996},
\begin{eqnarray}
n_{\rm eff} &=& -6{\deriv\ln\sigma\over\deriv\ln M}-3, \\
\omega &=& \sqrt{ca} {\delta_{\rm c}(z) \over \sigma}, \\
G_1 &=& \exp\left(-{1\over 2}\left[{(\log\omega-0.788)\over 0.6}\right]^2 \right), \\
G_2 &=& \exp\left(-{1\over 2}\left[{(\log\omega-1.138)\over 0.2}\right]^2 \right), \\
\end{eqnarray}
$A^\prime = 0.310$, $ca = 0.764$ and $p=0.3$ as in eqns.~(11) and (12) of \cite{reed_halo_2007}\footnote{With minor corrections to the published version (Reed, private communication).}. The mass variance, $\sigma^2(M)$, is computed using the cold dark matter transfer function of \cite{eisenstein_power_1999} together with a scale-free primordial power spectrum of slope $n_{\rm s}$ and normalization $\sigma_8$.

When constructing samples of dark matter halos we compute the number of halos, $N_{\rm halo}$, expected in some volume $V$ of the Universe within a logarithmic mass interval, $\Delta\ln M_{\rm V}$, according to this mass function, requiring that the number of halos in the interval never exceeds $N_{\rm max}$ and is never less than $N_{\rm min}$ to ensure a fair sample. We then generate halo masses at random using a Sobol' sequence \pcite{sobol__1967} drawn from a distribution which produces, on average, $N_{\rm halo}$ halos in each interval. This ensures a quasi-random, fair sampling of halos of all masses with no quantization of halo mass and with sub-Poissonian fluctuations in the number of halos in any mass interval.

\subsubsection{Merger Trees}\label{sec:MergerTrees}

Dark matter halo merger trees, which describe the hierarchical growth of structure in a cold dark matter universe, form the backbone of our model within which the process of galaxy formation proceeds. Merger trees are either constructed through a variant of the extended Press-Schechter Monte Carlo methodology, or are extracted from N-body simulations. 

\begin{figure}
 \includegraphics[width=80mm,viewport=0mm 55mm 200mm 245mm,clip]{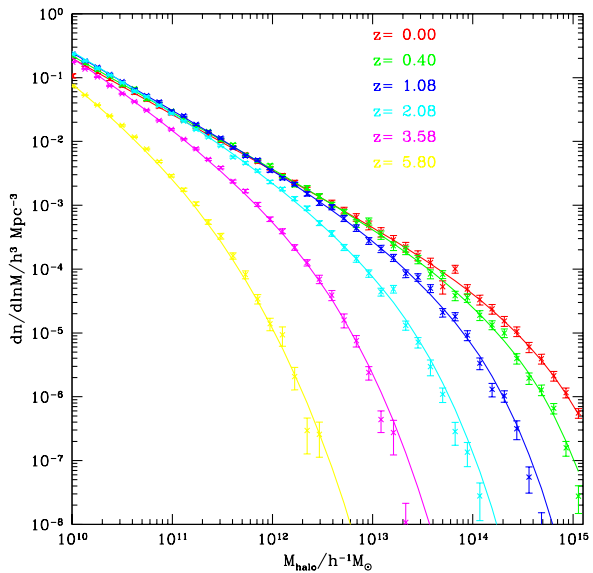}
 \caption{The dark matter halo mass function is shown at a number of different redshifts. Solid lines indicate the mass function expected from eqn.~(\protect\ref{eq:HaloMF}) while points with error bars indicate the mass function constructed using merger trees from our model. The trees in question were initiated at $z=0$ and grown back to higher redshifts using the methods of \protect\cite{parkinson_generating_2008}.}
 \label{fig:MCTrees}
\end{figure}

When constructing trees using Monte Carlo methods, we employ the merger tree algorithm described by \cite{parkinson_generating_2008} which is itself an empirical modification of that described by \cite{cole_hierarchical_2000}. We adopt the parameters $(G_0,\gamma_1,\gamma_2)=(0.57,0.38,-0.01)$ that \cite{parkinson_generating_2008} found provided the best fit\footnote{\protect\cite{benson_constraining_2008} found an alternative set of parameters which provided a better match to the evolution of the overall halo mass function but performed slightly less well (although still quite well) for the progenitor halo mass functions. We have chosen to use the parameters of \protect\cite{parkinson_generating_2008} as for the properties of galaxies we wish to get the progenitor masses as correct as possible.} to the statistics of halo progenitor masses measured from the Millennium Simulation by \cite{cole_statistical_2008}. We typically use a mass resolution (i.e. the lowest mass halo which we trace in our trees) of $5\times 10^9h^{-1}M_\odot$, which is sufficient to achieve resolved galaxy properties for all of the calculations considered in this work. An exception is when we consider Local Group satellites (see \S\ref{sec:LocalGroup}), for which we instead use a mass resolution of $10^7h^{-1}M_\odot$. Figure~\ref{fig:MCTrees} shows the resulting dark matter halo mass functions at several different redshifts and demonstrates that they are in good agreement with that expected from eqn.~(\ref{eq:HaloMF}).

\subsubsection{(Lack of) Halo Formation Events}\label{sec:FormEvents}

\cite{cole_hierarchical_2000} identified certain halos in each dark matter merger tree as being newly formed. ``Formation'' in this case corresponded to the point where a halo had doubled in mass since the previous formation event. The characteristic circular velocity and spin of halos was held fixed in between formation events and the time available for hot gas in a halo to cool was measured from the most recent formation event (such that the cooling radius was reduced to zero at each formation event). Additionally, any gas ejected by feedback was only allowed to begin recooling after a formation event, and any satellite halos that had not yet merged with the central galaxy of their host halo were assumed to have their orbits randomized by the formation event and consequently their merger timescales were reset.

While computationally useful, these formation events lack any solid physical basis. As such, we have excised them from our current implementation of \gf. Halo properties (virial velocity and spin) now change at each timestep in response to mass accretion. Additionally, the cooling and merging calculations no longer make use of formation events (see \S\ref{sec:Cooling} and \S\ref{sec:Merging} respectively).

\subsubsection{Density Profiles}\label{sec:HaloProfiles}

Recent N-body studies \pcite{navarro_inner_2004,merritt_universal_2005,prada_far_2006} indicate that the density profiles of dark matter halos in \CDM\ cosmologies are better described by the Einasto profile \pcite{einasto__1965} than the \NFW\ profile \pcite{navarro_universal_1997}. As such, we use the Einasto density profile,
\begin{equation}
\rho(r) = \rho_{-2} \exp\left( -{2\over \alpha} \left[ \left({r\over r_{-2}}\right)^\alpha- 1 \right] \right),
\end{equation}
where $r_{-2}$ is a characteristic radius at which the logarithmic slope of the density profile equals $-2$ and $\alpha$ is a parameter which controls how rapidly the logarithmic slope varies with radius. To fix the value of $\alpha$ we adopt the fitting formula of \cite{gao_redshift_2008}, truncated so that $\alpha$ never exceeds $0.3$,
\begin{equation}
\alpha = \left\{ \begin{array}{ll} 0.155 + 0.0095\nu^2 & \hbox{if } \nu < 3.907 \\ 0.3 & \hbox{if } \nu \ge 3.907,\end{array} \right.
\end{equation}
where $\nu=\delta_{\rm c}(a)/\sigma(M)$ which is a good match to halos in the Millennium Simulation\footnote{\protect\cite{gao_redshift_2008} were not able to probe the behaviour of $\alpha$ in the very high $\nu$ regime. Extrapolating their formula to $\nu > 4$ is not justified and we instead choose to truncate it at a maximum of $\alpha=0.3$.}. The value of $r_{-2}$ for each halo is determined from the known virial radius, $r_{\rm v}$, and the concentration, $c_{-2}\equiv r_{\rm v}/r_{-2}$. Concentrations are computed using the method of \cite{navarro_universal_1997} but with the best-fit parameters found by \cite{gao_redshift_2008}.

Various integrals over the density and mass distribution are needed to compute the enclosed mass, angular momentum, velocity dispersion, gravitational energy and so on of the Einasto profile. Some of these may be expressed analytically in terms of incomplete gamma functions \pcite{cardone_spherical_2005}. Expressions for the mass and gravitational potential are provided by \cite{cardone_spherical_2005}. One other integral, the angular momentum of material interior to some radius, can also be found analytically:
\begin{eqnarray}
J(r) &=& \pi^2 V_{\rm rot} \int_0^r r^{\prime (3+\alpha_{\rm rot})} \rho(r^\prime) \deriv r^\prime \nonumber \\
&=&\pi^2 V_{\rm rot} \rho_{-2} r_{-2}^{4+\alpha_{\rm rot}} {\e^{2/\alpha}\over \alpha} \left({\alpha\over 2}\right)^{4+\alpha_{\rm rot}} \nonumber \\ 
 & & \times\Gamma\left({4+\alpha_{\rm rot}\over \alpha},{2 (r/r_{-2})^\alpha\over\alpha}\right),
\end{eqnarray}
where the specific angular momentum at radius $r$ is assumed to be $r V_{\rm rot} (r/r_{\rm v})^{\alpha_{\rm rot}}$ and $\Gamma$ is the lower incomplete gamma function. Other integrals (e.g. gravitational energy) are computed numerically as needed.

\subsubsection{Angular momentum}\label{sec:HaloSpins}

As first suggested by \cite{hoyle_origin_1949}, and developed further by \cite{doroshkevich_space_1970}, \cite{peebles_origin_1969} and \cite{white_angular_1984}, the angular momenta of dark matter halos arises from tidal torques from surrounding large scale structure and is usually characterized by the dimensionless spin parameter,
\begin{equation}
 \lambda\equiv {J_{\rm v}|E_{\rm v}|^{1/2}\over\G M_{\rm v}^{5/2}},
 \label{eq:lambdaSpinDef}
\end{equation}
where $J_{\rm v}$ is the angular momentum of the halo and $E_{\rm v}$ its energy (gravitational plus kinetic). The distribution of $\lambda$ has been measured numerous times from N-body simulations \pcite{barnes_angular_1987,efstathiou_gravitational_1988,warren_dark_1992,cole_structure_1996,lemson_environmental_1999} and found to be reasonably well approximated by a log-normal distribution. More recent estimates by \cite{bett_spin_2007} using the Millennium Simulation show a somewhat different form for this distribution:
\begin{equation}
 P(\lambda) \propto \left({\lambda\over\lambda_0}\right)^3 \exp\left[-\alpha_\lambda\left({\lambda\over\lambda_0}\right)^{3/\alpha_\lambda}\right],
\end{equation}
where $\alpha_\lambda=2.509$ and $\lambda_0=0.04326$ are parameters.

\cite{cole_hierarchical_2000} assigned spins to dark matter halos by drawing them at random from the distribution of \cite{cole_structure_1996}. This approach has the disadvantage that spin is not influenced by the merging history of a given dark matter halo and, furthermore, spin can vary dramatically from one timestep to the next even if a halo experiences no (or only very minor) merging. This was not a problem for \cite{cole_hierarchical_2000}, who made use of the spin of each newly formed halo, ignoring any variation between formation events\footnote{As it seems reasonable to assume that the spins of a halo at two successive formation events, i.e. separated by a factor of two in halo mass, would be only weakly correlated.}. However, in our case, such behaviour would be problematic. We therefore revisit an idea first suggested by \citeauthor{vitvitska_origin_2002}~(\citeyear{vitvitska_origin_2002}; see also \citealt{maller_modelling_2002}). They followed the contribution to the angular momentum of each halo from its progenitor halos (which carry angular momentum in both their internal spin and orbit). Note that the angular momentum still arises via tidal torques (which are responsible for the orbital angular momenta of merging halos).

Halos in the merger tree which have no progenitors are assigned a spin by drawing at random from the distribution of \cite{bett_spin_2007}. For halos with progenitors, we proceed as follows:
\begin{enumerate}
\item Compute the internal angular momenta of all progenitor halos using their previously assigned spin and eqn.~(\ref{eq:lambdaSpinDef});
\item Select orbital parameters (specifically the orbital angular momentum) for each merging pair of progenitors by drawing at random from the distribution found by \cite{benson_orbital_2005};
\item Sum the internal and orbital angular momenta of all progenitors assuming no correlations between the directions of these vectors\footnote{Additionally, we are assuming that mass accretion below the resolution of the merger tree contributes the same mean specific angular momentum as accretion above the resolution.};
\item Determine the spin parameter of the new halo from this summed angular momentum and eqn.~(\ref{eq:lambdaSpinDef}).
\end{enumerate}
\cite{benson_orbital_2005} report orbital velocities for merging halos and give expressions for the angular momenta of those orbits assuming point mass halos. While this will be a reasonable approximation for high mass ratio mergers it will fail for mergers of comparable mass halos. In addition, halo mergers may not necessarily conserve angular momentum in the sense that some material, plausibly with the highest specific angular momentum, may be thrown out during the merging event leaving the final halo with a lower angular momentum. To empirically account for these two factors we divide the orbital angular momentum by a factor of $f_2\equiv 1+M_2/M_1$ (where $M_2<M_1$ are the masses of the dark matter halos). We find that this empirical factor leads to good agreement with the measured N-body spin distribution, but could be justified more rigorously by measuring the angular momentum (accounting for finite size effects) of the progenitor and remnant halos in N-body mergers.

\begin{figure}
 \includegraphics[width=80mm,viewport=0mm 60mm 200mm 245mm,clip]{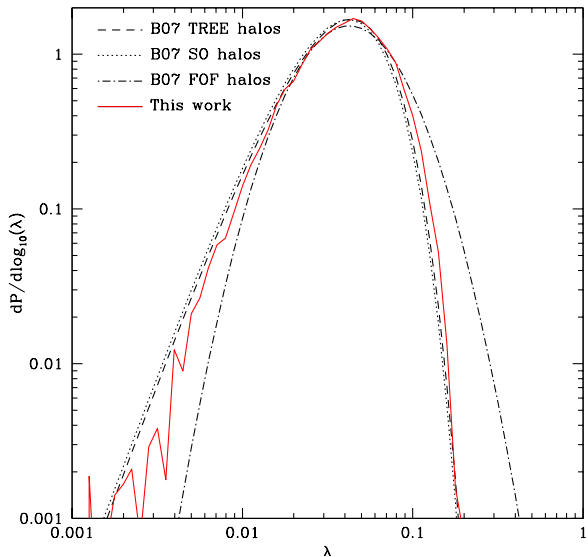}
 \caption{The distribution of dark matter halo spin parameters. Black lines show measurements of this distribution from the Millennium N-body simulation \protect\citeauthor{bett_spin_2007}~(\protect\citeyear{bett_spin_2007}; B07), for three different group finding algorithms. \protect\cite{bett_spin_2007} note that the ``{\sc tree}'' halos give the most accurate determination of the spin distribution. The red line shows the results of the Monte Carlo model described in this work, using \SpinModelNTree\ Monte Carlo realizations of merger trees spanning a range of masses identical to that used by \protect\cite{bett_spin_2007}.}
 \label{fig:SpinModel}
\end{figure}

To test the validity of this approach we generated \SpinModelNTree\ Monte Carlo realizations of merger trees drawn from a halo mass function consistent with that of the Millennium Simulation and with a range of masses consistent with that for which \cite{bett_spin_2007} were able to measure spin parameters and applied the above procedure. Figure~\ref{fig:SpinModel} shows the results of this test. We find remarkably good agreement between the distribution of spin measured by \cite{bett_spin_2007} and the results of our Monte Carlo model. It should be noted that our assumption of no correlation between the various angular momenta vectors of progenitor halos is not correct. However, \cite{benson_orbital_2005} shows that any such correlations are weak. Therefore, given the success of a model with no correlations, we choose to ignore them.

Our results are in good agreement with previous attempts to model the halo spin distribution in this way. \cite{maller_modelling_2002} found good agreement with N-body results using the same principles, although they found that introducing some correlation between the directions of spin and orbital angular momenta improved their fit. \cite{vitvitska_origin_2002} also found generally good agreement with N-body simulations using orbital parameters of halos drawn from an N-body simulation. Both of these earlier calculations relied on much less well calibrated orbital parameter distributions for merging halos and the simulations to which they compared their results had significantly poorer statistics than the Millennium simulation. Our results confirm that this approach to calculating halo spins from a merger history still works extremely well even when confronted with the latest high-precision measures of the spin distribution.

\subsection{Cooling Model}\label{sec:Cooling}

The cooling model described by \cite{cole_hierarchical_2000} determines the mass of gas able to cool in any timestep by following the propagation of the cooling radius in a notional hot gas density profile\footnote{We refer to this as a ``notional'' profile since it is taken to represent the profile before any cooling can occur. Once some cooling occurs presumably the actual profile adjusts in some way to respond to this and so will no longer look like the notional profile, even outside of the cooling radius.} which is fixed when a halo is flagged as ``forming'' and is only updated when the halo undergoes another formation event. The mass of gas able to cool in any given timestep is equal to the mass of gas in this notional profile between the cooling radius at the present step and that at the previous step. The cooling time is assumed to be the time since the formation event of the halo. Any gas which is reheated into or accreted by the halo is ignored until the next formation event, at which point it is added to the hot gas profile of the newly formed halo. The notional profile is constructed using the properties (e.g. scale radius, virial temperature etc.) of the halo at the formation event and retains a fixed metallicity throughout, corresponding to the metallicity of the hot gas in the halo at the formation event.

In this work we implement a new cooling model. We do away with the arbitrary ``formation'' events and instead use a continuously updating estimate of cooling time and halo properties. For the purposes of this calculation we define the following quantities:
\begin{itemize}
 \item $M_{\rm hot}$: The current mass of hot (i.e. as yet uncooled) gas remaining in the notional profile;
 \item $M_{\rm cooled}$: The mass of gas which has cooled out of the notional profile into the galaxy phase;
 \item $M_{\rm reheated}$: The mass of gas which has been reheated (by supernovae feedback) but has yet to be reincorporated back into the hot gas component.
 \item $M_{\rm ejected}$: The mass of gas which has been ejected beyond the virial radius of this halo, but which may later reaccrete into other, more massive halos.
\end{itemize}
The notional profile always contains a mass $M_{\rm total}=M_{\rm hot}+M_{\rm cooled}+M_{\rm reheated}$. The properties (density normalization, core radius) are reset, as described in \S\ref{sec:HotGasDist}, at each timestep. The previous infall radius (i.e. the radius within which gas was allowed to infall and accrete onto the galaxy) is computed by finding the radius which encloses a mass $M_{\rm cooled}+M_{\rm reheated}$ (i.e. the mass previously removed from the hot component) in the current notional profile.

We aim to compute a time available for cooling for the halo, $t_{\rm avail}$, from which we can compute a cooling radius in the usual way (i.e. by finding the radius in the notional profile at which $t_{\rm cool}=t_{\rm avail}$). In \cite{cole_hierarchical_2000} the time available for cooling is simply set to the time since the last formation event of the halo.

At any time, the rate of cooling per particle is just $\Lambda(T,\hbox{\boldmath $Z$},n_{\rm H},F_\nu) n_{\rm H}$ where $\Lambda(T,\hbox{\boldmath $Z$},n_{\rm H},F_\nu)$ is the cooling function, and $n_{\rm H}$ the number density of hydrogen, $\hbox{\boldmath $Z$}$ a vector of metallicity (such that the $i^{\rm th}$ component of $\hbox{\boldmath $Z$}$ is the abundance by mass of the $i^{\rm th}$ element) and $F_\nu$ the spectrum of background radiation. The total cooling luminosity is then found by multiplying by the number of particles, $N$, in some volume $V$ that we want to consider. If we take this volume to be the entire halo then $N\equiv M_{\rm total}/\mu m_{\rm H}$. If we integrate this luminosity over time, we find the total energy lost through cooling. The total thermal energy in our volume $V$ is just $3Nk_{\rm B}T/2$. The gas will have completely cooled once the energy lost via cooling equals the original thermal energy, i.e.:
\begin{equation}
 3Nk_{\rm B}T_{\rm v}/2 = \int_0^t \Lambda(t^\prime) n_{\rm H} N \deriv t^\prime,
\end{equation}
where for brevity we write $\Lambda(t)\equiv\Lambda[T_{\rm v}(t),\hbox{\boldmath $Z$}(t),n_{\rm H}(t),F_\nu(t)]$. We can write this as
\begin{equation}
 t_{\rm cool} = t_{\rm avail},
\end{equation}
where
\begin{equation}
 t_{\rm cool}(t) = {3k_{\rm B}T_{\rm v}(t) \over 2\Lambda(t) n_{\rm H}}
\end{equation}
is the usual cooling time and
\begin{equation}
 t_{\rm avail} = {\int_0^t \Lambda(t^\prime) n_{\rm H}(t^\prime) N \deriv t^\prime \over \Lambda(t) n_{\rm H}(t) N}
\end{equation}
is the time available for cooling. We can re-write this as
\begin{equation}
 t_{\rm avail} = {\int_0^t [T_{\rm v}(t^\prime) N / t_{\rm cool}(t^\prime)] \deriv t^\prime \over [T_{\rm v}(t) N / t_{\rm cool}(t)]}.
 \label{eq:tAvail}
\end{equation}
In the case of a static halo, where $T_{\rm v}$, $\hbox{\boldmath $Z$}$, $F_\nu$ and $N$ are independent of time, $t_{\rm avail}$ reduces to the time since the halo came into existence as we might expect. For a non-static halo the above makes more physical sense. For example, consider a halo which is below the $10^4$K cooling threshold from time $t=0$ to time $t=t_4$, and then moves above that threshold (with fixed properties after this time). Since $t_{\rm cool}=\infty$ (i.e. $\Lambda(t)=0$) before $t_4$ in this case we find that $t_{\rm avail}=t-t_4$ as expected. Note that since the number of particles, $N$, appears in both the numerator and denominator of eqn.~(\ref{eq:tAvail}) we can, in practice, replace $N$ by $M_{\rm total}$ without changing the resulting time.

The cooling time in the above must be computed for a specific value of the density. We choose to use the cooling time at the mean density of the notional profile at each timestep. This implicitly assumes that the density of each mass element of gas in the notional profile has the same time dependence as the mean density of the profile, i.e. that the profile evolves in a self-similar way and that $\Lambda(t)$ is independent of $n_{\rm H}$ (which will only be true in the collisional ionization limit). This may not be true in general, but serves as an approximation allowing us to describe the cooling of the entire halo with just a single integral\footnote{A more elaborate model could compute a separate integral for each shell of gas, following the evolution of its density as a function of time as the profile evolves due to continued infall and cooling.}.

Having computed the time available for cooling we solve for the cooling radius in the notional profile at which $t_{\rm cool}(r_{\rm cool})=t_{\rm avail}$ (as described in \S\ref{sec:CoolRadius}). We also estimate the largest radius from whch gas has had sufficient time to freefall to the halo centre (as described in \S\ref{sec:Freefall}). The current infall radius is taken to be the smaller of the cooling and freefall radii. Any mass between the current infall radius and that at the previous timestep is allowed to infall onto the galaxy during the current timestep---that is, it is transferred from $M_{\rm hot}$ to $M_{\rm cooled}$.

One refinement which must be introduced is to limit the integral
\begin{equation}
 {\mathcal E} = {3\over 2} k_{\rm B} \int_0^t [T_{\rm v}(t^\prime) N / t_{\rm cool}(t^\prime)] \deriv t^\prime,
 \label{eq:tAvailIntegral}
\end{equation}
so that the total radiated energy cannot exceed the total thermal energy of the halo. This limit is given by
\begin{equation}
 {\mathcal E}_{\rm max} = {3\over 2} k_{\rm B} T_{\rm v}(t) N {\bar{\rho}_{\rm total} \over \rho_{\rm total}(r_{\rm v})},
 \label{eq:tAvailIntegralMax}
\end{equation}
where $\bar{\rho}_{\rm total}$ is the mean density of the notional profile and $\rho_{\rm total}(r_{\rm v})$ is the density of the notional profile at the virial radius. For the entire halo (out to the virial radius) to cool takes longer than for gas at the mean density of the halo to cool, by a factor of  $\bar{\rho}_{\rm total} / \rho_{\rm total}(r_{\rm v})$. This is the origin of the ratio of densities in eqn.~(\ref{eq:tAvailIntegralMax}).

We must then consider two additional effects: accretion (\S\ref{sec:Accretion}) and reheating (\S\ref{sec:Reheating}). The cooling model is then fully specified once we specify the distribution of gas in the notional profile (\S\ref{sec:HotGasDist}), determine a cooling radius (\S\ref{sec:CoolRadius}) and freefall radius (\S\ref{sec:Freefall}), and consider how to compute the angular momentum of the infalling gas (\S\ref{sec:CoolAngMom}).

\subsubsection{Accretion}\label{sec:Accretion}

When a halo accretes another halo, we merge their notional gas profiles. Since the integral, ${\mathcal E} = \int (N T_{\rm v}/t_{\rm cool}) \deriv t$, that we are computing is the total energy lost we simply add ${\mathcal E}$ from the accreted halo to that of the halo it accretes into. This gives the total energy lost from the combined notional profile. However, we must consider the fact that only a fraction $M_{\rm hot}/M_{\rm total}$ of the gas from the accreted halo is added to the hot gas reservoir of the combined halo (the mass $M_{\rm cooled}$ from the accreted halo becomes the satellite galaxy while the mass $M_{\rm reheated}$ is added to the reheated reservoir of the new halo to await reincorporation into the hot component; see \S\ref{sec:Reheating}). We simply multiply the integral ${\mathcal E}$ of the accreted halo by this fraction before adding it to the new halo.

\begin{figure}
 \includegraphics[width=80mm,viewport=0mm 10mm 186mm 265mm,clip]{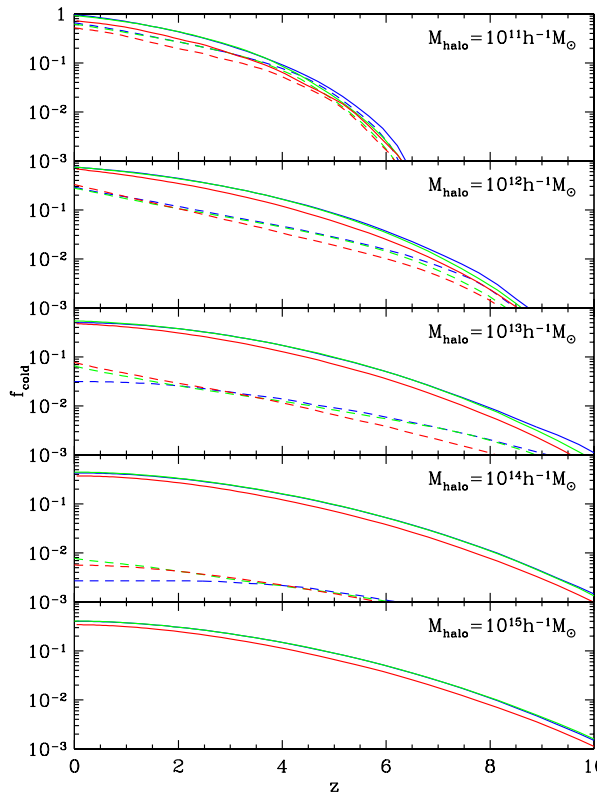}
 \caption{The mean cooled gas fractions in the merger trees of halos with masses $10^{11}$, $10^{12}$, $10^{13}$, $10^{14}$ and $10^{15}h^{-1}M_\odot$ at $z=0$ are shown by coloured lines. Green lines show results from the cooling model described in this work while red lines indicate the model of \protect\cite{cole_hierarchical_2000} and blue lines the cooling model of \protect\cite{bower_breakinghierarchy_2006}. Solid lines show the total cooled fraction in all branches of the merger trees, while dashed lines show the cooled fraction in the main branch of the trees. For the purposes of this figure, no star or black hole formation was included in these calculations, so consequently there is no reheating of gas, expulsion of gas from the halo or metal enrichment. Additionally, no galaxy merging was allowed. As such, the differences between models arises purely from their different implementations of cooling.}
 \label{fig:CoolingModels}
\end{figure}

Figure \ref{fig:CoolingModels} compares the mean cooled gas fractions in halos of different masses computed using the cooling model described here (green lines) and two previous cooling models used in \gf: that of \citeauthor{cole_hierarchical_2000}~(\citeyear{cole_hierarchical_2000}; red lines) and that of \citeauthor{bower_breakinghierarchy_2006}~(\citeyear{bower_breakinghierarchy_2006}; blue lines). The only significant difference between the cooling implementations of \cite{cole_hierarchical_2000} and \cite{bower_breakinghierarchy_2006} is that \cite{bower_breakinghierarchy_2006} allow reheated gas to gradually return to the hot component (and so be available for re-cooling) at each timestep (in the same manner as in the present work), while \cite{cole_hierarchical_2000} simply accumulated this reheated gas and returned it all to the hot component only at the next halo formation event (i.e. after a halo mass doubling). No star or black hole formation was included in these calculations, so consequently there is no reheating of gas, expulsion of gas from the halo or metal enrichment. Additionally, no galaxy merging was allowed. The thick lines show the total cooled fraction in all branches of the merger trees, while the thin lines show the cooled fraction in the main branch of the trees\footnote{We define the main branch of the merger tree as the set of progenitor halos found by starting from the final halo and repeatedly stepping back to the most massive progenitor of the current halo at each time step. It should be noted that definition is not unique, and can depend on the time resolution of the merger tree. It can also result in situations where the main branch does not correspond to the most massive progenitor halo at a given timestep.}.

The cooling model utilized by \cite{bower_breakinghierarchy_2006} was similar to that of \cite{cole_hierarchical_2000} except that it allowed accreted and reheated gas to rejoin the hot gas reservoir in a continuous manner rather than only at each halo formation event. Additionally, it used the current properties of the halo (e.g. virial temperature) to compute cooling rates rather than the properties of the halo at the previous formation event. As such, the \cite{bower_breakinghierarchy_2006} model contains many features of the current cooling model, but retains the fundamental division of the merger tree into discrete branches as in the \cite{cole_hierarchical_2000} model.

We find that, in general, the cooling model described here predicts a total cooled fraction very close to that predicted by the cooling model of \cite{bower_breakinghierarchy_2006}, the exception being at very early times in low mass halos where it gives a slightly lower value. The difference of course is that the new model does not contain artificial resets in the cooling calculation which, although they make little difference to this statistic, have a strong influence on, for example, calculations of the angular momentum of cooling gas. Both of these models predict somewhat more total cooled mass than the \cite{cole_hierarchical_2000} model. This is due entirely to the allowance of accreted gas to begin cooling immediately.

If we consider the cooled fraction in the main branch of each tree (i.e. the mass in what will become the central galaxy in the final halo) we see rather different behaviour. At early times, the new model tracks the \cite{bower_breakinghierarchy_2006} model. At late times, however, the \cite{bower_breakinghierarchy_2006} model shows a much lower cooling rate while the new model tracks the cooled fraction in the \cite{cole_hierarchical_2000} model quite closely. This occurs in massive halos where, in the \cite{bower_breakinghierarchy_2006} model the use of the current halo properties to determine cooling rates results in ever increasing cooling times as the virial temperature of the halo increases and the halo density (and hence hot gas density) decline. The \cite{cole_hierarchical_2000} model is less susceptible to this as it computes halo properties based on the halo at formation. The new cooling model produces results comparable to the \cite{cole_hierarchical_2000} model since, while it utilizes the present properties of the halo just as does the \cite{bower_breakinghierarchy_2006} model, it retains a memory of the early properties of the halo.

\subsubsection{Reheating}\label{sec:Reheating}

When gas is reheated (via feedback; \S\ref{sec:Feedback}) we assume that it is heated to the virial temperature of the current halo (i.e. the host halo for satellite galaxies) and is placed into a reservoir $M_{\rm reheated}$. Mass is moved from this reservoir back into the hot gas reservoir on a timescale of order the halo dynamical time, $\tau_{\rm dyn}$. Specifically, mass is returned to the hot phase at a rate
\begin{equation}
 \dot{M}_{\rm hot} =  \alpha_{\rm reheat} {M_{\rm reheated} \over \tau_{\rm dyn}}
\end{equation}
during each timestep. This effectively undoes the cooling energy loss which caused this gas to cool previously. The energy integral ${\mathcal E}$ is therefore modified by subtracting from it an amount $\Delta N_{\rm reheated} T_{\rm v}$ where $\Delta N_{\rm reheated}$ is the number of particles reheated.

Similarly, the notional profile is allowed to ``forget'' about any cooled gas on a timescale of order the dynamical time (i.e. we assume that the notional profile adjusts to the loss of this gas). This is implemented by removing mass from the cooled reservoir at a rate
\begin{equation}
 \dot M_{\rm cooled} = - \alpha_{\rm remove} {M_{\rm cooled} \over \tau_{\rm dyn}}.
\end{equation}

\subsubsection{Hot Gas Distribution}\label{sec:HotGasDist}

The hot gas is assumed to be distributed in a notional profile with a run of density consistent with that found in hydrodynamical simulations \pcite{sharma_angular_2005,stringer_formation_2007}. \cite{sharma_angular_2005} performed non-radiative cosmological \SPH\ simulations and studied the properties of the hot gas in dark matter halos. These simulations are therefore well suited to our purposes since they relate to the notional profile which is defined to be that in the absence of any cooling. The gas density profiles found by \cite{sharma_angular_2005} are well described by the expression:
\begin{equation}
 \rho(r) \propto {1 \over (r+r_{\rm core})^3},
\end{equation}
where $r_{\rm core}$ is a characteristic core radius for the profile. We choose to set $r_{\rm core} = a_{\rm core} r_{\rm v}$ where $a_{\rm core}$ is a parameter whose value is the same for all halos at all redshifts. The simulations suggest that $a_{\rm core}\approx 0.05$ \pcite{stringer_formation_2007}, but we will treat $a_{\rm core}$ as a free parameter to be constrained by observational data. The density profile is normalized such that
\begin{equation}
 \int_0^{r_{\rm v}} \rho(r) 4 \pi r^2 \deriv r = M_{\rm total},
\end{equation}
and the hot gas is assumed to be isothermal at the virial temperature
\begin{equation}
 T_{\rm v} = {1 \over 2} {\mu m_{\rm H} \over k} V_{\rm v}^2
\end{equation}
with a metallicity equal to $Z = M_{Z,{\rm hot}}/M_{\rm hot}$. Initially, $M_{Z,{\rm hot}}=0$ but can become non-zero due to metal production and outflows as a result of star formation and feedback.

\subsubsection{Cooling Radius}\label{sec:CoolRadius}

Given the time available for cooling from eqn.~(\ref{eq:tAvail}) the cooling radius is found by solving
\begin{equation}
t_{\rm avail} = {{3 \over 2} (n_{\rm tot}/n_{\rm H})k_{\rm B}T_{\rm V} \over n_{\rm H} (r_{\rm cool}) \Lambda(t)},
\end{equation}
where $n_{\rm tot}$ is the total number density of the atoms in the gas. Due to the dependence of $\Lambda(t)$ on density when a photoionizing background is present (see \S\ref{sec:PhotoEffect}) this equation must be solved numerically.

\subsubsection{Freefall Radius}\label{sec:Freefall}

To compute the mass of gas which can actually reach the centre of a halo potential well at any given time we require that not only has the gas had time to cool but also that it has had time to freefall to the centre of the halo starting from zero velocity at its initial radius. To estimate the maximum radius from which cold gas could have reached the halo centre through freefall we proceed as follows. We compute a time available for freefall in the halo, $t_{\rm avail,ff}$, using eqn.~(\ref{eq:tAvail}), but limit the integral ${\mathcal E}$ (defined in eqn.~\ref{eq:tAvailIntegral}) such that the time available can not exceed the freefall time at the virial radius. We then solve the freefall equation
\begin{equation}
 \int_0^{r_{\rm ff}} {\deriv r^\prime \over \sqrt{2[\Phi(r^\prime)-\Phi(r_{\rm ff})]}} = t_{\rm avail,ff},
\end{equation}
where $\Phi(r)$ is the gravitational potential of the halo, for the radius $r_{\rm ff}$ at which the freefall time equals the time available. Only gas within the minimum of the cooling and freefall radii at each timestep is allowed to reach the centre of the halo and become part of the forming galaxy.

\subsubsection{Angular Momentum}\label{sec:CoolAngMom}

The angular momentum of gas in the notional halo is tracked using a similar approach as for the mass. We define the following quantities:
\begin{description}
 \item [$J_{\rm hot}$:] The total angular momentum in the $M_{\rm hot}$ reservoir of the notional profile;
 \item [$J_{\rm cooled}$:] The total angular momentum in the $M_{\rm cooled}$ reservoir of the notional profile;
 \item [$J_{\rm reheated}$:] The total angular momentum in the $M_{\rm reheated}$ reservoir of the notional profile;
 \item [$j_{\rm new}$:] The specific angular momentum which newly accreted material must have in order to produce the correct change in angular momentum for this halo\footnote{The angular momentum of a halo differs from that of its main progenitor due to an increase in mass, change in virial radius and change in spin parameter. $j_{\rm new}$ is computed by finding the difference in the angular momentum of a halo and its main progenitor and dividing by their mass difference. Note that this quantity can therefore be negative.}.
\end{description}
$J_{\rm cooled}$ and $J_{\rm reheated}$ are initialized to zero at the start of the calculation. $J_{\rm hot}$ is initialized by assuming that any material accreted below the resolution of the merger tree arrives with the mean specific angular momentum of the halo. Angular momentum is then tracked using the following method:
\begin{enumerate}
\item At the start of a time step, all three angular momentum reservoirs from the most massive progenitor halo are added to those of the current halo.

\item We assume that the specific angular momentum of the gas halo is distributed according to the results of \cite{sharma_angular_2005} such that the differential distribution of specific angular momentum, $j$, is given by
\begin{equation}
 {1\over M}{\deriv M \over \deriv j} = {1 \over j_{\rm d}^{\alpha_j}\Gamma(\alpha_j)}j^{\alpha_j-1}{\rm e}^{-j/j_{\rm d}},
\end{equation}
where $\Gamma$ is the gamma function, $M$ is the total mass of gas, $j_{\rm d}=j_{\rm tot}/\alpha$ and $j_{\rm tot}$ is the mean specific angular momentum of the gas. The parameter $\alpha_j$ is chosen to be $0.89$, consistent with the median value found by \cite{sharma_angular_2005} in simulated halos. The fraction of mass with specific angular momentum less than $j$ is then given by
\begin{equation}
 f(<j) = \gamma\left(\alpha_j,{j\over j_{\rm d}}\right),
\end{equation}
where $\gamma$ is the incomplete gamma function. Once the mass of gas cooling in any given timestep is known the above allows the angular momentum of that gas to be found. This amount is added to the $J_{\rm cooled}$ reservoir.

\item If $J_{\rm reheated}>0$ then an angular momentum
\begin{equation}
 \Delta J_{\rm hot} = \left\{ \begin{array}{ll} J_{\rm reheated} \alpha_{\rm reheat} \Delta t/\tau_{\rm dyn} & \hbox{ if } \alpha_{\rm reheat} \Delta t < \tau_{\rm dyn} \\
                               J_{\rm reheated}& \hbox{ if } \alpha_{\rm reheat} \Delta t \ge \tau_{\rm dyn} \\
                              \end{array}
 \right.
\end{equation}
is transferred to back to the hot phase, consistent with the fraction of mass returned to the hot phase (see \S\ref{sec:Reheating}).

\item When a halo becomes a satellite of a larger halo, $J_{\rm hot}$ of the larger halo is increased by an amount, $j_{\rm new} M_{\rm hot,sat}$. This accounts for the orbital angular momentum of the gas in the satellite halo assuming that, on average, satellites have specific angular momentum of $j_{\rm new}$. We do the same for $J_{\rm reheated}$, assuming that the $M_{\rm reheated}$ reservoir of the satellite arrives with the same specific angular momentum.

\item When gas is ejected from a galaxy disk to join the reheated reservoir it is ejected with the mean specific angular momentum of the disk. Gas ejected during a starburst is also assumed to be ejected with the mean pseudo-specific angular momentum\footnote{As defined by \protect\citeauthor{cole_hierarchical_2000}~(\protect\citeyear{cole_hierarchical_2000}; their eqn.~C11) and equal to the product of the bulge half-mass radius and the circular velocity at that radius.} of the bulge.
\end{enumerate}

Because $j_{\rm new}$ can be negative on occasion it is possible that $J_{\rm hot}<0$ can occur. This, in turn, can lead to a galaxy disk with a negative angular momentum. We do not consider this to be a fundamental problem due to the vector nature of angular momentum. When computing disk sizes we simply consider the magnitude of the disk angular momentum, ignoring the sign.

\subsubsection{Cooling/Heating Rates of Hot Gas in Halos}\label{sec:Cloudy}

The cooling model described above requires knowledge of the cooling function, $\Lambda(T,\hbox{\boldmath $Z$},n_{\rm H},F_\nu)$. Given a gas metallicity and density and the spectrum of the ionizing background we can compute cooling and heating rates for gas in dark matter halos. Calculations were performed with version 08.00 of {\sc Cloudy}, last described by \cite{ferland_cloudy_1998}. In practice, we compute cooling/heating rates as a function of temperature, density and metallicity using the self-consistently computed photon background (\S\ref{sec:IGM}) after each timestep. The rates are computed on a grid which is then interpolated on to find the cooling/heating rate for any given halo.

Chemical abundances are assumed to behave as follows:
\begin{itemize}
 \item{} $Z=0$ : ``zero'' metallicity corresponding to the ``primordial'' abundance ratios as used by {\sc Cloudy} version 08.00 (see the \emph{Hazy} documentation of {\sc Cloudy} for details). 

 \item{} [Fe/H]$<-1$ : ``primordial'' abundance ratios from \cite{sutherland_cooling_1993};

 \item{} [Fe/H]$\ge 1$ : Solar abundance ratios as used by {\sc Cloudy} version 08.00 (see the \emph{Hazy} documentation of {\sc Cloudy} for details).
\end{itemize}
However, since our model can track the abundances of individual elements we know the abundances in each cooling halo. In principle, we could recompute a cooling/heating rate for each halo using its specific abundances as input into {\sc Cloudy}. This is computationally impractical however. Instead, we follow the approach of \citet{martinez-serrano_chemical_2008} who perform a \PCA\ to find the optimal linear combination of abundances which minimizes the variance between cooling/heating rates computed using that linear combination as a parameter and a full calculation using all abundances. The best linear combination turns out to be a function of temperature. We therefore track this linear combination of abundances at 10 different temperatures for all of the gas in our models and use it instead of metallicity when computing cooling/heating rates.

{\bf Compton Cooling:} \citet{cole_hierarchical_2000} allowed hot halo gas to cool via two-body collisional radiative processes. However, as we go to higher redshifts the effect of Compton cooling must be considered. The Compton cooling timescale is given by \pcite{peebles_recombination_1968}:
\begin{equation}
\tau_{\rm Compton} = {3 m_{\rm e}\clight (1+1/x_{\rm e}) \over 8\sigma_{\rm T}aT^4_{\rm CMB}(1-T_{\rm CMB}/T_{\rm e})},
\end{equation}
where $x_{\rm e}=n_{\rm e}/n_{\rm t}$, $n_{\rm e}$ is the electron number density, $n_{\rm t}$ is the number density of all atoms and ions, $T_{\rm CMB}$ is the \CMB\ temperature and $T_{\rm e}$ is the electron temperature of the gas.

The electron fraction, $x_{\rm e}$, is determined from photoionization equilibrium computed using {\sc Cloudy} (see above).\\

\noindent {\bf Molecular Hydrogen Cooling:} The molecular hydrogen cooling timescale is found by first estimating the abundance, $f_{{H_2},c}$, of molecular hydrogen that would be present if there is no background of $H_2$-dissociating radiation from stars. For gas with hydrogen number density $n_{\rm H}$ and temperature $T_{\rm V}$ the fraction is \pcite{tegmark_small_1997}:
\begin{eqnarray}
f_{{\rm H_2},c} &=& 3.5 \times 10^{-4}T_3^{1.52} [1+(7.4\times 10^8 (1+z)^{2.13} \nonumber \\
 & & \times \exp\left\{-3173/(1+z)\right\}/n_{\rm H 1})]^{-1},
\end{eqnarray}
where $T_3$ is the temperature $T_{\rm v}$ in units of 1000K and $n_{\rm H 1}$ is the hydrogen density in units of cm$^{-3}$. Using this initial abundance we calculate the final H$_2$ abundance, still in the absence of a photodissociating background, as
\begin{equation}
f_{\rm H_2} = f_{{\rm H_2},c}\exp\left({-T_{\rm v} \over 51920K}\right)
\end{equation}
where the exponential cut-off is included to account for collisional dissociation of H$_2$, as in \citet{benson_epoch_2006}. 

Finally, the cooling time-scale due to molecular hydrogen was computed using \pcite{galli_chemistry_1998}:
\begin{equation}
 \tau_{{\rm H}_2} = 6.56419 \time 10^{-33} T_{\rm e} f^{-1}_{{\rm H}_2} n^{-1}_{{\rm H} 1} \Lambda^{-1}_{{\rm H}_2},
\end{equation}
where
\begin{equation}
 \Lambda_{\rm H_2} = {\Lambda_{\rm LTE} \over 1+n^{\rm cr}/n_{\rm H}},
\end{equation}
where
\begin{equation}
 {n^{\rm cr}\over n_{\rm H}} = {\Lambda_{\rm H_2}({\rm LTE}) \over \Lambda_{\rm H_2}[n_{\rm H}\rightarrow0]},
\end{equation}
and
\begin{eqnarray}
\log_{10}\Lambda_{\rm H_2}[n_{\rm H}\rightarrow0] &=& -103+97.59 \ln(T) -48.05 \ln(T)^2 \nonumber \\
 & &+10.8 \ln(T)^3-0.9032 \ln(T)^4
\end{eqnarray}
is the cooling function in the low density limit (independent of hydrogen density) and we have used the fit given by \citet{galli_chemistry_1998},
\begin{equation}
 \Lambda_{\rm LTE} = \Lambda_r+\Lambda_v
\end{equation}
is the cooling function in local thermodynamic equilibrium and
\begin{eqnarray}
 \Lambda_r &=& {1\over n_{\rm H_1}}\left\{{9.5\times10^{-22} T_3^{3.76}\over1+0.12 T_3^{2.1}} \exp\left(-\left[{0.13\over T_3}\right]^3\right) \right. \nonumber \\
 & & \left. +3\times10^{-24} \exp\left(-{0.51\over T_3}\right) \right\} \hbox{ergs cm}^3\hbox{ s}^{-1}, \\
 \Lambda_v &=& {1\over n_{\rm H_1}}\left\{ 6.7\times 10^{-19} \exp\left(-{5.86\over T_3}\right) \right. \nonumber \\
 & & \left. +1.6\times 10^{-18} \exp\left(-{11.7\over T_3}\right)\right\} \hbox{ergs cm}^3\hbox{ s}^{-1}
\end{eqnarray}
are the cooling functions for rotational and vibrational transitions in H$_2$ \pcite{hollenbach_molecule_1979}.

The model also allows for an estimate of the rate of molecular hydrogen formation on dust grains using the approach of \citet{cazaux_molecular_2004}. In this case we have to modify equation (13) of \cite{tegmark_small_1997}, which gives the rate of change of the H$_2$ fraction, to account for the dust grain growth path. The molecular hydrogen fraction growth rate becomes:
\begin{equation}
 \dot{f} = k_{\rm d} f (1-x-2f) + k_{\rm m} n (1-x-2f) x,
\end{equation}
where $f$ is the fraction of H$_2$ by number, $x$ is the ionization fraction of H which has total number density $n$, 
\begin{equation}
k_{\rm d}=3.025\times 10^{-17}{\xi_{\rm d}\over0.01} S_{\rm H}(T) \sqrt{{T_{\rm g}\over100\hbox{K}}} \hbox{cm}^3 \hbox{s}^{-1}
\end{equation}
is the dust formation rate coefficient (\citealt{cazaux_molecular_2004}; eqn.~4), and $k_{\rm m}$ is the effective rate coefficient for H$_2$ formation (\citealt{tegmark_small_1997}; eqn.~14). We adopt the expression given by \citeauthor{cazaux_molecular_2004}~(\citeyear{cazaux_molecular_2004}; eqn.~3) for the H sticking coefficient, $S_{\rm H}(T)$ and $\xi_{\rm d}=0.53 Z$ for the dust-to-gas mass ratio as suggested by \cite{cazaux_molecular_2004} and which results in $\xi_{\rm d}\approx 0.01$ for Solar metallicity. This equation must be solved simultaneously with the recombination equation governing the ionized fraction $x$. The solution, assuming $x(t)=x_0/(1+x_0nk_1t)$ and $1-x-2f\approx 1$ as do \cite{tegmark_small_1997}, is
\begin{equation}
f(t) = f_0 {k_{\rm m} \over k_1} \exp\left[ {\tau_{\rm r} +t\over \tau_{\rm d}} \right] \left\{ {\rm E}_{\rm i}\left( {\tau_{\rm r} \over \tau_{\rm d}} \right) - {\rm E}_{\rm i}\left( {\tau_{\rm r} +t \over \tau_{\rm d}} \right) \right\} 
\end{equation}
where $\tau_{\rm r}=1/x_0/n_{\rm H}/k_1$, $\tau_{\rm d}=1/n_{\rm H}/k_{\rm d}$, $k_1$ is the hydrogen recombination coefficient and ${\rm E}_{\rm i}$ is the exponential integral.

\subsection{Sizes and Adiabatic Contraction}\label{sec:Sizes}

The angular momentum content of galactic components is tracked within our model, allowing us to compute sizes for disks and bulges. We follow the same basic methodology as \cite{cole_hierarchical_2000}---simultaneously solving for the equilibrium radii of disks and bulges under the influence of the gravity of the dark matter halo and their own self-gravity and including the effects of adiabatic contraction---but treat adiabatic contraction using updated methods.

For the bulge component with pseudo-specific angular momentum $j_{\rm b}$ the half-mass radius, $r_{\rm b}$, must satisfy
\begin{equation}
 j^2_{\rm b} = \G [M_{\rm h}(r_{\rm b})+M_{\rm d}(r_{\rm b})+M_{\rm b}(r_{\rm b})] r_{\rm b},
\end{equation}
where $M_{\rm h}(r)$, $M_{\rm d}(r)$ and $M_{\rm b}(r)$ are the masses of dark matter, disk and bulge within radius $r$ respectively, and which we can write as
\begin{equation}
 c_{\rm b} = [M_{\rm h}(r_{\rm b})+M_{\rm d}(r_{\rm b})+M_{\rm b}(r_{\rm b})] r_{\rm b},
\label{eq:cBulge}
\end{equation}
where $c_{\rm b} = j^2_{\rm b}/\G$. In the original \cite{blumenthal_contraction_1986} treatment of adiabatic contraction the right-hand side of eqn.~(\ref{eq:cBulge}) is an adiabatically conserved quantity allowing us to write
\begin{equation}
 c_{\rm b} = M_{\rm h}^0(r_{\rm b,0}) r_{\rm b,0},
\end{equation}
where $M_{\rm h}^0$ is the unperturbed dark matter mass profile and $r_{\rm b,0}$ the original radius in that profile. This allows us to trivially solve for $r_{\rm b,0}$ and $M_{\rm h}^0(r_{\rm b,0})$ and so, assuming no shell crossing, $M_{\rm h}(r_{\rm b}) = f_{\rm h} M_{\rm h}^0(r_{\rm b,0})$, where $f_{\rm h}$ is the fraction of mass that remains distributed like the halo. Given a disk mass and radius this allows us to solve for $r_{\rm b}$.

In the \cite{gnedin_response_2004} treatment of adiabatic contraction however, $M(r)r$ is no longer a conserved quantity. Instead, $M(\langle\overline{r}\rangle)r$ is the conserved quantity where $\langle\overline{r}\rangle/r_{\rm h} = A_{\rm ac} (r/r_{\rm h})^{w_{\rm ac}}$. In this case, we write
\begin{equation}
r_{\rm b}=\langle\overline{r_{\rm b}^\prime}\rangle = A_{\rm ac} r_{\rm h} (r_{\rm b}^\prime/r_{\rm h})^{w_{\rm ac}}.
\end{equation}
Equation~(\ref{eq:cBulge}) then becomes
\begin{equation}
 c_{\rm b}^\prime = [M_{\rm h}(\langle\overline{r_{\rm b}^\prime}\rangle)+M_{\rm d}(\langle\overline{r_{\rm b}^\prime}\rangle)+M_{\rm b}(\langle\overline{r_{\rm b}^\prime}\rangle)] r_{\rm b}^\prime,
 \label{eq:cBulgePrime}
\end{equation}
where
\begin{equation}
 c_{\rm b}^\prime = {c_{\rm b} \over A_{\rm ac}}\left({r_{\rm b}^\prime\over r_{\rm h}}\right)^{1-w_{\rm ac}}.
\end{equation}
The right-hand side of eqn.~(\ref{eq:cBulgePrime}) is now an adiabatically conserved quantity and we can write
\begin{equation}
 c_{\rm b}^\prime = M_{\rm h}^0(\langle\overline{r_{\rm b,0}^\prime}\rangle) r_{\rm b,0}.
\end{equation}
If we know $c_{\rm b}^\prime$ this expression allows us to solve for $r_{\rm b,0}$ and $M_{\rm h}^0(\langle\overline{r_{\rm b,0}^\prime}\rangle)$ which in turns gives $M_{\rm h}(r_{\rm b}) = f_{\rm h} M_{\rm h}^0(\langle\overline{r_{\rm b,0}^\prime}\rangle)$. Of course, to find $c_{\rm b}^\prime$ we need to know $r_{\rm b}$. This equation must therefore be solved iteratively. In practice, for a galaxy containing a disk and bulge, the coupled disk and bulge equations must be solved iteratively in any case, so this does not significantly increase computational demand.

The disk is handled similarly. We have
\begin{equation}
 {j^2_{\rm d} \over k^2_{\rm d}} = \G \left[M_{\rm h}(r_{\rm d})+{k_{\rm h}\over 2}M_{\rm d}+M_{\rm b}(r_{\rm d})\right] r_{\rm d},
\end{equation}
where $k_{\rm h}$ gives the contribution to the rotation curve in the mid-plane and $k_{\rm d}$ relates the total angular momentum of the disk to the specific angular momentum at the half-mass radius \pcite{cole_hierarchical_2000}. This becomes
\begin{equation}
 c_{\rm d,2}^\prime = M_{\rm h}^0(\langle\overline{r_{\rm d,0}^\prime}\rangle) r_{\rm d,0},
\end{equation}
where
\begin{equation}
 c_{\rm d,2}^\prime = {c_{\rm d,2} \over A_{\rm ac}}\left({r_{\rm b}^\prime\over r_{\rm h}}\right)^{1-w_{\rm ac}},
\end{equation}
and
\begin{equation}
 c_{\rm d,2} = {j^2_{\rm d} \over \G k^2_{\rm d}} - \left({k_{\rm h}\over 2}-{1\over 2}\right) r_{\rm d} M_{\rm d}.
\end{equation}
This system of equations must be solved simultaneously to find the radii of disk and bulge in a given galaxy. Once these are determined, the rotation curve and dark matter density as a function of radius are trivially found from the known baryonic distribution, pre-collapse dark matter density profile and the adiabatic invariance of $M(\langle\overline{r}\rangle)r$.

\subsection{Substructures and Merging}\label{sec:Merging}

N-body simulations of dark matter halos have convincingly shown that substructure persists within dark matter halos for cosmological timescales \pcite{moore_dark_1999}. Moreover, recent ultra-high resolution simulations \pcite{kuhlen_via_2008,springel_aquarius_2008,stadel_quantifyingheart_2009} demonstrate that multiple levels of substructure (e.g. sub-sub-halos) can exist. This ``substructure hierarchy'' is often neglected in semi-analytic models when merging is being considered. For example, \cite{cole_hierarchical_2000} and all other semi-analytic models to date\footnote{\protect\cite{taylor_evolution_2004}, who describe a model of the orbital dynamics of subhalos, do account for the orbital grouping of subhalos arriving as part of a pre-existing bound system (i.e. when a halo becomes a subhalo its own subhalos are given similar orbits in the new host). However, as noted by \cite{taylor_evolution_2005}, they do not include the self-gravity of subhalos and so sub-subhalos do not remain gravitationally bound to their subhalo. As such, sub-subhalos will gradually disperse and cannot merge with each other via dynamical friction.} consider only one level of substructure---a substructure in a group halo which merges into a cluster immediately becomes a substructure of the cluster for the purposes of merging calculations. This is unrealistic and may:
\begin{enumerate}
 \item neglect mergers between galaxies in substructures which \cite{angulo_fate_2009} have recently shown to be important for lower mass subhalos;
 \item bias the estimation of merging timescales for halos (and their galaxies).
\end{enumerate}
\cite{angulo_fate_2009} examine rates of subhalo-subhalo mergers in the Millennium Simulation and find that for subhalos with masses below 0.1\% the mass of the main halo mergers with other subhalos become equally likely as a merger with the central galaxy of the halo. They also find that subhalo-subhalo mergers tend to occur between subhalos that were physically associated before falling into the larger potential. This suggests that a treatment of subhalo-subhalo mergers must consider the interactions between subhalos and not simply consider random encounters as was done, for example, by \cite{somerville_semi-analytic_1999}.

We therefore implement a method to handle an arbitrarily deep hierarchy of substructure. We refer to isolated halos as $S^0$ substructures (i.e. not substructures at all), substructures of $S^0$ halos are called $S^1$ substructures and substructures of $S^n$ halos are $S^{n+1}$ substructures. When a halo forms it is an $S^0$ substructure, and when it first becomes a satellite it becomes a $S^1$ substructure.

For $S^n$ substructures with $n\ge2$ we check at the end of each timestep whether the substructure has been tidally stripped out of its $S^{n-1}$ host. If it has, it is promoted to being a $S^{n-1}$ substructure in the $S^{n-2}$ substructure which hosts its $S^{n-1}$ host.

\subsubsection{Orbital Parameters}

When a halo first becomes an $S^1$ subhalo it is assigned orbital parameters drawn from the distribution of \cite{benson_orbital_2005} which was measured from N-body simulations. This distribution gives the radial and tangential velocity components of the orbit. For later convenience, we compute from these velocities the radius of a circular orbit with the same energy as the actual orbit, $r_{\rm C}(E)$, and the circularity (the angular momentum of the actual orbit in units of the angular momentum of that circular orbit), $\epsilon$. These are computed using the gravitational potential of the host halo.

\subsubsection{Adiabatic Evolution of Host Potential}

As a subhalo orbits inside of a host halo the gravitational potential of that host halo will evolve due to continued cosmological infall. To model how this evolution affects the orbital parameters of each subhalo we assume that it can be well described as an adiabatic process\footnote{Halos are expected to grow on the Hubble time, while the characteristic orbital time is shorter than this by a factor of $\sqrt{\Delta}$ where $\Delta$ is the overdensity of dark matter halos. This expected validity of the adiabatic approximation has been confirmed in N-body simulations by \protect\cite{book_testingadiabatic_2010}.}. As such, the azimuthal and radial actions of the orbits:
\begin{equation}
 J_{\rm a} = {1\over 2\pi}\int_0^{2\pi} r^2 \dot{\phi}\deriv\phi,
\end{equation}
and
\begin{equation}
 J_{\rm r} = {1\over \pi}\int_{r_{\rm min}}^{r_{\rm max}} \dot{r}\deriv r,
\end{equation}
should be conserved (assuming a spherically symmetric potential). Therefore, at each timestep, we compute $J_{\rm a}$ and $J_{\rm r}$ for each satellite from the known orbital parameters in the current host halo potential. We assume these quantities are the same in the new host halo potential and convert them back into new orbital parameters $r_{\rm C}(E)$ and $\epsilon$.

\subsubsection{Tidal Stripping of Dark Matter Substructures}

Given orbital parameters, $r_{\rm C}(E)$ and $\epsilon$ we can compute the apocentric and pericentric distances of the orbit of each subhalo. At the end of each timestep, for each subhalo we find the pericentric distance and compute the tidal field of its host halo at that point:
\begin{equation}
 {\mathcal D}_{\rm t} = {\deriv \over \deriv r_{\rm h}}\left[ - {\G M_{\rm h}(r_{\rm h}) \over r_{\rm h}^2} \right] + \omega^2,
\end{equation}
where $\omega$ is the orbital frequency of the subhalo, and find the radius, $r_{\rm s}$, in the subhalo at which this equals
\begin{equation}
 {\mathcal D}_{\rm s} = {\G M_{\rm s}(r_{\rm s})\over r_{\rm s}^3}.
\end{equation}
This gives the tidal radius, $r_{\rm s}$, in the subhalo.

\subsubsection{Promotion through the hierarchy}

After computing tidal radii, for each $S^{\ge2}$ subhalo we compute the apocentric distance of its orbit and ask if this exceeds the tidal radius of its host. If it does, the subhalo is assumed to be tidally stripped from its host halo and promoted to an orbit in the host of its host: $S^n\rightarrow S^{n-1}$. To compute orbital parameters of the satellite in this new halo we determine its radius and velocity at the point where it crosses the tidal radius of its old host. These are added vectorially (assuming random orientations) to the position and velocity of its old host at pericentre in the new host. From this new position and velocity values of $r_{\rm C}(E)$ and $\epsilon$ are computed.

This approach can handle an arbitrarily deep hierarchy of substructure. In practice, the actual depth of the hierarchy will depend on both the mass resolution of the merger trees used and the efficiency of tidal forces to promote substructures through the hierarchy. Given the resolution of the trees used in our calculations we find that most substructures belonge to the $S^1$ and $S^2$ levels. However, the deepest substructure level that we have found at $z=0$ is $S^7$.

\subsubsection{Dynamical Friction}\label{sec:DynFric}

We adopt the fitting formula found by \cite{jiang_fitting_2008} to estimate merging timescales for dark matter substructures (and, consequently, the galaxies that they contain). The multiple levels of substructure hierarchy in our model allow for the possibility of satellite-satellite mergers. We intend to compare results from our model with N-body measures of this process in a future work.

When a halo first becomes a satellite, we set a dimensionless merger clock, $x_{\rm DF}=0$. On each subsequent timestep, $x_{\rm DF}$ is incremented by an amount $\Delta t / \tau_{\rm DF}$ where $\tau_{\rm DF}$ is the dynamical friction timescale for the satellite in the current host halo according to the expression of \cite{jiang_fitting_2008}, including the dependence on $r_{\rm C}(E)$. When $x_{\rm DF}=1$ the satellite is deemed to have merged with the central galaxy in the host halo.

When a satellite is tidally stripped out of its current orbital host and promoted to the host above it in the hierarchy the merging clock is reset so that dynamical friction calculations start anew in this new orbital host. This is something of an approximation since the dynamical friction timescale of \cite{jiang_fitting_2008} is calibrated using satellites that enter their halo at the virial radius. As such, it does not explore as a sufficiently wide range in $r_{\rm C}$ as is required for our models. Furthermore, when promoted to a new orbital host, a satellite will have already lost some mass due to tidal effects. This is not accounted for when computing a new dynamical friction timescale however, and so may cause us to underestimate merging timescales somewhat.

\begin{figure}
 \includegraphics[width=80mm,viewport=0mm 60mm 200mm 245mm,clip]{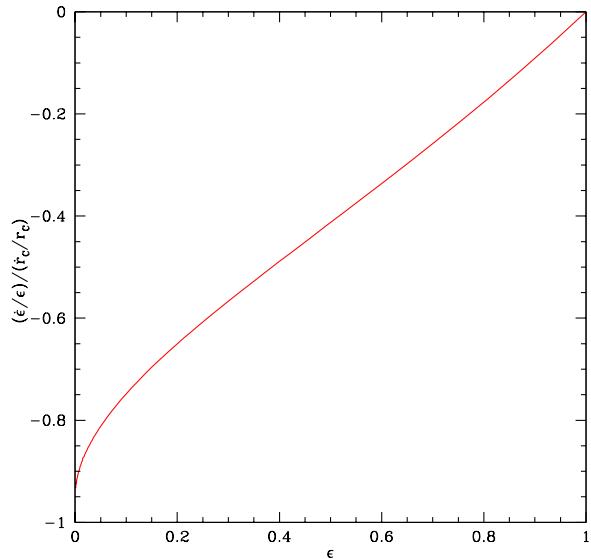}
 \caption{The ratio $(\dot{\epsilon}/\epsilon)/(\dot{r}_{\rm C}/r_{\rm C})$ for isothermal halos. This ratio is used in solving for the evolution of orbital circularity and orbital radius under the influence of dynamical friction as described in \S\protect\ref{sec:DynFric}.}
 \label{fig:Orbital_DynFric_Ratio}
\end{figure}

Dynamical friction also affects the orbital parameters of each subhalo. To simplify matters we follow \cite{lacey_merger_1993} and examine the evolution of these quantities in an isothermal dark matter halo. In such a halo, and for a circular orbit, $r_{\rm C}$ evolves as
\begin{equation}
 \left({r_{\rm C} \over r_{\rm C,0}}\right)^2 = 1-{t\over \tau_{\rm DF}}.
\end{equation}
Therefore, after each timestep we update
\begin{equation}
 r_{\rm C}^2 \rightarrow r_{\rm C}^2 - r_{\rm C,0}^2 {\Delta t \over \tau_{\rm DF}}.
\end{equation}
The fractional change in $\epsilon$ is assumed to be given by $(\dot{\epsilon}/\epsilon)/(\dot{r}_{\rm C}/r_{\rm C})$ as computed for the current orbit using the expressions of \cite{lacey_merger_1993}. This is a function of $\epsilon$ only and is plotted in Fig.~\ref{fig:Orbital_DynFric_Ratio}. Note that the timescale, $\tau_{\rm DF}$, used here is that from \cite{jiang_fitting_2008} and not the one from \cite{lacey_merger_1993}.

\subsection{Ram pressure and tidal stripping}\label{sec:Stripping}

We follow \cite{font_colours_2008} and estimate the extent to which ram pressure from the hot atmosphere of a halo may strip away the hot atmosphere of an orbiting subhalo. In addition, we also consider tidal stripping of this hot gas and both ram pressure and tidal stripping of material from galaxies.

Ram pressure and tidal forces are computed at the pericentre of each subhalo's orbit, which we now compute self-consistently with our orbital model (see \S\ref{sec:Merging}). For an $S^i$, where $i>1$, subhalo we compute the ram pressure force from all halos higher in the hierarchy and take the maximum of these to be the ram pressure force actually felt. The tidal field (i.e. the gradient in the gravitational force across the satellite) includes the centrifugal contribution at the orbital pericentre and is given by:
\begin{equation}
 {\mathcal F} = \omega^2 - {\deriv \over \deriv R} {\G M(<r) \over r^2} .
\end{equation}
The ram pressure is taken to be
\begin{equation}
 P_{\rm ram} = \rho_{\rm hot,host} V_{\rm orbit}^2
 \label{eq:RamPressure}
\end{equation}
where $\rho_{\rm hot,host}$ is the density of hot gas in the host halo at the pericentre of the orbit and $V_{\rm orbit}$ is the orbital velocity of the satellite at that position.

\subsubsection{Stripping of hot halo gas}\label{sec:HotStrip}

We find the ram pressure radius in the hot halo gas by solving
\begin{equation}
 P_{\rm ram} = \alpha_{\rm ram}{\G M_{\rm sat}(r_{\rm r}) \over r_{\rm r}} \rho_{\rm hot,sat}(r_{\rm r})
 \label{eq:HotRamPressure}
\end{equation}
for $r_{\rm r}$, where $\alpha_{\rm ram}$ is a parameter that we set equal to 2 as suggested by \cite{mccarthy_ram_2008}. Similarly, a tidal radius is found by solving
\begin{equation}
 {\mathcal F} = \alpha_{\rm tidal}^3 {\G M_{\rm sat}(r_{\rm t}) \over r_{\rm t}^3}
\end{equation}
for $r_{\rm t}$, where $\alpha_{\rm tidal}$ is a parameter that we set equal to unity. Once the minimum of the ram pressure and tidal stripping radii has been determined we follow \cite{font_colours_2008} and compute the cooling rate of the remaining, unstripped gas by cooling only the gas within the stripping radius and assuming that stripping does not alter the mean density of gas within this radius. We implement this by giving the satellite a nominal hot gas mass $M_{\rm hot}^\prime = M_{\rm hot} + M_{\rm strip}$ (where $M_{\rm hot}$ is the true hot gas content of the halo) and applying the same cooling algorithm as that used for central galaxies (except limiting the maximum cooling radius to $r_{\rm strip}$ rather than $R_{\rm v}$). This step ensures self-consistency in the treatment of the gas cooling between stripped and unstripped galaxies, and therefore that the colours of satellites are predicted correctly.

The initial stripping of re-heated gas is the same as for the hot gas, i.e. the same fraction is transferred from the re-heated gas of the satellite to the re-heated gas reservoir of the parent halo. We follow \cite{font_colours_2008} in modelling the time-dependence of the hot gas mass in the satellite halo and refer the reader to that paper for full details. This process introduces one free parameter, $\epsilon_{\rm strip}$ which represents the time averaged stripping rate after the initial pericentre. We treat $\epsilon_{\rm strip}$ as a free parameter which we will adjust to match observational constraints.

The stripping of satellites is also affected by the growth of the halo in
which the satellite is orbiting. \cite{font_colours_2008} took this effect into account by assigning each satellite galaxy new orbital parameters and deriving a new stripping factor every time the halo doubles in mass compared to the initial stripping event. In the present work we directly follow the evolution of the pericentric radius and velocity of each satellite due to both dynamical friction and host halo mass growth. For this reason, we take a different approach from \cite{font_colours_2008}, computing a new ram pressure radius in each timestep instead of only at every mass doubling event.

Any material stripped away from the subhalo is added to the halo which provided the greatest ram pressure force. For tidal forces, we consider only the contribution from the current orbital host as typically if this were exceeded by the tidal force from a parent higher up in the hierarchy the subhalo would have already been tidally stripped from this orbital host and promoted to a higher level in the hierarchy.

\subsubsection{Stripping of galactic gas and stars}

\begin{figure}
 \includegraphics[width=80mm,viewport=0mm 55mm 200mm 245mm,clip]{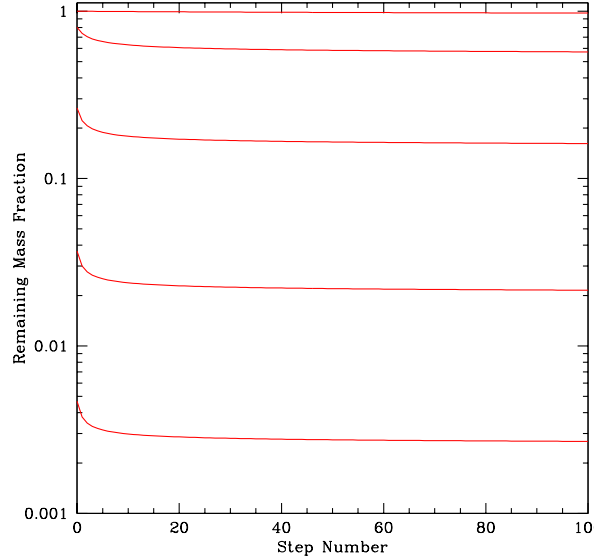}
 \caption{The remaining mass fraction in an exponential disk in a potential giving a flat rotation curve (and ignoring the disk self-gravity) subjected to tidal truncation at radius $r_{\rm t}/r_{\rm d,0}=0.1$, 0.3, 1.0, 3.0 and 10.0 (from lower to upper lines) after a given number of steps according to our model. The remaining mass fraction quickly converges to a near constant value.}
 \label{fig:MassLossSteps}
\end{figure}

The effective gravitational pressure that resists the ram pressure force in the disk plane is (for an exponential disk; \citealt{abadi_ram_1999}):
\begin{equation}
P_{\rm grav} = {\G M_{\rm d} M_{\rm g}\over 4\pi r_{\rm d}^4} x {\rm e}^{-x} \left[I_0\left({x\over 2}\right)K_1\left({x\over 2}\right)-I_1\left({x\over 2}\right)K_0\left({x\over 2}\right)\right],
\end{equation}
where $x=r/r_{\rm d}$ and $I_0$, $I_1$, $K_0$ and $K_1$ are Bessel functions. The ram pressure radius is found by solving for the radius at which $P_{\rm grav}=P_{\rm ram}$, where $P_{\rm ram}$ is given by eq.~(\ref{eq:RamPressure}). We assume that any stars in the galaxy which lie beyond the computed tidal radius and any gas which lies beyond the smaller of the tidal and ram pressure radii are instantaneously removed. Stars become part of the diffuse light component of the halo (i.e. that which is known as intracluster light in clusters of galaxies; see \S\ref{sec:ICL}), while gas is added to the reheated reservoir of the host halo. The remaining mass of each component (cold gas, disk and bulge stars) is computed and the specific angular momentum of the remaining material is computed assuming a flat rotation curve:
\begin{eqnarray}
 j_{\rm disk}&=&j_{\rm disk 0} \nonumber \\
&&\times
\left[
{
\int_0^{R_\star} \Sigma_\star(R) R^2 \deriv R
+
\int_0^{\rm R_{\rm g}} \Sigma_{\rm g}(R) R^2 \deriv R
\over
\int_0^\infty \Sigma(R) R^2 \deriv R
}
\right]\nonumber\\
&&\times
\left[
{
\int_0^{R_\star} \Sigma_\star(R) R \deriv R
+
\int_0^{\rm R_{\rm g}} \Sigma_{\rm g}(R) R \deriv R
\over
\int_0^\infty \Sigma(R) R \deriv R
}
\right]^{-1} \\
 &=& j_{\rm disk 0} \nonumber \\
&&\times
\left\{
f_\star\left[1-\left(1+x_\star+{x_\star^2\over 2}\right){\rm e}^{-x_\star}\right]\nonumber \right. \\
&&+
\left. f_{\rm g}\left[1-\left(1+x_{\rm g}+{x_{\rm g}^2\over 2}\right){\rm e}^{-x_{\rm g}}\right]
\right\}\nonumber\\
&&\times
\left\{
f_\star[1-(1+x_\star){\rm e}^{-x_\star}]
+
f_{\rm g}[1-(1+x_{\rm g}){\rm e}^{-x_{\rm g}}]
\right\}^{-1}
\end{eqnarray}
for the disk (the last line assuming an exponential disk) where $R_\star=r_{\rm tidal}$, $R_{\rm g}=\hbox{min}(r_{\rm tidal},r_{\rm ram})$, $x_\star=R_\star/R_{\rm d}$, $x_{\rm g}=R_{\rm g}/R_{\rm d}$, $f_\star=M_\star/(M_\star+M_{\rm g})$ and $f_{\rm g}=M_{\rm g}/(M_\star+M_{\rm g})$, and
\begin{equation}
  j_{\rm sph} = j_{\rm sph 0} {\left. \int_0^{r_{\rm tidal}} \rho_\star(R) R^3 \deriv R \right/ \int_0^\infty \rho_\star(R) R^3 \deriv R \over \left. \int_0^{r_{\rm tidal}} \rho_\star(R) R^2 \deriv R \right/ \int_0^\infty \rho_\star(R) R^3 \deriv R}
\end{equation}
for the bulge (and which must be evaluated numerically). Here, $j_{\rm disk 0}$ and $j_{\rm sph 0}$ are the pre-stripping specific angular momenta of disk and spheroid respectively, $\Sigma_\star(R)$ and $\Sigma_{\rm gas}$ are the surface density profiles of stars and gas in the disk prior to stripping and $\rho_\star(R)$ is the stellar density profile in the spheroid prior to stripping. Since \gf\ always assumes a de Vaucouler's spheroid and an exponential disk with stars tracing gas the stripped components will readjust to these configurations with their new masses and angular momenta. This is, therefore, an approximate treatment of stripping. In particular, some material will always ``leak'' back out beyond the stripping radius and so is easily stripped on the next timestep. Figure~\ref{fig:MassLossSteps} demonstrates that this is not a severe problem, with the remaining mass fraction asymptoting to a near constant value after just a few steps.

\subsection{IGM Interaction}\label{sec:IGM}

\cite{benson_effects_2002-1} introduced methods to simultaneously compute the evolution of the \IGM\ and the galaxy population in a self-consistent manner such that emission from galaxies ionized and heated the \IGM\ which in turn lead to suppression of future galaxy formation. A major practical limitation of \citeauthor{benson_effects_2002-1}'s~(\citeyear{benson_effects_2002-1}) method was that it required \gf\ to be run to generate an emissivity history for the Universe which was then fed into a model for the \IGM\ evolution. The \IGM\ evolution was used to predict the effects on galaxy formation and \gf\ run again. This loop was iterated around several times to find a converged solution. This problem was inherent in the implementation due to the fact that \gf\ was designed to evolve a single merger tree to $z=0$ then move onto the next one.

To circumvent this problem, we have adapted \gf\ to allow for multiple merger trees to be evolved simultaneously: Each tree is evolved for a single timestep after which the \IGM\ evolution for that same timestep is computed. This allows simultaneous, self-consistent evolution of the \IGM\ and galaxies without the need for iteration.

The model we adopt for the \IGM\ evolution is essentially identical to that of \cite{benson_effects_2002-1}, and consists of a uniform \IGM\ (with a clumping factor to account for enhanced recombination and cooling due to inhomogeneities) composed of hydrogen and helium and a photon background supplied by galaxies and \AGN. The reader is therefore referred to \cite{benson_effects_2002-1} for a full discussion. Here we will discuss only those aspects that are new or updated.

\subsubsection{Emissivity}

The two sources of photons in our model are quasars and galaxies. For \AGN\ we assume that the \SED\ has the following shape \pcite{haardt_radiative_1996}:
\begin{equation}
 f_\nu(\lambda) \propto \left\{ \begin{array}{ll}
                                 \lambda^{1.5} & \hbox{if } \lambda < 1216\hbox{\AA}; \\
                                 \lambda^{0.8} & \hbox{if } 1216\hbox{\AA} < \lambda < 2500\hbox{\AA}; \\
                                 \lambda^{0.3} & \hbox{if } \lambda > 2500\hbox{\AA}, \\
                                \end{array}
\right.
\end{equation}
where the normalization of each segment is chosen to give a continuous function and unit energy when integrated over all wavelengths. The emissivity per unit volume from \AGN\ is then
\begin{equation}
\epsilon_{\rm AGN} = f_{\rm esc,AGN} \epsilon_\bullet \dot{\rho}_\bullet \clight^2 f_\nu(\lambda),
\end{equation}
where $\epsilon_\bullet=0.1$ is an assumed radiative efficiency for accretion onto black holes, $\dot{\rho}_\bullet$ is the rate of black hole mass growth per unit volume computed by \gf\ and $f_{\rm esc,AGN}$ is an assumed escape fraction for \AGN photons which we fix at $10^{-2}$ to produce a reasonable epoch of HeII reionization.

The emissivity from galaxies was calculated directly by integrating the star formation rate per unit volume predicted by \gf\ over time and metallicity to give
\begin{equation}
\epsilon_{\rm gal} = \int_0^{t_{\rm now}} f_{\rm esc,gal}(t^\prime) \dot{M}_{\star}(t^\prime,Z)L_{\nu}(t_{\rm now}-t^\prime,Z[t^\prime]) \deriv t^{\prime},
\end{equation} 
where $\dot{M}_{\star}(t,Z)$ is the rate of star formation at metallicity $Z$, $L_\nu(t,Z)$ is the integrated luminosity per unit frequency and per Solar
mass of stars formed of a single stellar population of age $t$ and metallicity $Z$ and $f_{\rm esc,gal}$ is the escape fraction of ionizing photons from the galaxy.

The fraction of ionizing photons able to escape from the disk of each galaxy is computed using the expressions derived by \citet{benson_effects_2002} (their eqn.~A4) which is a generalization of the model of \cite{dove_photoionization_1994} in which OB associations with a distribution of luminosities ionize holes through the neutral hydrogen distribution through which their photons can escape.

The sum of $\epsilon_{\rm AGN}$ and $\epsilon_{\rm gal}$ gives the number of photons emitted from the galaxies and quasars in the model.

\subsubsection{IGM Ionization State}

The ionization state of the \IGM\ is computed just as in \cite{benson_effects_2002-1} except that we use effective photo-ionization cross-sections that account for the effects of secondary ionizations and are given by \citeauthor{shull_x-ray_1985} (\citeyear{shull_x-ray_1985}; as re-expressed by \citealt{venkatesan_heating_2001}):
\begin{eqnarray}
 \sigma_{\rm H}^\prime(E) &=& \left(1+\phi_{\hbox{\scriptsize H{\sc i}}} { E-E_{\rm H} \over E_{\rm H}} + \phi^*_{\hbox{\scriptsize He{\sc i}}} {E-E_{\rm H}\over 19.95\hbox{eV}}\right) \sigma_{\rm H}(E) \nonumber \\
 & & + \left(1+\phi_{\hbox{\scriptsize He{\sc i}}} {E-E_{\rm He}\over E_{\rm He}}\right) \sigma_{\rm He}(E) \\
 \sigma_{\rm He}^\prime(E) &=& \left(1+\phi_{\hbox{\scriptsize He{\sc i}}} {E-E_{\rm He} \over E_{\rm He}}\right) \sigma_{\rm He}(E) \nonumber \\
 & & + \left(\phi_{\hbox{\scriptsize He{\sc i}}} {E-E_{\rm H}\over24.6}\right) \sigma_{\rm H}(E)
\end{eqnarray}
where $\sigma(E)$ is the actual cross section \pcite{verner_analytic_1995} and
\begin{eqnarray}
 \phi_{\hbox{\scriptsize H{\sc i}}}   &=& 0.3908 (1-x_{\rm e}^{0.4092})^{1.7592}, \\
 \phi^*_{\hbox{\scriptsize He{\sc i}}} &=& 0.0246 (1-x_{\rm e}^{0.4049})^{1.6594}, \\
 \phi_{\hbox{\scriptsize He{\sc i}}}   &=& 0.0554 (1-x_{\rm e}^{0.4614})^{1.6660}.
\end{eqnarray}

\subsubsection{IGM Thermal State}

Heating of the \IGM\ is treated as in \citet{benson_effects_2002-1} with the exception that we account for heating by secondary electrons. Photoionization heats the \IGM\ at a rate of 
\begin{equation}
\Sigma_{\rm photo} = \int^{\infty}_0(E-E_i)\clight\sigma^\prime(E)n_in_{\gamma}(E) {\mathcal E} \deriv E
\end{equation}
where $E_i$ is the energy of the sampled photons which is associated with atom/ion number density $n_i$, $\clight$ is speed of light, $\sigma^\prime$ is the effective partial photo-ionization cross section (accounting for secondary ionizations) for the ionization stages of H and He, $n_{\gamma(E)}$ is the number density of photons of energy $E$, $E_i$ is the ionization potential of i and index i represent the different atoms and ions, H, H$^+$, He, He$^+$ and He$^{2+}$. In the above, ${\mathcal E}$ accounts for heating by secondary electrons and is given by \pcite{shull_x-ray_1985}:
\begin{equation}
 {\mathcal E} = 0.9971 [1-(1-x_{\rm e}^{0.2663})^{1.3163}].
\end{equation}

\subsubsection{Suppression of Baryonic Infall into Halos}\label{sec:BaryonSupress}

According to \citet{okamoto_mass_2008}, the mass of baryons which accrete from the \IGM\ into a halo after reionization is given by
\begin{equation}
 M_{\rm b} = M_{\rm b}^\prime+M_{\rm acc},
\end{equation}
where
\begin{equation}
 M_{\rm b}^\prime = \sum_{\rm prog} \exp\left(-{\delta t \over t_{\rm evp}}\right) M_{\rm b},
\end{equation}
and where the sum is taken over the progenitor halos of the current halo, $\delta t$ is the time since the previous timestep and $t_{\rm evp}$ is the timescale for gas to evaporate from the progenitor halo and is given by
\begin{equation}
 t_{\rm evp}=\left\{ \begin{array}{ll}
                      R_{\rm H}/c_{\rm s}(\Delta_{\rm evp}) & \hbox{if } T_{\rm v} < T_{\rm evp}, \\
                      \infty & \hbox{if } T_{\rm v} > T_{\rm evp}.
                     \end{array} \right.
\end{equation}
Here, $T_{\rm evp}$ is the temperature below which gas will be heated and evaporated from the halo. We follow \cite{okamoto_mass_2008} and compute $T_{\rm evp}$ by finding the equilibrium temperature of gas at an overdensity of $\Delta_{\rm evp}=10^6$. The accreted mass $M_{\rm acc}$ is given by
\begin{equation}
 M_{\rm acc} = \left\{ \begin{array}{ll}
                {\Omega_{\rm b}\over \Omega_0} M_{\rm v} - M_{\rm b}^\prime & \hbox{if } T_{\rm vir} > T_{\rm acc} \\
                0 & \hbox{if } T_{\rm vir} < T_{\rm acc}
               \end{array} \right.
\end{equation}
where $T_{\rm acc}$ is the larger of the temperature of \IGM\ gas adiabatically compressed to the density of accreting gas and the equilibrium temperature, $T_{\rm eq}$, at which radiative cooling balances photoheating for gas at the density expected at the virial radius. This ensures that a sensible temperature is used even when the photoionizing background is essentially zero.

The value of $T_{\rm acc}$ is computed at each timestep by searching for where the cooling function (see \S\ref{sec:Cloudy}) crosses zero for the density of gas just accreting at the virial radius (for which we use one third of the halo overdensity; \citealt{okamoto_mass_2008}).

\subsection{Recycling and Chemical Evolution}\label{sec:NonInstGasEq}

In \cite{cole_hierarchical_2000} the instantaneous recycling approximation for chemical enrichment was used. While this is a reasonable approximation for $z=0$ it fails for high redshifts (where the main sequence lifetimes of the stars which do the majority of the enrichment become comparable to the age of the Universe). It also prevents predictions for abundance ratios (e.g. [$\alpha$/Fe]) from being made and ignores any metallicity dependence in the yield.

\citeauthor{nagashima_metal_2005}~(\citeyear{nagashima_metal_2005}; see also \citealt{nagashima_metal_2005-1}, \citealt{arrigoni_galactic_2009}) previously implemented a non-instantaneous recycling calculation in \gf. We implement a similar model here, following their general approach, but with some specific differences.

The fraction of material returned to the \ISM\ by a stellar population as a function of time is given by
\begin{equation}
 R(t) = \int_{M(t;Z)}^\infty [M-M_{\rm r}(M;Z)]\phi(M) {\deriv M \over M}
\end{equation}
where $\phi(M)$ is the \IMF\ normalized to unit stellar mass, $M_{\rm r}(M)$ is the remnant mass of a star of initial mass $M$. Here, $M(t)$ is the mass of a star with lifetime $t$. Similarly, the yield of element $i$ is given by
\begin{equation}
 p_i(t) = \int_{M(t;Z)}^\infty M_i(M_0;Z)\phi(M_0) {\deriv M_0\over M_0}
\end{equation}
where $M_i(M_0;Z)$ is the mass of metals produced by stars of initial mass $M_0$. For a specified \IMF\ we compute $R(t;Z)$ and $y_i(t;Z)$ for all times and elements of interest. This means that, unlike most previous implementations of \gf, the recycled fraction and yield are not free parameters of the model, but are fixed once an \IMF\ is chosen. However, it should be noted that significant uncertainties remain in calculations of stellar yields, which may therefore influence our calculations. Note that, unlike \citet{nagashima_metal_2005}, we include the full metallicity dependence in these functions. Stellar data are taken from \citet{portinari_galactic_1998} for low and intermediate mass stars and \citet{marigo_chemical_2001} for high mass stars.

In \gf\ the evolution of gas and stellar masses in a galaxy are controlled by the following equations\footnote{These are identical to those given in \protect\citeauthor{cole_hierarchical_2000}~(\citeyear{cole_hierarchical_2000}; their equations 4.6 and 4.8) except for the explicit inclusion of the recycling terms---\protect\cite{cole_hierarchical_2000} included these using the instantaneous recycling approximation.}:
\begin{eqnarray}
 \dot{M}_\star & = & {M_{\rm gas} \over \tau_\star} - \dot{M}_{\rm R} \\
 \dot{M}_{\rm gas} & = & -(1+\beta^\prime){M_{\rm gas} \over \tau_\star} + \dot{M}_{\rm R} + \dot{M}_{\rm infall}.
\end{eqnarray}
where
\begin{equation}
 \tau_\star = \left\{ \begin{array}{ll} \epsilon_\star^{-1} \tau_{\rm disk} \left({V_{\rm disk} \over 200\hbox{km s}^{-1}}\right)^{\alpha_\star} & \hbox{for disks} \\
               f_{\rm dyn} \tau_{\rm bulge} & \hbox{for bursts},
              \end{array} \right.
\end{equation}
is the star formation timescale, $\tau_{\rm disk}$ is the dynamical time at the disk half-mass radius, $\tau_{\rm bulge}$ is the dynamical time at the bulge half-mass radius, $f_{\rm dyn}=2$ and $\beta^\prime$ quantifies the strength of supernova feedback (see \S\ref{sec:Feedback}). In \cite{cole_hierarchical_2000}, the instantaneous recycling approximation implies that $\dot{M}_{\rm R}\propto M_{\rm gas} / \tau_\star$, and the cosmological infall term $\dot{M}_{\rm infall}$ is approximated as being constant over each short timestep. This permits a simple solution to these equations. In our case, we retain the assumption of constant $\dot{M}_{\rm infall}$ and further assume that the mass recycling rate, $\dot{M}_{\rm R}$, can be approximated as being constant throughout the timestep\footnote{This will be approximately true if the timestep is sufficiently short that $\ddot{R}\Delta t \ll \dot{R}$.}. We therefore write
\begin{equation}
 \dot{M}_{\rm R} = {M_{\rm R,past} + M_{\rm R,now} \over \Delta t},
\end{equation}
where $\Delta t$ is the timestep,
\begin{equation}
 M_{\rm R,past} = \int_{t_0}^{t_0+\Delta t} \deriv t^{\prime\prime} \int_0^{t_0} \deriv t^\prime  \dot{M}_\star(t^\prime) \dot{R}(t^{\prime\prime}-t^\prime) 
\end{equation}
is the mass of gas returned to the \ISM\ from populations of stars formed in previous timesteps (and is trivially computed from the known star formation rate of the galaxy on past timesteps) and 
\begin{equation}
 M_{\rm R,now} = \int_{t^\prime}^{t_0+\Delta t} \deriv t^{\prime\prime} \int_{t_0}^{t_0+\Delta t} \deriv t^\prime  \dot{M}_\star(t^\prime) \dot{R}(t^{\prime\prime}-t^\prime),
\end{equation}
is the mass returned to the \ISM\ by star formation during the current timestep. With these approximations, the gas equations always have the solution
\begin{eqnarray}
 M_{\rm gas}(t) = M_{\rm gas 0} \exp\left(-{t\over \tau_{\rm eff}}\right) + \dot{M}_{\rm input} \tau_{\rm eff}\left[1-\exp\left(-{t\over \tau_{\rm eff}}\right)\right],
\end{eqnarray}
where $M_{\rm gas 0}$ is the mass of gas at time $t=0$ (measured from the start of the timestep and
\begin{eqnarray}
 \dot{M}_{\rm input} &=& \dot{M}_{\rm infall} \nonumber \\
 & & + \left\{\left[{M_{\rm gas 0} \over \tau_{\rm eff}}-{M_{\rm R,past}\over\Delta t}\right]I_{\rm R1}(\Delta t,\tau_{\rm eff}) \right. \nonumber \\
 & & \left. + {M_{\rm R,past}\over\Delta t} I_{\rm R0}(\Delta t)\right\} \nonumber \\
 & & \times \left\{ (1+\beta) + [I_{\rm R1}(\Delta t,\tau_{\rm eff}) -I_{\rm R0}(\Delta t)]/\Delta t \right\}^{-1}
\end{eqnarray}
where
\begin{eqnarray}
 I_{\rm R0}(t) &=& \int_0^t R(t-t^\prime) \deriv t^\prime, \\
 I_{\rm R1}(t,\tau) &=& \int_0^t \exp(-t^\prime/\tau) R(t-t^\prime) \deriv t^\prime.
\end{eqnarray}
In the above, the effective e-folding timescale for star formation (accounting for \SNe\ driven outflows), $\tau_{\rm eff}$, is given by
\begin{equation}
 \tau_{\rm eff} = {\tau_\star \over 1 + \beta^\prime}
\end{equation}
where $\beta^\prime$ measures the strength of \SNe\ feedback and is defined below in eqn.~(\ref{eq:Beta}).

The evolution of the metal mass is treated in a similar way, assuming a constant rate of input of metals from infall, star formation from previous timesteps and star formation from the current timestep. Metals in the cold gas reservoir of a galaxy are assumed to be uniformly mixed into the gas, such that the reservoir has a uniform metallicty. Metals then flow from the cold gas reservoir into the stellar phase and out into the reheated reservoir at a rate proportional to the star formation rate and mass outflow rate respectively, with the constant of proportionality being the cold gas metallicity. Material recycled from stars to the cold phase carries with it metals corresponding to the original metallicity of those stars, augmented by the appropriate metal yield. Finally, gas infalling from the surrounding halo may have been enriched in metals by previous galaxy formation and so deposits metals into the cold phase gas at a rate proportional to the mass infall rate, with proportionality equal to the (assumed uniform) metallicity of the notional profile gas. Apart from the fact that metals from stellar recycling and yields are not added instantaneously to the cold reservoir this treatment of metals remains identical to that of \cite{cole_hierarchical_2000}. The net rate of metal mass input to the cold phase (from both cosmological infall and returned from stars) is
\begin{eqnarray}
 \dot{M}_{Z_i {\rm input}} &=& \dot{M}_{Z_i {\rm infall}} \nonumber \\
  & &+ {[{M_{Z_i {\rm gas 0}}\over\tau_{\rm eff}}-{M_{Z_i {\rm R}}^{\rm past}\over\Delta t}]I_{\rm R1}(\Delta t,\tau_{\rm eff}) + {M_{Z_i {\rm R}}^{\rm past}\over\Delta t} I_{\rm R0}(\Delta t) \over \Delta t [(1+\beta) + (I_{\rm R1}(\Delta t,\tau_{\rm eff}) -I_{\rm R0}(\Delta t))\Delta t]} \nonumber \\
 & & + {[{M_{\rm gas 0}\over\tau_{\rm eff}}-{M_{\rm R}^{\rm past}\over\Delta t}]I_{\rm p1}(\Delta t,\tau_{\rm eff}) + {M_{\rm R}^{\rm past}\over\Delta t} I_{\rm p0}(\Delta t) \over \Delta t [(1+\beta) + (I_{\rm p1}(\Delta t,\tau_{\rm eff}) -I_{\rm p0}(\Delta t))\Delta t]},
\end{eqnarray}
where $M_{Z_i {\rm R,past}}$ is the mass of metal $i$ recycled from star formation in previous timesteps and
\begin{eqnarray}
 I_{\rm p0}(t) &=& \int_0^t p(t-t^\prime) \deriv t^\prime, \\
 I_{\rm p1}(t,\tau) &=& \int_0^t \exp(-t^\prime/\tau) p(t-t^\prime) \deriv t^\prime.
\end{eqnarray}

\subsubsection{Star Bursts}

In previous implementations of \gf\ star bursts were assumed to have an exponentially declining star formation rate. Such a rate results from assuming an instantaneous star formation rate of
\begin{equation}
 \dot{M}_\star = {M_{\rm cold}\over \tau_\star},
\label{eq:BurstSFLaw}
\end{equation}
where $\tau_\star$ is a star formation timescale (fixed throughout the duration of the burst), an outflow rate proportional to the star formation rate and a rate of recycling given by $R\dot{M}_\star$. The resulting differential equations have a solution with an exponentially declining star formation rate.

When the instantaneous recycling approximation is dropped the rate of recycling is no longer proportional to the star formation rate and the differential equations no longer have an exponential solution. We choose to retain the original star formation law (eqn.~\ref{eq:BurstSFLaw}) and solve the differential equations to determine the star formation rate, outflow rate etc. as a function of time in the burst. The resulting set of equations have solutions identical to those in \S\ref{sec:NonInstGasEq} but with zero cosmological infall terms. Recycled material and the effects of feedback (see \S\ref{sec:Feedback}) are applied to the gas in the burst during the lifetime of the burst. Any recycling and feedback occurring after the burst is finished are applied to the disk.

In \citet{cole_hierarchical_2000} while bursts were treated as having finite duration for the purposes of computing the luminosity of their stellar populations at some later time, the change in the mass of the galaxy due to the burst occurred instantaneously. We drop this approximation and correctly follow the change in mass of each component (gas, stars, outflow) during each timestep.

\subsection{Feedback}\label{sec:Feedback}

Feedback from supernovae is also modified to account for the delay between star formation and supernova. In \cite{cole_hierarchical_2000} the outflow rate due to supernovae feedback was
\begin{equation}
 \dot{M}_{\rm out} = \beta \dot{M}_\star,
\end{equation}
where
\begin{equation}
\beta = \left({V_{\rm hot}\over V_{\rm galaxy}}\right)^{\alpha_{\rm hot}},
\end{equation}
$V_{\rm hot}$ and $\alpha_{\rm hot}$ are parameters of the model (we allow for two different values of $V_{\rm hot}$, one for quiescent star formation in disks and one for bursts of star formation) and $V_{\rm galaxy}$ is the circular velocity at the half-mass radius of the galaxy, determines the strength of feedback and is a function of the depth of the galaxy's potential well. We modify this to
\begin{equation}
 \dot{M}_{\rm out} = \beta^\prime \dot{M}_\star,
\end{equation}
where
\begin{equation}
 \beta^\prime = \beta { \int_0^t \dot{\phi}_\star(t^\prime) \dot{N}_{\rm SNe}(t-t^\prime)\d t^\prime \over \dot{\phi}_\star(t) N^{\rm (II)}_{\rm SNe}(\infty) }
 \label{eq:Beta}
\end{equation}
where $N_{\rm SNe}(t)$ is the total number of \SNe\ (of all types) arising from a single population of stars after time $t$, such that the outflow rate scales in proportion to the current rate of supernovae but produces the same net mass ejection after infinite time (for constant $\beta$). In fact, we compute $\beta$ using the present properties of the galaxy at each timestep. The qualifier ``(II)'' appearing in the quantity $N^{\rm (II)}_{\rm SNe}(\infty)$ in the denominator of eqn.~(\ref{eq:Beta}) indicates that we normalize the outflow rate by reference to the number of supernovae from our adopted Population II \IMF\ (see \S\ref{sec:StellarPop}). This results in the outflow correctly encapsulating any differences in the effective number of supernovae between Population II and III stars. For supernovae rates, we assume that all stars with initial masses greater than $8M_\odot$ will result in a Type II supernova allowing the rate to be found from the lifetimes of these stars and the adopted \IMF. We adopt the calculations of \cite{nagashima_metal_2005} to compute the Type Ia \SNe\ rate.

Since $\beta^\prime$ appears in the gas equations of \S\ref{sec:NonInstGasEq} but also depends on the star formation rate during the current timestep we must iteratively seek a solution for $\beta^\prime$ which is self-consistent with the star formation rate. We find that a simple iterative procedure, with an initial guess of $\beta^\prime=\beta$ quickly converges.

When gas is driven out of a galaxy in this way it can be either reincorporated into the $M_{\rm reheated}$ reservoir in the notional hot gas profile of the current halo, or it can be expelled from the halo altogether and allowed to reaccrete only further up the hierarchy once the potential well has become deeper. 

We assume that the expelled fraction is given by
\begin{equation}
 f_{\rm exp}=\exp\left( - {\lambda_\phi V^2 \over \langle e \rangle} \right),
\end{equation}
such that the rate of mass input to the reheated reservoir is
\begin{equation}
 \dot{M}_{\rm reheated} = (1 - f_{\rm exp})  \beta^\prime   \dot{M}_\star.
\end{equation}
Here, $\lambda_\phi$ is a dimensionless parameter relating the depth of the potential well to $V^2$ (we set $\lambda_\phi=1$ always), $V$ is the circular velocity of the galaxy disk or bulge (for quiescent or bursting star formation respectively) and $\langle e \rangle$ is the mean energy per unit mass of the outflowing material. We further assume
\begin{equation}
  \langle e \rangle ={1\over 2} \lambda_{\rm expel} V^2,
\end{equation}
where $\lambda_{\rm expel}$ is a parameter of order unity relating the energy of the outflowing gas to  the potential of the host galaxy, and will be treated as a free parameter to be constrained from observations (we actually allow for $\lambda_{\rm expel}$ to have different values for quiescent and bursting star formation; see \S\ref{sec:Selection}). We then proceed to the parent halo and allow a fraction
\begin{equation}
f_{\rm acc} = \exp\left( - {V_{\rm max}^2 \over \langle e \rangle} \right) - \exp\left( - {V_{\rm v}^2 \over \langle e \rangle } \right)
\end{equation}
to be reaccreted into the hot gas reservoir of the notional profile, where $V_{\rm max}$ is the maximum of $\sqrt{\lambda_{\rm expel}} V$ and any parent halo $V_{\rm v}$ yet found. We then proceed to the parent's parent and repeat the accretion procedure, continuing until the base of the tree is reached. In this way, all of the gas will be reaccreted if the potential well becomes sufficiently deep.

\subsection{AGN feedback}\label{sec:AGNFeedback}

In recent years, the possibility that feedback from \AGN plays a significant role in shaping the properties of a forming galaxy has come to the forefront \pcite{croton_many_2006,bower_breakinghierarchy_2006,somerville_semi-analytic_2008}. We adopt the black hole growth model of \cite{malbon_black_2007} and the \AGN\ feedback model of \cite{bower_breakinghierarchy_2006} as modified by \cite{bower_flip_2008}. The reader is referred to those papers for a full description of our implementation of \AGN\ feedback.

\subsection{Stellar Populations}\label{sec:StellarPop}

\begin{figure}
 \includegraphics[width=80mm,viewport=0mm 10mm 186mm 265mm,clip]{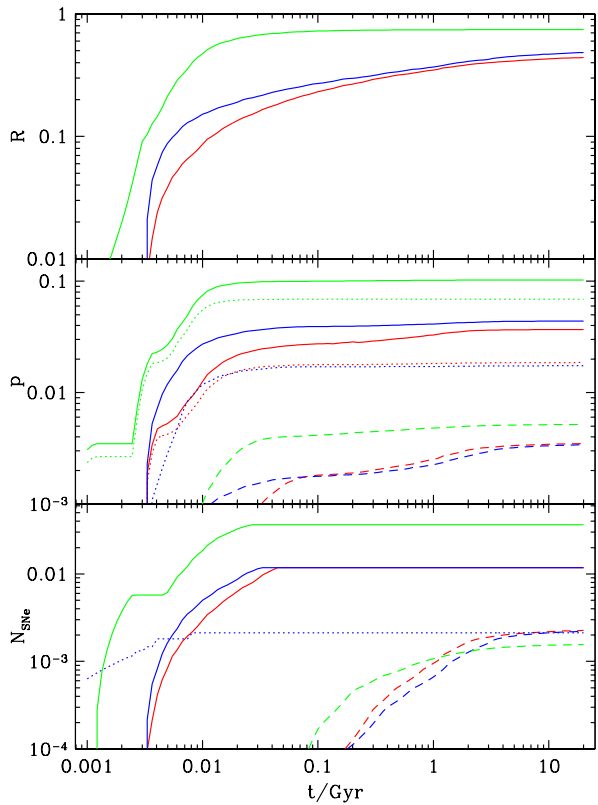}
 \caption{Upper, middle and lower panels show the recycled fraction, yield and effective number of supernovae respectively for a Chabrier \protect\IMF\ (two metallicities, defined as the mass fraction of heavy elements, are shown: \ChabrierIMFiZa\ as red lines and \ChabrierIMFiZb\ as blue lines) and for metal free Population III stars with type ``A'' \protect\IMF\ from \protect\cite{tumlinson_chemical_2006} (green lines). \emph{Top-panel:} The fraction of mass from a single stellar population, born at time $t=0$, recycled to the interstellar medium after time $t$. \emph{Middle-panel:} The total metal yield from a single stellar population born at time, $t=0$, after time $t$ is shown by the solid lines. Dotted and dashed lines show the yield of oxygen and iron respectively. \emph{Lower-panel:} Cumulative energy input into the interstellar medium, expressed as the number of equivalent supernovae, per unit mass of stars formed as a function of time. The dotted line indicates the contribution from stellar winds, the solid line the contribution from Type~II supernovae and the dashed line the contribution from Type~Ia supernovae.}
 \label{fig:Chabrier_NonInstant}
\end{figure}

We consider both Pop~II and Pop~III stars. To compute luminosities of Population II stellar populations we employ the most recent version\footnote{Specifically, {\tt v2.0} downloaded from {\tt http://www.astro.princeton.edu/$\sim$cconroy/SPS/} with bug fixes up to January 7, 2010.} of the Conroy, Gunn \& White spectral synthesis library \pcite{conroy_propagation_2009}\footnote{For calculations of \protect\IGM\ evolution we \emph{do not} use the \protect\cite{conroy_propagation_2009} spectra because they assign stars hotter than $5\times 10^4$K pure blackbody spectra. This leads to an unrealistically large ionizing flux for young, metal rich populations. We therefore instead use the \protect\cite{bruzual_stellar_2003} spectral synthesis library for \protect\IGM\ evolution calculations.}. We adopt a Chabrier \IMF\ \pcite{chabrier_galactic_2003}
\begin{equation}
 \phi(M) \propto \left\{ \begin{array}{ll}\exp\left( - {1\over 2}{ [\log_{10} M/M_{\rm c}]^2  \over \sigma^2} \right) & \hbox{for} M\le 1M_\odot \\
            M^{-\alpha} & \hbox{for} M>1M_\odot,
           \end{array} \right.
\end{equation}
where $M_{\rm c}=0.08M_\odot$ and $\sigma=0.69$ and the two expressions are forced to coincide at $1M_\odot$. Recycled mass fractions, yield and supernovae rates are computed self-consistently from this \IMF\ as described in \S\ref{sec:NonInstGasEq} and \S\ref{sec:Feedback} and are shown in Fig.~\ref{fig:Chabrier_NonInstant}.

For Population~III stars (which we assume form below a critical metallicity of $Z_{\rm crit}=10^{-4}Z_\odot$) we adopt \IMF\ ``A'' from \cite{tumlinson_chemical_2006}. Spectral energy distributions for this \IMF\ as a function of population age were kindly provided by J.~Tumlinson. Lifetimes for these stars are taken from the tabulation given by \cite{tumlinson_cosmological_2003}. Recycled fractions and yields and energies from pair instability supernovae are computed using the data given by \cite{heger_nucleosynthetic_2002}. Recycled mass fractions, yield and supernovae rates are computed self-consistently from these Population III stars as shown in Fig.~\ref{fig:Chabrier_NonInstant} by green lines.

\subsubsection{Extinction by Dust}\label{sec:DustModel}

\citet{cole_hierarchical_2000} introduced a model for dust extinction in galaxies which significantly improved upon earlier ``slab'' models. In \citet{cole_hierarchical_2000} the mass of dust is assumed to be proportional to the mass and metallicity of the \ISM\ and to be mixed homogeneously with the \ISM\ (possibly with a different scale height from the stars) and to have 
properties consistent with the extinction law observed in the
Milky Way. To compute the extinction of any galaxy, a random inclination angle is selected and the extinction computed using the results of radiative transfer calculations carried out by \citet{ferrara_atlas_1999}.

Following \citet{gonzalez-perez_massive_2009}, we extend this model\footnote{An alternative method for rapidly computing dust extinction and re-emission within the \gf +\grasil\ frameworks based on artificial neural networks is described by \protect\cite{almeida_modellingdusty_2009}.} by assuming that some fraction, $f_{\rm cloud}$, of the dust is in the form of dense molecular clouds where the stars form (see \citealt{baugh_canfaint_2005,lacey__2010}). Stars are assumed to form in these clouds and to escape on a timescale of $\tau_{\rm quies}$ (for quiescent star formation in disks) or $\tau_{\rm burst}$ (for star formation in bursts), which is a parameter of the dust model \pcite{granato_infrared_2000}, so these stars spend a significant fraction of their lifetime inside the clouds. Since massive, short-lived stars dominate the \UV\ emission of a galaxy this enhances the extinction at short wavelengths. 

To compute emission from dust we assume a far infrared opacity of
\begin{equation}
   \kappa = \left\{ \begin{array}{ll}
          \kappa_1 (\lambda/\lambda_1)^{-\beta_1} & \hbox{for } \lambda<\lambda_{\rm break} \\
        \kappa_1 (\lambda_{\rm break}/\lambda_1)^{-\beta_1} (\lambda/\lambda_{\rm break})^{-\beta_2} & \hbox{for } \lambda>\lambda_{\rm\
 break},
\end{array} \right.
\end{equation}
where the opacity normalization at $\lambda_1=30\mu$m is chosen to be $\kappa_1=140$cm$^2$/g  to reproduce the dust opacity model used in 
\grasil, as described in \cite{silva_modelingeffects_1998}. The dust grain 
model in \grasil\ is a slightly modified version of that proposed by \cite{draine_optical_1984}. Both the \cite{draine_optical_1984} and \grasil\ dust models have 
been adjusted to fit data on dust extinction and emission in the local \ISM\ 
(with much more extensive \ISM\ dust emission data being used by \citealt{silva_modelingeffects_1998}).
The normalization is set at 30$\mu$m because the dust opacity in the \cite{draine_optical_1984} and \grasil\ 
models is well fit by a power-law longwards of that wavelength, but not shortwards. The dust luminosity is then assumed to be
\begin{equation}
 L_\nu = 4\pi\kappa(\nu)B_\nu(T) M_{\rm Z,gas},
 \end{equation}
where $B_\nu(T) = [2{\rm h}\nu^3/{\rm c}^2]/[\exp({\rm h}\nu/{\rm k}T)-1]$ is the Planck blackbody spectrum and $M_{\rm Z,gas}$ is the mass of metals in gas. The dust temperature, $T$, is chosen such that the bolometric dust luminosity equals the luminosity absorbed by dust.

Values of the parameters used in dust model are given in Table~\ref{tb:DustParams} and were found by \citet{gonzalez-perez_massive_2009} to give the best match to the results of the full \grasil\ model. 

\begin{table}
 \caption{Parameters of the dust model used throughout this work. The parameters are defined in \S\protect\ref{sec:DustModel}.}
 \label{tb:DustParams}
 \begin{center}
 \begin{tabular}{lc}
 \hline
 {\bf Parameter} & {\bf Value} \\
 \hline
 $f_{\rm cloud}$ & 0.25 \\
 $r_{\rm burst}$ & 1.0 \\
 $\tau_{\rm quies}$ & 1~Myr \\
 $\tau_{\rm burst}$ & 1~Myr \\
 $\lambda_{\rm 1, disk}$ & 30$\mu$m \\
 $\lambda_{\rm break, disk}$ & 10000$\mu$m \\
 $\beta_{\rm 1, disk}$ & 2.0 \\
 $\beta_{\rm 2, disk}$ & 2.0 \\
 $\lambda_{\rm 1, burst}$ & 30$\mu$m \\
 $\lambda_{\rm break, burst}$ & 100$\mu$m \\
 $\beta_{\rm 1, burst}$ & 1.6 \\
 $\beta_{\rm 2, burst}$ & 1.6 \\
 \hline
 \end{tabular}
\end{center}
\end{table}

This extended dust model, including diffuse and molecular cloud dust components, provides a better match to the detailed radiative transfer calculation of dust extinction carried out by the spectrophotometric code \grasil\ \pcite{silva_modelingeffects_1998,baugh_predictions_2004,baugh_canfaint_2005,lacey_galaxy_2008} while being orders of magnitude faster, although it does not capture details such as \PAH\ features. 

\cite{fontanot_evaluating_2009} have explored similar models which aim to reproduce the results of \grasil\ using simple, analytic prescriptions. They found that by fitting the results from \grasil\ they were able to obtain a better match to the extinction in galaxies than previous, simplistic models of dust extinction had been able to attain. In this respect, our conclusions are in agreement with theirs---the model we describe here provides a significantly better match to the results of the full \grasil\ model than, for example, the dust extinction model described by \cite{cole_hierarchical_2000}.

At high redshifts model galaxies often undergo periods of near continuous bursting as a result of experiencing disk instabilities on each subsequent timestep. This rather chaotic period of evolution is not well modelled presently---it is treated as a sequence of quiescent gas accretion periods punctuated by instability-triggered bursts while in reality we expect it to correspond more closely to a near continuous, high star formation rate mode somewhere in between the quiescent and bursting behaviour. While our model probably estimates the total amount of star formation during this period reasonably well (as it is controlled primarily by the cosmological infall rate and degree of outflow due to supernovae) we suspect that it does a rather poor job of accounting for dust extinction. After each burst the gas (and hence dust) content of each galaxy is reduced to zero, resulting in no extinction. Our model therefore tends to contain too many dust-free galaxies at high redshifts. To counteract this effect we force galaxies in this regime to be observed during a bursting phase, so that they always experience some dust extinction.

Dust remains one of the most challenging aspects of galaxies to model. We will return to aspects of our model related to dust (utilizing the more detailed \grasil\ model) in a future work, but note that even this is unlikely to be sufficient---what is needed is a better understanding of the complicated distribution of dust within galaxies, particularly during these early, chaotic phases.

Indeed the distribution of star formation within galaxies at $z=3$ to 5 has recently become within reach of observational studies \pcite{stark_formation_2008,elmegreen_bulge_2009,lehnert_physical_2009,swinbank_???_2009}. It seems that this aspect of the model is indeed supported by observational data. A future project will be to compare the internal properties of observed galaxies at these redshifts with those predicted by the model.

\subsection{Absorption by the IGM}

Where necessary, we model the attenuation of galaxy \SED s by neutral hydrogen in the intervening \IGM\ using the model of \cite{meiksin_colour_2006}.

\section{Model Selection}\label{sec:Selection}

The model described above has numerous free parameters which reflect our ignorance of the details of certain physical processes or order unity uncertainties in (e.g. geometrical) coefficients. To determine suitable values for these parameters we appeal to a broad range of observational data and search the model parameter space to find the best fit model.

The problem of how to implement the computationally challenging problem of fitting a complicated semi-analytic model with numerous free parameters to observational data has been considered before by \cite{henriques_monte_2009} and \cite{bower_parameter_2010}. To constrain model parameters in this work we use the ``Projection Pursuit'' method of \cite{bower_parameter_2010}. We give a brief description of that method here and refer the reader to \cite{bower_parameter_2010} for complete details.

Running a single set of model parameters, including all of the redshifts and wavelengths required for our analysis, is a relatively slow process. In particular, running a model with self-consistently computed \IGM\ evolution is entirely impractical for a parameter space search. We therefore chose to run models without a self-consistently computed \IGM\ or photoionizing background. Even then, each model takes around 2 hours to run on a fast computer. To mimic the effects of a photoionizing background we adopt the ``$V_{\rm cut}$--$z_{\rm cut}$'' model described by \cite{font_modelingmilky_2009} and which they show to reproduce quite well the results of the self-consistent calculation. Briefly, this model inhibits cooling of gas in halos with virial velocities below $V_{\rm cut}$ at redshifts below $z_{\rm cut}$. We then include $V_{\rm cut}$ and $z_{\rm cut}$ as parameters in our fitting process.

This approach is not ideal, but is required due to computational limitations. \cite{bower_parameter_2010} show that local (i.e. low redshift) properties of the model are not significantly affected by the inclusion of self-consistent reionization (i.e. those data do not constrain $V_{\rm cut}$ or $z_{\rm cut}$), and, where they are, the ``$V_{\rm cut}$--$z_{\rm cut}$'' model provides a reasonable approximation \cite{font_modelingmilky_2009}. In any case, as we will discuss below, some manual tuning of parameters is still required after the automated search of parameter space is completed. This manual search is then conducted using the fully self-consistent \IGM\ calculation.

We envision the problem in terms of a multi-dimensional parameter space into which constraints from observational data are mapped. Given the large number of model parameters and the fact that running a single realization of the model requires a significant amount of computer time, we can not perform a simple grid-search of the parameter space on a sufficiently fine grid. Instead, we begin by specifying plausible ranges for model parameters. The ranges considered for each parameter are listed in Table~\ref{tb:PCAranges}---for some parameters we choose to consider the logarithm of the parameter as the variable in our parameter space, to allow for efficient exploration of several decades of parameter value. We scale each model parameter such that it varies between 0 and 1 across this allowed range. We then generate a set of points in this limited and scaled model parameter space using Latin hypercube sampling \pcite{mckay_comparison_1979}, thereby ensuring an efficient coverage of the parameter space. A model is run for each set of parameters and a goodness of fit measure computed. 

\begin{table}
 \caption{The allowed ranges for each parameter in our fitting parameter space. For some parameters, we choose to use the logarithm of the parameter to allow efficient exploration of several decades of parameter value.}
 \label{tb:PCAranges}
 \begin{center}
 \begin{tabular}{lr@{.}lr@{.}l}
  \hline
  {\bf Parameter} & \multicolumn{2}{c}{{\bf Minimum}} & \multicolumn{2}{c}{{\bf Maximum}} \\
  \hline
  \input{Data/PCA_Table_Ranges}
  \hline
 \end{tabular}
 \end{center}
\end{table}

The choice of a goodness of fit measure is important and non-trivial (see \citealt{bower_parameter_2010}). We do not expect our model to fit all of the constraints in a statistically rigorous manner, as the model is clearly approximate. The Bayesian approach to this is issue is to assign a prior assessment of the reliability of the model to each of the data set comparisons and to define a correlation matrix reflecting the a priori connections between datasets. This concept (referred to as ``model discrepancy'' in the statistical literature) is discussed in detail for $z=0$ luminosity function constraints in \cite{bower_parameter_2010}. However, in the present paper, we needed a simpler approach to the problem. We therefore adopted a non-Bayesian methodology of simply summing $\chi^2$ for each dataset that we used. This has the advantage of simplicity, but clearly there may be more appropriate choices for the relative weighting of different data sets: we will explore this issue in a future paper. There is little doubt that a better measure of goodness of fit could be found. In particular, the relative weightings given to each dataset should really reflect how well we think the model performs in that particular quantity, how accurately we think that we have been able to match any observational selection and, inevitably, how much we believe the data itself. These are extremely thorny issues to which, at present, we do not have a good answer.

Specifically, in this work, the goodness of fit measure is taken to be
\begin{equation}
 \widetilde{\chi}^2 = \sum_i w_i {\chi^2_i \over N_i},
\end{equation}
where $\chi^2_i$ is the usual goodness of fit measure for dataset $i$, $N_i$ is the number of degrees of freedom in that dataset and $w_i$ is a weight assigned to each dataset. The sum is taken over all datasets shown in \S\ref{sec:Results} and, additionally, cosmological parameters were allowed to vary within the $2\sigma$ intervals permitted by the \cite{dunkley_five-year_2009} constraints, and were included in the goodness of fit measure using a Gaussian prior. When computing $\chi^2$ for each dataset we estimate the error in each datum to be the sum in quadrature of the experimental error and any statistical error present in the model due to the finite number of Monte Carlo merger tree realizations that we are able to carry out. This ensures that two models which differ by an amount comparable to the random noise in the models have similar values of $\chi^2$. The specific datasets used, along with the weights assigned to them (estimated using our best judgement of the reliability of each dataset and the \gf's ability to model it) are listed in Table~\ref{tb:constraints}.

\begin{table*}
 \caption{The set of datasets used as constraints on our model, together with a reference to where the dataset is shown in this paper and the value of the weight, $w_i$, assigned to each constraint.}
 \label{tb:constraints}
 \begin{tabular}{lcr@{.}l}
 \hline
 {\bf Constraint} & {\bf Reference} & \multicolumn{2}{c}{Weight ($w_i$)} \\
 \hline
 \input{Data/Constraints_Table}
 \hline
 \end{tabular}
\end{table*}

Once a set of models have been run, a principal components analysis is performed on the goodness of fit values of those models with $\widetilde{\chi}^2$ values in the lower $10^{\rm th}$ percentile of all models to find which linear combinations of parameters provide the \emph{minimum} variance in goodness of fit. These are the parameter combinations that are most tightly constrained by the observational data. A principal component with low variance implies that this particular combination of the parameters is tightly constrained if the model is likely to produce an acceptable fit. Of course, even if this constraint is satisfied, a good model is not guaranteed; rather we can be confident that if it is not satisfied the fit will not be good\footnote{This is only strictly true if the relationships between $\widetilde{\chi}^2$ and the parameters are approximately linear and unimodal. If there exists a separate small island of good values somewhere, our \PCA+Latin Hypercube method might happen to miss the region, or it might not exert sufficient pull on the \PCA\ compared to the large region and might be
subsequently ignored. The advantage of the emulator approach used by \protect\cite{bower_parameter_2010} is that it gives an estimate of the error made by excluding regions from further evaluations.}. When analysing the acceptable region in this way, we also need to bear in mind that the \PCA\ assumes that the relationships are linear, whereas \cite{bower_parameter_2010} show that the actual acceptable space is curved. This will prevent any of the suggested projections being arbitrarily thin and limit the accuracy of constraints. Nevertheless, the procedure substantially cuts down the volume of parameter space where model evaluations need to be run. These linear combinations are used to define rotated axes in the parameter space within which we select a new set of points again using Latin hypercube sampling. The process is repeated until a suitably converged model is found\footnote{In practice these calculations were run on distributed computing resources (including machines at the ICC in Durham, TeraGrid and Amazon EC2. Each machine was given an initial small set of models to run. After running each model, the results were transferred back to a central server. Periodically, the server would collate all available results, perform the \PCA\ and generate a new set of models which it then distributed to all active computing resources.}. This process is not fast, requiring around 150,000~CPU hours\footnote{The authors, feeling the need to help preserve our own small region of one realization of the Universe, purchased carbon offsets to counteract the carbon emissions resulting from this large investment of computing time.}, but does produce a model which is a good match to the input data.

\begin{figure*}
 \begin{center}
 \begin{tabular}{cc}
 \includegraphics[width=85mm,viewport=0mm 55mm 205mm 250mm,clip]{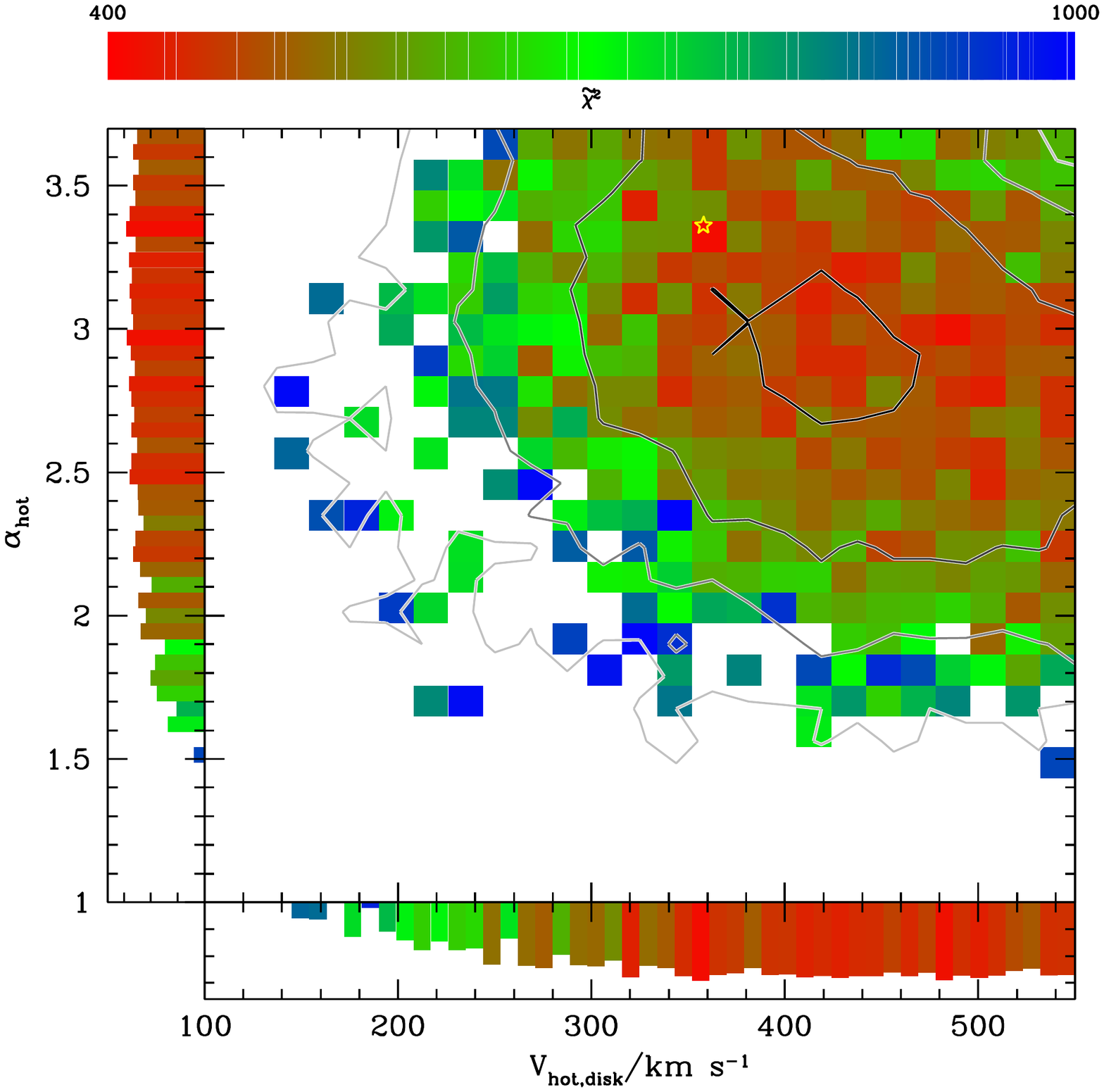} &
 \includegraphics[width=85mm,viewport=0mm 55mm 205mm 250mm,clip]{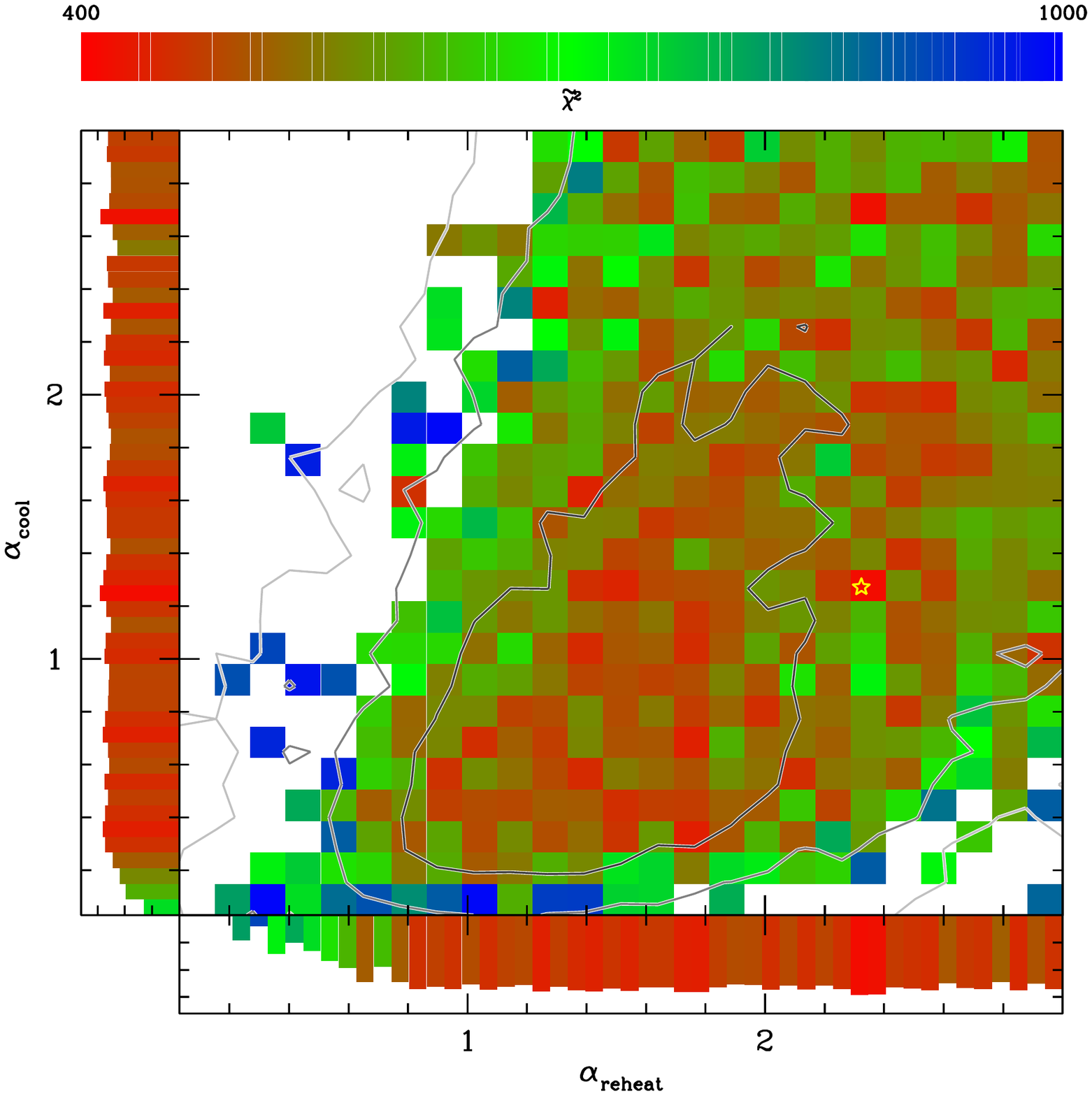} \\
 \includegraphics[width=85mm,viewport=0mm 55mm 205mm 250mm,clip]{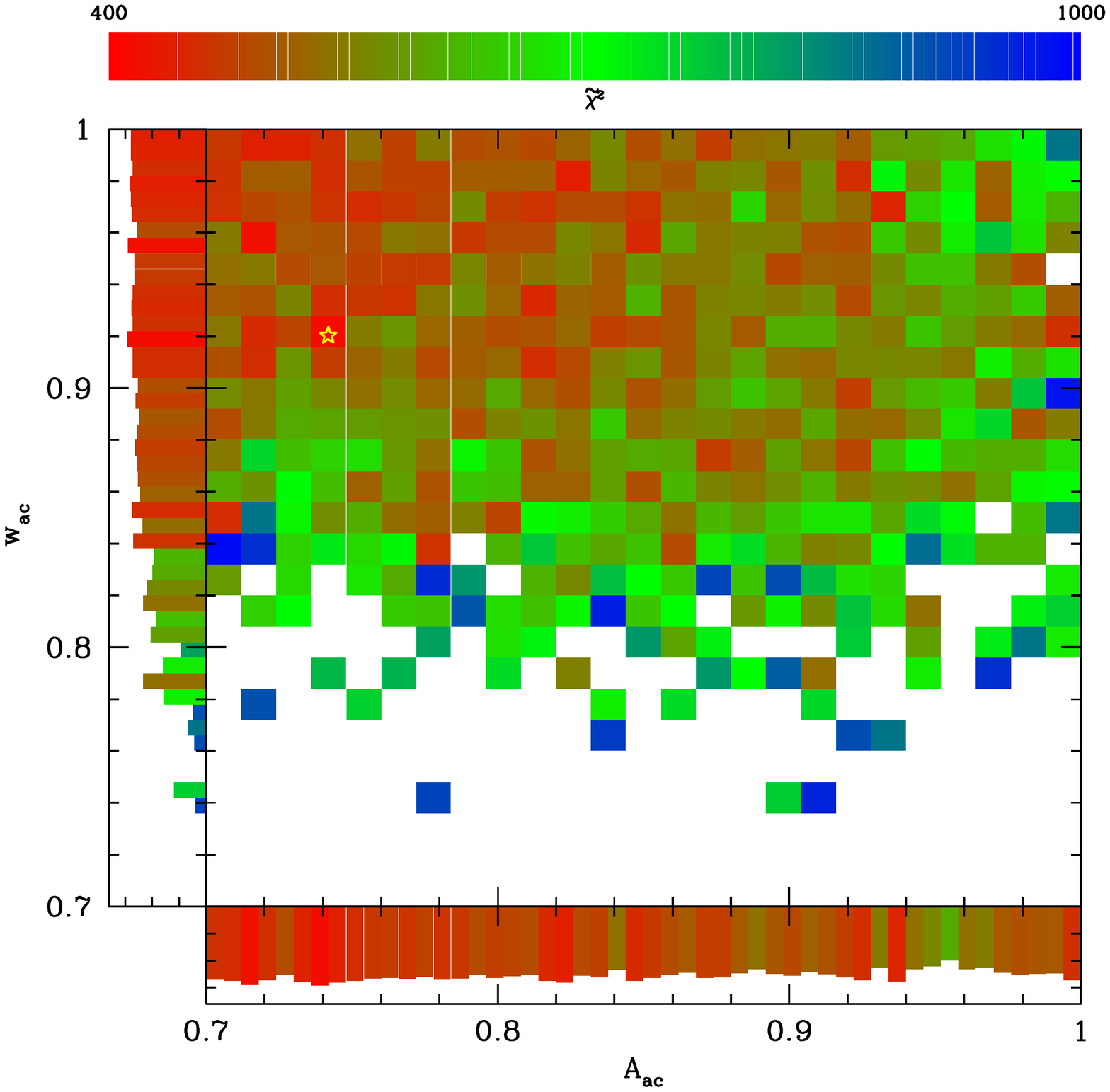} &
 \includegraphics[width=85mm,viewport=0mm 55mm 205mm 250mm,clip]{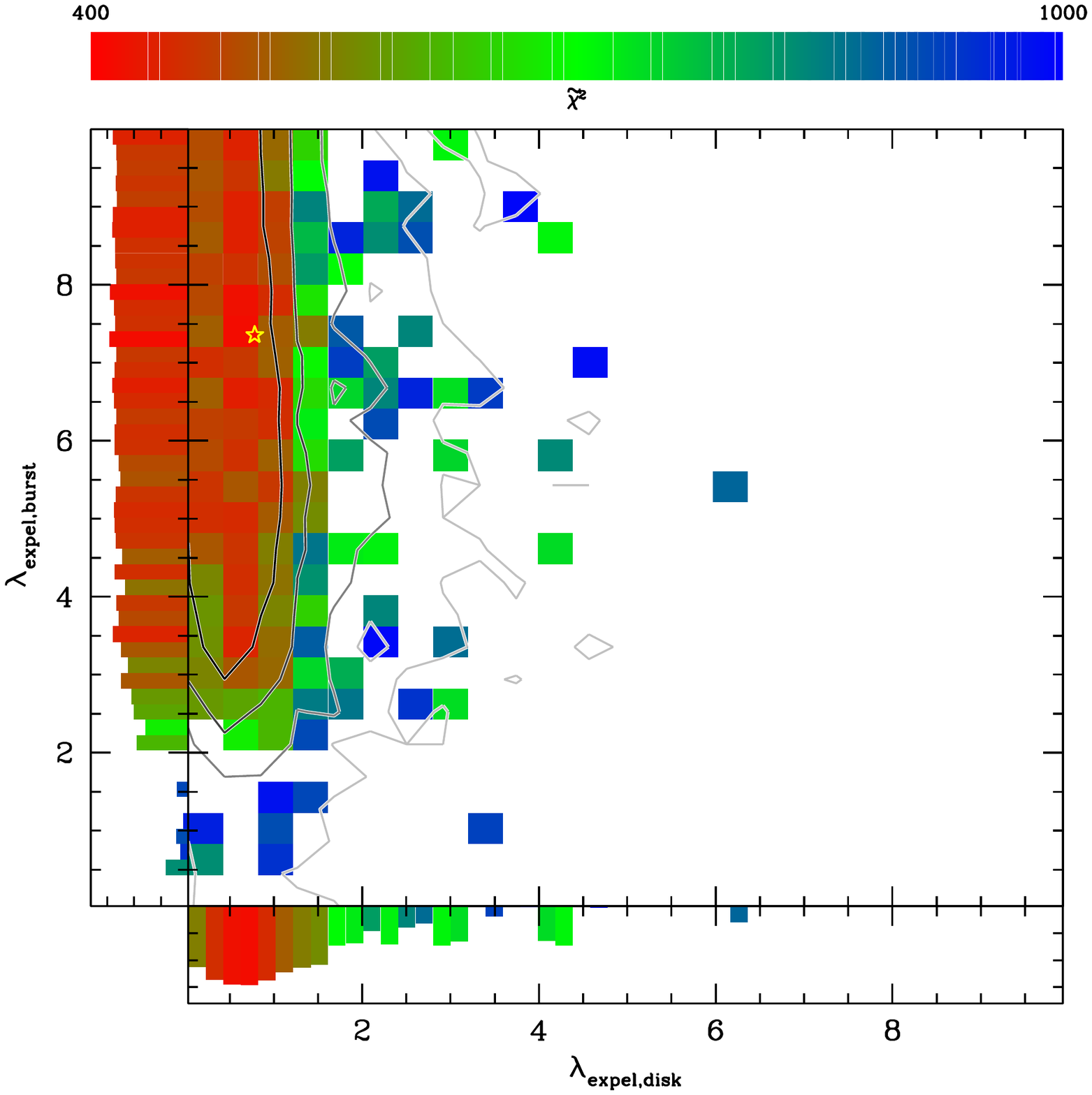}
 \end{tabular}
 \end{center}
 \caption{Constraints on model parameters shown as 2D slices through the multi-dimensional parameter space. In each panel, the colour scale indicates the value of $\widetilde{\chi}^2$ as shown by the bar above the panel, with the yellow star indicating the best-fit model. Each point in the plane is coloured to correspond to the minimum value of $\widetilde{\chi}^2$ found when projecting over all other dimensions of the parameter space. Contours illustrate the relative number of model evaluations at each point in the plane---from lightest to darkest line colour they correspond to 10, 30, 100 and 300 evaluations per grid cell. Most evaluations are carried out when the best model fits are found, indicating that our method is efficient in concentrating resources where good models are most likely to be found. To each side of the plane, the distribution of $\widetilde{\chi}^2$ is projected over one of the remaining dimensions to show constraints on the indicated parameter. \emph{Top left panel:} Shows the main parameters of the \protect\SNe\ feedback model, $V_{\rm hot,disk}$ and $\alpha_{\rm hot}$. \emph{Top right panel:} Shows critical parameters controlling the cooling and \protect\AGN\ feedback models, $\alpha_{\rm reheat}$ and $\alpha_{\rm cool}$. \emph{Lower left panel:} Show parameters of the adiabatic contraction model which have important consequences for the sizes of galaxies, $A_{\rm ac}$ and $w_{\rm ac}$. \emph{Lower right panel:} Shows parameters of the \protect\SNe\ feedback model that control the amount of material expelled from halos, $\lambda_{\rm expel,disk}$ and $\lambda_{\rm expel,burst}$.}
 \label{fig:Constraints}
\end{figure*}

Figure~\ref{fig:Constraints} demonstrates the efficacy of our method using four 2D slices through the multi-dimensional parameter space. The colour scale in each panel shows constraints on two of the model parameters, while the projections below and to the left of the panel indicate the constraints on the indicated single parameter. Contours illustrate the relative number of model evaluations which were performed at each point in the plane. It can be clearly seen that our ``Projection Pursuit'' methodology concentrates model evaluations in those regions which are most likely to provide a good fit. The nominal best-fit model is indicated by a yellow star in each panel. Despite the large number of models run we do not believe that this precise point should be considered as the ``best'' model---the dimensionality of the parameter space is so large that we do not believe that it has been sufficiently well mapped to draw this conclusion. Additionally, we also need a model discrepancy matrix---without this, we can not say whether a model is acceptable (in the sense that it should only agree with the data as well as we expect given the level of approximation in the model). Without the discrepancy term, we will tend to overfit the model. Instead, we utilize these results to suggest the region of parameter space in which the best model is likely to be found. We then adjust parameters manually to find the final model (utilizing our intuition of how the model will respond to changes in parameters).

Interesting constraints and correlations can be seen in Figure~\ref{fig:Constraints}. For example, the combination $\alpha_{\rm hot}$--$V_{\rm hot,disk}$ is quite well constrained and somewhat anti-correlated (such that an increase in $\alpha_{\rm hot}$ can be played off against a decrease in $V_{\rm hot,disk}$. It is immediately clear, for example, that no good model can be found with $\lambda_{\rm expel,disk}\gsim1.5$ while $\lambda_{\rm expel,bulge}$ is much less well constrained, but must be larger then about $1.5$.

The principal component vectors from the final set of \NPCASearch\ models are shown in Table~\ref{tb:PCAmatrix}. We note here that these vectors are quite different from those found by \cite{bower_parameter_2010}. This is not too surprising as our implementation of \gf\ is quite different from theirs and we constrain our model to a much broader collection of datasets. We will examine the \PCA\ vectors in greater detail in a future paper, and so restrict ourselves to a brief discussion here. Taking the first \PCA\ vector for example, we see that it is dominated by $z_{\rm cut}$, $\alpha_\star$ and $\alpha_{\rm hot}$. These parameters all have strong effects on the faint end of luminosity functions. Luminosity functions are abundant in our set of constraints and have been well measured. As such, they provide some of the strongest constraints on the model. It can be seen that an increase in $\alpha_{\rm hot}$, which will flatten the slope of the faint of a luminosity function, has a similar effect as a decrease in $\alpha_\star$, which will preferentially reduce rates of star formation in low-mass galaxies and so also flatten the faint end slope. The second \PCA\ component shows a strong but opposite dependence on $\Omega_{\rm b}$ and $\lambda_{\rm expel,burst}$. Increasing $\Omega_{\rm b}$ results in more fuel for galaxy formation, while increasing $\lambda_{\rm expel,burst}$ causes material to be lost by being expelled from halos. As we continue to further \PCA\ vectors the parameter combinations they represent become more complicated and difficult to interpret---the advantage of our methodology is that these complex interactions can be taken into account when exploring the model parameter space.

The differences between our results and those of \mbox{\cite{bower_parameter_2010}} are interesting in their own right. For example, \cite{bower_parameter_2010} found two ``islands'' of good fit in the supernovae feedback parameter space ($V_{\rm hot,disk}$ and $V_{\rm hot,burst}$): a strong feedback island (corresponding approximately to what we find in this work) and a weak feedback island (which we do not find). The weak feedback island is ruled out in the present work as, while a good fit to the galaxy luminosity function can be found in it (as demonstrated by \citealt{bower_parameter_2010}), no good fit to, for example, galaxy sizes can be found.

\begin{sidewaystable*}
 \caption{: The principal components, rank ordered by their contribution to the variance, $\sigma^2$, from our models. In each row, the dominant elements (those with an absolute value in excess of $0.33$, are shown in bold.}
 \label{tb:PCAmatrix}
 \begin{tabular}{lr@{.}lr@{.}lr@{.}lr@{.}lr@{.}lr@{.}lr@{.}lr@{.}lr@{.}lr@{.}lr@{.}lr@{.}lr@{.}lr@{.}lr@{.}lr@{.}l}
  \hline
  \input{Data/PCA_Table_Part1}
 \end{tabular}
\end{sidewaystable*}

\begin{sidewaystable*}
 \addtocounter{table}{-1}
 \caption{: \emph{(cont.)} The principal components, rank ordered by their contribution to the variance, $\sigma^2$, from our models. In each row, the dominant elements (those with an absolute value in excess of $0.33$, are shown in bold.}
 \begin{tabular}{lr@{.}lr@{.}lr@{.}lr@{.}lr@{.}lr@{.}lr@{.}lr@{.}lr@{.}lr@{.}lr@{.}lr@{.}lr@{.}lr@{.}lr@{.}l}
  \hline
  \input{Data/PCA_Table_Part2}
 \end{tabular}
\end{sidewaystable*}

\section{Results}\label{sec:Results}

In this section we will begin by identifying the best-fit model and will then show results from that model compared to the observational data that was used to constrain the model parameters. With the exception of results shown in \S\ref{sec:Predictions} all of the data shown in this section were used to constrain the model and, as such, the results do not represent predictions of the model. (In \S\ref{sec:GasPhases} we examine the distribution of gas between different phases as a function of halo mass, while in \S\ref{sec:ICL} we explore the fraction of stellar mass in the intracluster light component of halos. The data shown in these comparisons were \emph{not} used as constraints when searching for the best-fit model.) The overall best-fit model (i.e. that which best describes the union of all datasets) is shown by blue lines. Additionally, we show as magenta lines the best-fit model to each individual dataset (as described in the figure captions) for comparison. We do not claim that the following represents a complete census of the observational data that \emph{could} be used to constrain our galaxy formation model. Instead, we have selected data which spans a range of physical characteristics and redshifts that we think best constrains the physics of our model, while remaining within the limited (although substantial) computational resources at our disposal.

In addition to these best-fit models, we will, where possible, compare our current results with those from the previous implementation of \gf\ described by \cite{bower_breakinghierarchy_2006}. Results from the \cite{bower_breakinghierarchy_2006} model are shown by green lines in each figure. We have not included figures for every constraint used in this work---specifically, in many cases we show examples of the constraints only for a limited number of magnitude or redshift ranges. However, all of the constraints used are listed in Table~\ref{tb:constraints} and are discussed in the text.

\subsection{Best Fit Model}

The resulting set of best-fit parameters are listed in Table~\ref{tb:BestFitParams}. We will not investigate the details of these results here, leaving an exploration of which data constrain which parameters and the possibility of alternative, yet acceptable, parameter sets to a future work. The best fit model turns out to be a reasonably good match to local luminosity data, galaxy colours, metallicities, gas content, supermassive black hole masses and constraints on the epoch of reionization, but to perform less well in matching galaxy sizes, clustering and the Tully-Fisher relation. In addition, luminosity functions become increasingly more discrepant with the data as we move to higher redshifts. In the remainder of this section we will briefly discuss some important aspects of the best fit parameter set.

\begin{table}
 \caption{Parameters of the best fit model used in this work and of the \protect\cite{bower_breakinghierarchy_2006} model. Note that the best-fit model listed here is one that includes self-consistent reionization and evolution of the \protect\IGM\ (see \S\ref{sec:IGM}) and which has been adjusted to also produce a reasonable reionization history (see \S\ref{sec:IGMResults}). It therefore does not correspond to the location of the best-fit model indicated in Fig.~\ref{fig:Constraints}. Where appropriate, references are given to the article, or section of this work, in which the parameter is described.}
 \label{tb:BestFitParams}
 \begin{center}
 \begin{tabular}{cr@{.}lr@{.}lc}
  \hline
                  & \multicolumn{4}{c}{{\bf Value}} & \\
  \cline{2-5}
  {\bf Parameter} & \multicolumn{2}{c}{{\bf This Work}} &  \multicolumn{2}{c}{{\bf Bower06}} & {\bf Reference} \\
  \hline
  \hline
  \multicolumn{6}{c}{\emph{Cosmological}} \\
  $\Omega_0$ & \BestFitomega0 & 0&250\\
  $\Lambda_0$ & \BestFitlambda0 & 0&750 \\
  $\Omega_{\rm b}$ & \BestFitomegab & 0&04500\\
  $h_0$ & \BestFith0 & 0&730\\
  $\sigma_8$ & \BestFitsigma8 & 0&900 \\
  $n_{\rm s}$ & \BestFitnspec & 1&000 \\
  \hline
  \multicolumn{6}{c}{\emph{Gas Cooling Model}} \\
  $\alpha_{\rm reheat}$ & \BestFitalphareheat & 1&260 & \S\ref{sec:Reheating} \\
  $\alpha_{\rm cool}$ & \BestFitalphacool & 0&580 & \S\ref{sec:AGNFeedback} \\
  $\alpha_{\rm remove}$ & \BestFitalphaCooledRemove & \multicolumn{2}{c}{N/A} & \S\ref{sec:Reheating} \\
  $a_{\rm core}$ & \BestFitcore & 0&100 & \S\ref{sec:HotGasDist} \\
  \hline
  \multicolumn{6}{c}{\emph{Adiabatic Contraction}} \\
  $A_{\rm ac}$ & \BestFitAGnedin & 1&000 & \S\ref{sec:Sizes} \\
  $w_{\rm ac}$ & \BestFitwGnedin & 1&000 & \S\ref{sec:Sizes} \\
  \hline
  \multicolumn{6}{c}{\emph{Star Formation}} \\
  $\epsilon_\star$ & \BestFitepsilonStar & 0&0029 & \cite{cole_hierarchical_2000} \\
  $\alpha_\star$ & \BestFitalphastar & -1&50 & \cite{cole_hierarchical_2000} \\
  \hline
  \multicolumn{6}{c}{\emph{Disk Stability}} \\
  $\epsilon_{\rm d,gas}$ & \BestFitstabledisk & 0&800\footnotemark & \S\ref{sec:MinorChanges} \\
  \hline
  \multicolumn{6}{c}{\emph{Supernovae Feedback}} \\
  $V_{\rm hot,disk}$ & \BestFitvhotdisk~km/s & 485&0~km/s & \S\ref{sec:Feedback} \\
  $V_{\rm hot,burst}$ & \BestFitvhotburst~km/s & 485&0~km/s & \S\ref{sec:Feedback} \\
  $\alpha_{\rm hot}$ & \BestFitalphahot & 3&20 & \S\ref{sec:Feedback} \\
  $\lambda_{\rm expel,disk}$ & \BestFitlambdaExpelDisk & \multicolumn{2}{c}{N/A} & \S\ref{sec:Feedback} \\
  $\lambda_{\rm expel,burst}$ & \BestFitlambdaExpelBurst & \multicolumn{2}{c}{N/A} & \S\ref{sec:Feedback} \\
  \hline
  \multicolumn{6}{c}{\emph{Ram Pressure Stripping}} \\
  $\epsilon_{\rm strip}$ & \BestFitRamPressureTransferFraction & \multicolumn{2}{c}{N/A} & \S\ref{sec:HotStrip} \\
  \hline
  \multicolumn{6}{c}{\emph{Merging}} \\
  $f_{\rm ellip}$ & \BestFitfellip & 0&3000 & \cite{cole_hierarchical_2000} \\
  $f_{\rm burst}$ & \BestFitfburst & 0&100 & \cite{cole_hierarchical_2000} \\
  $f_{\rm gas,burst}$ & \BestFitfgasburst & 0&100 & \S\ref{sec:Unchanged} \\
  $B/T_{\rm burst}$ & \BestFitbtburst & \multicolumn{2}{c}{N/A} &  \S\ref{sec:Unchanged} \\
  \hline
  \multicolumn{6}{c}{\emph{Black Hole Growth}} \\
  $\epsilon_\bullet$ & \BestFitepsilonSMBHEddington & 0&0398 & \S\ref{sec:AGNFeedback} \\
  $\eta_\bullet$ & \BestFitetaSMBH & \multicolumn{2}{c}{N/A} & \S\ref{sec:AGNFeedback} \\
  $F_{\rm SMBH}$ & \BestFitFSMBH & 0&00500 & \cite{malbon_black_2007} \\
  \hline
 \end{tabular}
 \end{center}
\end{table}

\footnotetext{The \protect\cite{bower_breakinghierarchy_2006} model used a single value of $\epsilon_{\rm d}$ for both gaseous and stellar disks.}

\begin{table*}
 \caption{Parameters of the overall best fit model compared to those of models which best fit individual datasets (as indicated by column labels). Parameters which play a key role (as discussed in the relevant subsections of \S\ref{sec:Results}) in helping to obtain a good fit to each dataset are shown in bold type.}
 \label{tb:BestModelParams}
 \setlength\tabcolsep{1pt}
 \begin{center}
 \begin{tabular}{lr@{.}l@{\hspace{2pt}}r@{.}l@{\hspace{2pt}}r@{.}l@{\hspace{2pt}}r@{.}l@{\hspace{2pt}}r@{.}l@{\hspace{2pt}}r@{.}l@{\hspace{2pt}}r@{.}l@{\hspace{2pt}}r@{.}l@{\hspace{2pt}}r@{.}l@{\hspace{2pt}}r@{.}l@{\hspace{2pt}}r@{.}l@{\hspace{2pt}}r@{.}l@{\hspace{2pt}}r@{.}l@{\hspace{2pt}}r@{.}l@{\hspace{2pt}}r@{.}l@{\hspace{2pt}}r@{.}l@{\hspace{2pt}}r@{.}l@{\hspace{2pt}}r@{.}l}
 \input{Data/BestModelsTable0}
 \end{tabular}
 \end{center}
 \setlength\tabcolsep{5pt}
\end{table*}

\begin{table*}
 \addtocounter{table}{-1}
 \caption{\emph{(cont.)} Parameters of the overall best fit model compared to those of models which best fit individual datasets (as indicated by column labels). Parameters which play a key role in helping to obtain a good fit to each dataset are shown in bold type.}
 \setlength\tabcolsep{1pt}
 \begin{center}
 \begin{tabular}{lr@{.}l@{\hspace{2pt}}r@{.}l@{\hspace{2pt}}r@{.}l@{\hspace{2pt}}r@{.}l@{\hspace{2pt}}r@{.}l@{\hspace{2pt}}r@{.}l@{\hspace{2pt}}r@{.}l@{\hspace{2pt}}r@{.}l@{\hspace{2pt}}r@{.}l@{\hspace{2pt}}r@{.}l@{\hspace{2pt}}r@{.}l@{\hspace{2pt}}r@{.}l@{\hspace{2pt}}r@{.}l@{\hspace{2pt}}r@{.}l@{\hspace{2pt}}r@{.}l@{\hspace{2pt}}r@{.}l@{\hspace{2pt}}r@{.}l@{\hspace{2pt}}r@{.}l}
 \input{Data/BestModelsTable1}
 \end{tabular}
 \end{center}
 \setlength\tabcolsep{5pt}
\end{table*}

The cosmological parameters are all close to the \WMAP\ five-year expectations (by construction). The parameters of the gas cooling model are all quite reasonable: the three parameters $\alpha_{\rm reheat}$ and $\alpha_{\rm cool}$ are all of order unity as expected, $\alpha_{\rm remove}$ is somewhat smaller but still plausible, while the core radius $a_{\rm core}$ is around 22\% of the virial radius. The parameters of the adiabatic contraction model differ from those proposed by \cite{gnedin_response_2004} but are within the range of values found by \cite{gustafsson_baryonic_2006} when fitting the profiles of dark matter halos in simulations including galaxy formation with feedback. The disk stability parameter, $\epsilon_{\rm d,gas}$ is close to, albeit lower than, the value of $0.9$ suggested by the theoretical work of \cite{christodoulou_new_1995}. The stripping parameter, $\epsilon_{\rm strip}$, is of order unity as expected.

The star formation parameters are reasonable, implying a low efficiency of star formation. The feedback parameters, $V_{\rm hot,disk|burst}$ are much lower than the value of 485~km/s required by \cite{bower_breakinghierarchy_2006} and significantly closer to the value of 200~km/s adopted by \cite{cole_hierarchical_2000}. This is desirable as values around 200~km/s already stretch the \SNe\ energy budget. We also note that the value of $\alpha_{\rm hot}$ is lower than that required by \cite{bower_breakinghierarchy_2006} and closer to the ``natural'' value of $2$, which would imply an efficiency of supernovae energy coupling into feedback that was independent of galaxy properties. The expulsion parameters, $\lambda_{\rm expel,disk|burst}$, are close to unity as expected.

The parameters of the merging model imply that mass ratios of 1:10 or greater are required for a major merger, a little low, but within the range of plausibility, while only 1:5 or greater mergers trigger a burst. Minor mergers in which the primary galaxy has at least 34\% gas by mass and at least 34\% of its mass in a disk can also lead to bursts.

Finally, the black hole growth parameters are quite reasonable: black holes radiate at about 9\% of the Eddington luminosity, 5\% of cooling gas reaches the black hole during radio mode feedback and around 0.5\% of gas in a merging event is driven into the black hole.

Overall, the parameters of the best fit model seem reasonable on physical grounds. Given the large dimensionality of the parameter space, the complexity of the model and the various assumptions used in modelling complex physical processes we would not consider these values to be either precise or accurate (which is why we do not quote error bars here), but to merely represent the most plausible values within the context of the \gf\ semi-analytic model of galaxy formation.

In addition to this overall best-fit model, we show in Table~\ref{tb:BestModelParams} the parameters which produced the best-fit to subsets of the data (as indicated). We caution that these models were selected from runs without self-consistent reionization and also with relatively few realizations of merger trees, making them noisy. This means that, after re-running these models with many more merger tree realizations it is possible that they will not be such good fits to the data. We do, in fact, find such cases as we will highlight below. Nevertheless, we will refer to this table in the remainder of this section when exploring the ability of our model to match each dataset. We also point out that there is no guarantee that any of these models that provide a good match to an individual dataset are good matches overall---for example, the model which best matches galaxy sizes may produce entirely unacceptable $z=0$ luminosity functions.

\subsection{Star Formation History}
\label{sec:SFH}

\begin{figure}
 \includegraphics[width=80mm,viewport=0mm 60mm 200mm 245mm,clip]{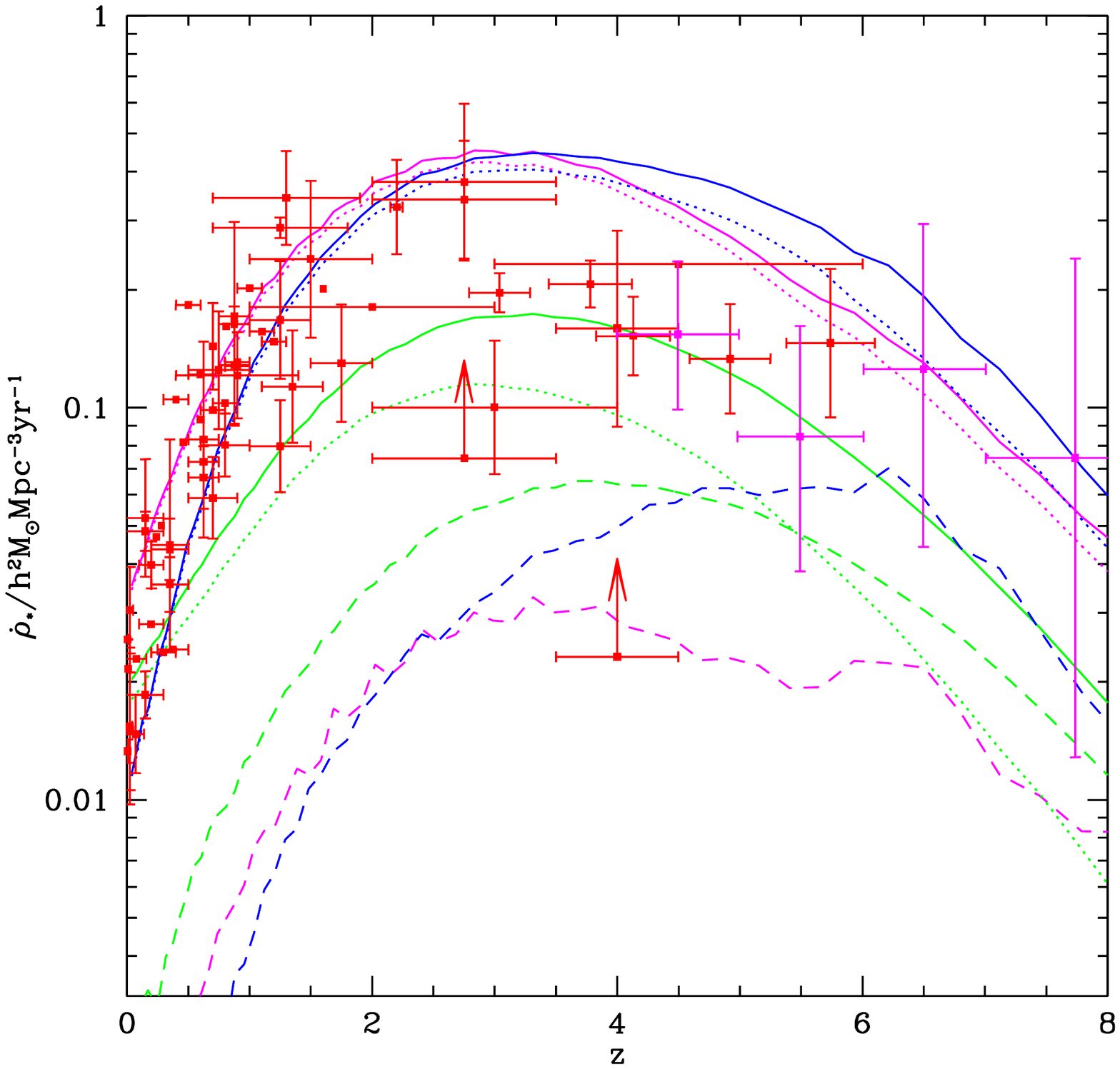}
 \caption{The star formation rate per unit comoving volume in the Universe as a function of redshift. Red points show observational estimates from a variety of sources as compiled by \protect\cite{hopkins_evolution_2004} while magenta points show the star formation rate inferred from gamma ray bursts by \protect\cite{kistler_star_2009}. The solid lines show the total star formation rate density from our models, while the dotted and dashed lines show the contribution to this from quiescent star formation in disks and starbursts respectively. Blue lines show the overall best-fit model, while magenta lines indicate the best-fit model to this dataset and the green lines show results from the \protect\cite{bower_breakinghierarchy_2006} model.}
 \label{fig:SFH}
\end{figure}

Figure~\ref{fig:SFH} shows the star formation rate per unit volume as a function of redshift, with symbols indicating observational estimates and lines showing results from our model. Dotted and dashed lines show quiescent star formation in disks and bursts of star formation respectively, while solid lines indicate the sum of these two. The quiescent mode dominates at all redshifts, although we note that at high redshifts model disks are typically unstable and undergo frequent instability events. These galaxies may therefore not look like typical low redshift disk galaxies. The best fit model is in excellent agreement with the star formation rate data from $z=1$ to $z=8$, reproducing the sharp decline in star formation below $z=2$ while maintaining a relatively high star formation rate out to the highest redshifts. Our model lies below the data at $z\lsim 1$ despite being a good match to the b$_{\rm J}$-band luminosity function (see \S\ref{sec:z0LFResults}). This suggests some inconsistency in the data analysis, perhaps related to the choice of \IMF\ or the calibration of star formation rate indicators. Indeed, the model which best fits this particular dataset (shown as magenta lines in Fig.~\ref{fig:SFH}) does so by virtue of having a large value of $\epsilon_\star$ (see Table~\ref{tb:BestModelParams}; this increases star formation rates overall) and a small value of $\alpha_{\rm cool}$ (which alters the critical mass scale for \AGN\ feedback and thereby delays the truncation of star formation at low redshifts). While these changes result in a better fit to the star formation rate, they produce very unacceptable fits to the luminosity functions (which have too many bright galaxies) and galaxies which are far too depleted of gas.

The \cite{bower_breakinghierarchy_2006} model has a much lower star formation rate density than our best-fit model at $z>0.5$, although it shows a comparable amount of star formation in bursts. (The \cite{bower_breakinghierarchy_2006} model still manages to obtain a good match to the K-band luminosity function at $z=0$ however by virtue of the fact that at $z\lsim1$, where much of the build up of stellar mass occurs, the two models have comparable average star formation rates, and because it uses a different \IMF\ which results in a different mass-to-light ratio. Our best-fit model produces has 65\% more mass in stars at $z=0$ than the \cite{bower_breakinghierarchy_2006} model, but produces only 35\% more K-band luminosity density, as will be shown in Fig.~\ref{fig:K_LF}, mostly from faint galaxies.) Our best-fit model can be seen to be in significantly better agreement with the data than the \cite{bower_breakinghierarchy_2006} model and nicely reproduces the sharp decline in star formation rate at low redshifts.

\subsection{Luminosity Functions}\label{sec:z0LFResults}

Luminosity functions have traditionally represented an important constraint for galaxy formation models. We therefore include a variety of luminosity functions, spanning a range of redshifts in our constraints.

\begin{figure}
 \includegraphics[width=80mm,viewport=0mm 55mm 200mm 245mm,clip]{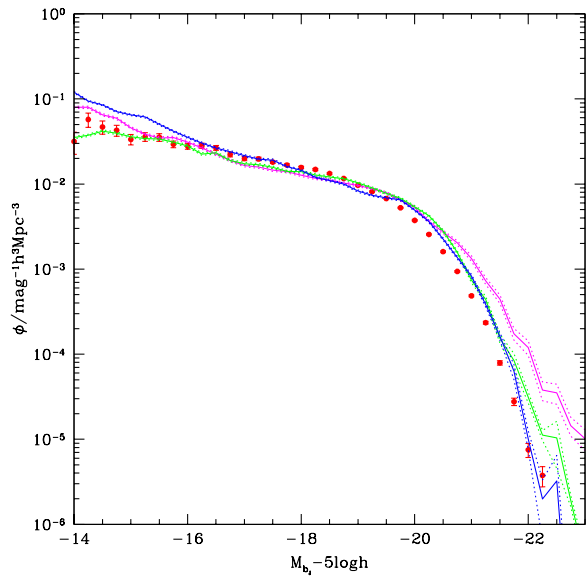}
 \caption{The $z=0$ b$_{\rm J}$-band luminosity function from our models: the solid lines show the luminosity function after dust extinction is applied while the dotted lines show the statistical error on the model estimate. Red points indicate the observed luminosity function from the \protect\TdF\ \protect\pcite{norberg_2df_2002}. Blue lines show the overall best-fit model, while magenta lines indicate the best-fit model to this dataset and the $z=0$ K-band luminosity function (see Fig.~\protect\ref{fig:K_LF}; note that the requirement that this model be a good match to the $z=0$ K-band luminosity function is the reason why the fit here is not as good as that of the overall best-fit model) and green lines show results from the \protect\cite{bower_breakinghierarchy_2006} model.}
 \label{fig:bJ_LF}
\end{figure}

\begin{figure}
 \includegraphics[width=80mm,viewport=0mm 55mm 200mm 245mm,clip]{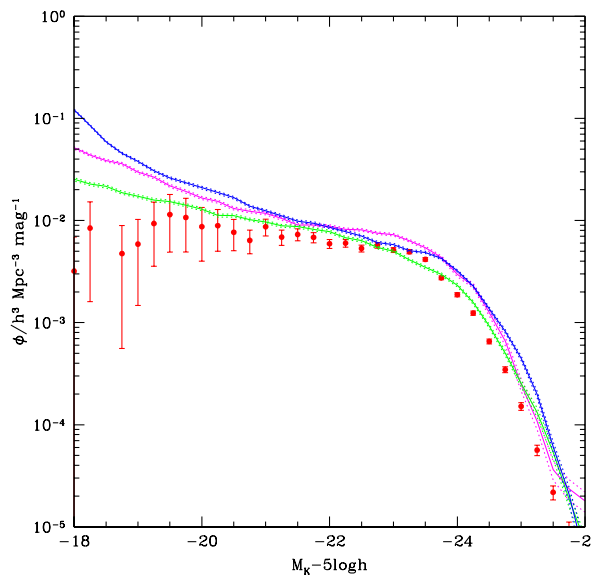}
 \caption{The $z=0$ K-band luminosity function from our models: the solid lines show the luminosity function after dust extinction is applied while the dotted lines show the statistical error on the model estimate. Red points indicate data from the \protect\TdF +\protect\TMASS\ \protect\pcite{cole_2df_2001}. Blue lines show the overall best-fit model, while magenta indicate the best-fit model to this dataset and the $z=0$ b$_{\rm J}$-band luminosity function (see Fig.~\protect\ref{fig:bJ_LF}) and green show results from the \protect\cite{bower_breakinghierarchy_2006} model.}
 \label{fig:K_LF}
\end{figure}

Figures~\ref{fig:bJ_LF} and \ref{fig:K_LF} show local ($z\approx 0$) luminosity functions from the \TdF\ (\citealt{norberg_2df_2002}; b$_{\rm J}$ band) and the \TMASS\ (\citealt{cole_2df_2001}; $K$ band) respectively together with model predictions.

It is well established that the faint end slope of the luminosity function, which is flatter than would be naively expected from the slope of the dark matter halo mass function, requires some type of feedback in order to be reproduced in models. The supernovae feedback present in our model is sufficient to flatten the faint end slope of the local luminosity functions and bring it into good agreement with the data in the b$_{\rm J}$ band, except perhaps at the very faintest magnitudes shown. The K-band shows an even flatter faint end slope and this is not as well reproduced by our model.

Both our best-fit model and the \cite{bower_breakinghierarchy_2006} model produce good fits to these luminosity functions (although our best fit model produces a break which is slightly too bright in the K-band, indicating that the galaxy colours are not quite right---see \S\ref{sec:Colours}). This is not surprising of course as these were primary constraints used to find parameters for the \cite{bower_breakinghierarchy_2006} model. The \cite{bower_breakinghierarchy_2006} model does give a noticeably better match to the faint end of the K-band luminosity function (although it is far from perfect), due to the higher value of $\alpha_{\rm hot}$ that it adopts (see Table~\ref{tb:BestFitParams}). Unfortunately, this large value of $\alpha_{\rm hot}$ adversely affects the agreement with other datasets and so our best-fit model is forced to adopt a lower value. The important point here is that the \cite{bower_breakinghierarchy_2006} model was designed to fit just these luminosity functions, while the current model is being asked to simultaneously fit a much larger compilation of datasets. This point is further illustrated by the magenta lines in Figs.~\ref{fig:bJ_LF} and \ref{fig:K_LF} which show the model that best matches these two datasets. It achieves a flatter faint end slope by virtue of having quite large values of $\alpha_{\rm hot}$ and $\alpha_{\rm cool}$. This improved match to the faint end is at the expense of the bright end though ($\chi^2$ fitting gives more weight to the faint end, which has more data points with smaller error bars).

\begin{figure}
 \includegraphics[width=80mm,viewport=0mm 55mm 200mm 245mm,clip]{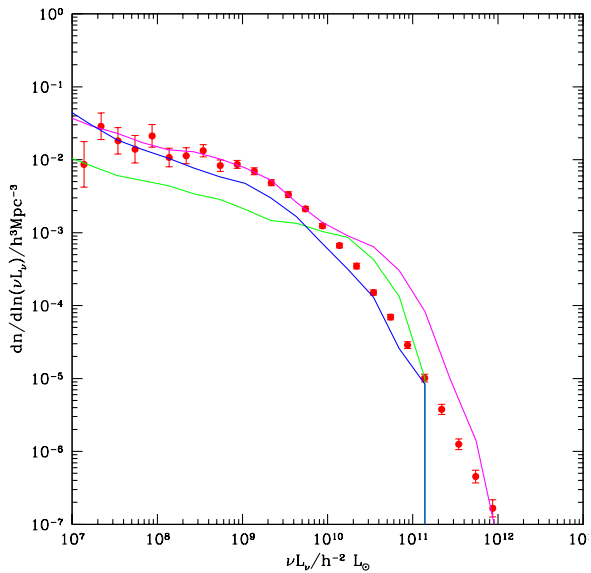}
 \caption{The $z=0$ 60$\mu$m luminosity functions from our models are shown by the solid lines. Red points indicate data from \protect\cite{saunders_60-micron_1990}. Blue lines show the overall best-fit model, while magenta lines indicate the best-fit model to this dataset and the green lines show results from the \protect\cite{bower_breakinghierarchy_2006} model.}
\label{fig:60mu_z0_LF}
\end{figure}

Figure~\ref{fig:60mu_z0_LF} shows the 60$\mu$m infrared luminosity function from \cite{saunders_60-micron_1990} (red points) and the corresponding model results (lines). The 60$\mu$m luminosity function constrains the dust absorption and reemission in our model and so is complementary to the optical and near-\IR\ luminosity functions discussed above. Our best fit model produces a very good match to the data at low luminosities---the sharp cut off at $10^{11}h^{-2}L_\odot$ is artificial and due to the limited number of merger trees which we are able to run and the scarcity of these galaxies (which are produced by massive bursts of star formation). The \cite{bower_breakinghierarchy_2006} model matches well at high luminosities but underpredicts the number of faint galaxies. This is due to the higher frequency of starbursts at low redshifts in the \cite{bower_breakinghierarchy_2006} model (see Fig.~\ref{fig:SFH}), which populate the bright end of the 60$\mu$m luminosity function. It must be kept in mind that absorption and re-emission of starlight by dust is one of the most challenging processes to model semi-analytically, and we expect that approximations made in this work may have significant effects on emission at $60\mu$m. A more detailed study, utilizing \grasil, will be presented in a future work. The best fit model to this specific dataset is a good fit to the data although it has somewhat too many  $60\mu$m-bright galaxies. This is achieved by adopting a much lower value of $f_{\rm gas,burst}$ which lets minor mergers trigger bursts more easily. This increases the abundance of bursting galaxies with high star formation rates and fills in the bright end of the $60\mu$m luminosity function.

\begin{figure}
\begin{center}
$z=1.0$\\
\vspace{-5mm}\includegraphics[width=80mm,viewport=0mm 55mm 200mm 245mm,clip]{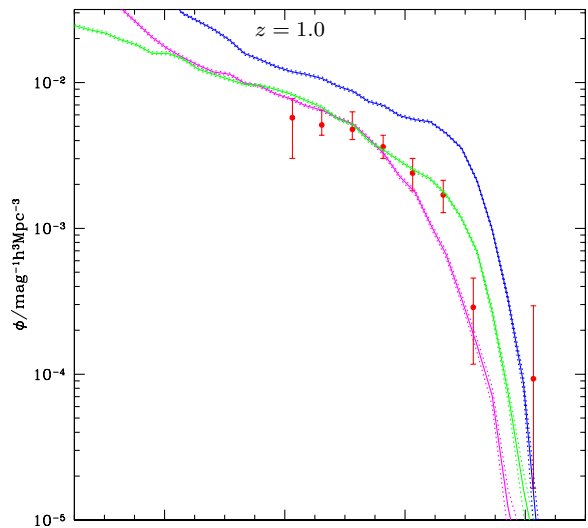}
\end{center}
\caption{The $z=1$ K$_{\rm s}$-band luminosity function from our models is shown by the solid lines with dotted lines indicating the statistical uncertainty on the model estimates. Red points indicate data from \protect\cite{pozzetti_k20_2003}. Blue lines show the overall best-fit model, while magenta lines indicate the best-fit model to this dataset and the green lines show results from the \protect\cite{bower_breakinghierarchy_2006} model.}
\label{fig:K20_Ks_LF}
\end{figure}

\begin{figure*}
 \begin{tabular}{ccc}
 \includegraphics[width=55mm,viewport=0mm 55mm 205mm 245mm,clip]{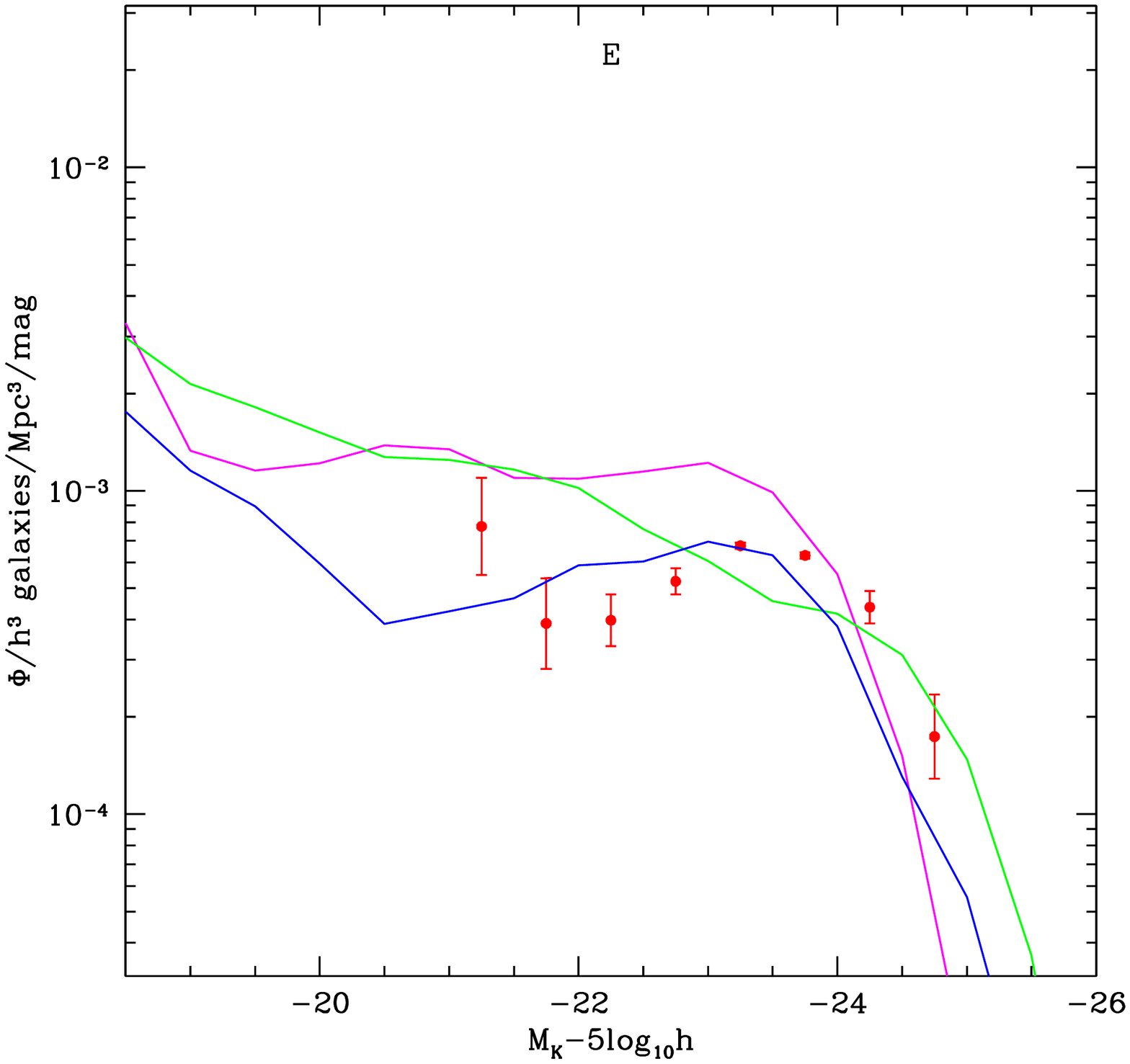} &
 \includegraphics[width=55mm,viewport=0mm 55mm 205mm 245mm,clip]{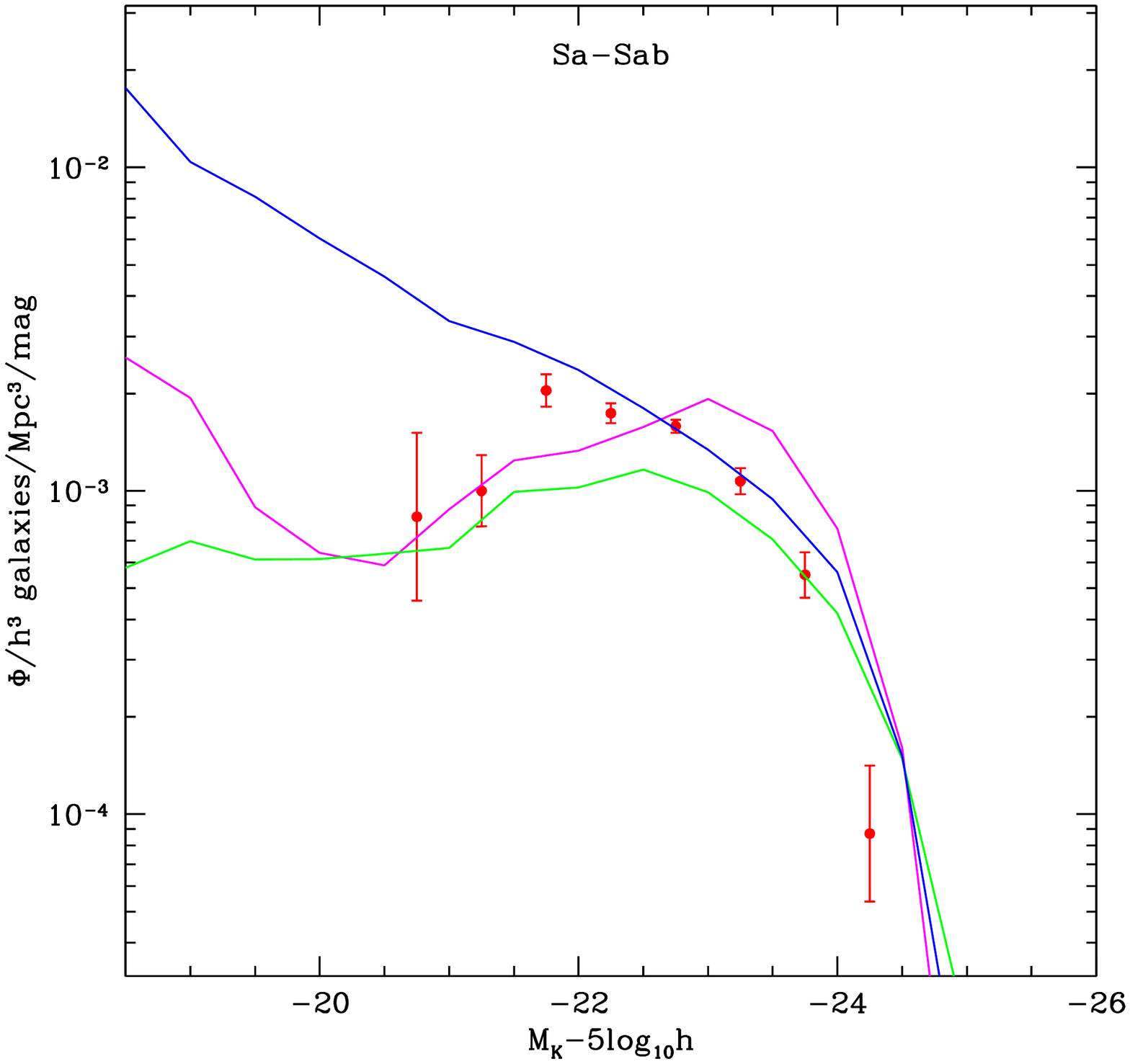} &
 \includegraphics[width=55mm,viewport=0mm 55mm 205mm 245mm,clip]{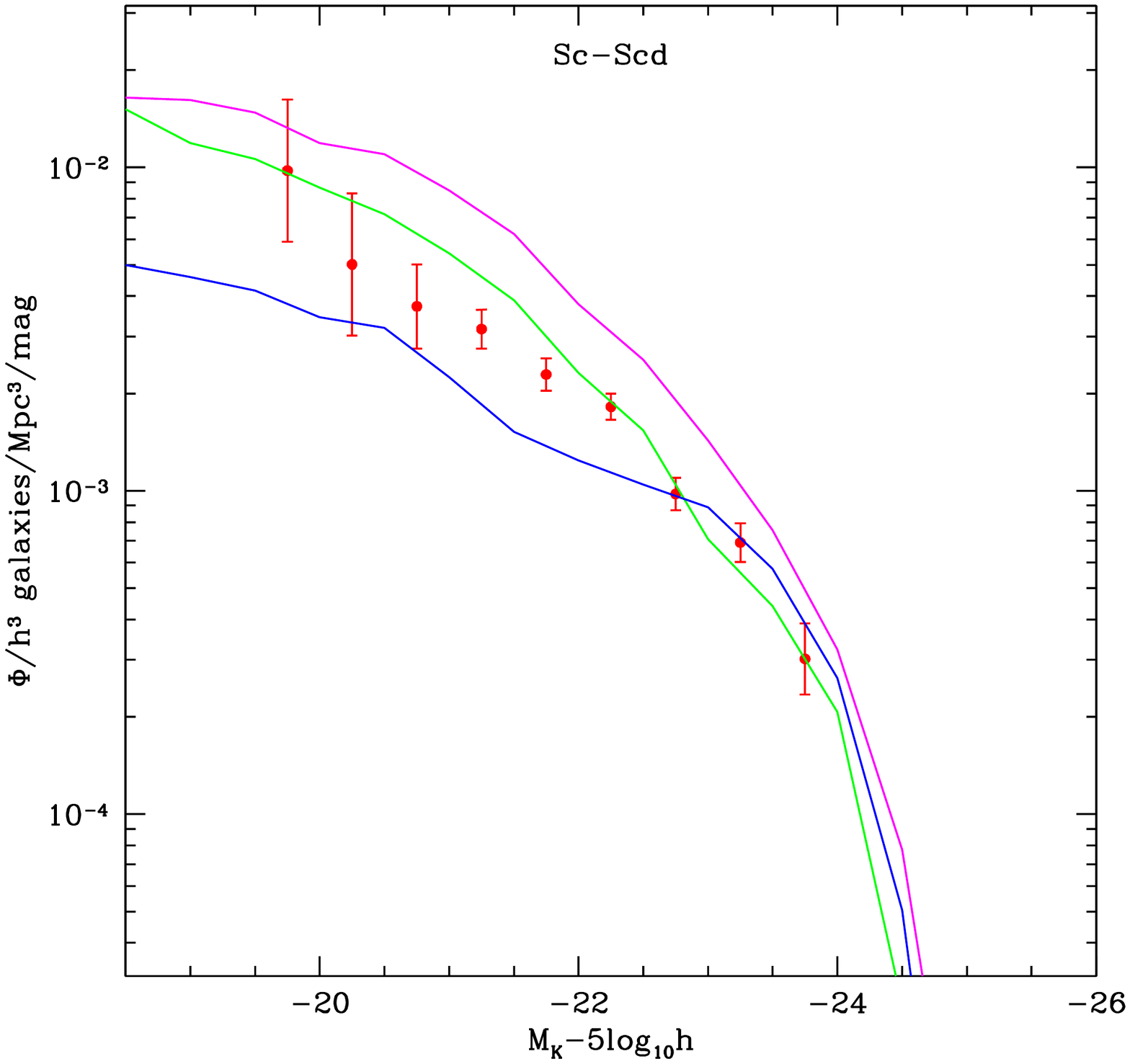}
 \end{tabular}
 \caption{The $z=0$ morphologically segregated K-band luminosity functions from our models. Points indicate the observed luminosity function from \protect\cite{devereux_morphological_2009} for morphological classes as indicated in each panel. Blue lines show the overall best-fit model, while magenta lines indicate the best-fit model to this dataset and the green lines show results from the \protect\cite{bower_breakinghierarchy_2006} model.}
 \label{fig:K_Morpho_LF}
\end{figure*}

Figure~\ref{fig:K20_Ks_LF} shows the K$_{\rm s}$-band luminosity function from the K20 survey \pcite{pozzetti_k20_2003} at $z=1.0$. (The data at $z=0.5$ and $1.5$ were used as constraints also.) The model traces the evolution of the luminosity function quite well but overpredicts the abundance at all redshifts. This is in contrast to the \cite{bower_breakinghierarchy_2006} model which matches these luminosity functions quite well. This is partly due to the tension between luminosity functions and the star formation rate density of Fig.~\ref{fig:SFH} which would be better fit if the model produced an even higher star formation rate density. This constraint forces our best-fit model to build up more stellar mass than the \cite{bower_breakinghierarchy_2006} model, consequently, to overpredict the abundance of galaxies at these redshifts. This tension between luminosity function and star formation rate constraints may in part be due to the difficulties involved with estimating the latter observationally (due to uncertainties in the \IMF\, calibration of star formation rate indicators and so on; see \cite{hopkins_normalization_2006} for a detailed examination of these issues). The best-fit model to this specific dataset successfully matches the data at all three redshifts. It achieves this through a combination of relatively high (i.e. less negative) $\alpha_\star$ and a high value of $\alpha_{\rm hot}$. Together, this combination allows for a flatter faint-end slope while maintaining the normalization of the bright end.

In addition to these luminosity functions that include all galaxy types, in Fig.~\ref{fig:K_Morpho_LF} we show the morphologically selected luminosity function of \cite{devereux_morphological_2009} overlaid with model results. We base morphological classification of model galaxies on \BT\ in dust-extinguished K-band light. We determine the mapping between \BT\ and morphology by requiring that the relative abundance of each type in the model agrees with the data in the interval $-23.5 < M_{\rm K}-5\log_{10}h \le -23.0$ but the morphological mix is not enforced outside this magnitude range. Our best-fit model reproduces the broad trends seen in this data---although we find that too many Sb-Sbc galaxies are produced at the highest luminosities. The \cite{bower_breakinghierarchy_2006} gives a better match to this data overall. The best fit to the particular dataset (magenta lines in Fig.~\ref{fig:K_Morpho_LF}) has a relatively large value of $f_{\rm ellip}$, but is not significantly better than our best-fit model.

\begin{figure}
 \includegraphics[width=80mm,viewport=0mm 55mm 205mm 245mm,clip]{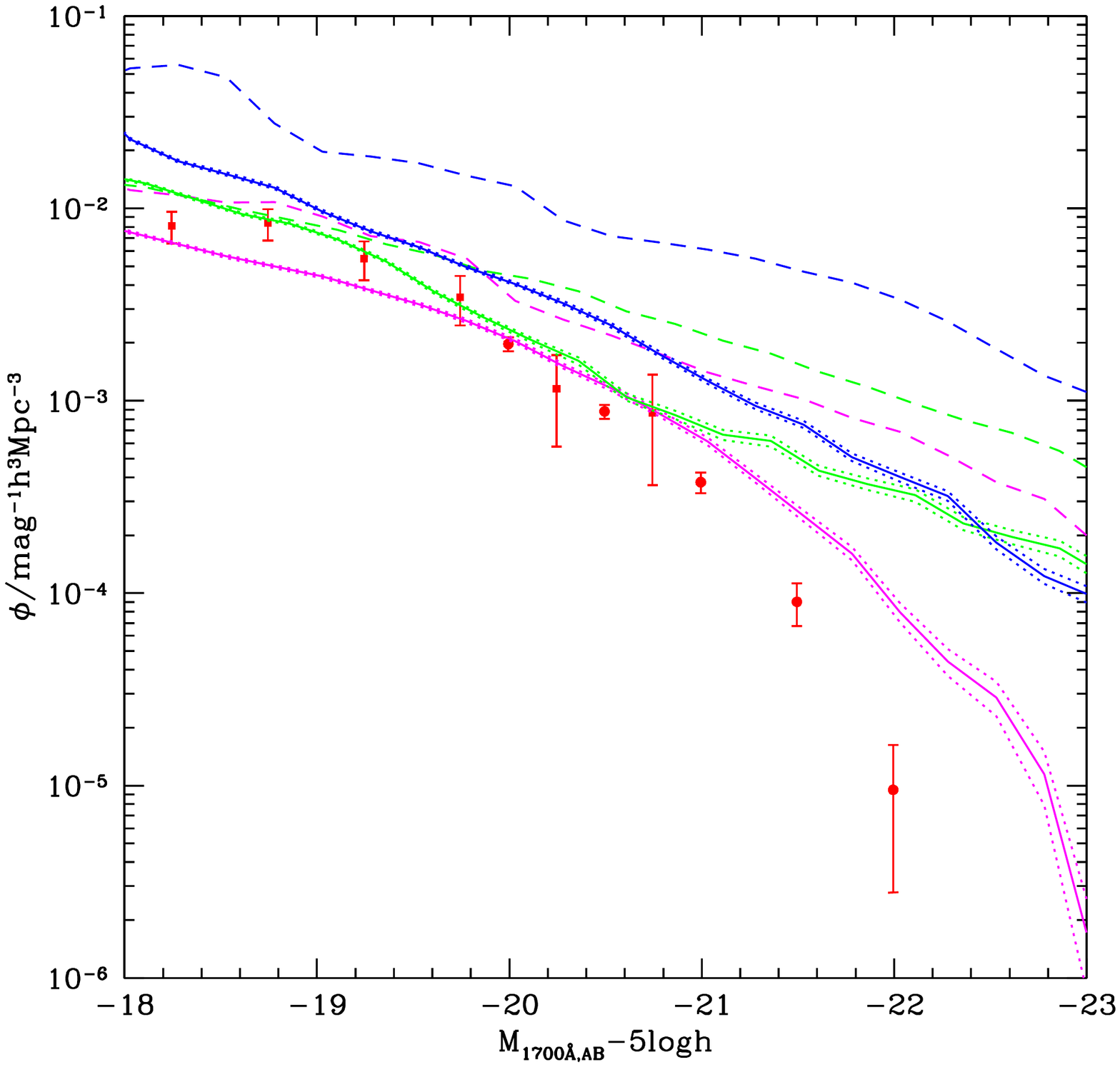}
 \caption{The $z=3$ 1700\AA\ luminosity functions from our models are shown by the solid lines with dotted lines showing the statistical uncertainty on the model estimates. The dashed lines indicate the luminosity function when the effects of dust extinction are neglected. Red points indicate the observed luminosity function from \protect\citeauthor{steidel_lyman-break_1999}~(\protect\citeyear{steidel_lyman-break_1999}; circles) and \protect\citeauthor{dickinson_color-selected_1998}~(\protect\citeyear{dickinson_color-selected_1998}; squares). Blue lines show the overall best-fit model, while magenta lines indicate the best-fit model to this dataset and the green lines show results from the \protect\cite{bower_breakinghierarchy_2006} model.}
\label{fig:LyBreakLF}
\end{figure}

In addition to these relatively low redshift constraints we are particularly interested here in examining constraints from the highest redshifts currently observable. Therefore, Fig.~\ref{fig:LyBreakLF} shows the luminosity function of $z\approx 3$ Lyman-break galaxies together with the expectation from our best fit model (blue line). Model galaxies are drawn from the entire sample of galaxies at $z=3$ found in the model. The model significantly overpredicts the number of luminous galaxies even when internal dust extinction is taken into account (the dashed line in Fig.~\ref{fig:LyBreakLF} shows the luminosity function without the effects of dust extinction). The \cite{bower_breakinghierarchy_2006} model gives a similarly bad match to this data at the bright-end (although is slightly better at low luminosities), producing too many highly luminous galaxies. The best-fit model to this specific dataset turns out to be not such a good fit, although it is better than either of the other models shown. The problem here is one of noise. The models run for our parameter space search utilized relatively small numbers of merger tree realizations (to permit them to run in a reasonable amount of time). In this particular case, the model run during the parameter space search looked like a good match to the $z\approx 3$ Lyman-break galaxy luminosity function, but, when re-run with many more merger trees, it turned out that the apparently good fit was partly a result of fortuitous noise. This luminosity function is particular sensitive to such effects, as the bright end is dominated by rare starburst galaxies.

\begin{figure}
 \begin{tabular}{cc}
 \includegraphics[width=80mm,viewport=0mm 55mm 200mm 245mm,clip]{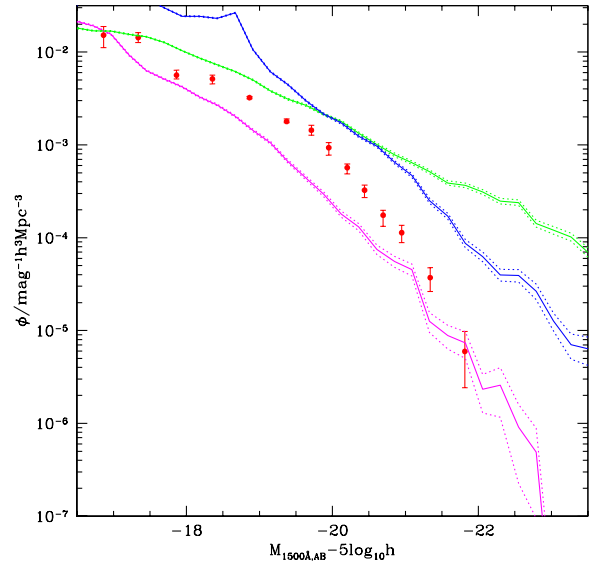}
 \end{tabular}
 \caption{The $z=5$ rest-frame 1500\AA\ luminosity function from our models are shown by the solid lines, with statistical errors indicated by the dotted lines. Red points indicate data from \protect\cite{mclure_luminosity_2009}. Blue lines show the overall best-fit model, while magenta lines indicate the best-fit model to this dataset and the green lines show results from the \protect\cite{bower_breakinghierarchy_2006} model.}
\label{fig:z5_6_LF}
\end{figure}

Finally, at the highest redshifts for which we presently have statistically useful data, Fig~\ref{fig:z5_6_LF} shows rest-frame \UV\ luminosity function at $z=5$ from \cite{mclure_luminosity_2009}. These highest redshift luminosity functions in principle place a strong constraint on the model. However, the effects of dust become extremely important at these short wavelengths and so our model predictions are less reliable. As such, these constraints are less fundamental than most of the others which we consider. We use our more detailed dust modelling for the \cite{bower_breakinghierarchy_2006} model here even though the original \cite{bower_breakinghierarchy_2006} used the simpler dust model of \cite{cole_hierarchical_2000}. However, as noted in \protect\S\ref{sec:DustModel}, in our current model we ensure that high-$z$ galaxies which are undergoing near continuous instability driven bursting are observed during the dust phase of the burst. In the \cite{bower_breakinghierarchy_2006} model shown here this is not the case---such systems are almost always observed in a gas and dust free state, making them appear much brighter. It is clear that the treatment of these galaxies in terms of punctuated equilibrium of disks is inadequate and we will return to this issue in more detail in a future work.

The best fit model again overpredicts the number and/or luminosities of galaxies at these redshifts. The \cite{bower_breakinghierarchy_2006} model performs much worse here however---drastically overpredicting the number of luminous galaxies. The majority of this difference is due to the treatment of dust in bursts in our current model. Additionally, however, this difference simply reflects the fact that high-$z$ constraints were not considered when selecting the parameters of the \cite{bower_breakinghierarchy_2006} model---the improved agreement here illustrates the benefits of considering a wide range of datasets when constraining model parameters. The best fit model to these specific datasets shows a steeper decline at high luminosities and a lower normalization over all luminosities. Once again, the best fit here is not particularly good, for the same reasons that the $z=3$ \UV\ luminosity function is not too well fit (i.e. that the models run to search parameter space use relatively few merger trees, leading to significant noise in these luminosity functions which depend on galaxies that form in rare halos). This is achieved through a combination of strong feedback (i.e. high $V_{\rm hot,disk}$) and highly efficient star formation with a very strong dependence on galaxy circular velocity. However, this achieves only a relatively small improvement over the overall best fit model, at the expense of significantly worse fits to other datasets.

\subsection{Colours}\label{sec:Colours}

The bimodality of the galaxy colour-magnitude diagram has long been understood to convey important information regarding the evolutionary history of different types of galaxy. Recently, semi-analytic models have paid close attention to this diagnostic \pcite{croton_many_2006,bower_breakinghierarchy_2006}. In particular, \cite{font_colours_2008} found that the inclusion of detailed modelling of ram pressure stripping of hot gas from satellite galaxy halos is crucial for obtaining an accurate determination of the colour-magnitude relation. That same model of ram pressure stripping is included in the present work.

\begin{figure*}
 \begin{tabular}{ccc}
   $-22 < ^{0.1}M_{\rm g}-5\log h \le -21$ &  $-20 < ^{0.1}M_{\rm g}-5\log h \le -19$ & $-18 < ^{0.1}M_{\rm g}-5\log h \le -17$ \\
   \includegraphics[width=55mm,viewport=0mm 50mm 200mm 245mm,clip]{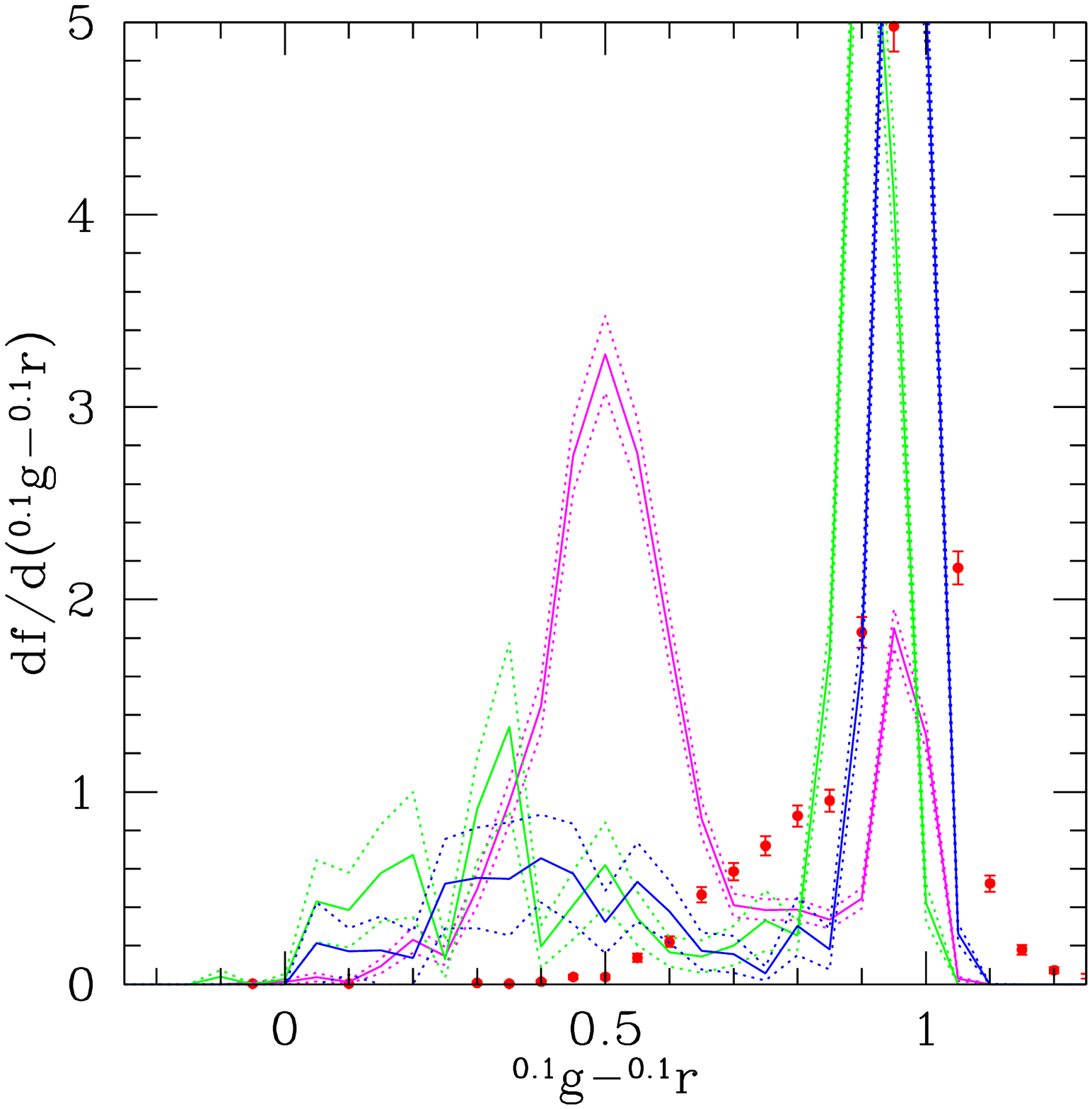} &
   \includegraphics[width=55mm,viewport=0mm 50mm 200mm 245mm,clip]{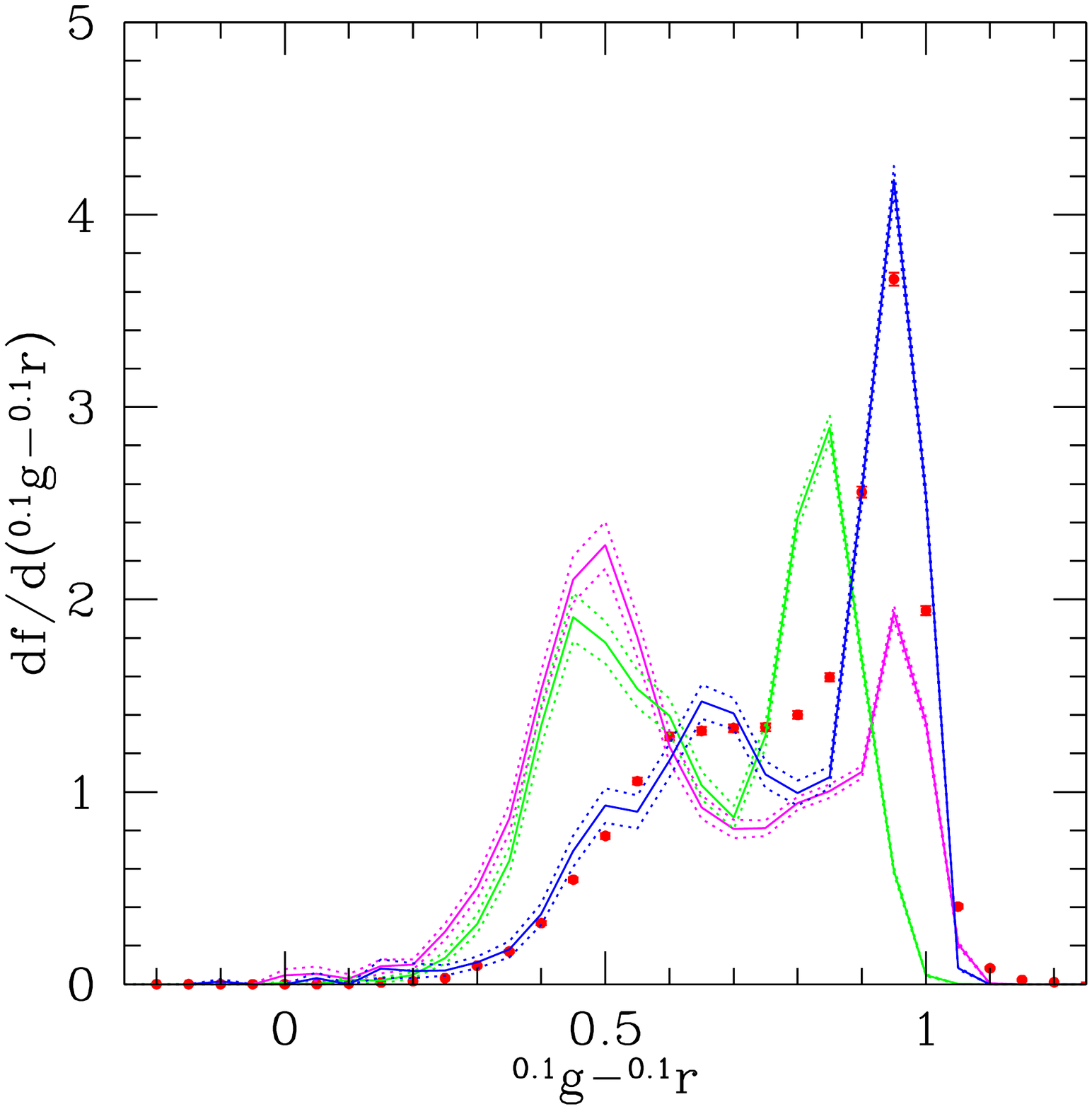} &
   \includegraphics[width=55mm,viewport=0mm 50mm 200mm 245mm,clip]{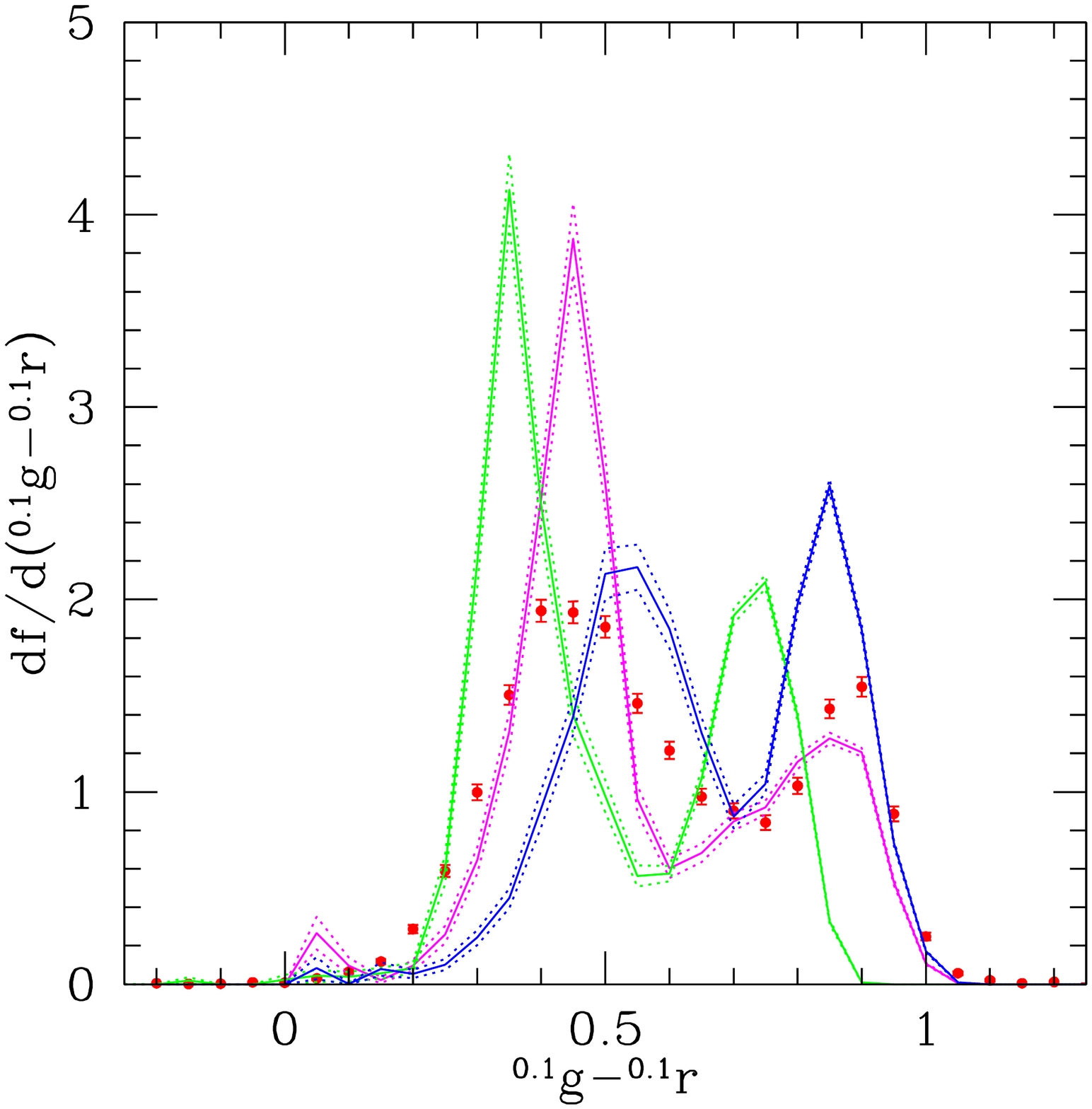}
 \end{tabular}
 \caption{$^{0.1}$g$-^{0.1}$r colour distributions for galaxies at $z=0.1$ split by g-band absolute magnitude (see above each panel for magnitude range). Solid lines indicate the distributions from our models while the red points show data from the \protect\SDSS\ \protect\pcite{weinmann_properties_2006}. Blue lines show the overall best-fit model, while magenta lines indicate the best-fit model to this dataset and the green lines show results from the \protect\cite{bower_breakinghierarchy_2006} model. Note that the magenta model is selected on the basis on more panels than are shown here.}
 \label{fig:SDSS_Colours}
\end{figure*}

Figure~\ref{fig:SDSS_Colours} shows slices of constant magnitude through the colour magnitude diagram of \cite{weinmann_properties_2006}, overlaid with results from our model. The model is very successful in matching these data, showing that at bright magnitudes the red galaxy component dominates, shifting to a mix of red and blue galaxies at fainter magnitudes. The median colours of the blue and red components of the galaxy population are reproduced better in our current model than by that of \cite{bower_breakinghierarchy_2006}, although there is clearly an offset in the blue cloud at faint magnitudes (model galaxies in the blue cloud are slightly too red). Our model reproduces the colours of galaxies reasonably well, so this offset may be partly due to the limitations of stellar population synthesis models. This problem with the \cite{bower_breakinghierarchy_2006} model was noted by \cite{font_colours_2008} who demonstrated that a combination of a higher yield of $p=0.04$ in the instantaneous recycling approximation (\cite{bower_breakinghierarchy_2006} assumed a yield of $p=0.02$) and ram pressure stripping of cold gas in galaxy disks lead to a much better match to galaxy colours. The yield is not a free parameter in our model, instead it is determined from the \IMF\ and stellar metal yields directly (see Fig.~\ref{fig:Chabrier_NonInstant}), potentially rising as high as $p=0.04$ after several Gyr. This is very close to the value adopted by \cite{font_colours_2008}, and our model is able to produce a good match to the colours. As we will see later (in \S\ref{sec:GasMetals}), the \cite{bower_breakinghierarchy_2006} model has more serious problems with galaxy metallicities which are somewhat rectified in our present model thereby helping us obtain a better match to the galaxy colours. The best-fit model to this specific dataset is a better match than our overall best-fit model for fainter galaxies, although it performs less well at brighter magnitudes. At faint magnitudes it produces a bluer blue-cloud which better matches that which is observed. It achieves this success by having a much larger value (i.e. less negative) of $\alpha_\star$. This parameter controls how star formation rates scale with galaxy mass, with this model having less dependence than any other. This improves the match to galaxy colours (at the expense of steepening the faint end slope of the luminosity function), particularly for fainter galaxies.

\subsection{Scaling Relations}\label{sec:TF}

\begin{figure*}
 \begin{tabular}{cccc}
  $-20 \le M_{i} < -19$ & $-21 \le M_{i} < -20$ & $-22 \le M_{i} < -21$ & $-23 \le M_{i} < -22$\vspace{-6mm} \\
  \includegraphics[width=40mm,viewport=0mm 55mm 205mm 245mm,clip]{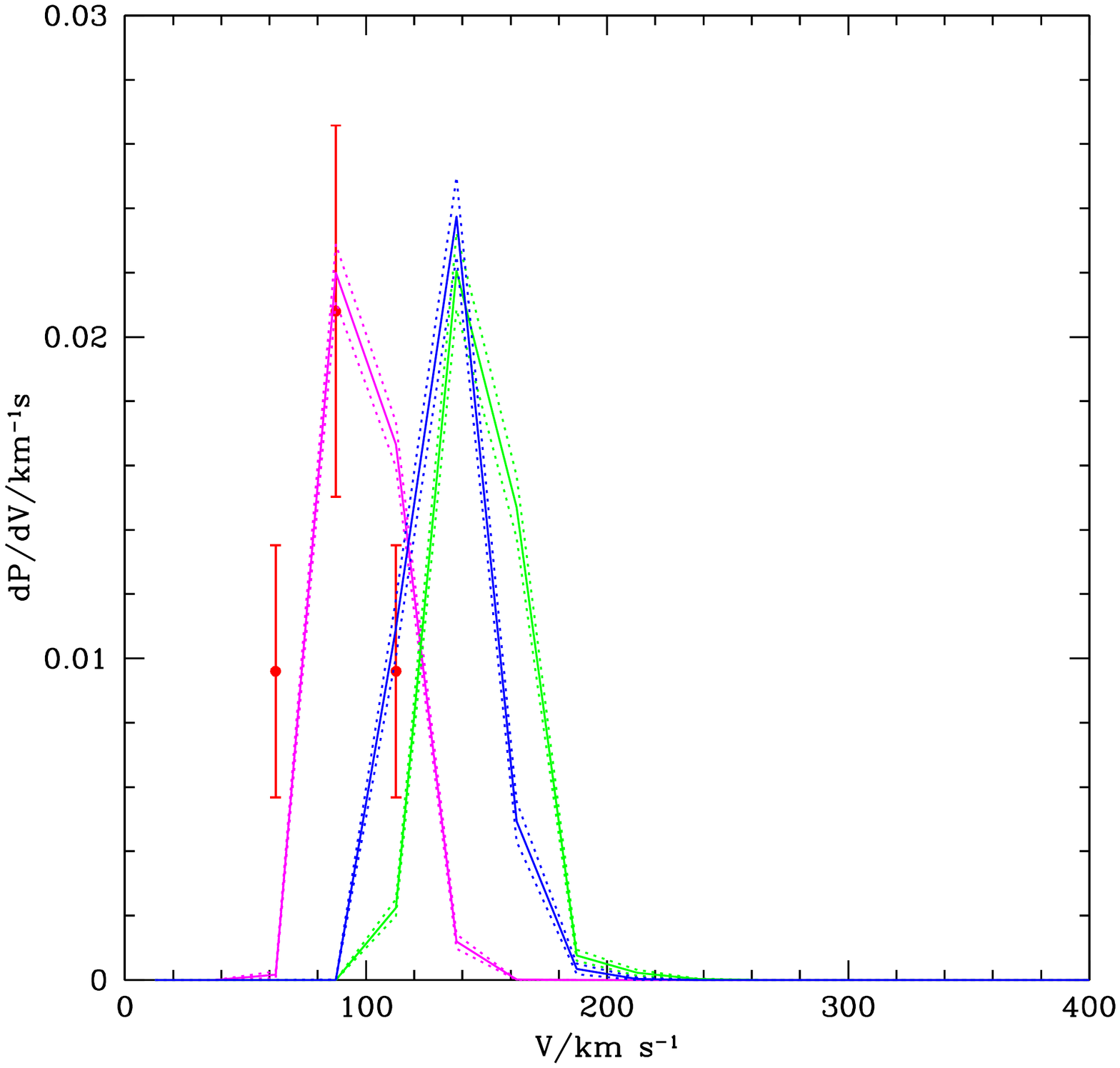} &
 \includegraphics[width=40mm,viewport=0mm 55mm 205mm 245mm,clip]{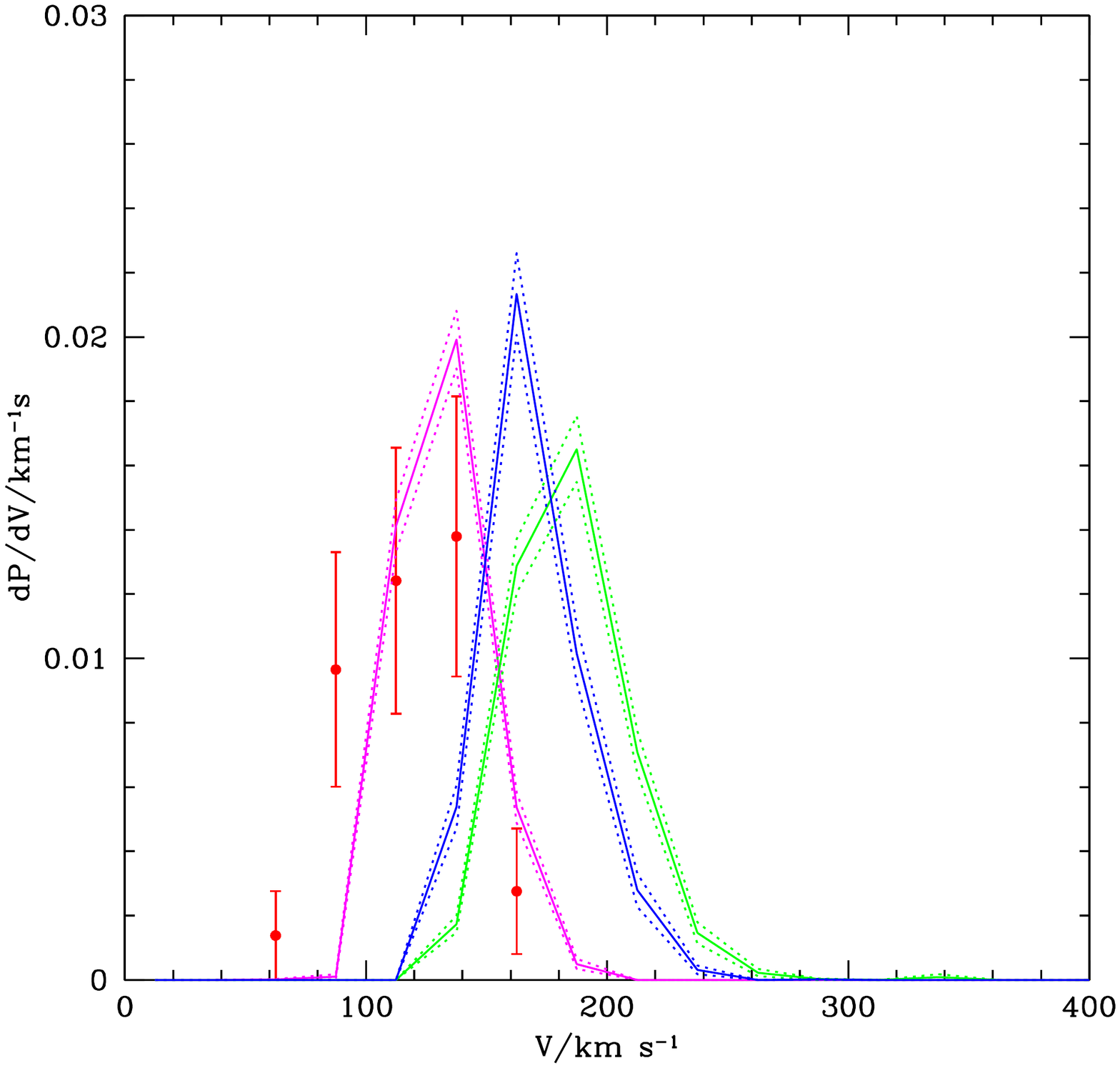} &
  \includegraphics[width=40mm,viewport=0mm 55mm 205mm 245mm,clip]{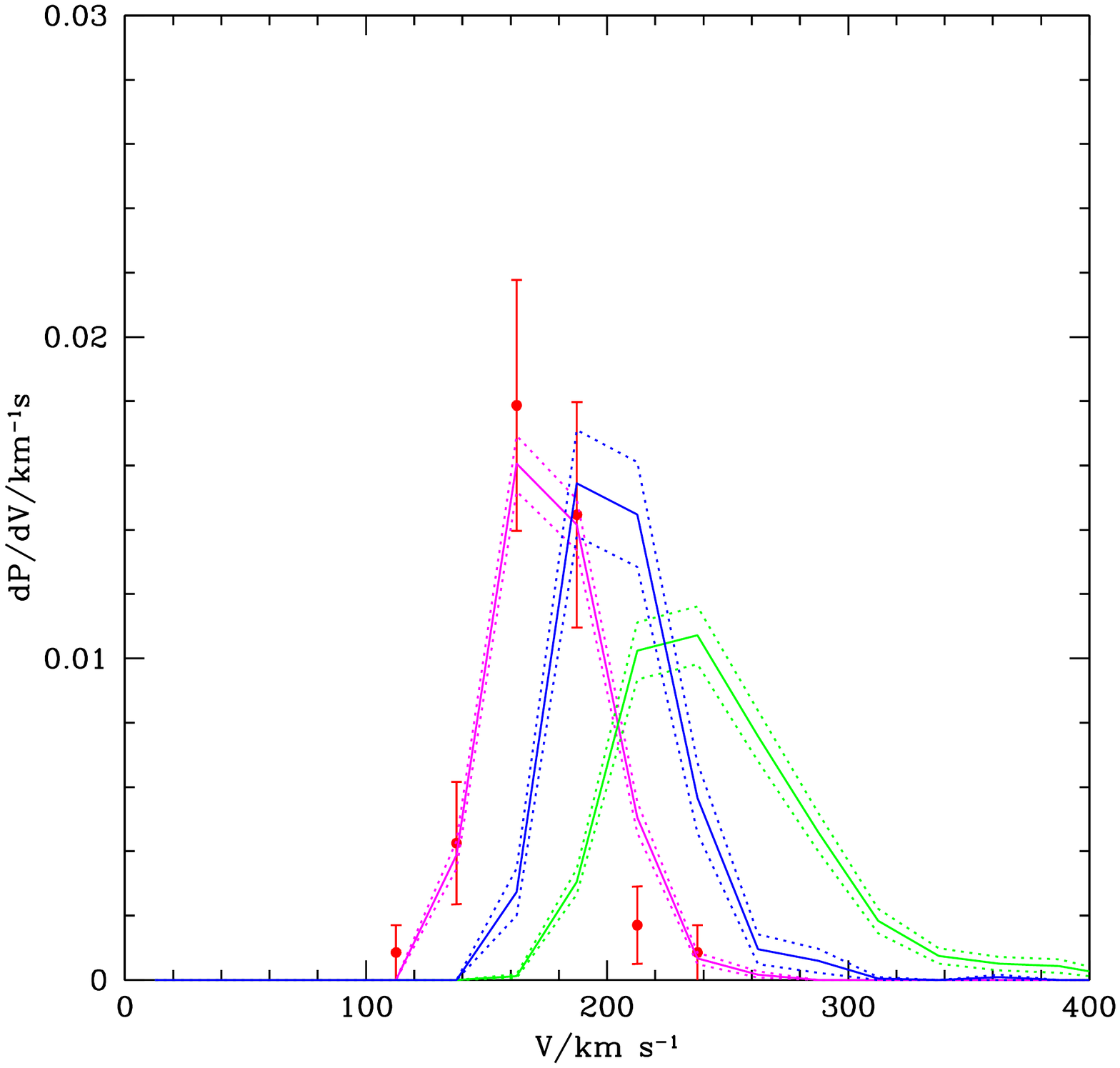} &
 \includegraphics[width=40mm,viewport=0mm 55mm 205mm 245mm,clip]{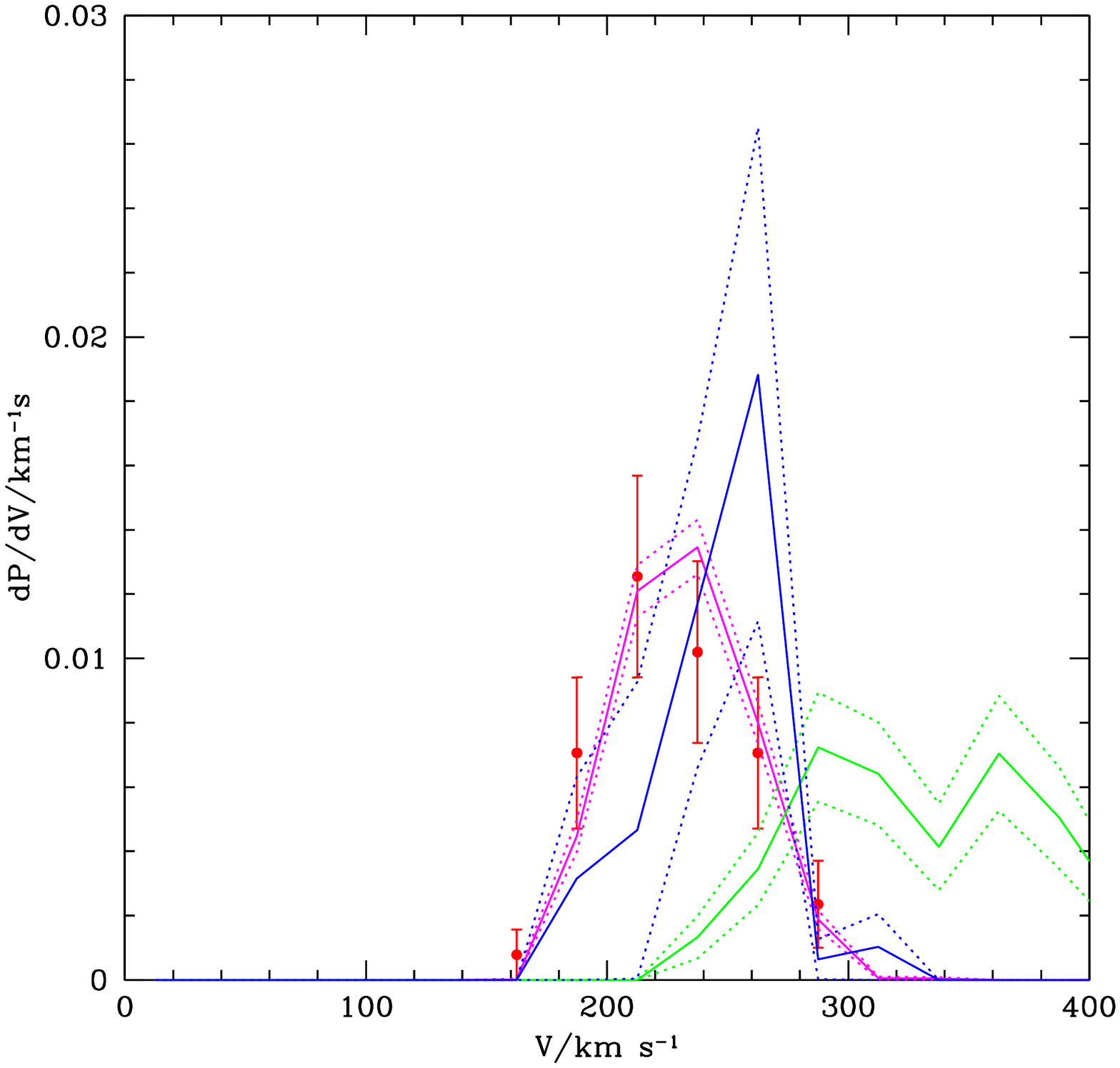}
 \end{tabular}
 \caption{Slices though the i-band Tully-Fisher relation from the \protect\SDSS\ \protect\pcite{pizagno_tully-fisher_2007} at constant absolute magnitude are shown by red points. Solid lines show results from our models with dotted lines indicating the statistical error on the model estimate. Blue lines show the overall best-fit model, while magenta lines indicate the best-fit model to this dataset and the green lines show results from the \protect\cite{bower_breakinghierarchy_2006} model.}
 \label{fig:SDSS_TF}
\end{figure*}

Fitting the Tully-Fisher relation simultaneously with the luminosity function has been a long-standing challenge for models of galaxy formation (see \cite{dutton_tully-fisher_2008} and references therein). Figure~\ref{fig:SDSS_TF} shows the Tully-Fisher relation from the \SDSS\ as measured by \cite{pizagno_tully-fisher_2007} together with the result from out best-fit model. The model is in reasonable agreement with zero point, although somewhat offset to higher velocities, and in good agreement with the luminosity dependence and width of the Tully-Fisher relation. Our new model is a significantly better match to the Tully-Fisher relation than that of \cite{bower_breakinghierarchy_2006}, which produces galaxies with rotation speeds that are systematically too large (particularly for the brightest galaxies). For example, for the most luminous galaxies shown the \cite{bower_breakinghierarchy_2006} predicts a population of galaxies with circular velocities of 300--400km/s or greater---strongly ruled out by observations. The new model on the other hand predicts essentially no galaxies in this velocity range. The best-fit model to this particular dataset is a significantly better match than our overall best-fit model. No single parameter is responsible for the improvement, but $\lambda_{\rm expel,burst}$ plays an important role---it is much lower in the best-fit model to the Tully-Fisher data.

\subsection{Sizes}

\begin{figure*}
 \begin{tabular}{ccc}
$-18.5 > \hspace{1mm}^{0.1}\hspace{-1mm}M_{\rm r}-5\log h \ge -18.0$ &
$-20.5 > \hspace{1mm}^{0.1}\hspace{-1mm}M_{\rm r}-5\log h \ge -21.0$ &
$-22.5 > \hspace{1mm}^{0.1}\hspace{-1mm}M_{\rm r}-5\log h \ge -23.0$ \\
   \includegraphics[width=55mm,viewport=0mm 5mm 200mm 270mm,clip]{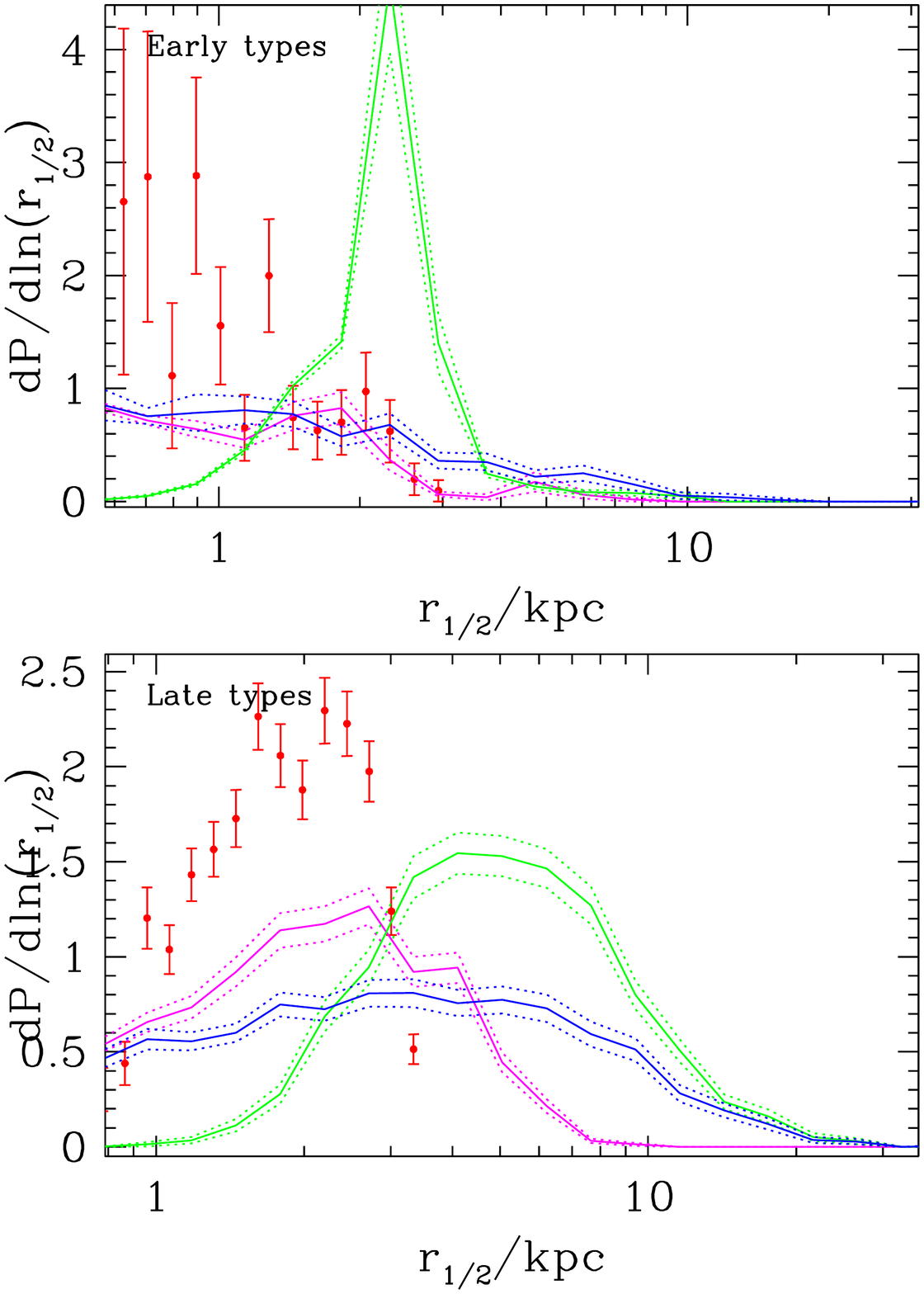} &
   \includegraphics[width=55mm,viewport=0mm 5mm 200mm 270mm,clip]{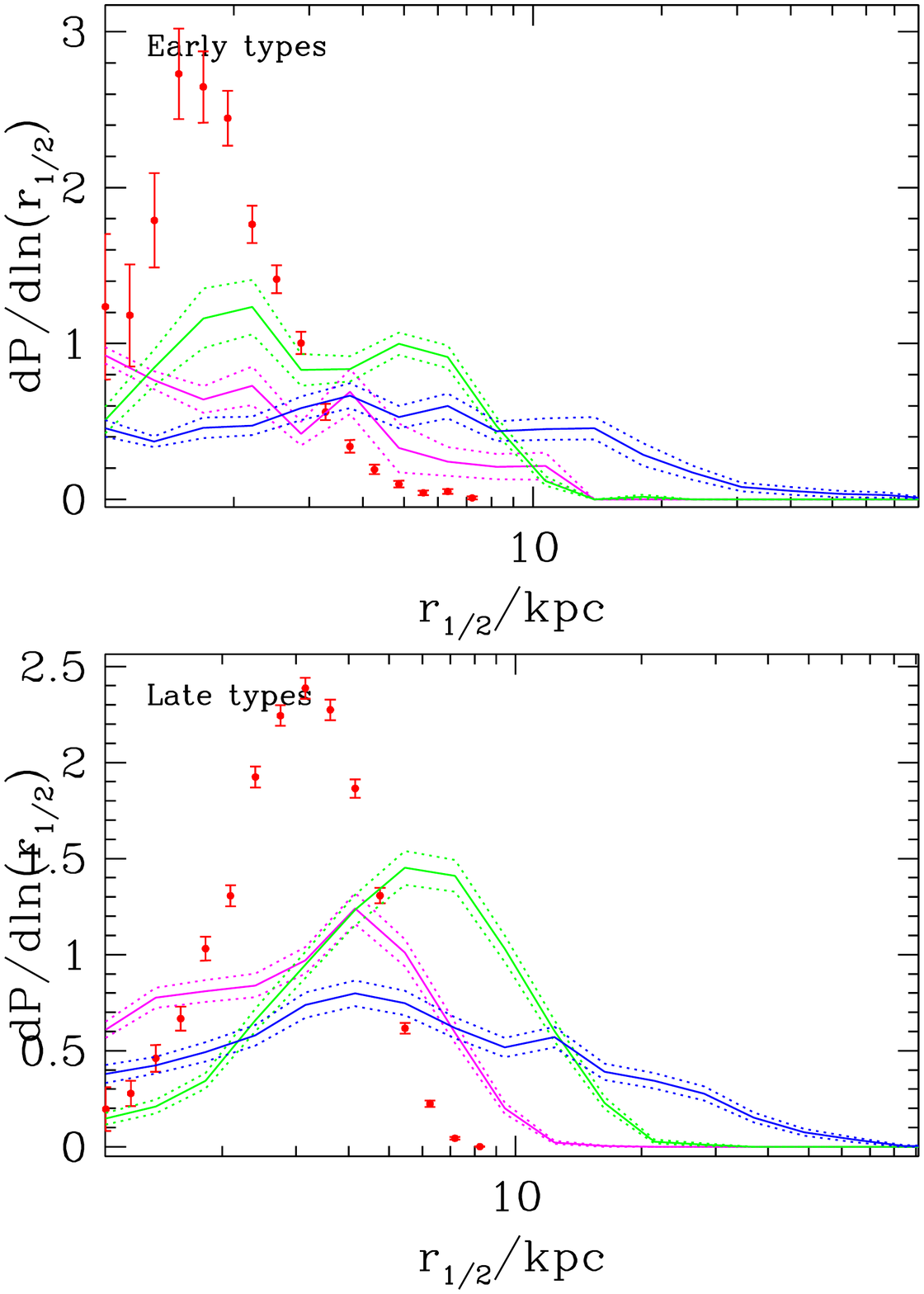} &
   \includegraphics[width=55mm,viewport=0mm 5mm 200mm 270mm,clip]{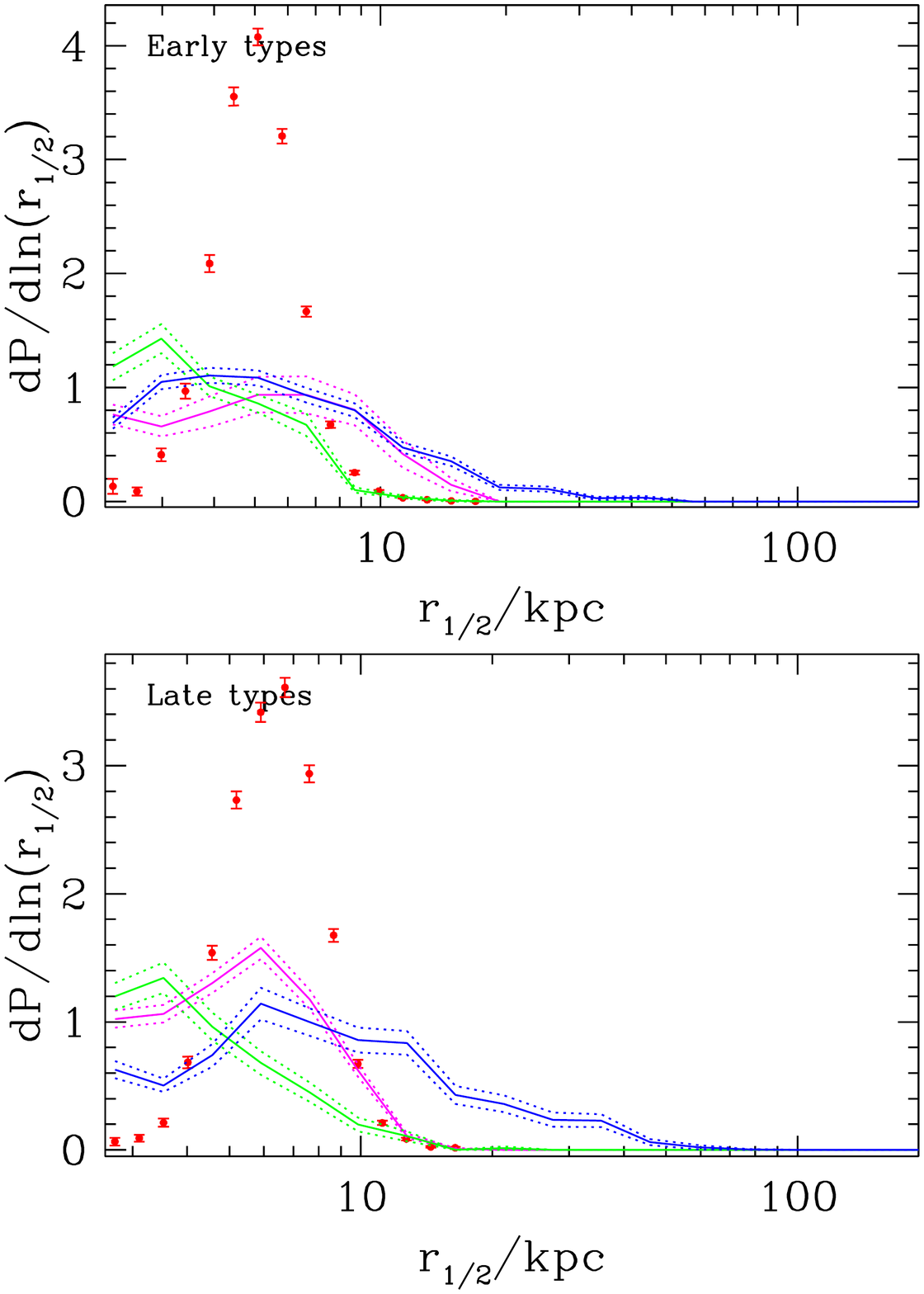} \\
 \end{tabular}
 \caption{Distributions of galaxy half-light radii (measured in the dust-extinguished face-on r-band light profile) at $z=0.1$ segregated by r-band absolute magnitude and by morphological class. Solid lines show results from our models while dotted lines show the statistical error on the model estimates. Red points data from the \protect\SDSS\ \protect\pcite{shen_size_2003}. Blue lines show the overall best-fit model, while magenta lines indicate the best-fit model to this dataset and the green lines show results from the \protect\cite{bower_breakinghierarchy_2006} model.}
 \label{fig:SDSS_Sizes}
\end{figure*}

Figure~\ref{fig:SDSS_Sizes} shows the distribution of galaxy sizes, split by morphological type and magnitude, from the \SDSS\ \pcite{shen_size_2003}. To morphologically classify model galaxies we utilize the bulge-to-total ratio in dust-extinguished $^{0.1}$r-band light. From the K-band morphologically segregated luminosity function (see \S\ref{sec:z0LFResults}) we find that E and S0 galaxies are those with B/T$>0.714$ for the best fit model. There is no convincing reason to expect this value to correspond precisely to the morphological selection used by \cite{shen_size_2003}, but it is currently our best method to choose a division between early and late types in our model. For simplicity, we employ the same morphological cut for all three models plotted in Fig.~\ref{fig:SDSS_Sizes}. Model results are overlaid as lines. Model galaxies are too large compared to the data, by factors of about two, and the distribution of model galaxy sizes is too broad. This problem is more significant for the fainter galaxies.

\begin{figure*}
 \begin{tabular}{ccc}
  \includegraphics[width=55mm,viewport=0mm 45mm 200mm 255mm,clip]{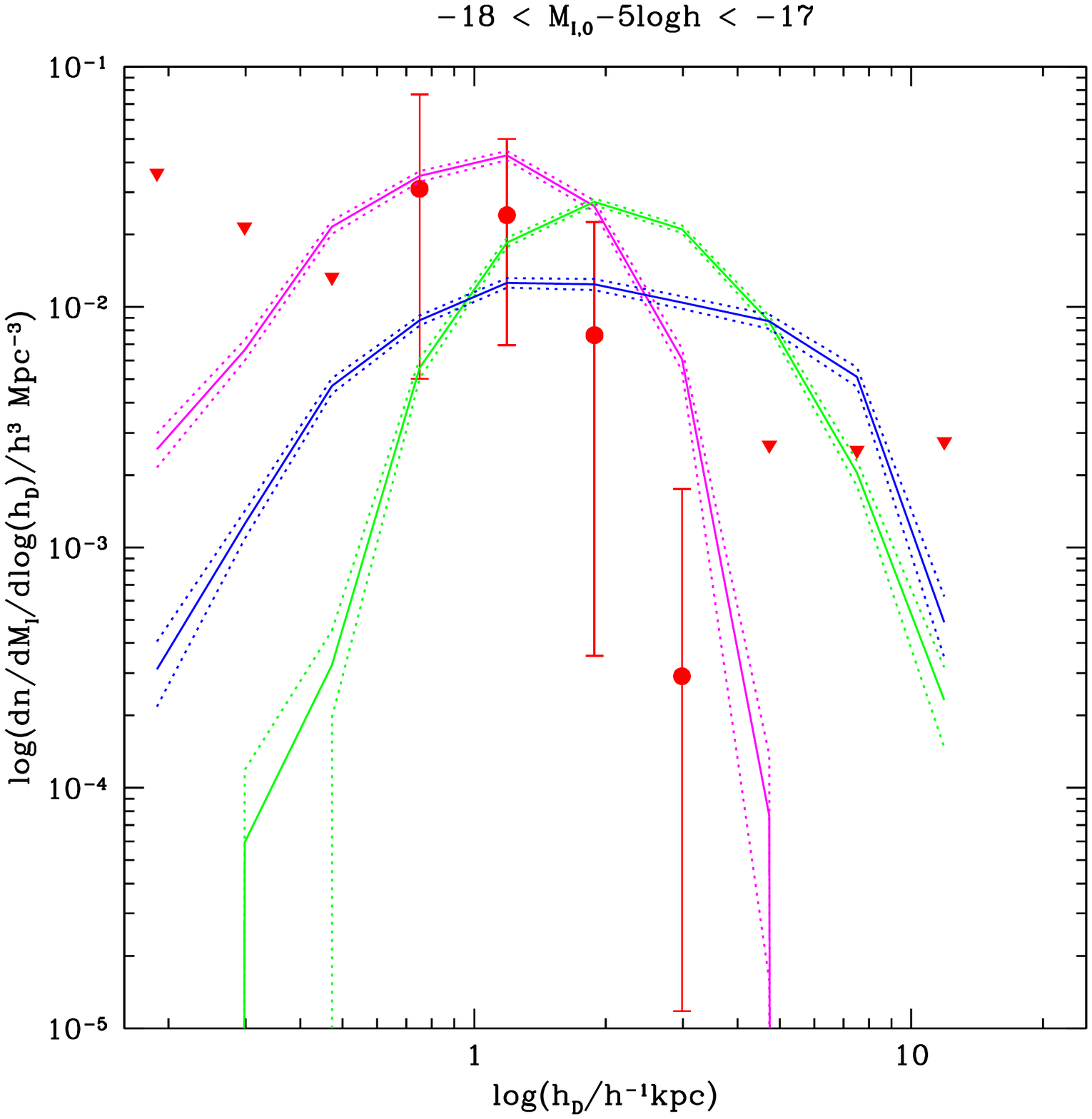} &
  \includegraphics[width=55mm,viewport=0mm 45mm 200mm 255mm,clip]{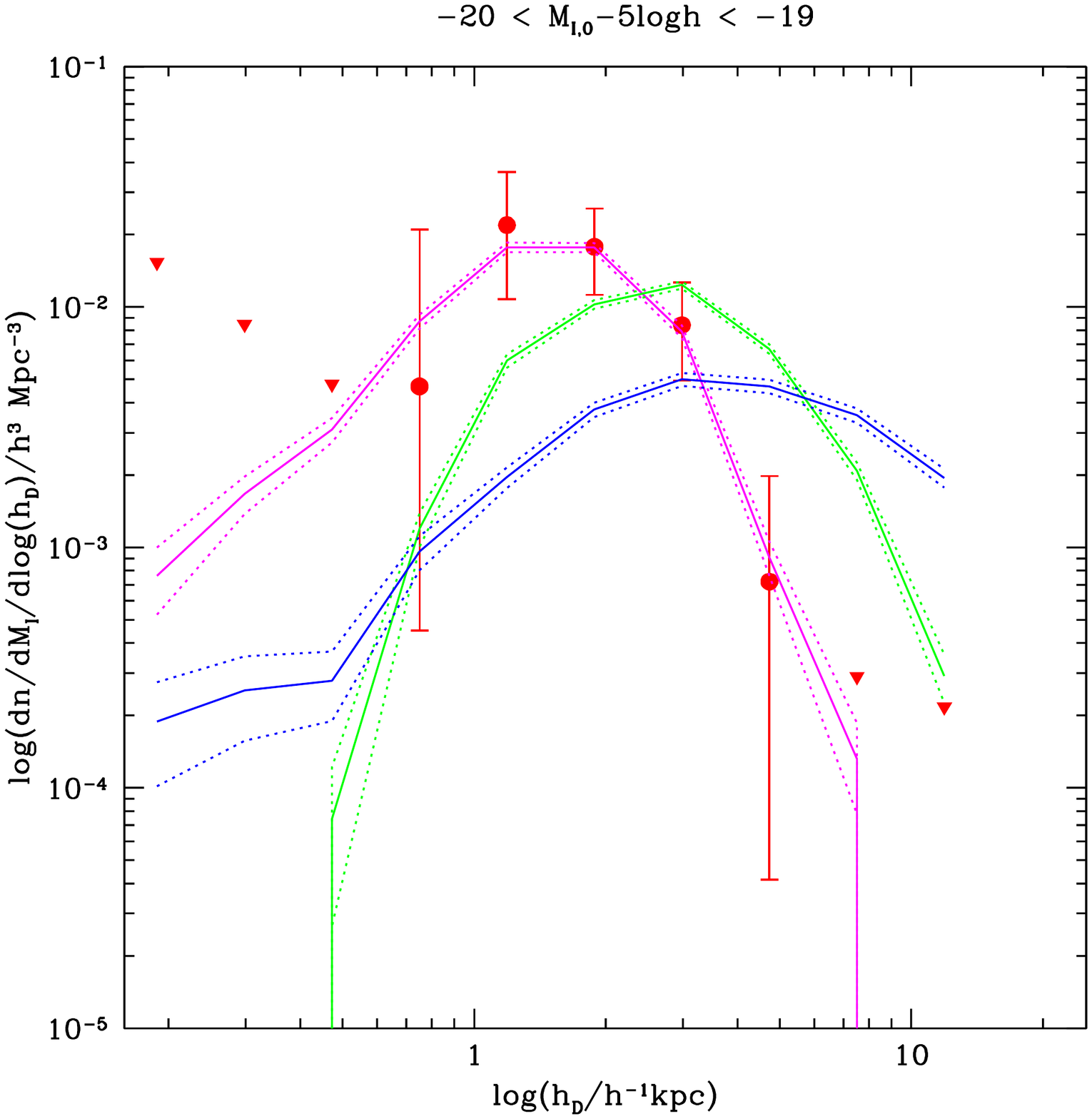} &
  \includegraphics[width=55mm,viewport=0mm 45mm 200mm 255mm,clip]{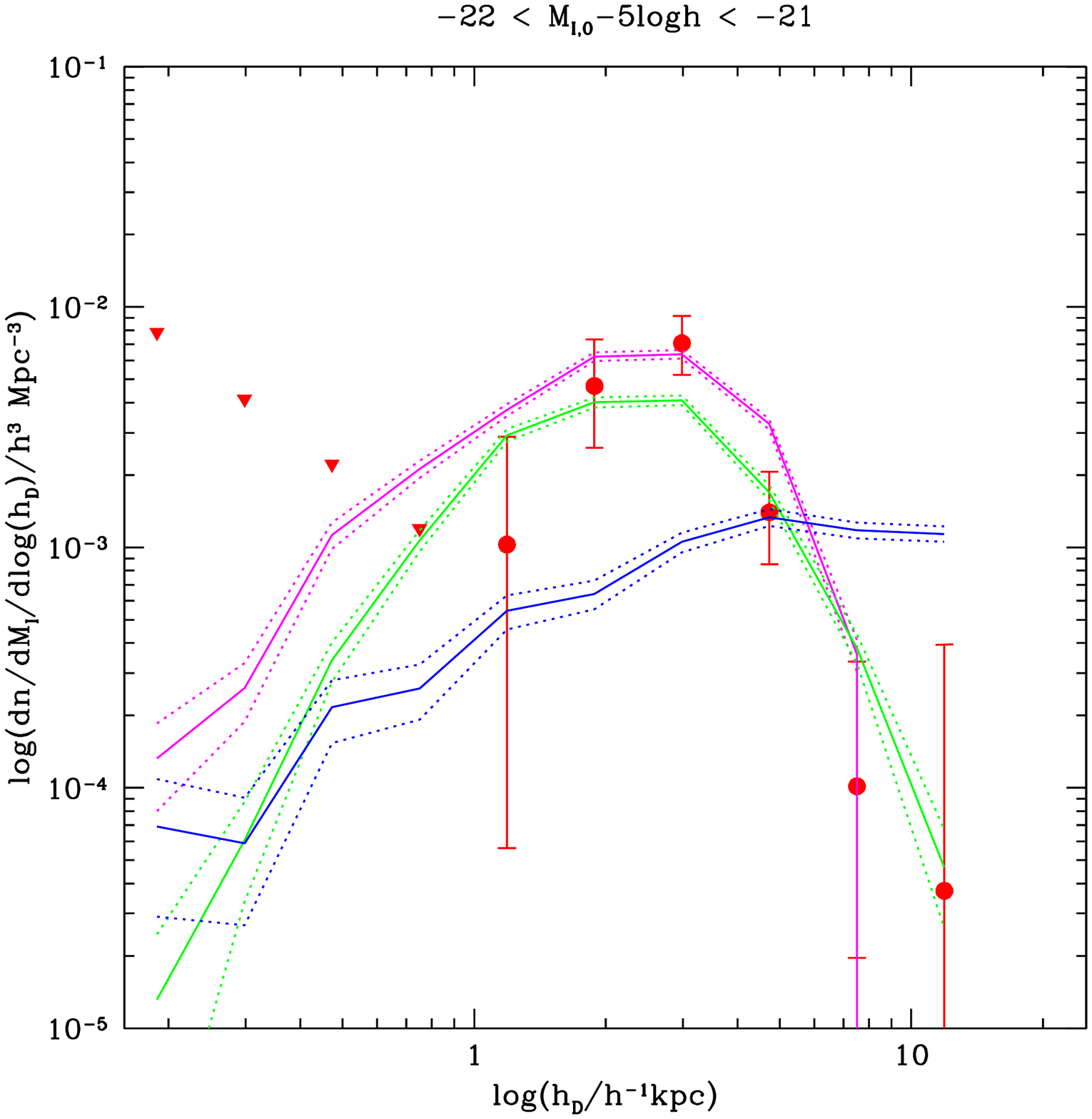}
 \end{tabular}
 \caption{Distribution of disk scale lengths for galaxies at $z=0$ segregated by face-on I-band absolute magnitude. Solid lines show results from our models while dotted lines indicate the statical uncertainty on the model estimates. Red circles show data from \protect\cite{de_jong_local_2000} with upper limits indicated by red triangles. Blue lines show the overall best-fit model, while magenta lines indicate the best-fit model to this dataset and the green lines show results from the \protect\cite{bower_breakinghierarchy_2006} model.}
 \label{fig:Other_Sizes}
\end{figure*}

Figure~\ref{fig:Other_Sizes} shows the distribution of disk sizes from \cite{de_jong_local_2000} with model results overlaid as lines. This permits a more careful comparison with the model as it does not require us to assign morphological types to model galaxies. Model disks are somewhat too large in all luminosity bins considered, and the width of the distribution of disk sizes is broader than that observed.

The \cite{bower_breakinghierarchy_2006} model produces galaxies which are systematically smaller than those in our current best-fit model at bright magnitudes, but larger at faint magnitudes. It also produces a narrower distribution of disk sizes. Our best fit model to these combined size datasets is a rather poor match to the distribution of disk sizes. We find that it is challenging to obtain realistic sizes for disks in our model while simultaneously matching other observational constraints. This problem, which may reflect inaccuracies in the angular momentum of cooling gas, angular momentum loss during cooling or merging, or internal processes which transfer angular momentum out of galaxies, will be addressed in greater detail in a future work.

\subsection{Gas and Metal Content}\label{sec:GasMetals}

\begin{figure}
 \includegraphics[width=80mm,viewport=0mm 10mm 195mm 265mm,clip]{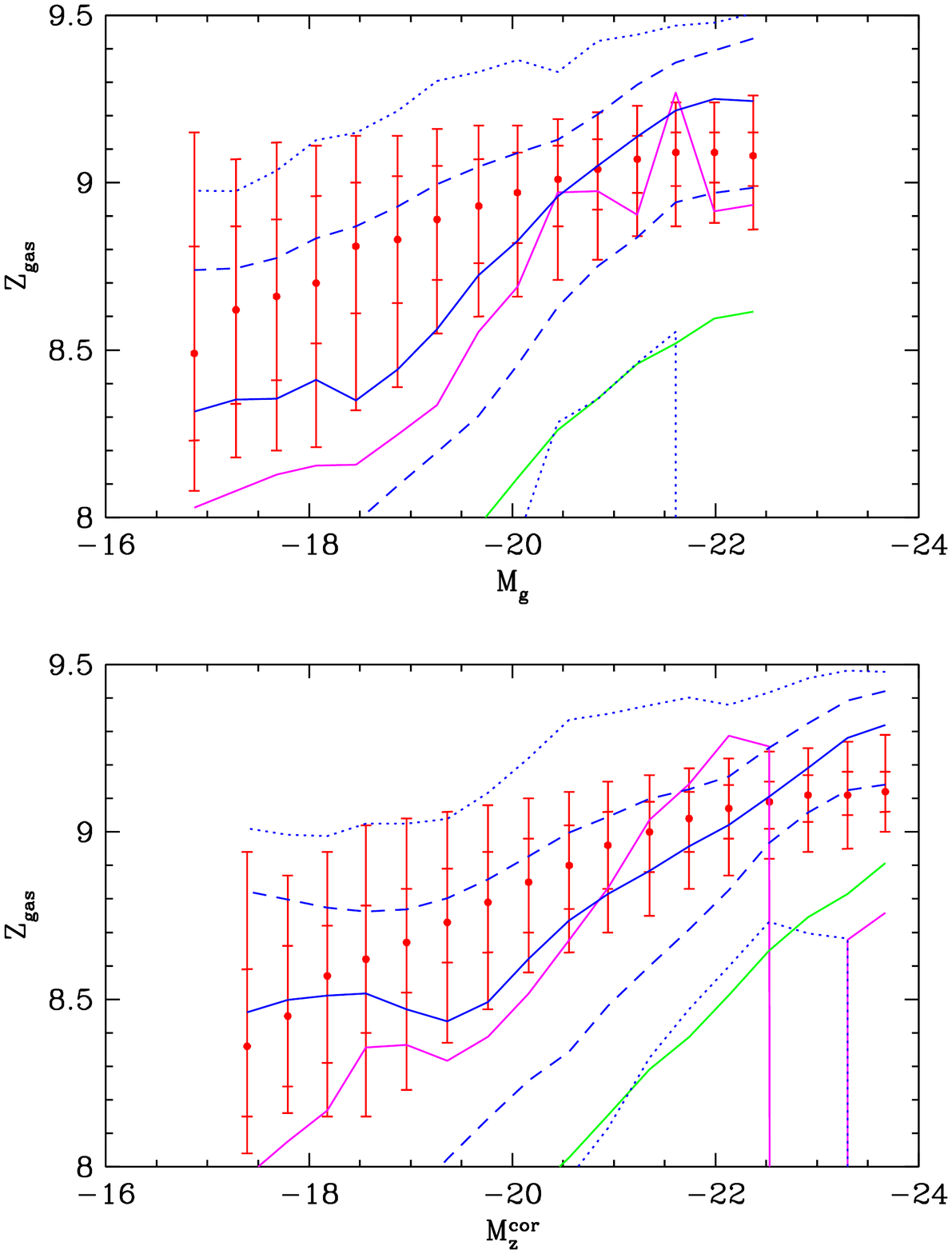}
 \caption{Gas-phase metallicity as a function of absolute magnitude from the \protect\SDSS\ \protect\pcite{tremonti_origin_2004} is shown by the red points. Points show the median value, while error bars indicate the 2.5, 16, 84 and 97.5 percentiles of the distribution. Lines indicate results form our best-fit model. Solid lines indicate the median model relation, dashed lines the 16 and 84 percentiles and dotted lines the 2.5 and 97.5 percentiles, corresponding to the error bars on the data. Blue lines show the overall best-fit model, while magenta lines indicate the best-fit model to this dataset and the green lines show results from the \protect\cite{bower_breakinghierarchy_2006} model. (Note that dashed and dotted lines are shown only for the best-fit model for clarity.)}
 \label{fig:SDSS_Zgas}
\end{figure}

\begin{figure*}
 \begin{tabular}{cccc}
{\tiny $-15 < M_{\rm B} -5\log h\le -14$} &
{\tiny $-17 < M_{\rm B} -5\log h\le -16$} &
{\tiny $-19 < M_{\rm B} -5\log h\le -18$} &
{\tiny $-21 < M_{\rm B} -5\log h\le -20$} \\
  \includegraphics[width=40mm,viewport=5mm 50mm 200mm 245mm,clip]{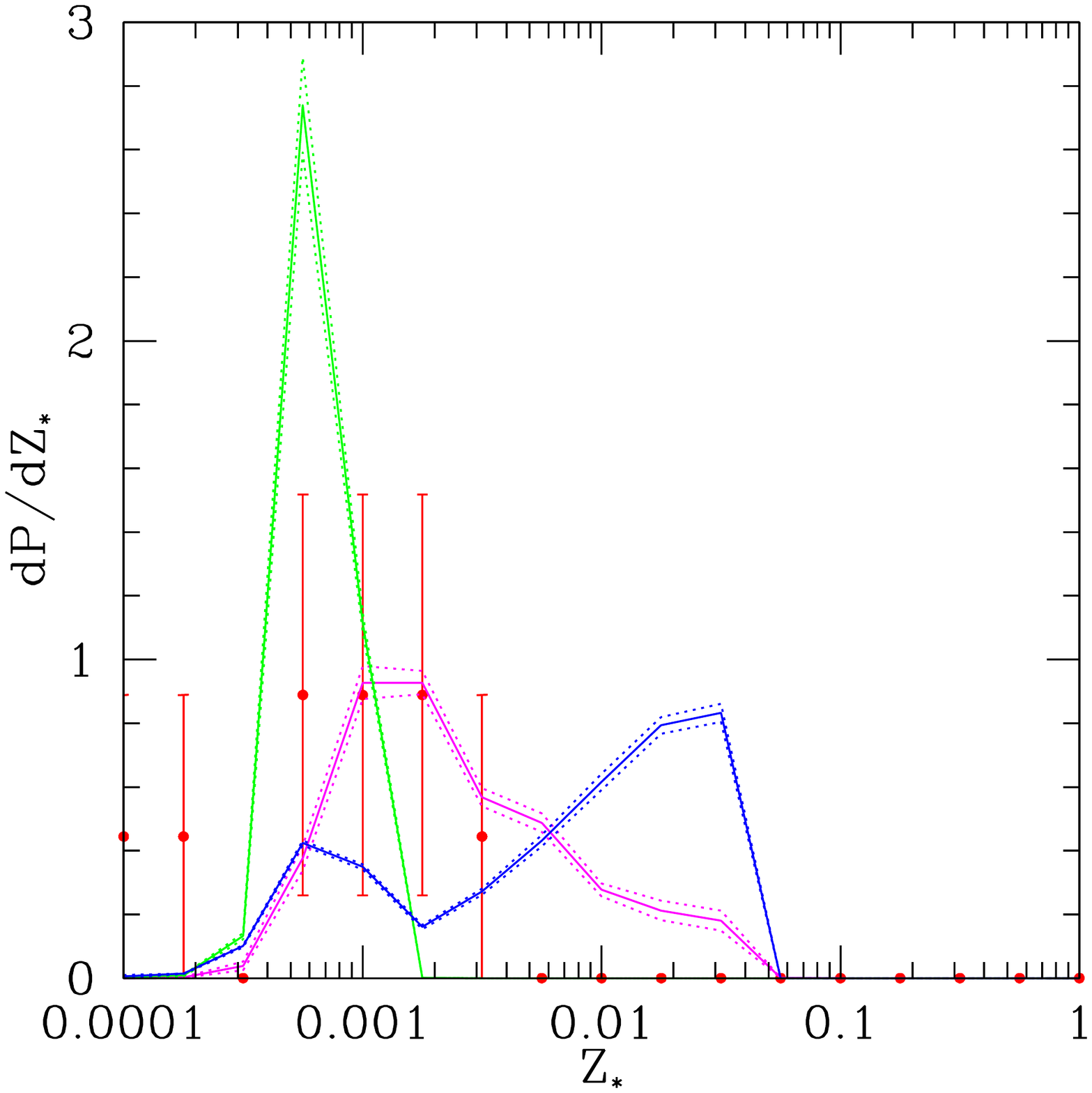} &
  \includegraphics[width=40mm,viewport=5mm 50mm 200mm 245mm,clip]{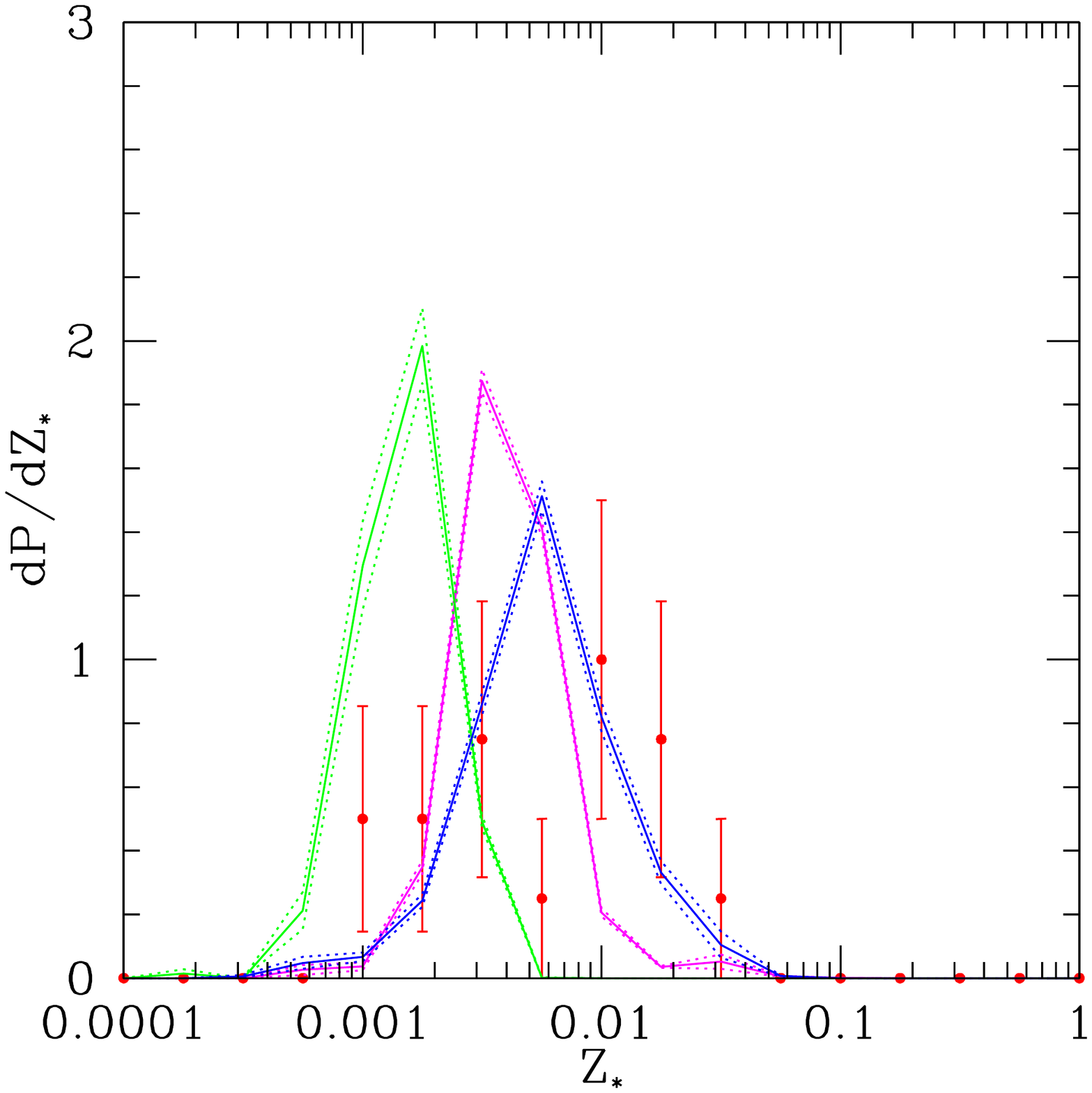} &
  \includegraphics[width=40mm,viewport=5mm 50mm 200mm 245mm,clip]{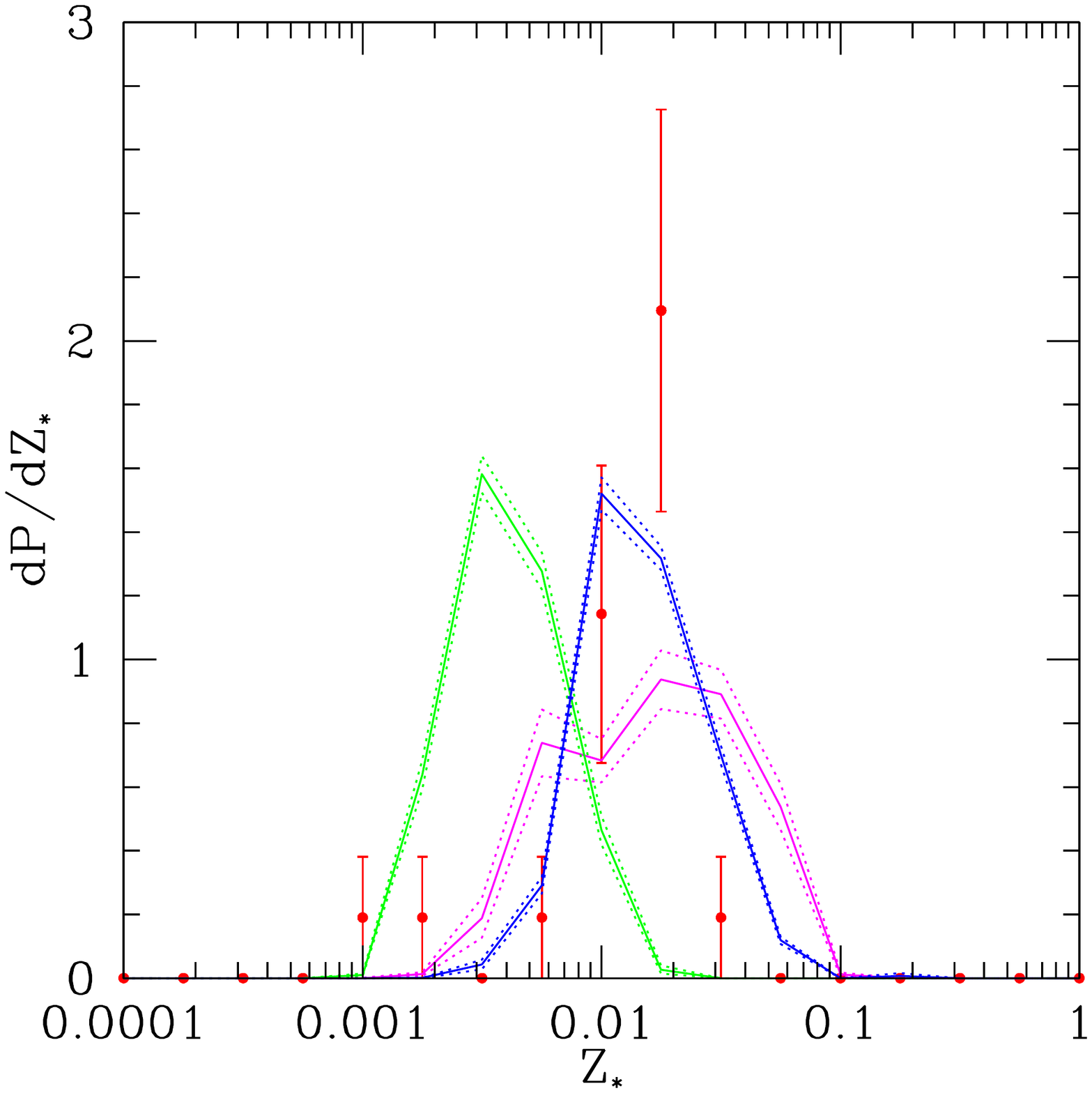} &
  \includegraphics[width=40mm,viewport=5mm 50mm 200mm 245mm,clip]{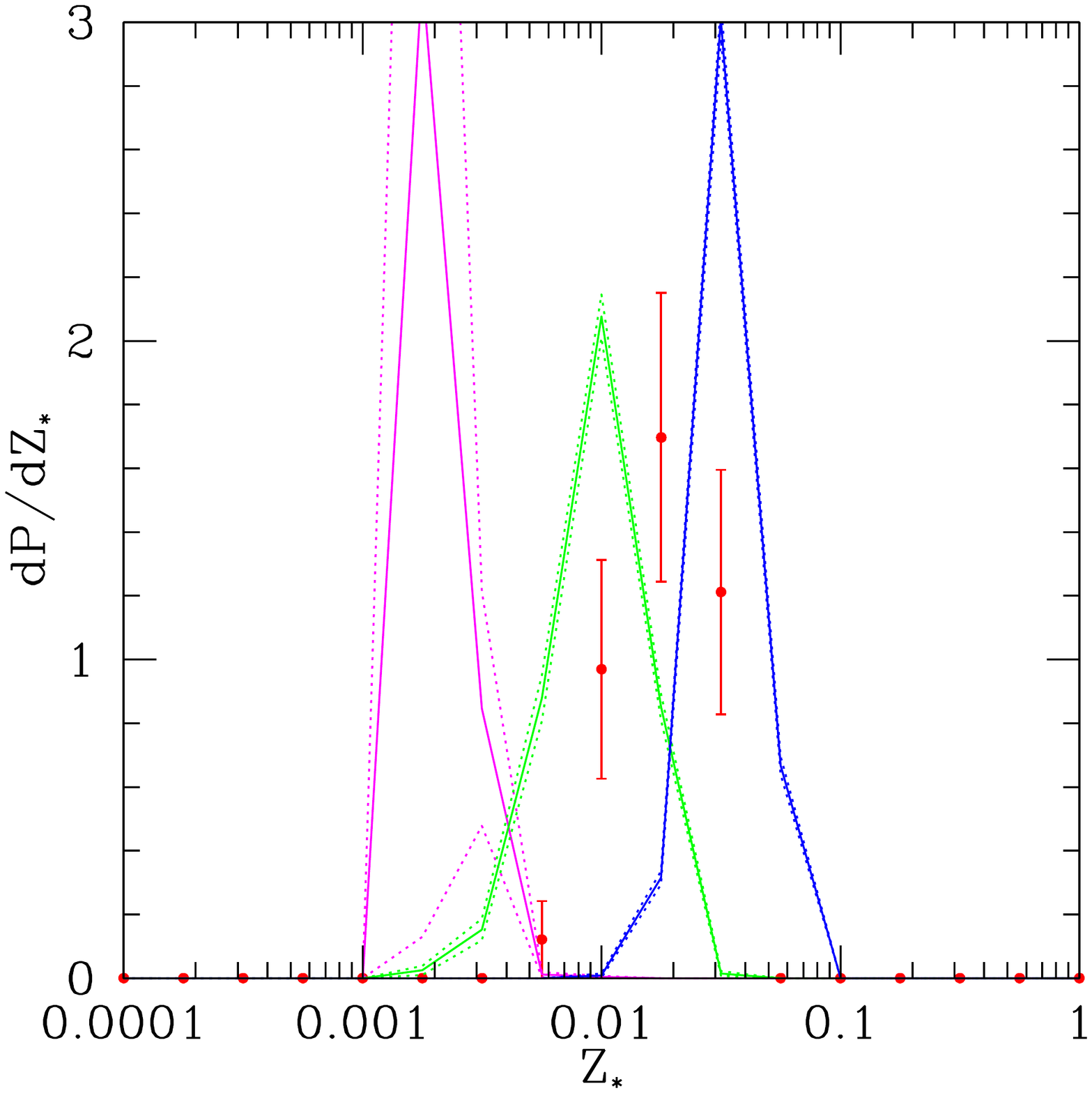}
 \end{tabular}
 \caption{Distributions of mean stellar metallicity at different slices of absolute magnitude. Red points show observational data compiled by \protect\cite{zaritsky_h_1994}. Solid lines indicate results from our models while dotted lines show the statistical uncertainty on the model estimate. Blue lines show the overall best-fit model, while magenta lines indicate the best-fit model to this dataset and the green lines show results from the \protect\cite{bower_breakinghierarchy_2006} model.}
 \label{fig:Zstar}
\end{figure*}

\begin{figure}
 \includegraphics[width=80mm,viewport=0mm 55mm 205mm 245mm,clip]{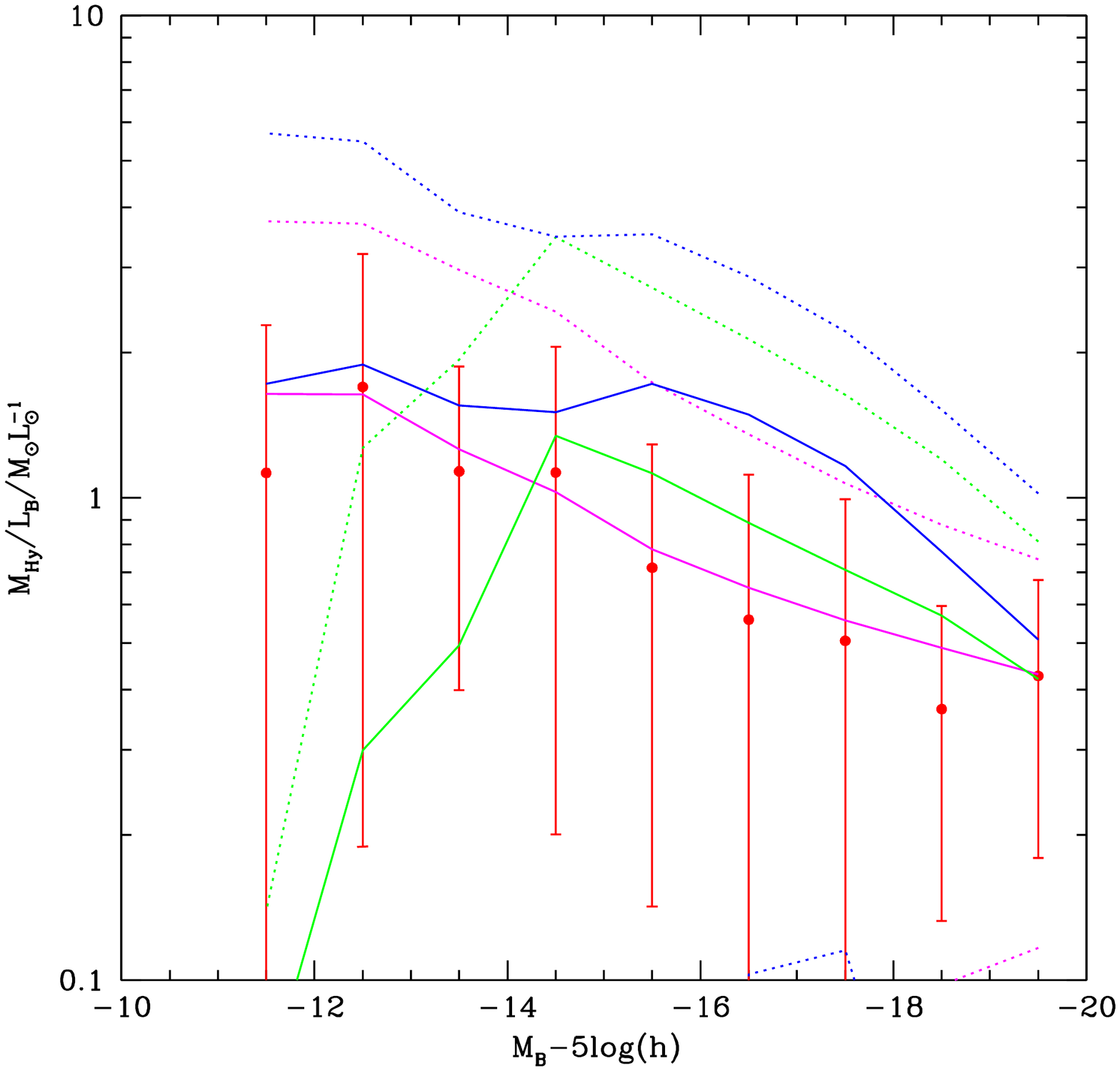}
 \caption{Gas (hydrogen) to B-band light ratios at $z=0$ as a function of B-band absolute magnitude. The solid lines show the mean ratio from our models while the dotted lines show the dispersion around the mean. Red points show the mean ratio from a compilation of data from \protect\cite{huchtmeier_h_1988} and \protect\cite{sage_molecular_1993} with error bars indicating the dispersion in the distribution. Blue lines show the overall best-fit model, while magenta lines indicate the best-fit model to this dataset and the green lines show results from the \protect\cite{bower_breakinghierarchy_2006} model. Model galaxies were selected to have bulge-to-total ratios in B-band light of $0.4$ or less and gas fractions of 3\% or more in order to attempt to match the morphological selection (Sa and later types) in the observations.}
 \label{fig:Gas2Light}
\end{figure}

The star formation and supernovae feedback prescriptions in our model can be constrained by measurements of the gas and metal content of galaxies. Figure~\ref{fig:SDSS_Zgas} shows the distribution of gas-phase metallicities from the \SDSS\ \pcite{tremonti_origin_2004} compared with results from our best-fit model. Model galaxies are drawn from the entire population of galaxies at $z=0.1$. \cite{tremonti_origin_2004} select star forming galaxies---essentially those with well detected H$\beta$, H$\alpha$ and [N{\sc ii}] $\lambda$6584 lines---and also reject galaxies with a significant \AGN\ component. We have not attempted to reproduce these observational selection criteria here\footnote{Both because we can not, at present, include the \protect\AGN\ component in the spectra and because it would involve constructing mock catalogues which is too expensive during our parameter space search.}, but note that excluding galaxies with very low star formation rates makes negligible difference to our results. The model clearly produces a strong trend of increasing metallicity with increasing luminosity, just as is observed, although the relation is somewhat too steep, resulting in metallicities which are around a factor of two too low at the lowest luminosities plotted. This relation is driven, in the model, by supernovae feedback: in low luminosity galaxies feedback is more efficient at ejecting material from a galaxy making it less efficient at self-enriching. The trend is somewhat steeper in the model than is observed and therefore underpredicts the metallicity of low luminosity galaxies. The spread in metallicity at fixed luminosity is larger than that which is observed. The best fit model to the metallicity datasets presented in this subsection can be seen to actually be a worse fit to the gas phase metallicity, a consequence of tensions between fitting this data and stellar metallcities and gas fractions.

Figure~\ref{fig:Zstar} shows distributions of mean stellar metallicity in various bins of absolute B-band magnitude. Data, shown by points, are taken from \cite{zaritsky_h_1994}, while results from our best-fit model are shown by lines. For model galaxies, we plot the luminosity-weighted mean metallicity of all stars (i.e. both disk and bulge stars). Although the data are quite noisy, there is, in general, good agreement of the model with this data. The \cite{bower_breakinghierarchy_2006} model fails to match the scaling of metallicity with stellar mass seen in these data. An increase in the yield in this model (from $p=0.02$ to $p=0.04$ as required to better match galaxy colours; \citealt{font_colours_2008}) would improve this situation significantly, but some reduction in the dependence of \SNe\ feedback on galaxy mass is likely still required to obtain the correct scaling.

Finally, Fig.~\ref{fig:Gas2Light} shows the distribution of gas-to-light ratios from a compilation of data compared to results from our best-fit model. Model galaxies are selected to have bulge-to-total ratios in B-band light of $0.4$ or less and gas fractions of 3\% or more in order to attempt to match the morphological selection (Sa and later types) in the observations. The results are somewhat sensitive to the morphological criteria used, a fact which must be taken into account when considering the comparison with the observational data. The model ratio is somewhat too high (too much gas per unit light), but displays approximately the correct dispersion. The \cite{bower_breakinghierarchy_2006} model gets closer to the observed mean for bright galaxies, but shows a dramatic downturn at low luminosities (a result of its very strong supernovae feedback). The best fit model to this specific dataset is an excellent match to both the mean and dispersion in the gas fraction data. This is achieved primarily via a very low efficiency of star formation (allowing gas fractions to stay high) coupled with strongly velocity dependent feedback which helps obtain the measured slope in this relation.

Overall, the \cite{bower_breakinghierarchy_2006} performs much less well in matching metallicity and gas content properties. This problem can be traced to the very strong scaling of supernovae feedback strength with galaxy circular velocity adopted in the \cite{bower_breakinghierarchy_2006} model and the low yield. This strongly suppresses the effective yield in low mass galaxies, resulting in them being too metal poor, and likewise strongly suppresses the gas content of those same low mass galaxies. These constraints are among the primary drivers causing our best fit model to adopt a lower value of $\alpha_{\rm hot}$.

\subsection{Clustering}\label{sec:Clustering}

\begin{figure*}
 \begin{tabular}{ccc}
{$-19.0 < M_{\rm b_{\rm J}} -5\log h\le -18.5$} &
{$-19.5 < M_{\rm b_{\rm J}} -5\log h\le -19.0$} &
{$-20.0 < M_{\rm b_{\rm J}} -5\log h\le -19.5$} \\
   \includegraphics[width=55mm,viewport=0mm 50mm 200mm 245mm,clip]{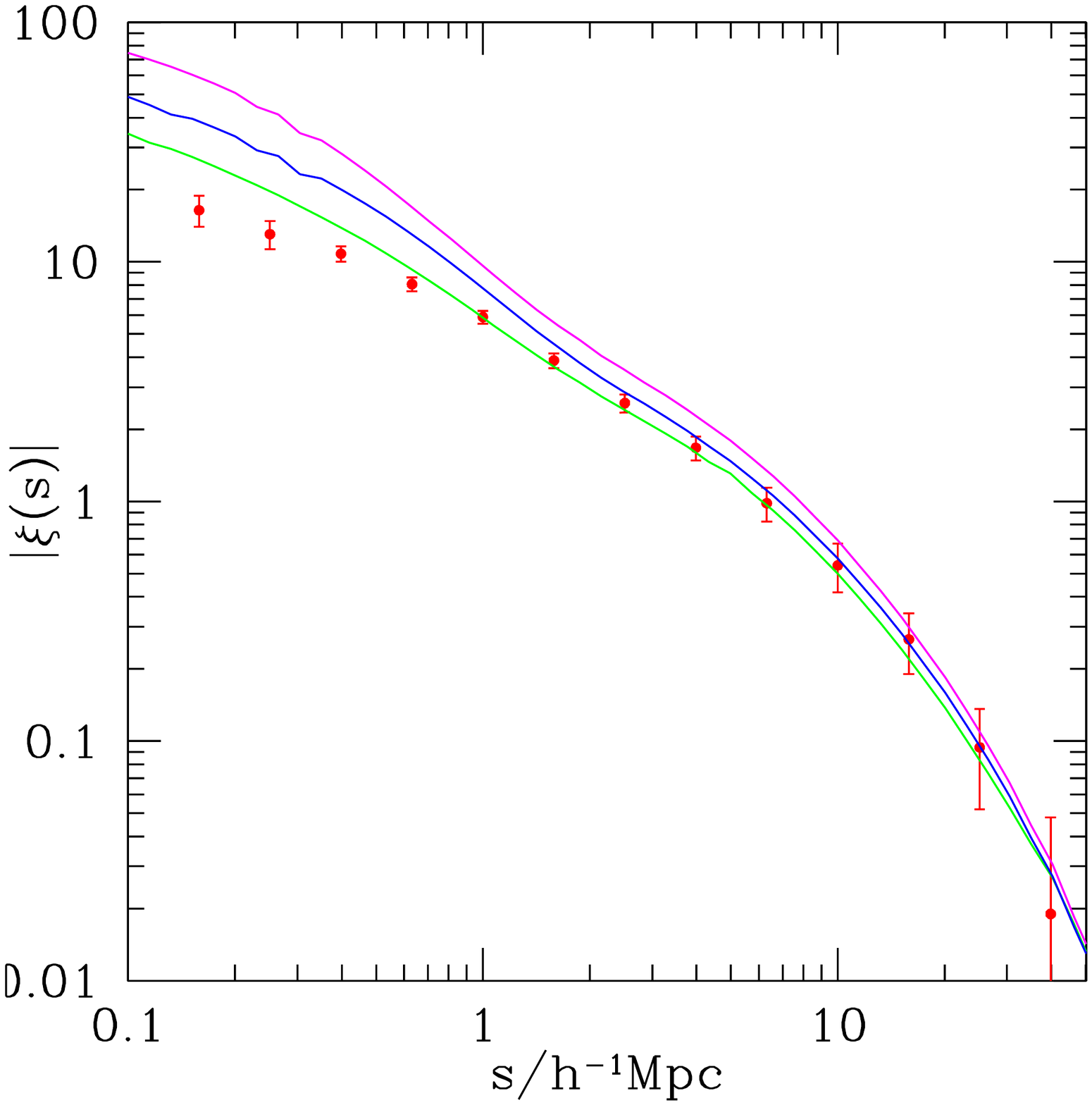} &
   \includegraphics[width=55mm,viewport=0mm 50mm 200mm 245mm,clip]{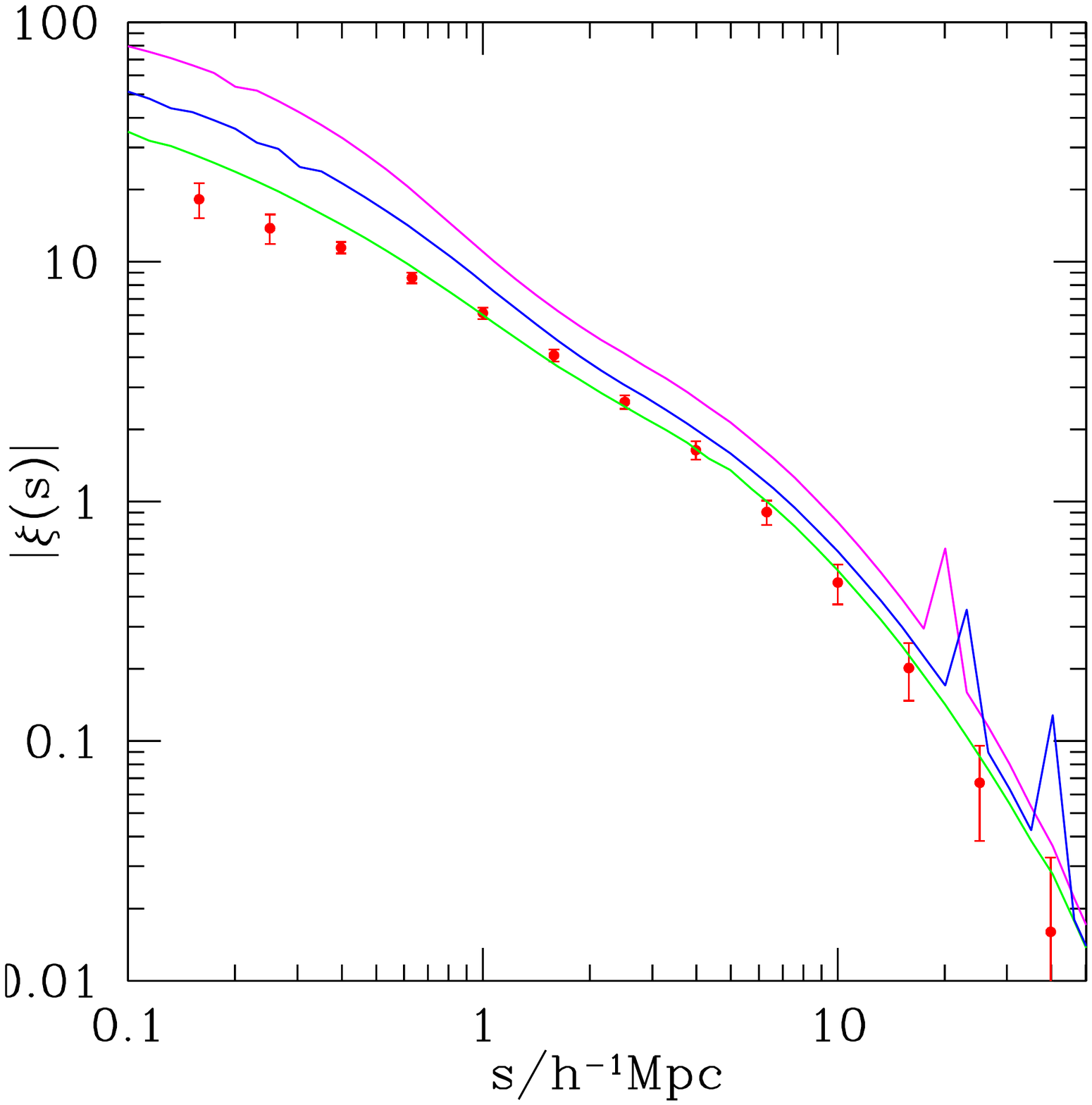} &
   \includegraphics[width=55mm,viewport=0mm 50mm 200mm 245mm,clip]{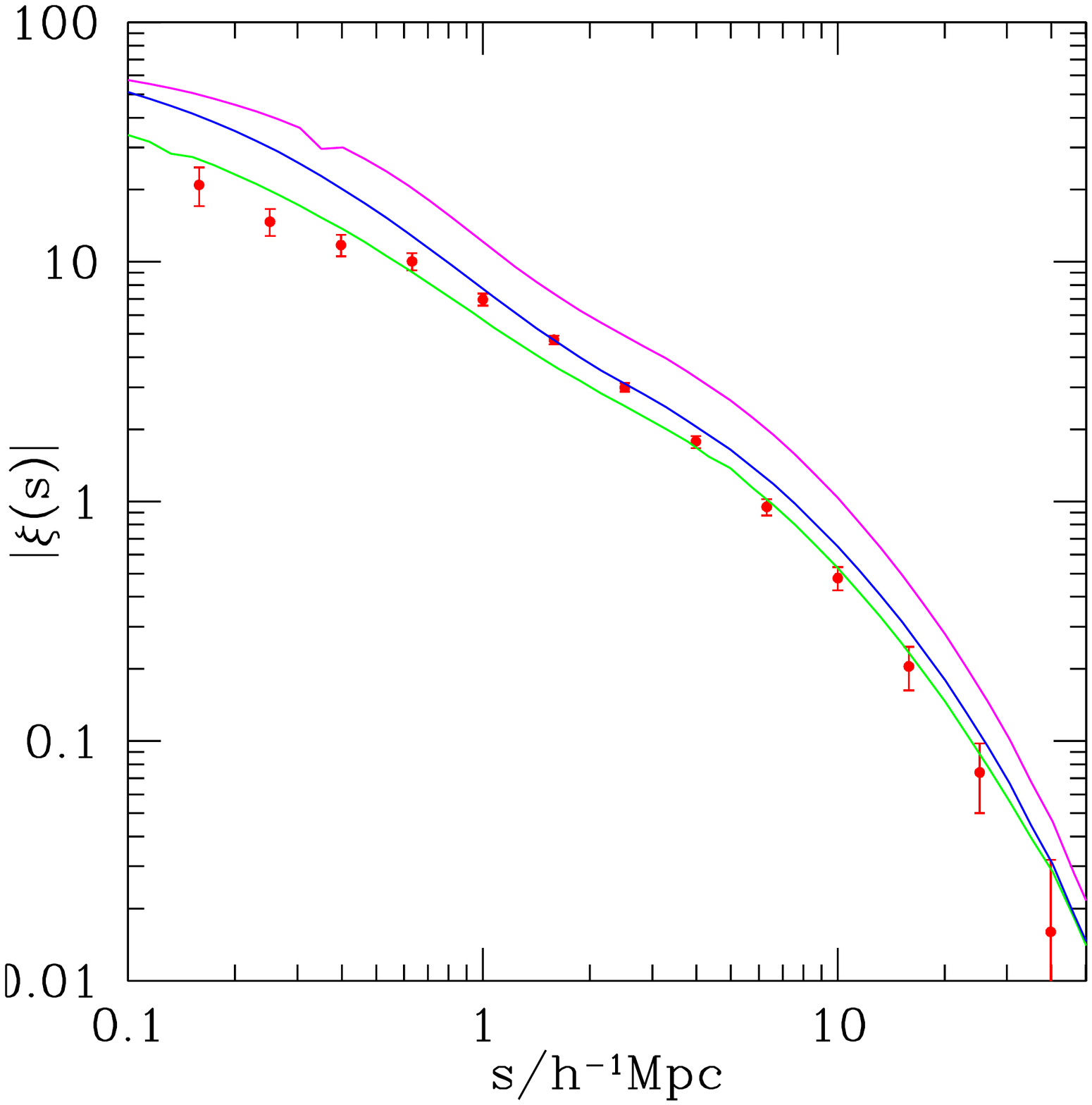}
 \end{tabular}
 \caption{Redshift space two-point correlation functions of galaxies selected by their b$_{\rm J}$ band absolute magnitude. Solid lines show results from our models while red points indicate data from the \protect\TdF\ \protect\pcite{norberg_2df_2002}. Model correlation functions are computed using the halo model of clustering \protect\pcite{cooray_halo_2002} with the input halo occupation distributions computed directly from our best-fit model. Blue lines show the overall best-fit model, while magenta lines indicate the best-fit model to this dataset and the green lines show results from the \protect\cite{bower_breakinghierarchy_2006} model.}
 \label{fig:2dFGRS_Clustering}
\end{figure*}

Galaxy clustering places strong constraints on the occupancy of galaxies within dark matter halos and, therefore, the merger rate (amongst other things). To compute the clustering properties of galaxies we make use of the fact that halo occupation distributions are naturally predicted by the \gf\ model. We therefore extract halo occupation distributions directly from our best fit model. We then employ the halo model of galaxy clustering \pcite{cooray_halo_2002} to compute two-point correlation functions in redshift space. These are compared to measured redshift-space correlation functions from the \TdF\ \pcite{norberg_2df_2002} in Fig.~\ref{fig:2dFGRS_Clustering}.

There is excellent agreement between the model and data on large scales (where the two halo term dominates). On small scales, in the one halo regime, the model systematically overestimates the correlation function. This discrepancy, which is due to the model placing too many satellite galaxies in massive halos, has been noted and discussed previously by \cite{seek_kim_modelling_2009}. In their study, \cite{seek_kim_modelling_2009} demonstrated that this problem might be resolved by invoking destruction of satellite galaxies by tidal forces and by accounting for satellite-satellite mergers (both processes reduce the number of satellites). The current model includes both of these processes and treats them in a significantly more realistic way than did \cite{seek_kim_modelling_2009}. We find that they are not enough to bring the model correlation function into agreement with the data on small scales (although they do help), in our particular model. This may indicate that these processes have not been modelled sufficiently accurately, or that our model simply begins with too many satellites. We note that the \cite{bower_breakinghierarchy_2006} model performs similarly well on large scales and somewhat better on small scales (the stronger feedback in this model helps reduce the number of satellite galaxies of a given luminosity in high mass halos), although it still overpredicts the small scale clustering, as has been noted by \cite{seek_kim_modelling_2009}. The best-fit model to the clustering data alone is not very successful. This is again due to the difficulty of computing accurate correlation functions using the relatively small sets of merger trees that we are able to utilize for parameter space searches, and serves as an excellent example of the need to include better estimates of the model uncertainty (i.e. the variance in predictions from the model due to the limited number of merger trees utilized) when computing goodness of fit measures.

\subsection{Supermassive Black Holes}\label{sec:SMBH}

\begin{figure*}
 \begin{tabular}{ccc}
$9 \le \log_{10}(M_{\rm bulge}/h^{-1}M_\odot) < 10$ &
$10 \le \log_{10}(M_{\rm bulge}/h^{-1}M_\odot) < 11$ &
$11 \le \log_{10}(M_{\rm bulge}/h^{-1}M_\odot) < 12$ \\
  \includegraphics[width=55mm,viewport=7mm 55mm 205mm 245mm,clip]{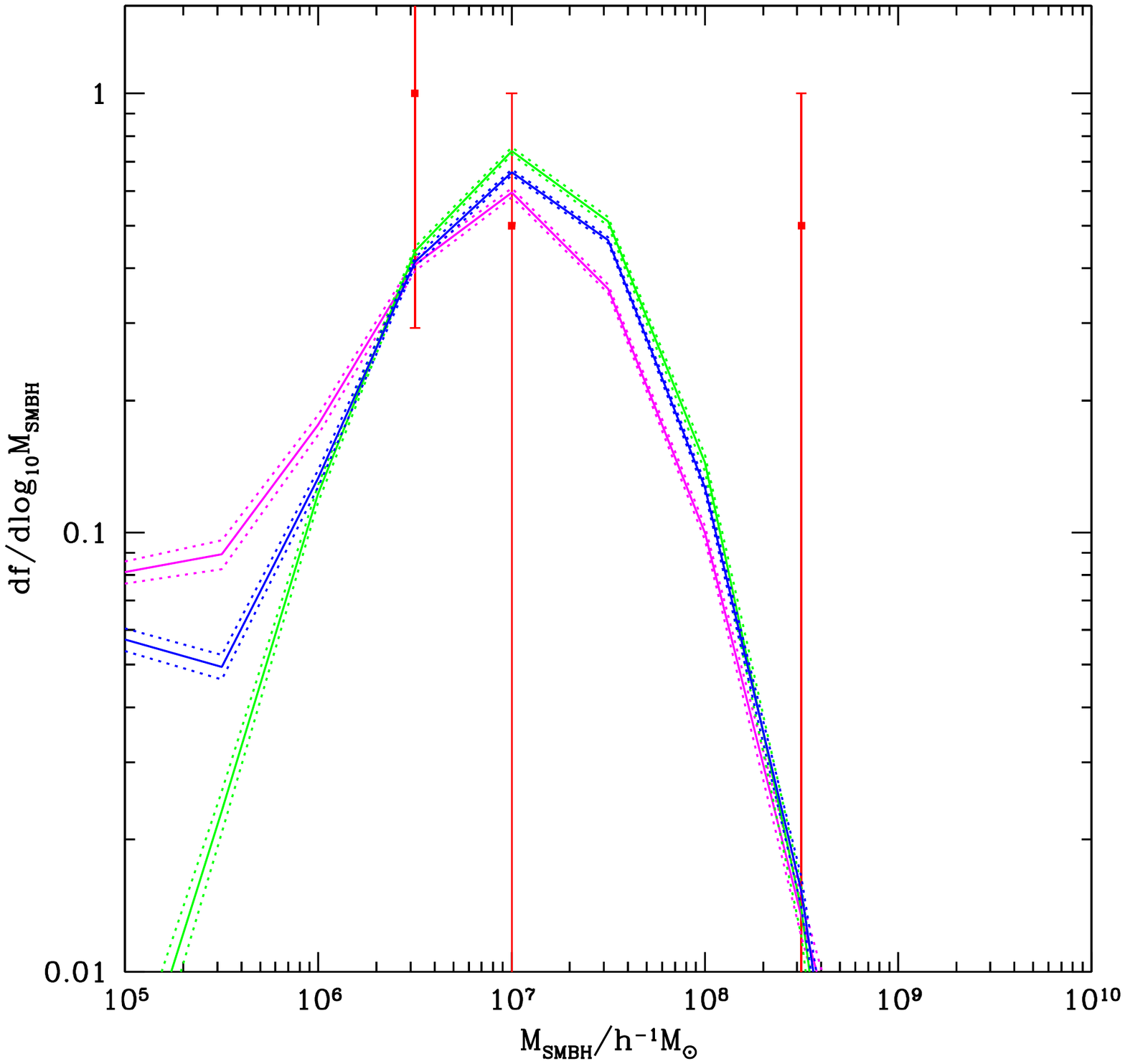} &
  \includegraphics[width=55mm,viewport=7mm 55mm 205mm 245mm,clip]{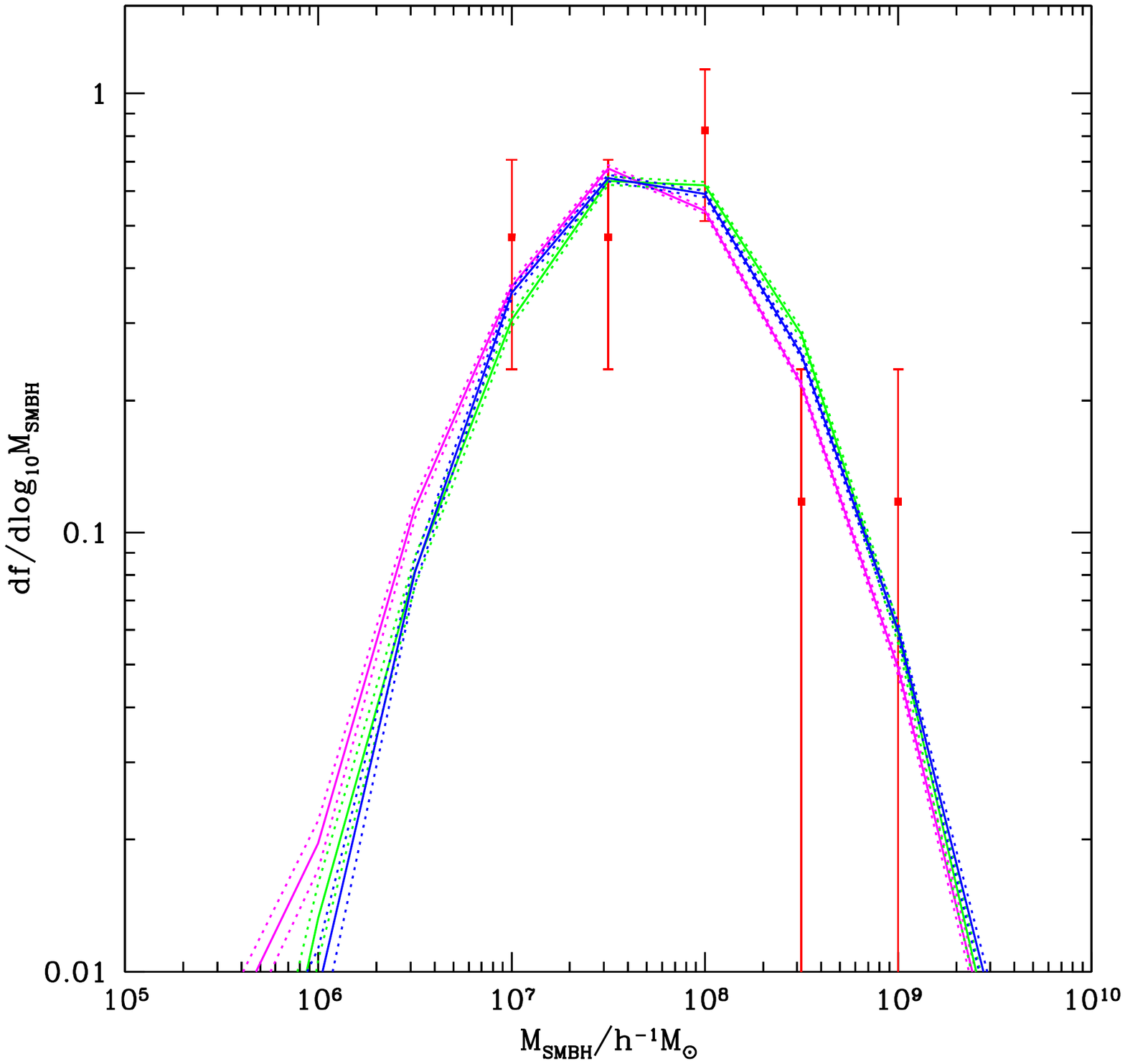} &
  \includegraphics[width=55mm,viewport=7mm 55mm 205mm 245mm,clip]{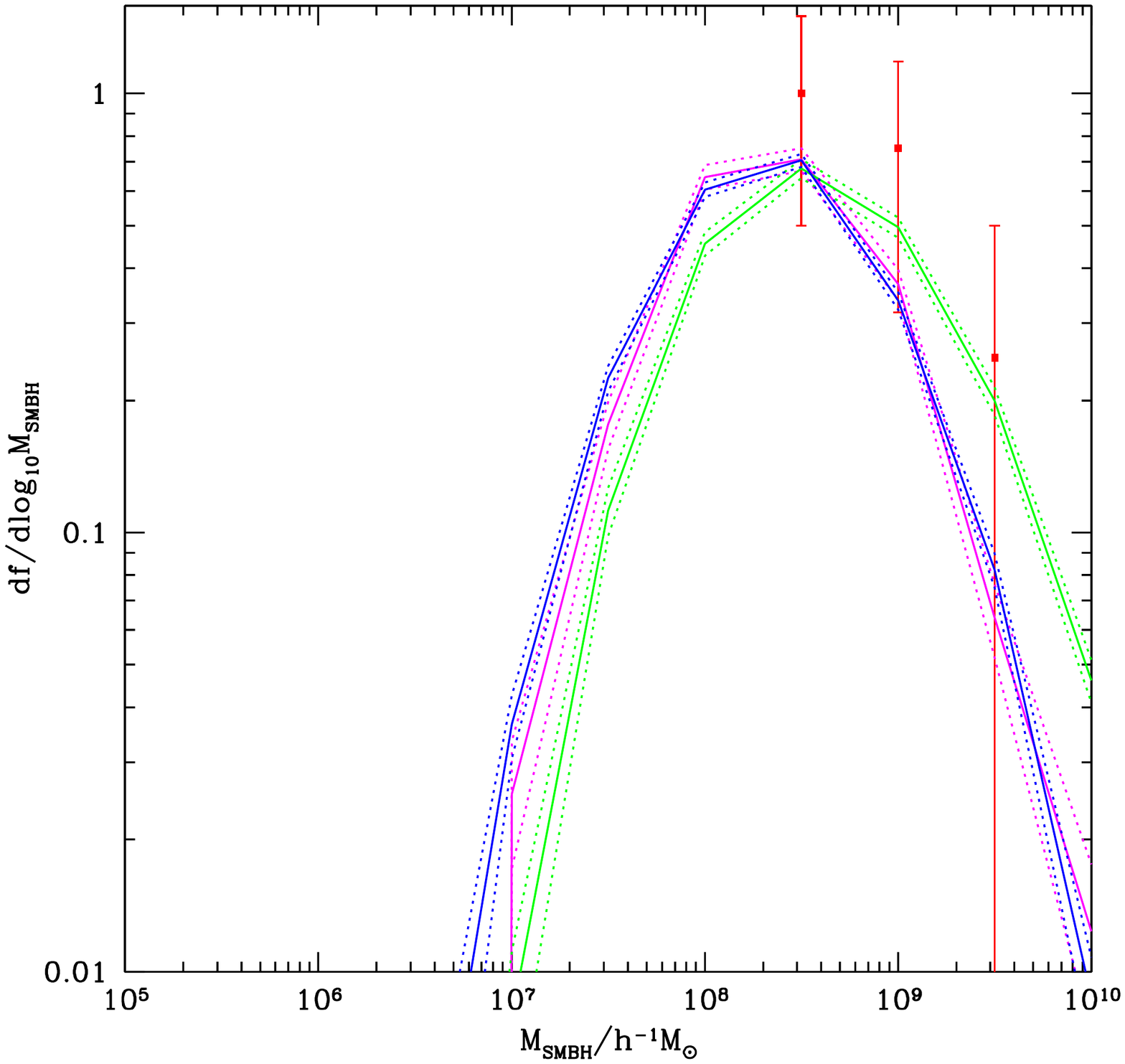}
 \end{tabular}
 \caption{The distribution of supermassive black hole mass in three slices of galaxy bulge mass. Data are taken from \protect\cite{haring_black_2004} and are shown by red points. Solid lines indicate results from our models with dotted lines showing the statistical uncertainty on the model estimate. Blue lines show the overall best-fit model, while magenta lines indicate the best-fit model to this dataset and the green lines show results from the \protect\cite{bower_breakinghierarchy_2006} model.}
 \label{fig:MSMBH}
\end{figure*}

The inclusion of \AGN\ feedback in semi-analytic models of galaxy formation necessitates the inclusion of the supermassive black holes that are responsible for that feedback. As such, it is important to constrain the properties of these black holes to match those that are observed. Figure~\ref{fig:MSMBH} shows the distribution of supermassive black hole masses in three slices of galaxy bulge mass. Points show observational data from \cite{haring_black_2004} while lines show results from our best-fit model. The model is in excellent agreement with the current data. The \cite{bower_breakinghierarchy_2006} model produces nearly identical results for the black hole masses. This is not surprising since, as pointed out by \cite{bower_parameter_2010}, the $F_\bullet$ parameter can be adjusted to achieve a good fit here without significantly affecting any other predictions. For this same reason, the best fit model to these black hole data in not significantly better than either the \cite{bower_breakinghierarchy_2006} or the overall best fit model.

\subsection{Local Group}\label{sec:LocalGroup}

\begin{figure}
   \includegraphics[width=80mm,viewport=7mm 55mm 205mm 255mm,clip]{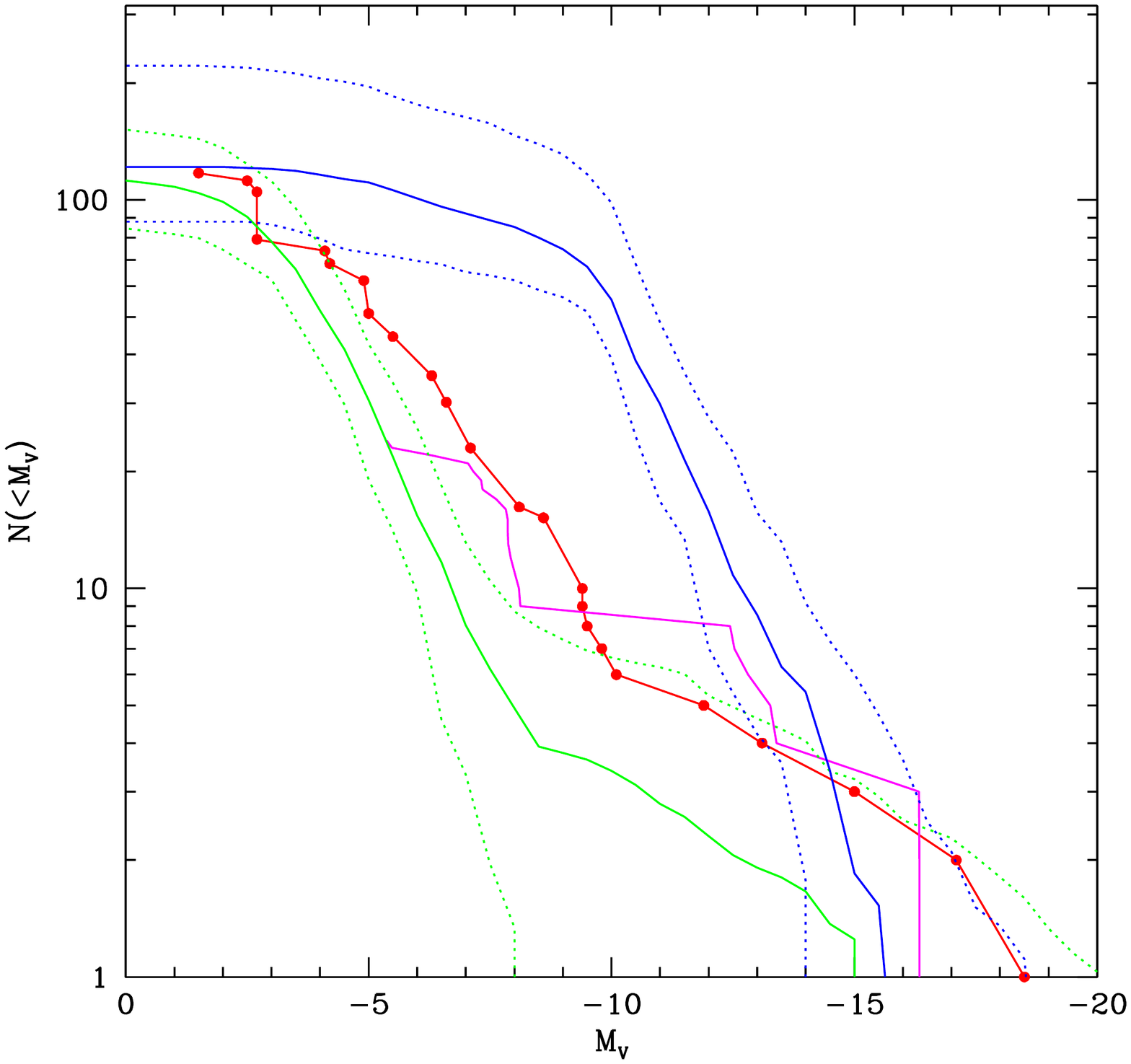}
 \caption{The luminosity function of Local Group satellite galaxies in our models. Red points show current observational estimates of the luminosity function from \protect\cite{koposov_luminosity_2008} including corrections for sky coverage and selection probability from \protect\cite{tollerud_hundreds_2008}. Solid lines show the median luminosity functions of model satellite galaxies located in Milky Way-hosting halos, while dotted lines indicate the $10^{\rm th}$ and $90^{\rm th}$ percentiles of the distribution of model luminosity functions. Blue lines show the overall best-fit model, while magenta lines indicate the best-fit model to this dataset and the green lines show results from the \protect\cite{bower_breakinghierarchy_2006} model.}
 \label{fig:LocalGroup_LF}
\end{figure}

The recent discovery of several new satellite galaxies of the Milky Way has lead to their abundance and properties being more robustly known and therefore acting as a strong constraint on models of galaxy formation and has attracted significant attention recently \pcite{bullock_reionization_2000,benson_effects_2002,somerville_can_2002,gnedin_fossils_2006,madau_dark_2008,madau_fossil_2008,munoz_probingepoch_2009,bovill_pre-reionization_2009,busha_impact_2009,macci`o_luminosity_2009}. Our model is the only one of which we are aware that follows the formation of these galaxies within the context of a self-consistent model of the \IGM\ and the global galaxy population which fits a broad range of experimental constraints on galaxies and the \IGM.

To compute the expected properties of Milky Way satellites in our model we simulate a large number of dark matter halos with masses at $z=0$ in the range $2\times 10^{11}$--$3\times 10^{12}h^{-1}M_\odot$. From these, we select only those halos with a virial velocity in the range 125--180km/s (consistent with recent estimates; \citealt{dehnen_velocity_2006,xue_milky_2008}) and which contain a central galaxy with a bulge-to-total ratio between 5 and 20\% to approximately match the properties of the Milky Way. This step is potentially important, as it ensures that the satellite populations that we consider are consistent with the formation of a Milky Way-like galaxy\footnote{The merging history of a halo will affect both the properties of the central galaxy and the population of satellite galaxies. By selecting only halos whose merger history was suitable to produce a Milky Way we ensure that we are looking only at satellite populations consistent with the presence of such a galaxy.}. In practice, we find that the morphological selection has little effect on the satellite luminosity function. However, the selection of suitable halos based on virial velocity produces a significant reduction (by about a factor of 2) in the number of satellites compared to the common practice of selecting halos with masses of approximately $10^{12}h^{-1}M_\odot$. Halo selection is clearly of great importance when addressing the missing satellite problem. We prefer to use a selection on halo virial velocity here rather than a selection on galaxy stellar mass, as was used by \cite{benson_effects_2002} for example, since we know that the Tully-Fisher relation in our model is incorrect (see \S\ref{sec:TF}) and so selecting on galaxy mass would result in an incorrect sample of halo masses.

Figure~\ref{fig:LocalGroup_LF} shows the V-band luminosity function of Milky Way satellite galaxies from our best fit model compared with the latest observational estimate. Our model is able to produce a sufficient number of the brightest satellites in a small fraction of realizations, although the median lies below the observed luminosity function for the Milky Way. At lower luminosities our best fit model overpredicts the observed number of satellites by factors of up to 5. It has recently been pointed out \pcite{busha_impact_2009,font_modelingmilky_2009} that inhomogeneous reionization (namely the reionization of the Lagrangian volume of the Milky Way halo by Milky Way progenitors) is an important consideration when computing the abundance of Local Group satellites. In particular, \cite{font_modelingmilky_2009} find a similar level of discrepancy in the luminosity function when they ignore this effect (as we do here) and use a similar feedback model, but demonstrate that consideration of inhomogeneous reionization can reconcile the predicted and observed abundance of satellites. We do not consider inhomogeneous reionization here, but will return to it in greater detail in a future work. It must be noted, however, that this may have an impact on the luminosity function of Local Group satellites. The \cite{bower_breakinghierarchy_2006} model gives a reasonably good match to the data, producing slightly fewer satellites than are observed at all luminosities. The best fit model to this specific dataset is in good agreement with the observations down to $M_{\rm V}=-5$, but fails to produce fainter satellites. (It also produces very few halo/galaxy pairs which meet our criteria to be deemed ``Milky Way-like'', resulting in poor statistics for this model. The models utilized during the parameter space search happened to produce more faint galaxies, resulting in them being judged a good fit---this is another example of where understanding the model uncertainty is of crucial importance.)

\begin{figure*}
 \begin{tabular}{cc}
 \vspace{-3mm} $-15 < M_{\rm V} \le -10$ & $-10 < M_{\rm V} \le -5$ \\
   \includegraphics[width=80mm,viewport=7mm 55mm 205mm 255mm,clip]{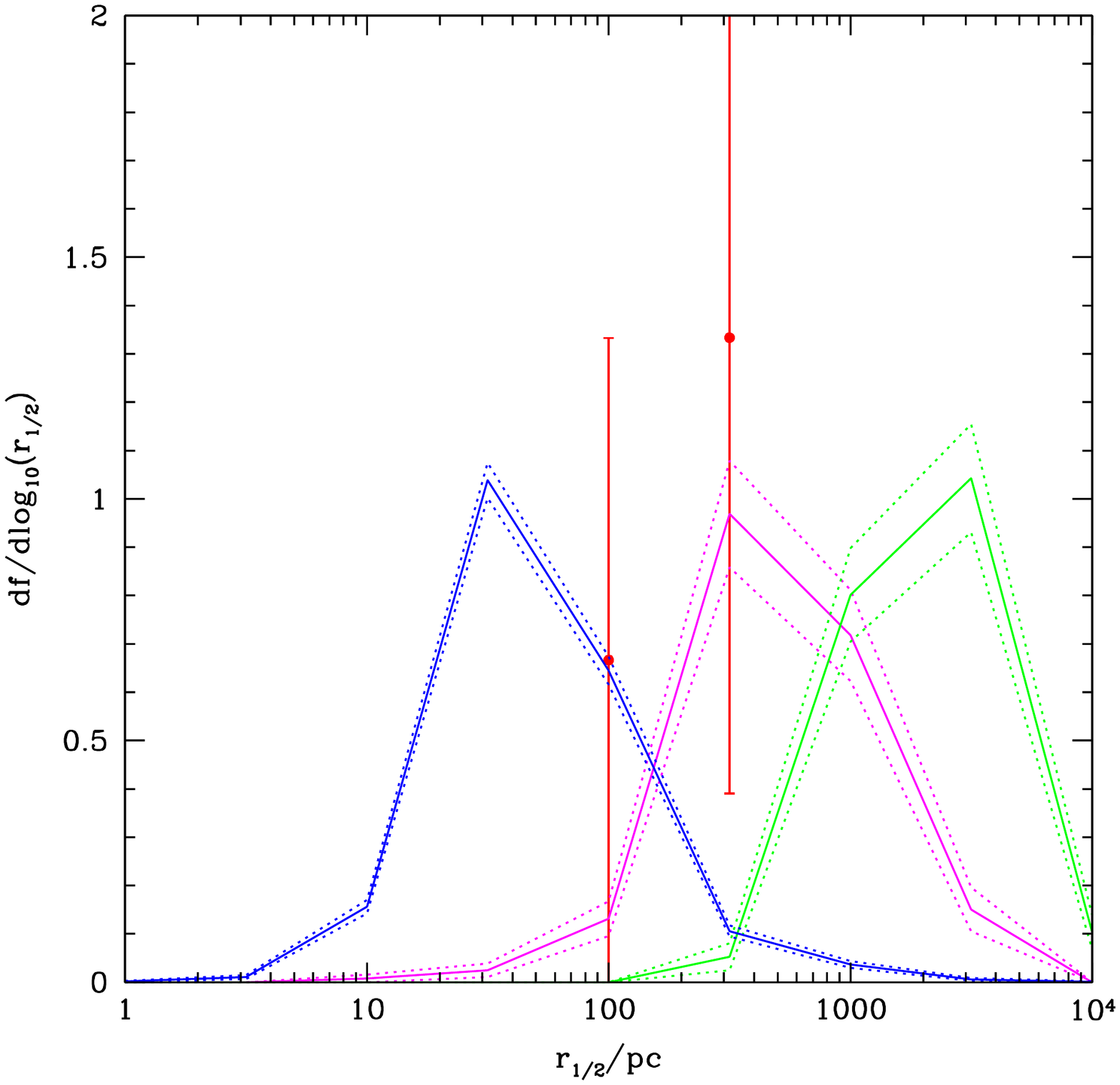} &
   \includegraphics[width=80mm,viewport=7mm 55mm 205mm 255mm,clip]{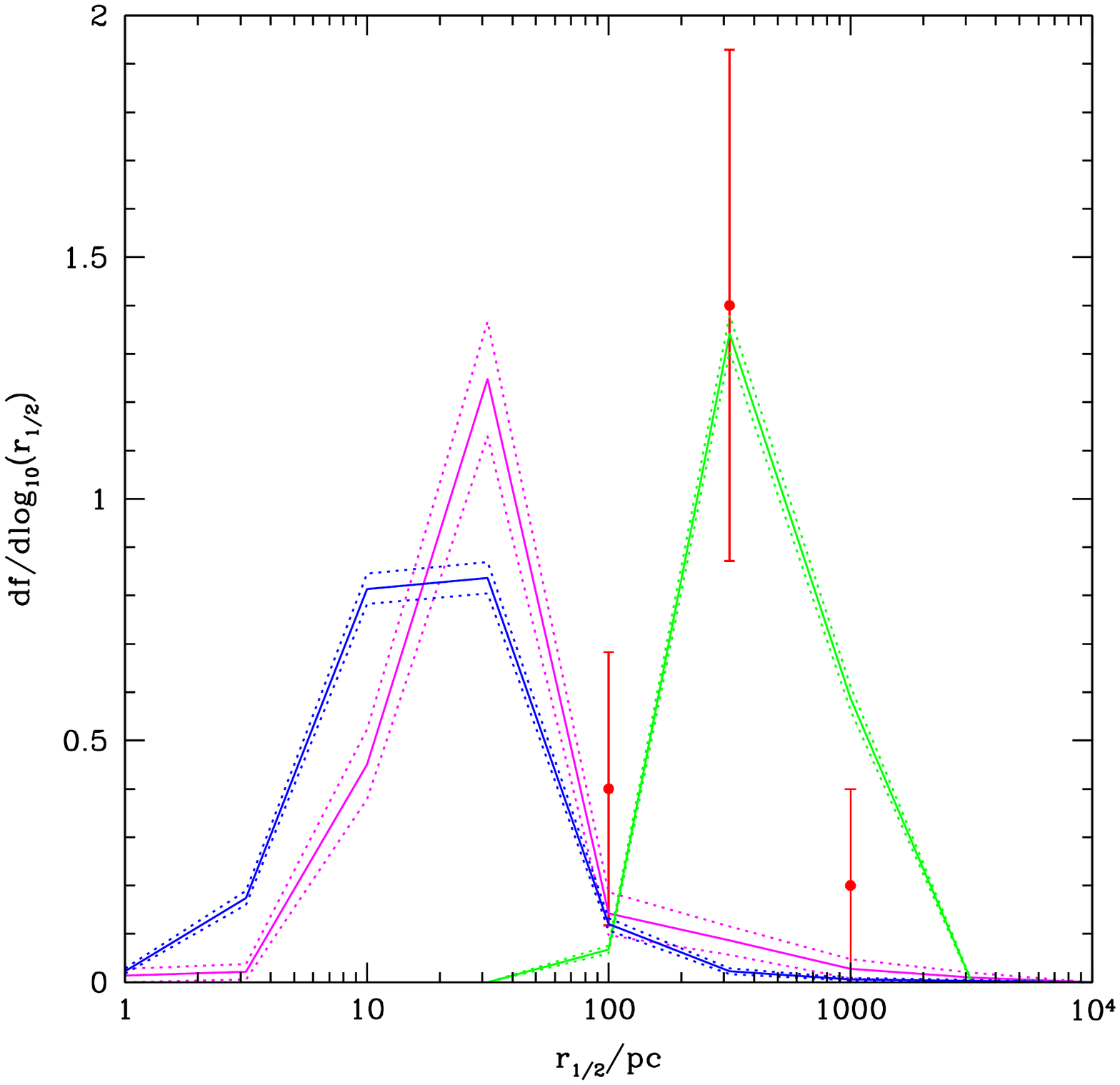}
 \end{tabular}
 \caption{The size distribution of Local Group satellite galaxies in our models. Red points show current observational estimates of the size distribution from \protect\cite{tollerud_hundreds_2008}. Solid lines show the size distribution of model satellite galaxies located in Milky Way-hosting halos with dotted lines showing the statistical uncertainty on the model estimate. Blue lines show the overall best-fit model, while magenta lines indicate the best-fit model to this dataset and the green lines show results from the \protect\cite{bower_breakinghierarchy_2006} model.}
 \label{fig:LocalGroup_Sizes}
\end{figure*}

Figure~\ref{fig:LocalGroup_Sizes} shows the distribution of half-mass radii for Milky Way satellites split into four bins of V-band absolute magnitude (only two of the bins are shown). The data are sparse, but the model produces galaxies that are too small compared to the observed satellites by factors of around 3--6. The \cite{bower_breakinghierarchy_2006} model has the opposite problem, producing faint satellites that are too large but doing well at matching the sizes of brighter satellites. The best fit model to the Local Group size data alone is not significantly better than the overall best fit model---the sizes tend to be rather insensitive to most parameters.

\begin{figure*}
 \begin{tabular}{cc}
  \vspace{-3mm} $-15 < M_{\rm V} \le -10$ & $-10 < M_{\rm V} \le -5$ \\
   \includegraphics[width=80mm,viewport=7mm 55mm 205mm 255mm,clip]{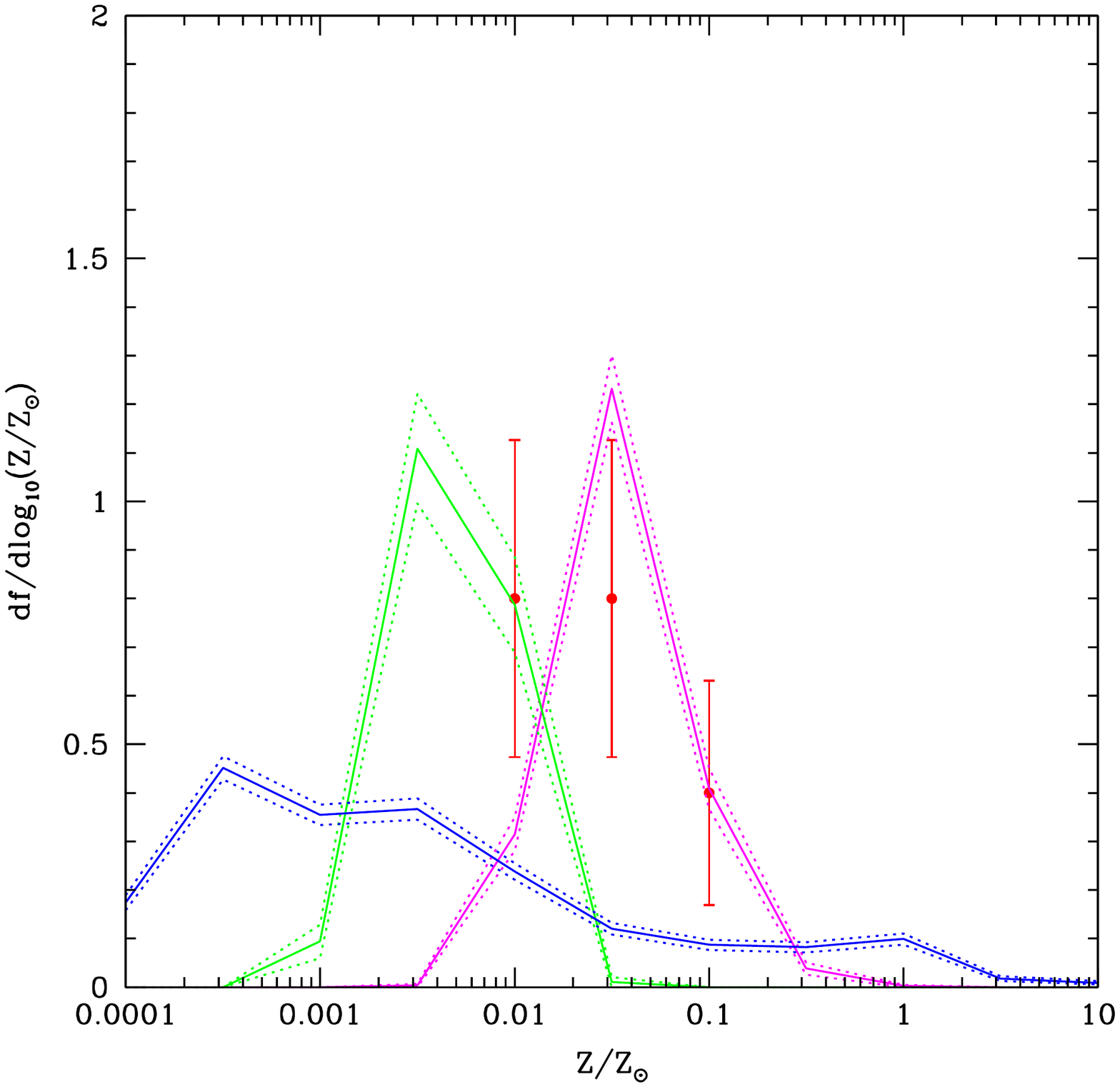} &
   \includegraphics[width=80mm,viewport=7mm 55mm 205mm 255mm,clip]{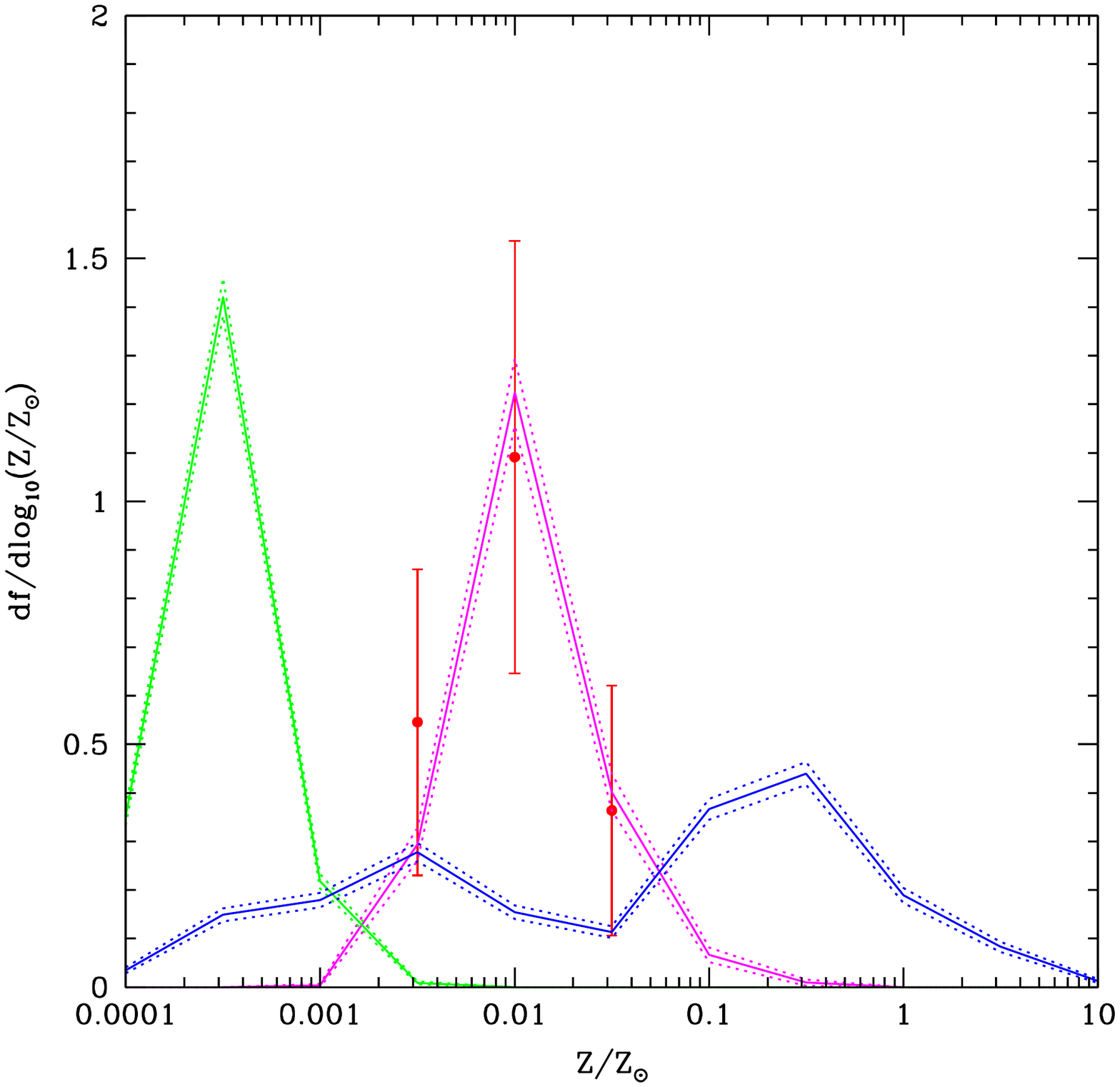}
 \end{tabular}
 \caption{The metallicity distribution of Local Group satellite galaxies in our models. Red points show current observational estimates of the metallicity distribution from the compilation of \protect\cite{mateo_dwarf_1998} and from \protect\cite{kirby_uncovering_2008}. Solid lines show the metallicity distribution of model satellite galaxies located in Milky Way-hosting halos with dotted lines showing the statistical uncertainty on the model estimate. Blue lines show the overall best-fit model, while magenta lines indicate the best-fit model to this dataset and the green lines show results from the \protect\cite{bower_breakinghierarchy_2006} model.}
 \label{fig:LocalGroup_Metallicities}
\end{figure*}

Figure~\ref{fig:LocalGroup_Metallicities} shows the distribution of stellar metallicities for Milky Way satellites split into the same four bins of V-band absolute magnitude (of which only two are shown). Once again, the data are sparse, but the model is seen to predict distributions of metallicity that are too broad compared to those observed. The \cite{bower_breakinghierarchy_2006} model performs poorly here, significantly underestimating the metallicities of the fainter satellites. This problem can be directly traced to the high value of $\alpha_{\rm hot}$ used by the \cite{bower_breakinghierarchy_2006} model which results is exceptionally strong supernovae feedback, and consequently very low effective yields, for low mass galaxies. The best fit model to the Local Group metallicity data alone performs much better than the \cite{bower_breakinghierarchy_2006} and significantly better than the overall best fit model in reproducing both the trend with luminosity and scatter at fixed luminosity. This is achieved through a combination of relatively weakly velocity dependent feedback (i.e. a low value of $\alpha_{\rm hot}$) and a weak scaling of star formation efficiency with velocity. Together, these parameters determine the trend of effective yield with mass and the degree of self-enrichment in these galaxies. However, this weaker feedback and low $\alpha_{\rm hot}$ also result in a steeper faint end slope for the global luminosity function compared to \cite{bower_breakinghierarchy_2006}, thereby giving less success in matching the data in that particular statistic.

\subsection{IGM Evolution}\label{sec:IGMResults}

As described in \S\ref{sec:IGM}, our model self-consistently evolves the properties of the intergalactic medium along with those of galaxies. In this section we discuss basic properties of the \IGM\ (and related quantities) from our best-fit model.

Photoheating of the \IGM\ begins to raise its temperature above the adiabatic expectation at $z\approx 25$, reaching a peak temperature of approximately 15,000K when hydrogen becomes fully reionized before cooling to around 2,000K by $z=0$.  Hydrogen is fully reionized by $z=8$. Helium is singly ionized at approximately the same time. There follows an extended period during which helium is partially doubly ionized, but is not fully doubly ionized until much later, around $z=4$.

\begin{figure*}
 \begin{tabular}{cc}
 \includegraphics[width=80mm,viewport=7mm 55mm 205mm 255mm,clip]{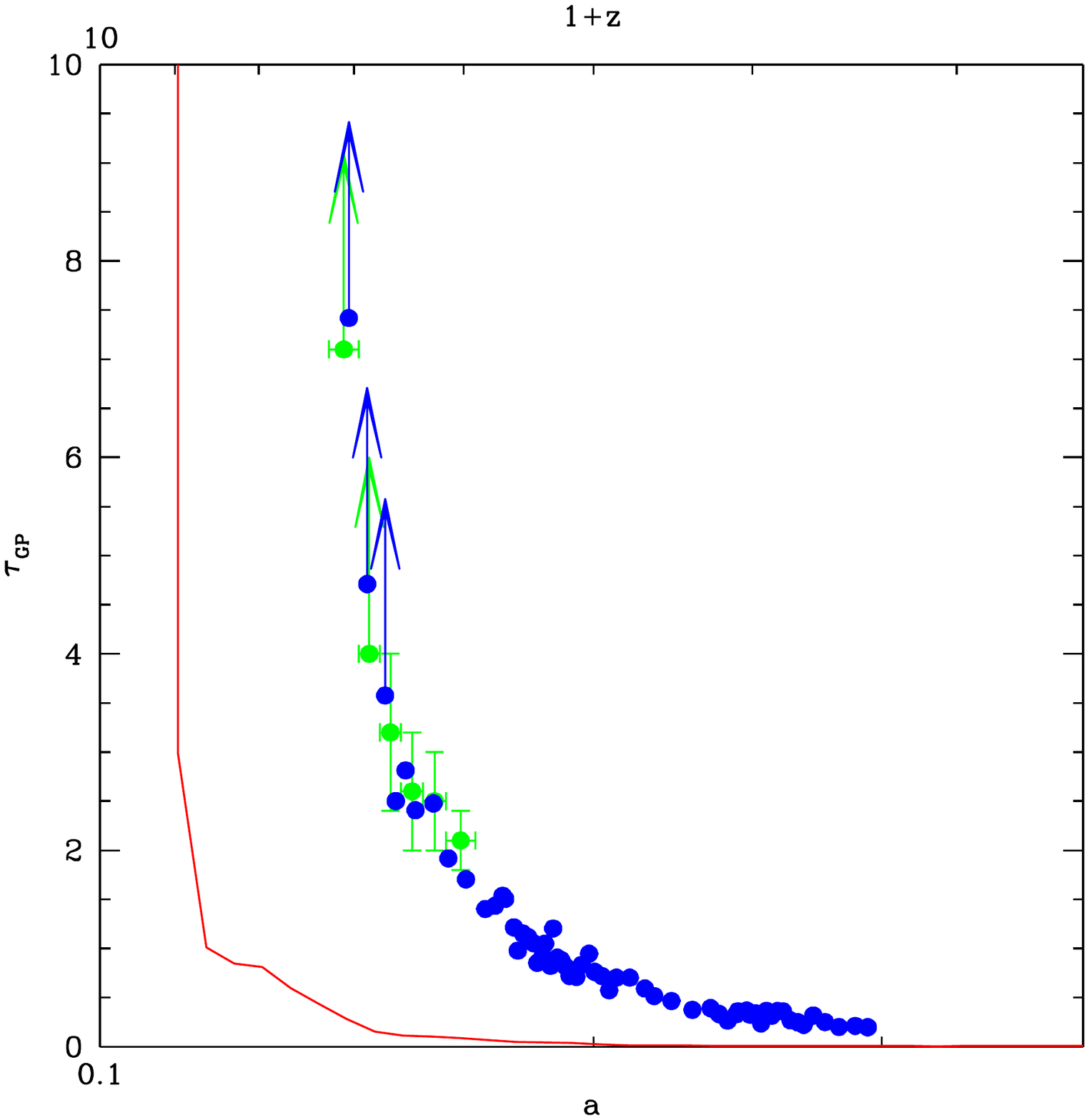} &
 \includegraphics[width=80mm,viewport=7mm 55mm 205mm 255mm,clip]{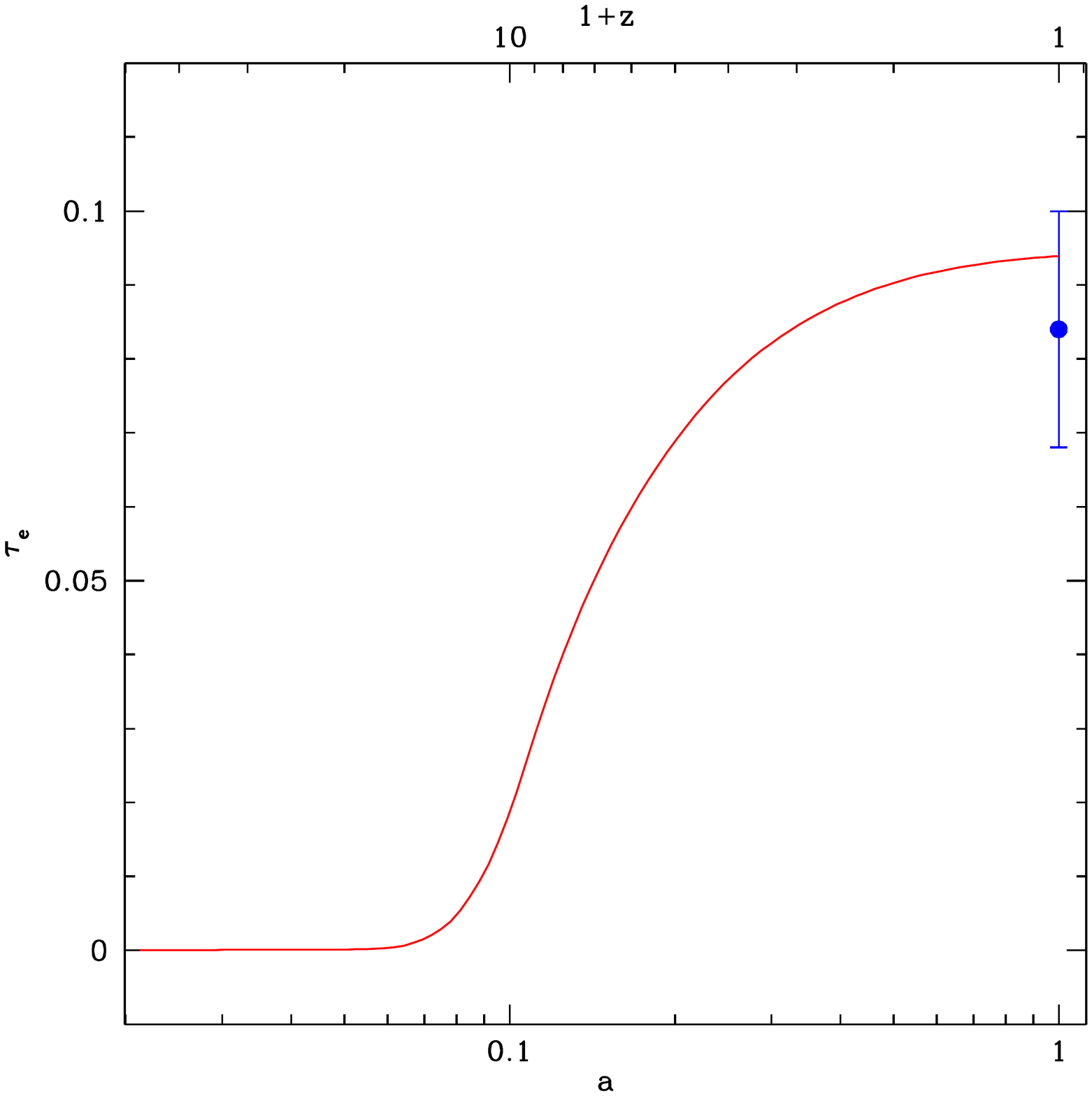}
 \end{tabular}
 \caption{\emph{Left-hand panel:} The Gunn-Peterson \protect\pcite{gunn_density_1965} optical depth as a function of expansion factor and redshift in our best-fit model. Points show observational constraints from \protect\citeauthor{songaila_evolution_2004} (\protect\citeyear{songaila_evolution_2004}; blue points) and \protect\citeauthor{fan_constrainingevolution_2006} (\protect\citeyear{fan_constrainingevolution_2006}; green points). \emph{Right-hand panel:} The electron scattering optical depth to the \protect\CMB\ as a function of redshift in our best-fit model. The blue point shows the \protect\WMAP 5 constraint \protect\pcite{dunkley_five-year_2009}.}
 \label{fig:IGM_tau}
\end{figure*}

Figure~\ref{fig:IGM_tau} shows the Gunn-Peterson \pcite{gunn_density_1965} and electron scattering optical depths as a function of redshift. The Gunn-Peterson optical depth rises sharply at the epoch of reionization becoming optically thick at $z=8$. The rise in Gunn-Peterson optical depth is offset from that seen in observations of high redshift quasars, suggesting that reionization of hydrogen occurs somewhat too early in our model, although \cite{becker_evolution_2007} have argued that this trend in optical depth does not necessarily coincide with the epoch of reionization, but is instead consistent with a smooth extrapolation of the Lyman-$\alpha$ forest from lower redshifts (our model does not include the Lyman-$\alpha$ forest). The electron scattering optical depth is an excellent match to that inferred from \WMAP\ observations of the cosmic microwave background (i.e. consistent within the errors) suggesting that our model reionizes the Universe at the correct epoch.

One of the key effects of the reionization of the Universe is to suppress the formation of galaxies in low mass dark matter halos. We find that the accretion temperature, $T_{\rm acc}$, remains approximately constant at around 30,000K below $z=3$, corresponding to a mass scale increasing with time. The filtering mass rises sharply during reionization and remains large until the present day.

We note that the model predicts too much flux at 912\AA\ in the photon background. We suspect that this is due to the fact that our \IGM\ model is uniform. Inclusion of a non-uniform \IGM\ (i.e. the Lyman-$\alpha$ forest) would result in a greater mean optical depth and would reduce the model flux.

\subsection{Additional Results}\label{sec:Predictions}

In this section we present two additional results that were not used to constrain the model, and therefore represent predictions.

\subsubsection{Gas Phases}\label{sec:GasPhases}

\begin{figure*}
 \begin{tabular}{cc}
 \includegraphics[width=80mm,viewport=7mm 55mm 205mm 255mm,clip]{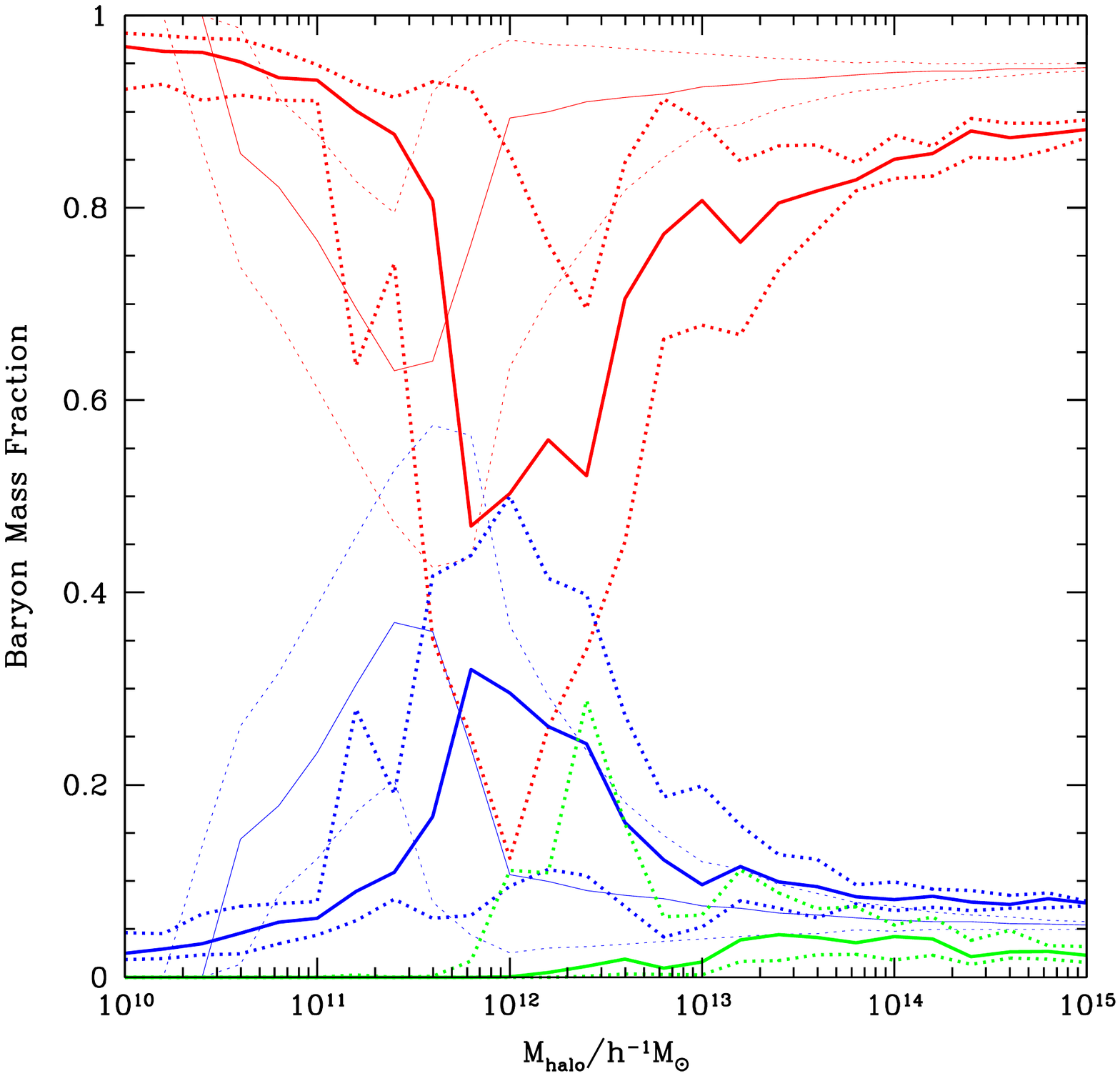} &
 \includegraphics[width=80mm,viewport=7mm 55mm 205mm 255mm,clip]{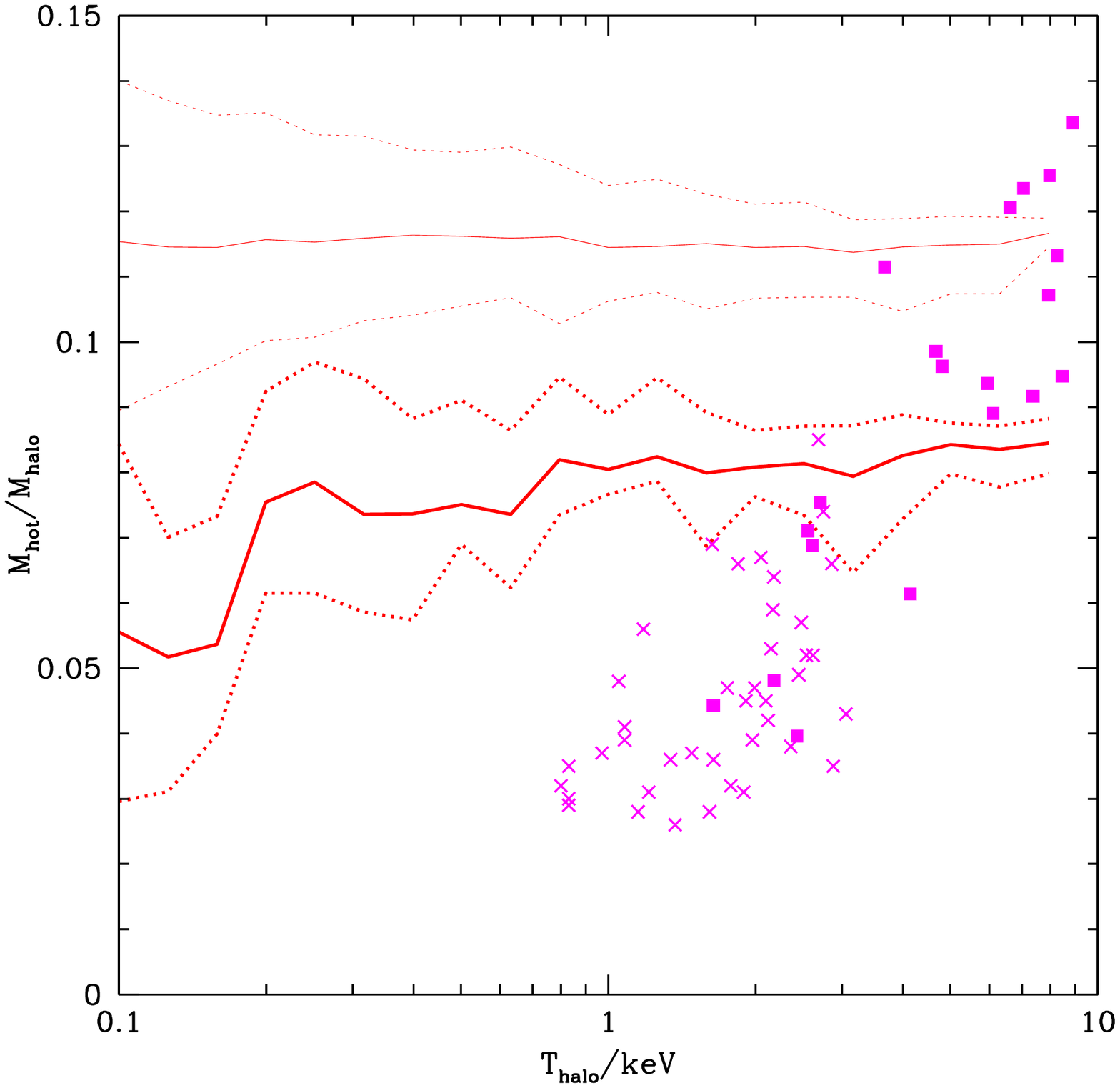}
 \end{tabular}
 \caption{\emph{Left panel:} Solid lines show the median fraction of baryons in different phases as a function of halo mass, while dotted lines indicate the $10^{\rm th}$ and $90^{\rm th}$ percentiles of the distribution. Red lines show gas in the hot phase (which includes any gas in the $M_{\rm reheated}$ reservoir), blue lines gas in the galaxy phase and green lines gas which has been ejected from the halo. Thin lines indicate results from the \protect\cite{bower_breakinghierarchy_2006} model while thick lines show results from the best fit model used in this work. \emph{Right panel:} The ratio of hot gas mass to total halo mass as a function of halo virial temperature is shown by the solid read line. Magenta points show data from \protect\cite{sun_chandra_2009} (crosses) and \protect\cite{vikhlinin_chandra_2009} (squares). Both the observed data and the model results are measured within $r_{2500}$ (the radius enclosing an overdensity of 2500). These data were not included as constraints in our search of the model parameter space.}
 \label{fig:GasPhases}
\end{figure*}

While not included in our fitting procedure, it is interesting to examine the distribution of gas between different phases as a function of dark matter halo mass. Figure~\ref{fig:GasPhases} shows the fraction of baryons in hot (including reheated gas), galaxy (cold gas in disks plus stars in disks and spheroids) and ejected (lost from the halo) phases. The \cite{bower_breakinghierarchy_2006} model (which has no ejected material) shows a peak in galaxy phase fraction at $M_{\rm halo}\approx 2\times 10^{11}h^{-1}M_\odot$ with a rapid decline to lower mass and asymptoting to a constant fraction of 5\% in higher mass halos. This follows the general trend found in semi-analytic models (see, for example, \citealt{benson_nature_2000}) in which supernovae feedback suppresses galaxy formation in low mass halos, while inefficient cooling and \AGN\ feedback does the same in the highest mass halos. In contrast, our best-fit model shows modest ejection of gas in massive halos and a corresponding suppression in the hot gas fraction, although the trends are qualitatively the same as in \cite{bower_breakinghierarchy_2006}. This is different from the dependence of hot gas fraction on halo mass found by \cite{bower_flip_2008}---our current model produces less ejection than found by \cite{bower_flip_2008} resulting in the hot gas fraction being too high in intermediate mass halos. In particular, the right-hand panel of Fig.~\ref{fig:GasPhases}, shows the gas fraction in model halos as a function of hot gas temperature. Model gas fractions were computed within a radius enclosing an overdensity of 2500, just as were the observed data. This radius, and the gas fraction within it, is computed using the dark matter and gas density profiles described in \S\ref{sec:HaloProfiles} and \S\ref{sec:HotGasDist} respectively. Compared to the data (magenta points), the \cite{bower_breakinghierarchy_2006} model is a very poor match, showing almost no trend with temperature. Our best fit model also performs poorly, and it is clear that the suppression in hot gas fraction does not have the correct dependence on halo mass\footnote{Given the hot gas profile assumed in our model and the baryon fraction, the largest ratio of hot gas to dark matter mass we could find here in massive halos is $0.10$ (since the gas profile is cored, but the dark matter profile is not).}. In contrast, the \cite{bower_flip_2008} model produced an excellent match to these data (as it was designed to do). We therefore expect that our best-fit model will not give a good match to the X-ray luminosity-temperature relation, and would instead require more efficient ejection, with a stronger dependence on halo mass in the relevant range, to achieve a good fit. We reiterate that these data were not included as a constraint when searching parameter space for the best-fit model. We will return to this issue in future work, including these constraints directly.

\subsubsection{Intrahalo Light}\label{sec:ICL}

\begin{figure}
 \includegraphics[width=80mm,viewport=7mm 55mm 205mm 255mm,clip]{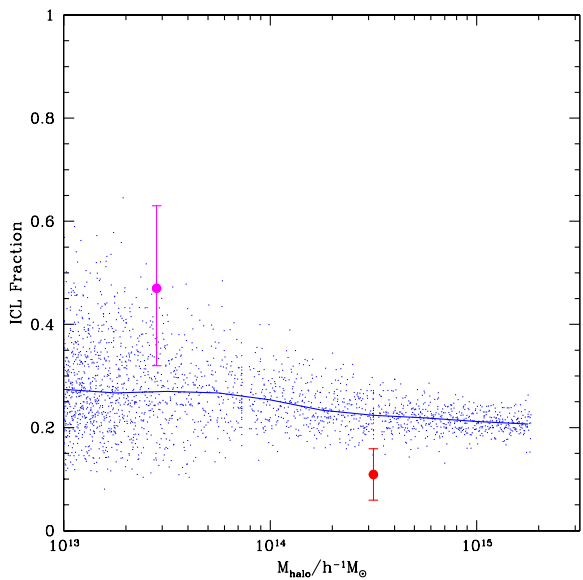} 
 \caption{The fraction of stars which are part of the intrahalo light as a function of halo mass. Blue points show individual model halos, while the blue line shows the running median of this distribution. The magenta and red points indicate the observational determinations of \protect\cite{mcgee_constraintsintragroup_2009} and \protect\cite{zibetti_intergalactic_2005} for groups and clusters respectively.}
 \label{fig:ICL}
\end{figure}

Stars that are tidally stripped from model galaxies become part of a diffuse intrahalo component which we assumes fills the host halo. We can therefore predict the fraction of stars which are found in this intrahalo light as a function of halo mass and compare it to measurements of this quantity. \cite{zibetti_intergalactic_2005} have measured this quantity for clusters, while \cite{mcgee_constraintsintragroup_2009} have measured it for galaxy groups. In Fig.~\ref{fig:ICL} we show their results overlaid on results from our model. Blue points show individual model halos, while the blue line shows the running median of this distribution. The magenta and red points indicate the above mentioned observational determinations for groups and clusters respectively. Our model predicts an intrahalo light fraction which is a very weak function of halo mass, remaining at 20--25\% over two orders of magnitude in halo mass. At fixed halo mass, there is significant scatter, particularly for the lower mass halos. Our predictions are in agreement with the current observational determinations, given their rather large errors bars, and it is clear that in the future such measurements have the potential to provide valuable constraints on models of tidal stripping.

\section{Effects of Physical Processes}\label{sec:Effects}

In the previous section we have explored the effects of varying parameters of the model and their effect on key galaxy properties. We will now instead briefly explore the effects of certain physical processes (those which are either new to this work or have not been extensively examined in the past) on the results of our galaxy formation model. The intent here is not to assess whether these models are ``better'' than our standard model---they all utilize less realistic physical models---but to examine the effects of ignoring certain physical processes or of making certain assumptions. This emphasises one of the key strengths of the semi-analytic approach: the ability to rapidly investigate the importance of different physical processes on the properties of galaxies. Rather than showing all model results in each case, we will show a small selection of model results which best demonstrate the effects of the updated model.

\subsection{Reionization and Photoheating}\label{sec:PhotoEffect}

Our standard model includes a fully self-consistent treatment of the evolution of the \IGM\ and its back reaction on galaxy formation. Two key physical processes are at work here. The first is the suppression of baryonic infall into halos due to the heating of the \IGM\ by the photoionizing background (see \S\ref{sec:BaryonSupress}). The second is the reduction in cooling rates of gas in halos as a result of photoheating by the same background (see \S\ref{sec:Cloudy}). Here, we compare this standard model to a model with identical parameters, but with these two physical processes switched off. (We retain Compton cooling and molecular hydrogen cooling, but revert to collisional ionization equilibrium cooling curves since there is no photon background in this model.)

\begin{figure*}
 \begin{tabular}{cc}
 \vspace{-10mm}\hspace{65mm}a & \hspace{65mm}b\\
 \vspace{10mm}\includegraphics[width=80mm,viewport=7mm 55mm 205mm 255mm,clip]{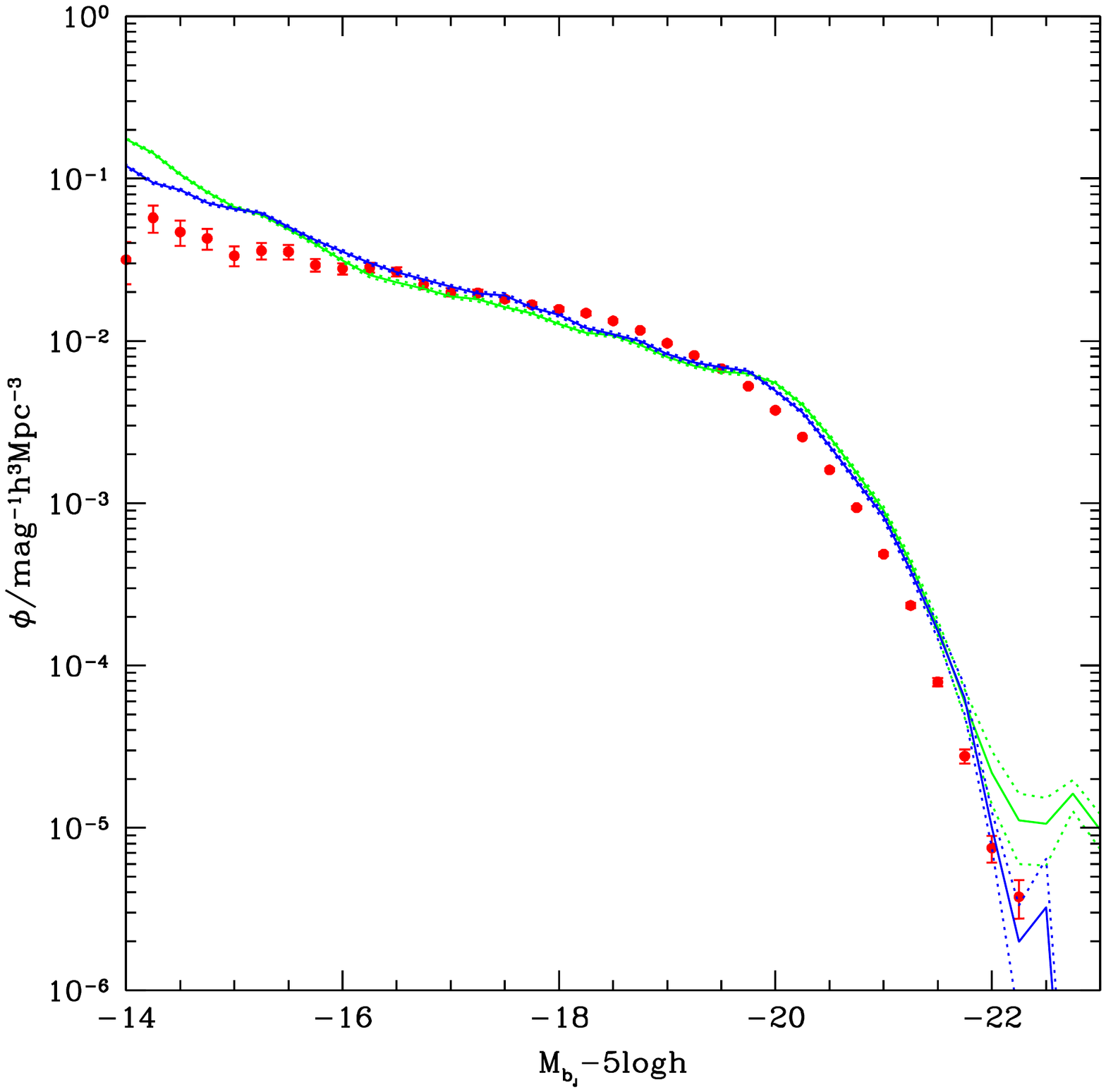} &
 \includegraphics[width=80mm,viewport=7mm 55mm 205mm 255mm,clip]{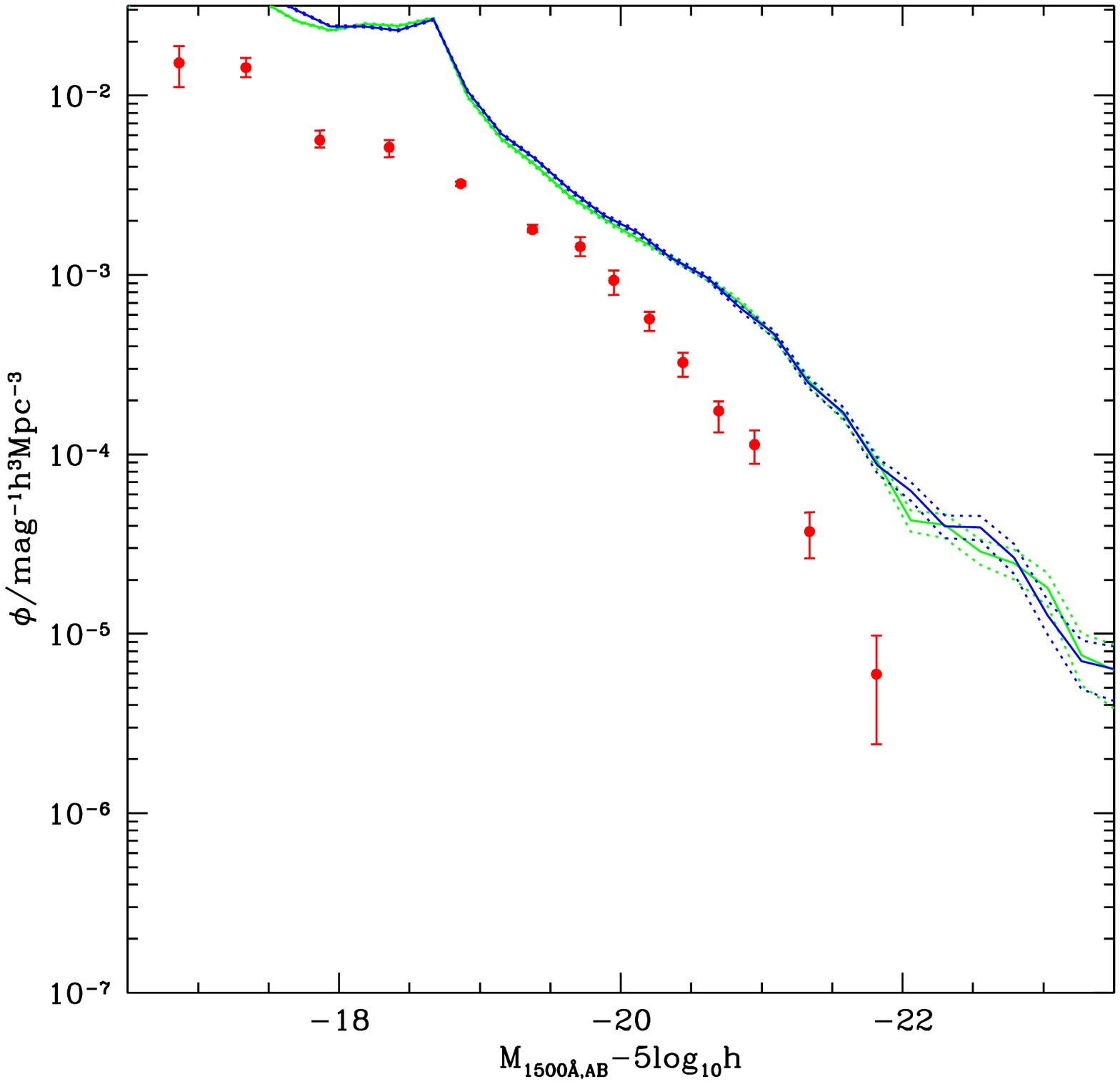} \\
 \vspace{-10mm}\hspace{65mm}c & \hspace{65mm}d\\
 \includegraphics[width=80mm,viewport=7mm 55mm 205mm 255mm,clip]{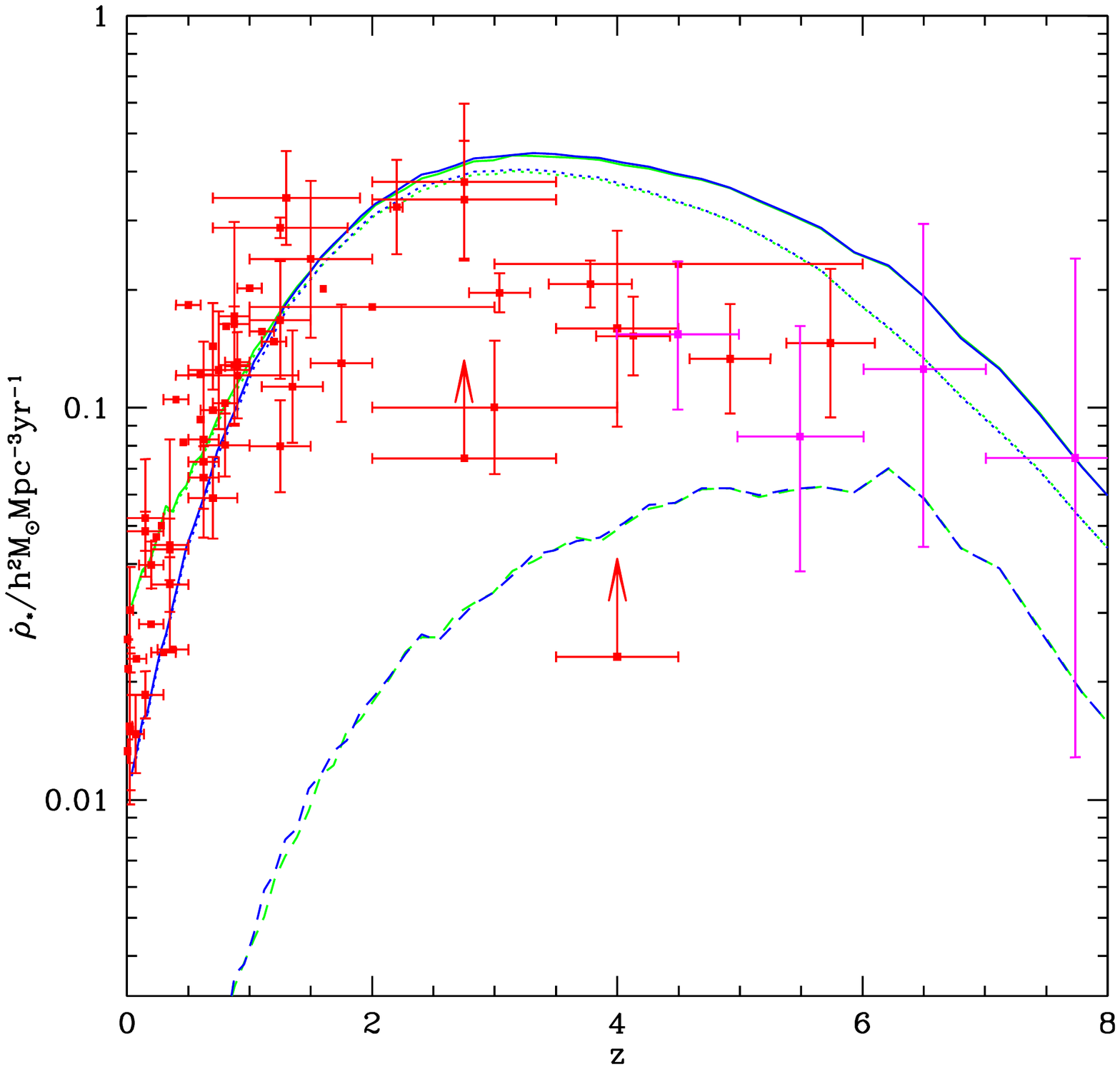} &
 \includegraphics[width=80mm,viewport=7mm 55mm 205mm 255mm,clip]{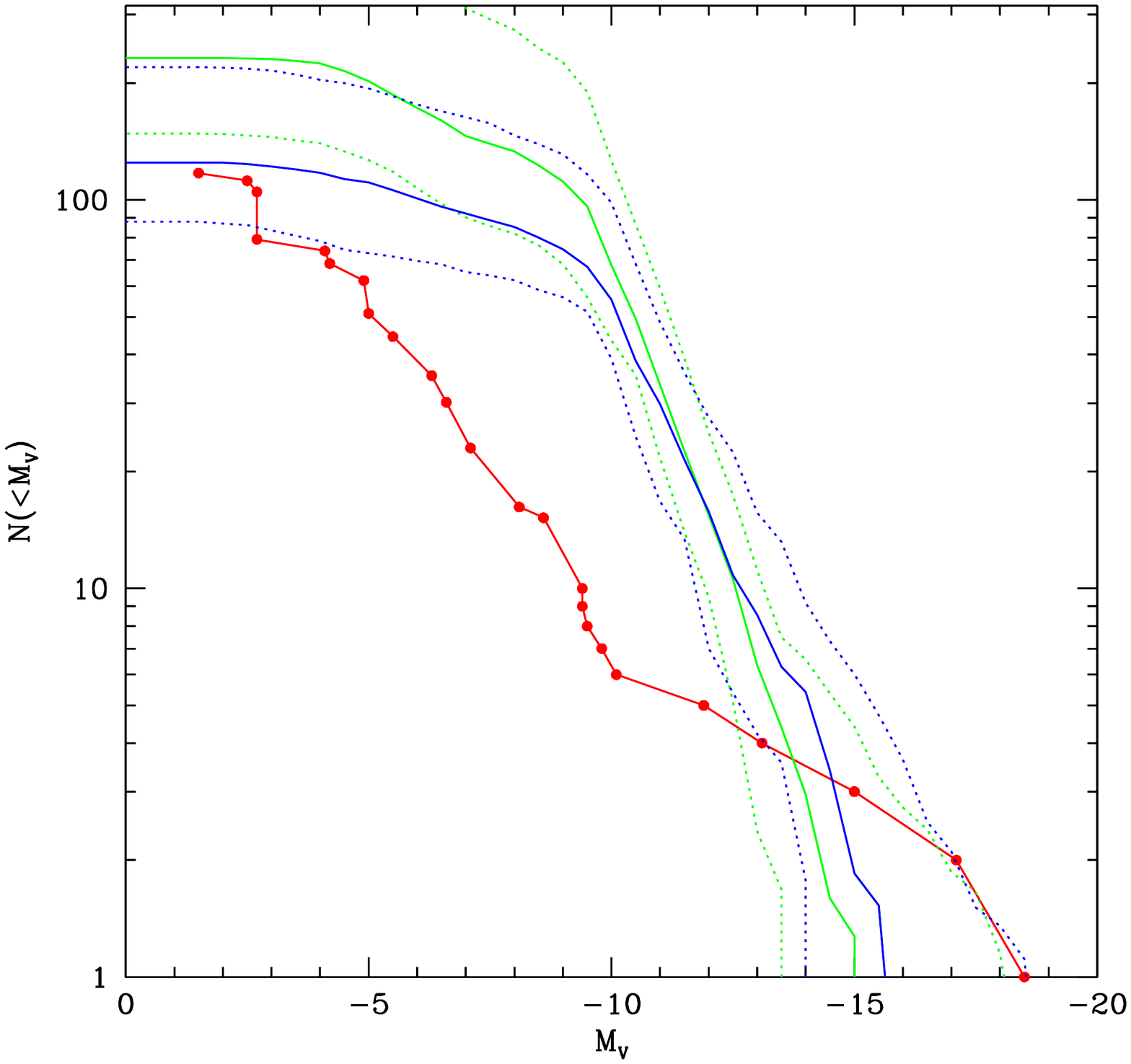} \\
 \end{tabular}
 \caption{Comparisons between our best-fit model (blue lines) and the same model without the effects of suppression of baryonic accretion or photoionization equilibrium cooling (green lines). \emph{Panel a:} The $z=0$ b$_{\rm J}$-band luminosity function as in Fig.~\protect\ref{fig:bJ_LF}. \emph{Panel b:} The $z=5$ 1500\AA\ luminosity function as in Fig.~\protect\ref{fig:z5_6_LF}. \emph{Panel c:} The mean star formation rate density in the Universe as a function of redshift as in Fig.~\protect\ref{fig:SFH}. \emph{Panel d:} The luminosity function of Local Group satellite galaxies as in Fig.~\protect\ref{fig:LocalGroup_LF}.}
 \label{fig:NoReion}
\end{figure*}

Figure~\ref{fig:NoReion} shows some of the key effects of making these changes to our best-fit model. In panel ``a'' we show the $z=0$ b$_{\rm J}$-band luminosity function. The model with no baryonic accretion suppression or photoheating (green line) shows a small excess of very bright galaxies relative to the best-fit model (blue line) due to slightly different cooling rates in this model which affect the efficiency of \AGN\ feedback. As shown in panel ``b'' of Fig.~\ref{fig:NoReion}, the $z=5$ and $z=6$ \UV\ luminosity functions are almost identical in this variant model and our best-fit model. At these higher redshifts \AGN\ feedback has yet to become a significant factor in galaxy evolution. A small excess of galaxies is seen in the model with no baryonic accretion suppression or photoheating at the faintest magnitudes plotted. This is as expected---those mechanisms preferentially suppress the formation of very low mass galaxies.

The effects of this change in the \AGN\ feedback can be seen also in panel ``c'', where we show the star formation history of the Universe. At high redshifts, the two models are nearly identical. However, below $z\approx 1.5$ when \AGN\ feedback begins to come into play, the two models diverge (primarily due to differences in their quiescent star formation rates---the rates of bursting star formation remain quite similar), due to the weakened \AGN\ feedback in this variant model.

Finally, in panel ``d'', we show the luminosity function of Local Group satellites. There is little difference between this variant model and the best-fit model for satellites brighter than about $M_{\rm V}=-10$---photoheating and baryonic suppression play only a minor role in shaping the properties of these brighter satellites. At fainter magnitudes, the variant model predicts more satellites than the best-fit model---by about a factor of two. Suppression of baryonic accretion and photoheating are clearly then important mechanisms for determining the number of satellites in the Local Group, but other baryonic effects (namely \SNe\ feedback) are clearly at work in reducing the number of satellites below the number of dark matter subhalos.

\subsection{Orbital Hierarchy}\label{sec:HierarchyEffect}

\begin{figure*}
 \begin{tabular}{cc}
 \vspace{-10mm}\hspace{65mm}a & \hspace{65mm}b\\
 \vspace{10mm}\includegraphics[width=80mm,viewport=7mm 50mm 205mm 255mm,clip]{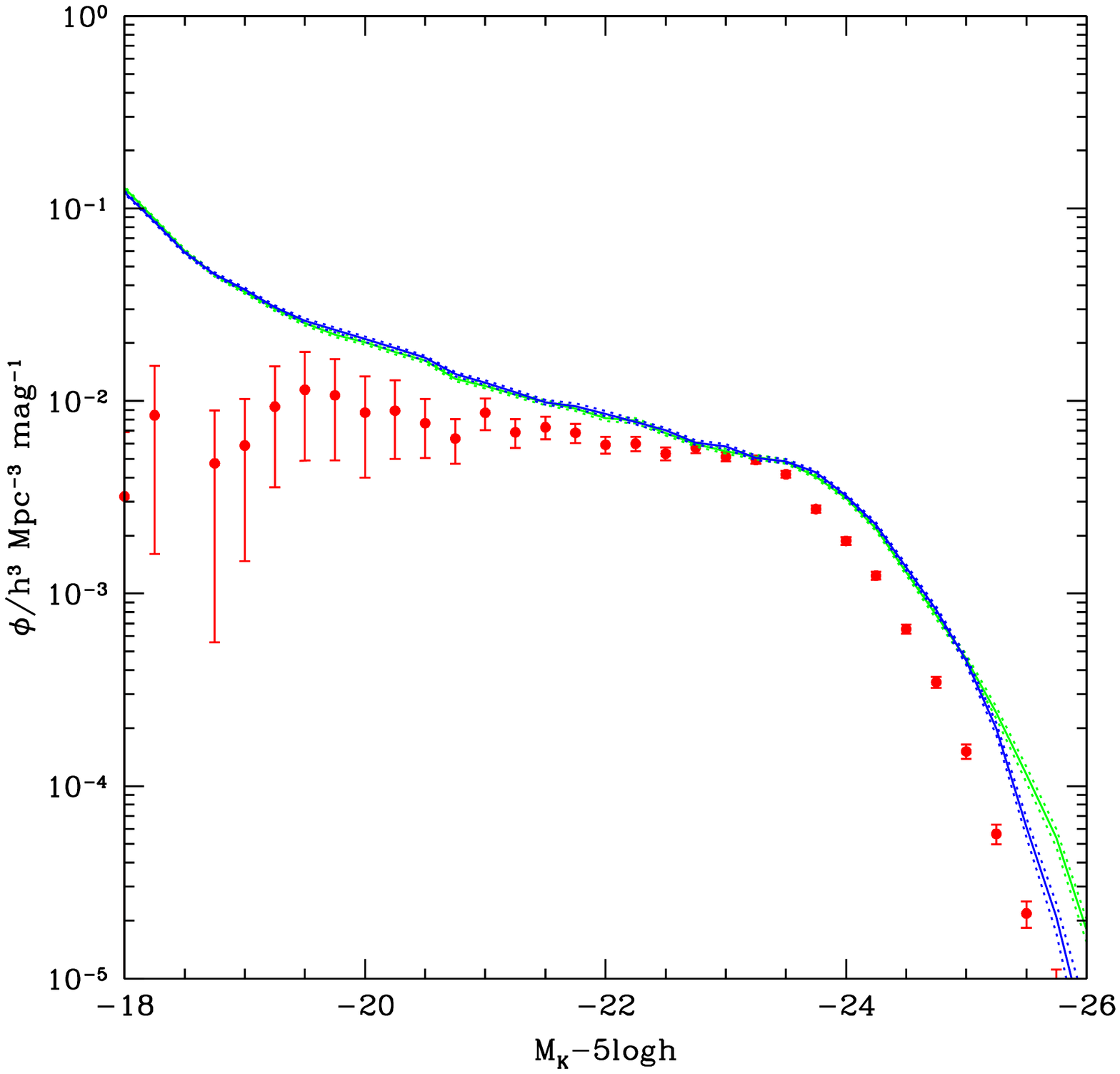} &
 \includegraphics[width=80mm,viewport=7mm 50mm 205mm 255mm,clip]{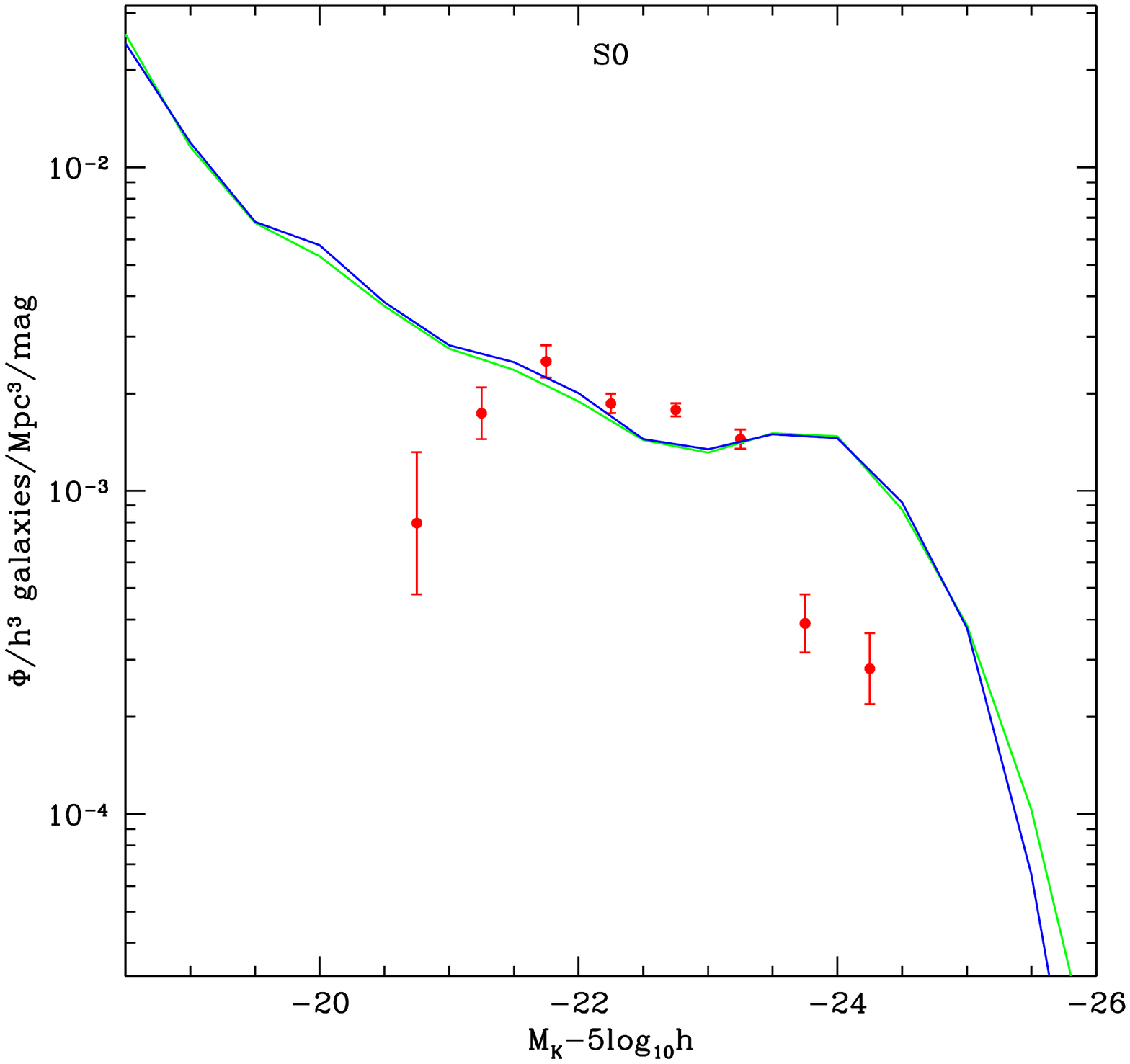}\\
 \vspace{-10mm}\hspace{65mm}c & \hspace{65mm}d\\
 \includegraphics[width=80mm,viewport=7mm 50mm 205mm 255mm,clip]{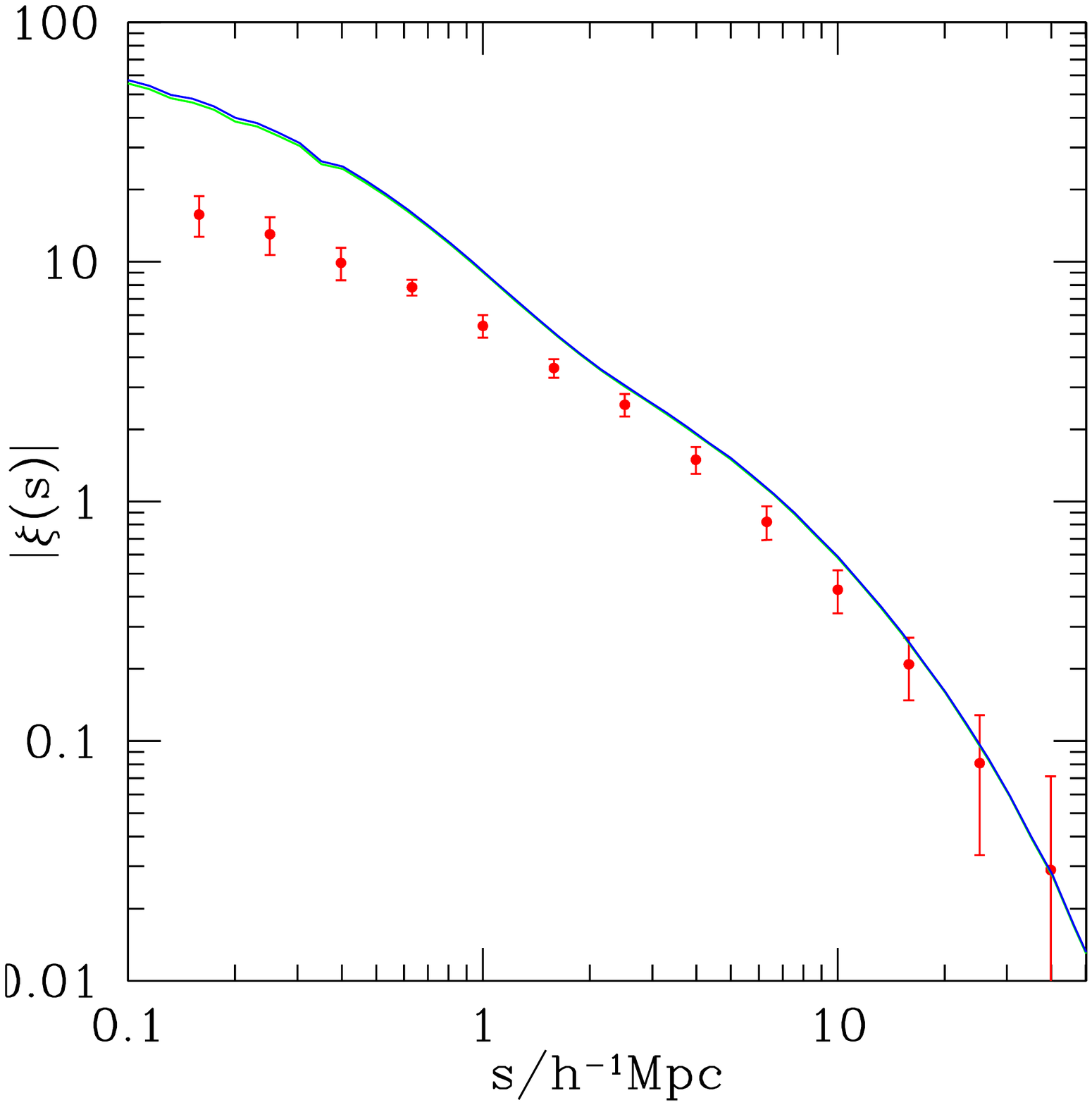} &
  \includegraphics[width=80mm,viewport=7mm 50mm 205mm 255mm,clip]{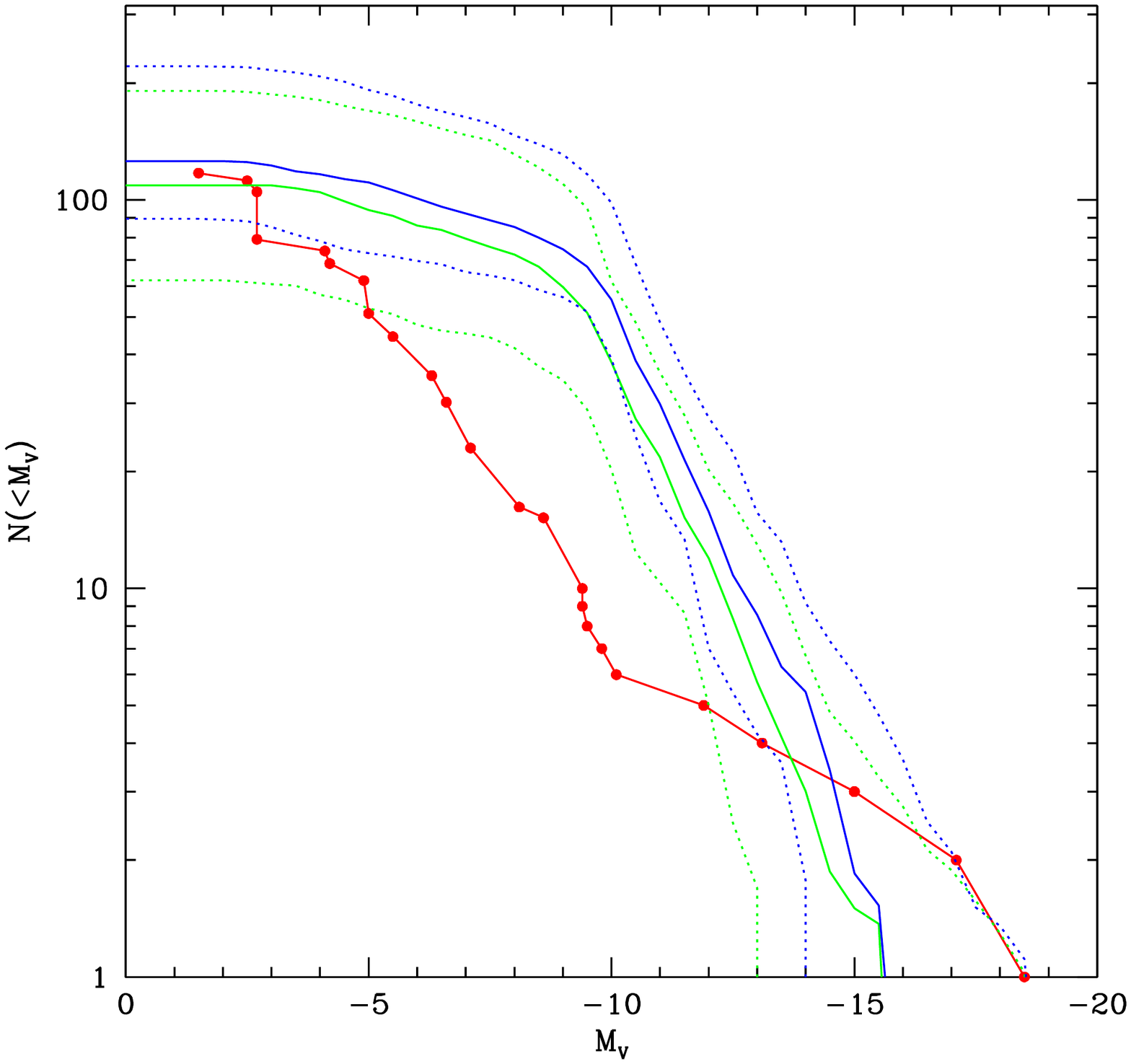}\\
 \end{tabular}
 \caption{Comparisons between our best-fit model (blue lines) and the same model without a full hierarchy of substructures (green lines). \emph{Panel a:} The $z=0$ b$_{\rm J}$-band luminosity function as in Fig.~\protect\ref{fig:bJ_LF}. \emph{Panel b:} The K-band $z=0$ luminosity function of S0 galaxies as in Fig.~\protect\ref{fig:K_Morpho_LF}. \emph{Panel c:} The redshift space two-point correlation function of galaxies with $-18.5<{\rm b}_{\rm J}\le-17.5$ as in Fig.~\protect\ref{fig:2dFGRS_Clustering}. \emph{Panel d:} The luminosity function of Local Group satellite galaxies as in Fig.~\protect\ref{fig:LocalGroup_LF}.}
 \label{fig:NoOrbitalHierarchy}
\end{figure*}

In our standard model, the full hierarchy of substructures (i.e. halos within halos within halos\ldots) is followed (see \S\ref{sec:Merging}). This is in contrast to all previous semi-analytic treatments, in which only the first level of the hierarchy has been considered (i.e. only subhalos, no sub-subhalos etc.). Figure~\ref{fig:NoOrbitalHierarchy} compares results from this variant model (green lines) with those from our best-fit standard model (blue lines). Panel ``a'' of this figure shows the $z=0$ b$_{\rm J}$-band luminosity function of galaxies. Without a hierarchy of substructures we find that this luminosity function is unchanged over most of the range of luminosities shown. The exception is for the brightest galaxies, which become slightly brighter when no hierarchy of substructures is used. These galaxies grow primarily through merging, and this suggests therefore that including a hierarchy of substructures reduces the rate of merging onto these galaxies. At first sight, this seems counter intuitive as galaxies should have more opportunity to merge as they pass through each level of the hierarchy. In fact, this is not the case. A subhalo may sink within the potential well of a halo and then be tidally stripped, releasing any sub-subhalos it may contain into the halo. These sub-subhalos (which become subhalos in their new host) are placed onto new orbits consistent with their orbital position and velocity at the time at which their subhalo was disrupted. The merging timescale for these orbits plus the time they have already spent orbiting with a subhalo can be longer than the merging timescale they would have received if they had been made subhalos as soon as they crossed the virial radius of the host halo. This is due in part to the relatively weak dependence of merging timescale on $r_{\rm C}(E)$ in the \cite{jiang_fitting_2008} fitting formula\footnote{We note that this formula has not been well-tested in the regime in which we are employing it. A more detailed study of the merging timescales and orbits of sub-subhalos is clearly warranted.} and partly due to the fact that sub-subhalos are ejected onto relatively energetic orbits (since they effectively gain a kick in velocity as their subhalo no longer holds them in place).

\begin{figure*}
 \begin{tabular}{cc}
 \vspace{-10mm}\hspace{65mm}a & \hspace{65mm}b\\
 \vspace{10mm}\includegraphics[width=80mm,viewport=7mm 50mm 205mm 255mm,clip]{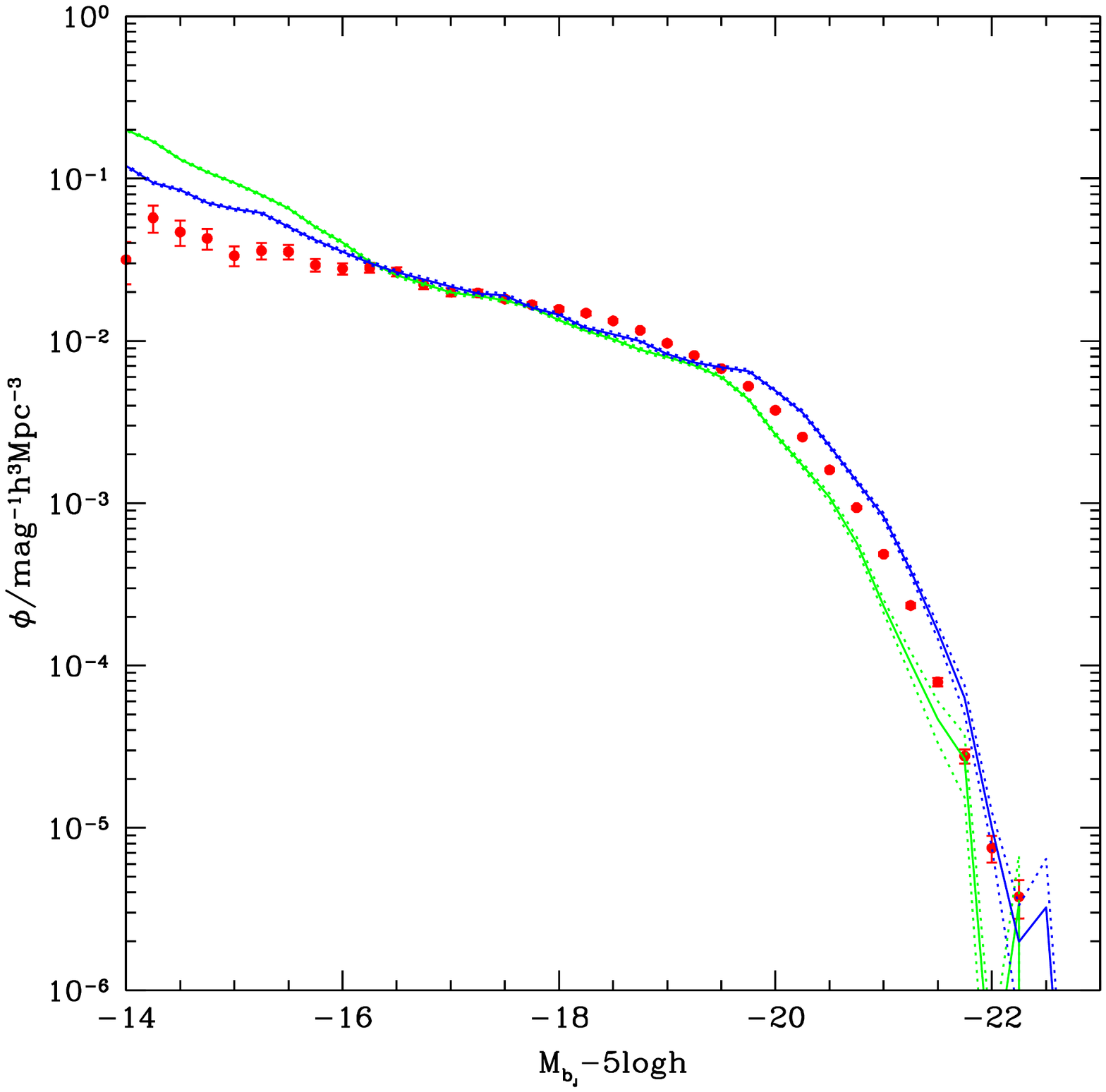} &
 \includegraphics[width=80mm,viewport=7mm 50mm 205mm 255mm,clip]{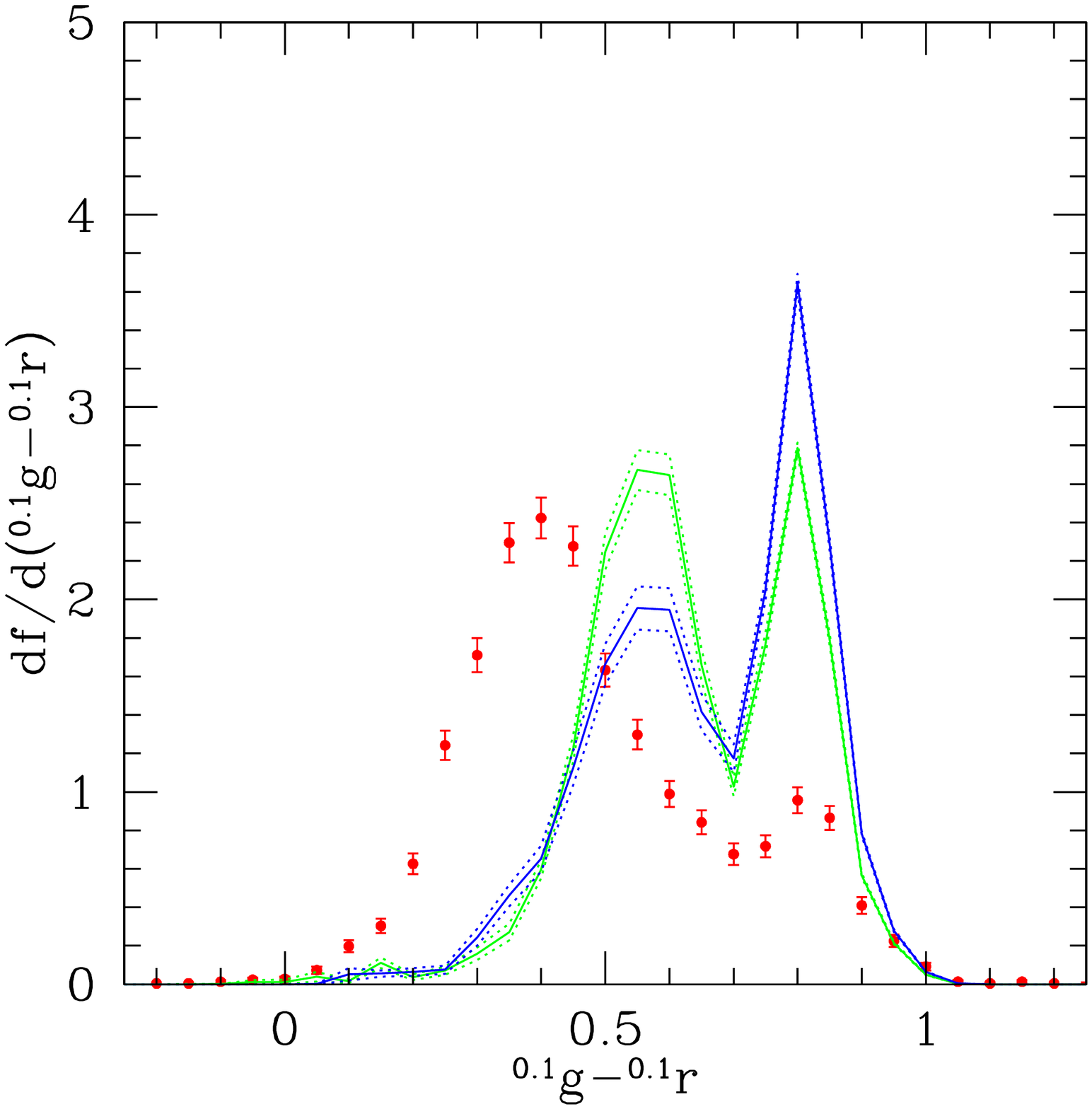}\\
 \vspace{-10mm}\hspace{65mm}c & \hspace{65mm}d\\
 \includegraphics[width=80mm,viewport=7mm 50mm 205mm 255mm,clip]{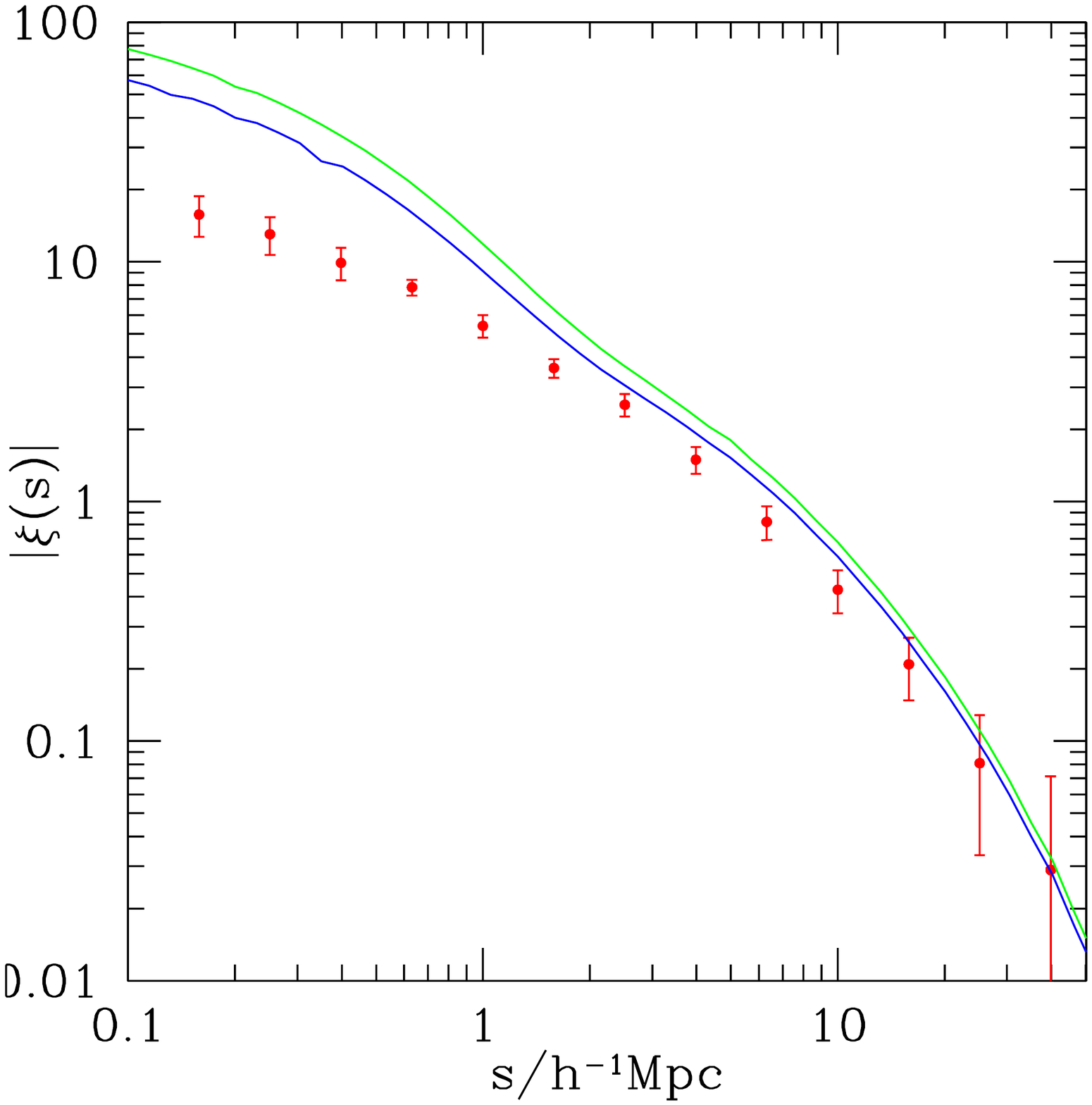} &
  \includegraphics[width=80mm,viewport=7mm 50mm 205mm 255mm,clip]{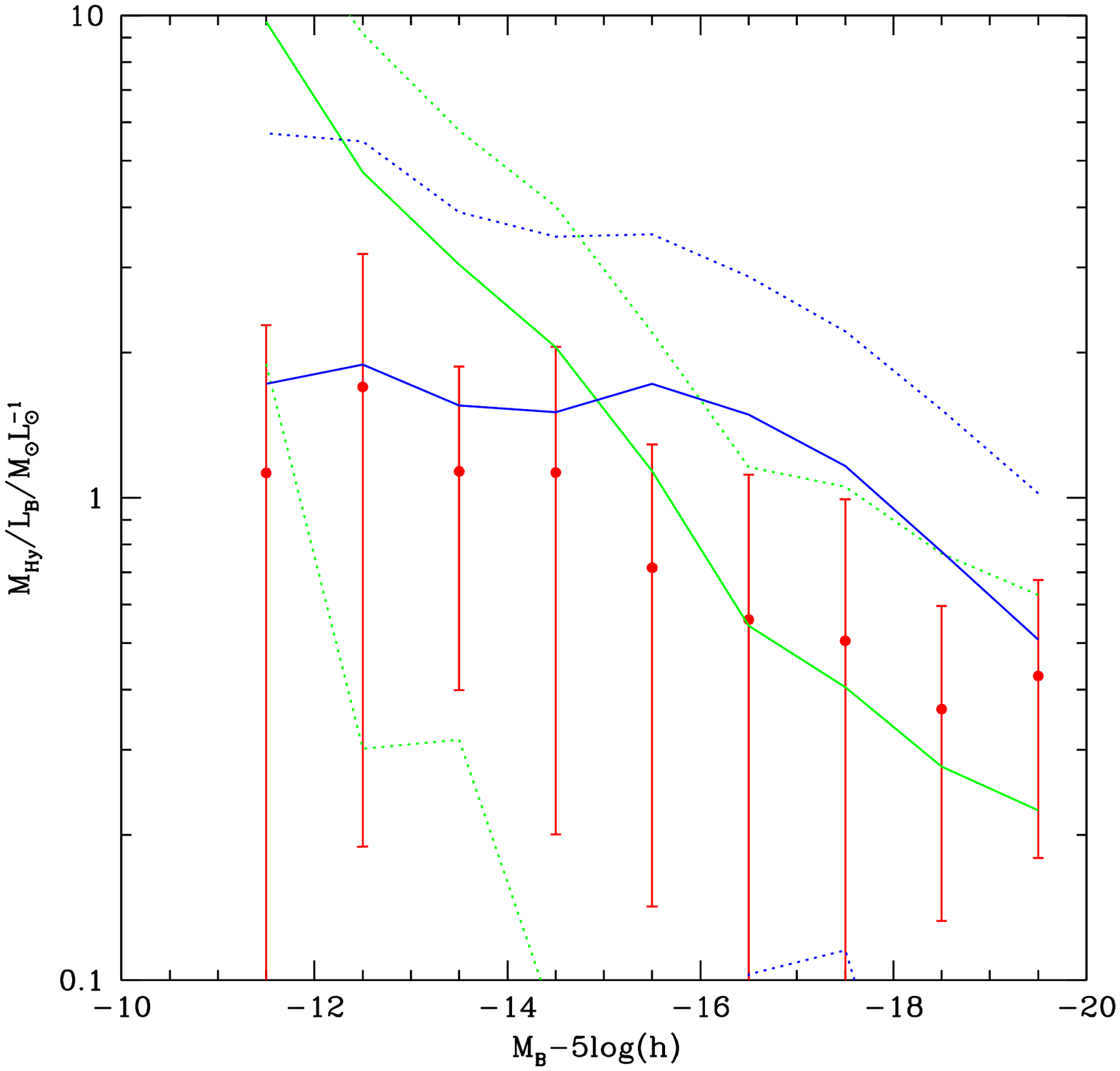}\\
 \end{tabular}
 \caption{Comparisons between our best-fit model (blue lines) and the same model without the effects of tidal or ram pressure stripping of gas and stars from galaxies and their hot atmospheres (green lines). \emph{Panel a:} The $z=0$ b$_{\rm J}$-band luminosity function as in Fig.~\protect\ref{fig:bJ_LF}. \emph{Panel b:} The $^{0.1}$g$-^{0.1}$r colour distribution for galaxies at $z=0.1$ with $-17<M_{^{0.1}\rm g}\le-16$ as in Fig.~\protect\ref{fig:SDSS_Colours}. \emph{Panel c:} The redshift space two-point correlation function of galaxies with $-18.5<{\rm b}_{\rm J}\le-17.5$ as in Fig.~\protect\ref{fig:2dFGRS_Clustering}. \emph{Panel d:} Gas (hydrogen) to B-band light ratios at $z=0$ as a function of B-band absolute magnitude as in Fig.~\protect\ref{fig:Gas2Light}.}
 \label{fig:NoTidalOrRam}
\end{figure*}

Panel ``b'' in Fig.~\ref{fig:NoOrbitalHierarchy} shows that most of the increase in luminosity when the orbital hierarchy is ignored occurs in the S0 morphological class, which, in this model, makes up a significant part of the bright end of the luminosity function. Panel ``c'' shows that the inclusion of the orbital hierarchy makes little difference to the correlation function of galaxies. Mergers between galaxies remain dominated by subhalo-halo interactions, such that this new physics has little impact on the number of pairs of galaxies in massive halos. Finally, panel ``d'' shows the luminosity function of Local Group galaxies. Their numbers are slightly reduced when the orbital hierarchy is ignored, a direct consequence of the slightly increased merger rate.

\subsection{Tidal and Ram Pressure Stripping}\label{sec:StripEffect}

Our standard model incorporates both ram pressure and tidal stripping of gas and stars from galaxies and their hot gaseous atmospheres. We compare this standard model to one in which both of these stripping mechanisms have been switched off. In general, tidal stripping of stars will reduce the luminosity of satellite galaxies. Ram pressure or tidal stripping of gas from galaxies or their hot atmospheres will also reduce the luminosity of satellites and, additionally, may increase the luminosity of central galaxies (since the stripped gas is added to their supply of potential fuel).

\begin{figure*}
 \begin{tabular}{cc}
 \vspace{-10mm}\hspace{65mm}a & \hspace{65mm}b\\
 \vspace{10mm}\includegraphics[width=80mm,viewport=7mm 50mm 205mm 255mm,clip]{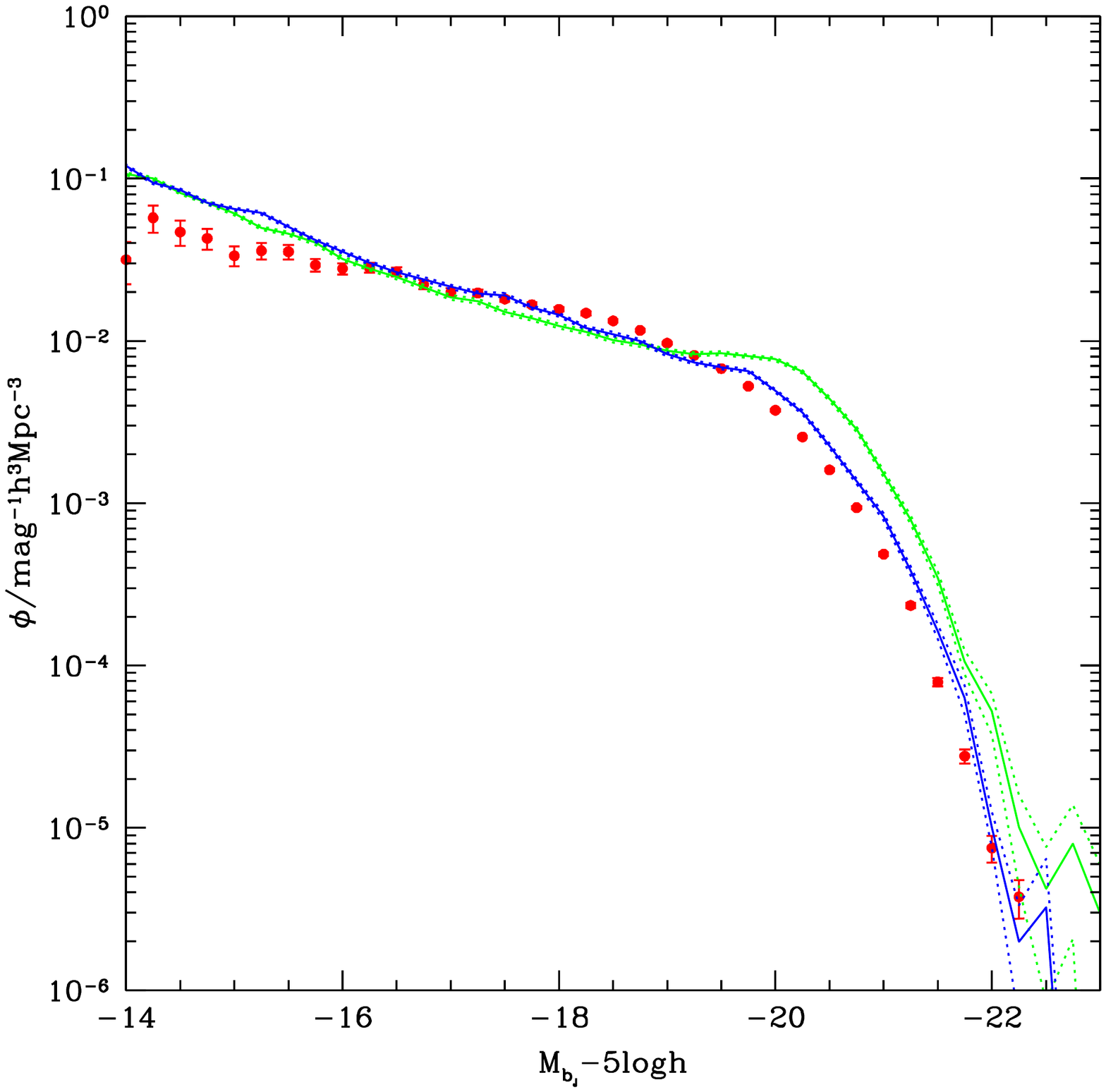} &
 \includegraphics[width=80mm,viewport=7mm 50mm 205mm 255mm,clip]{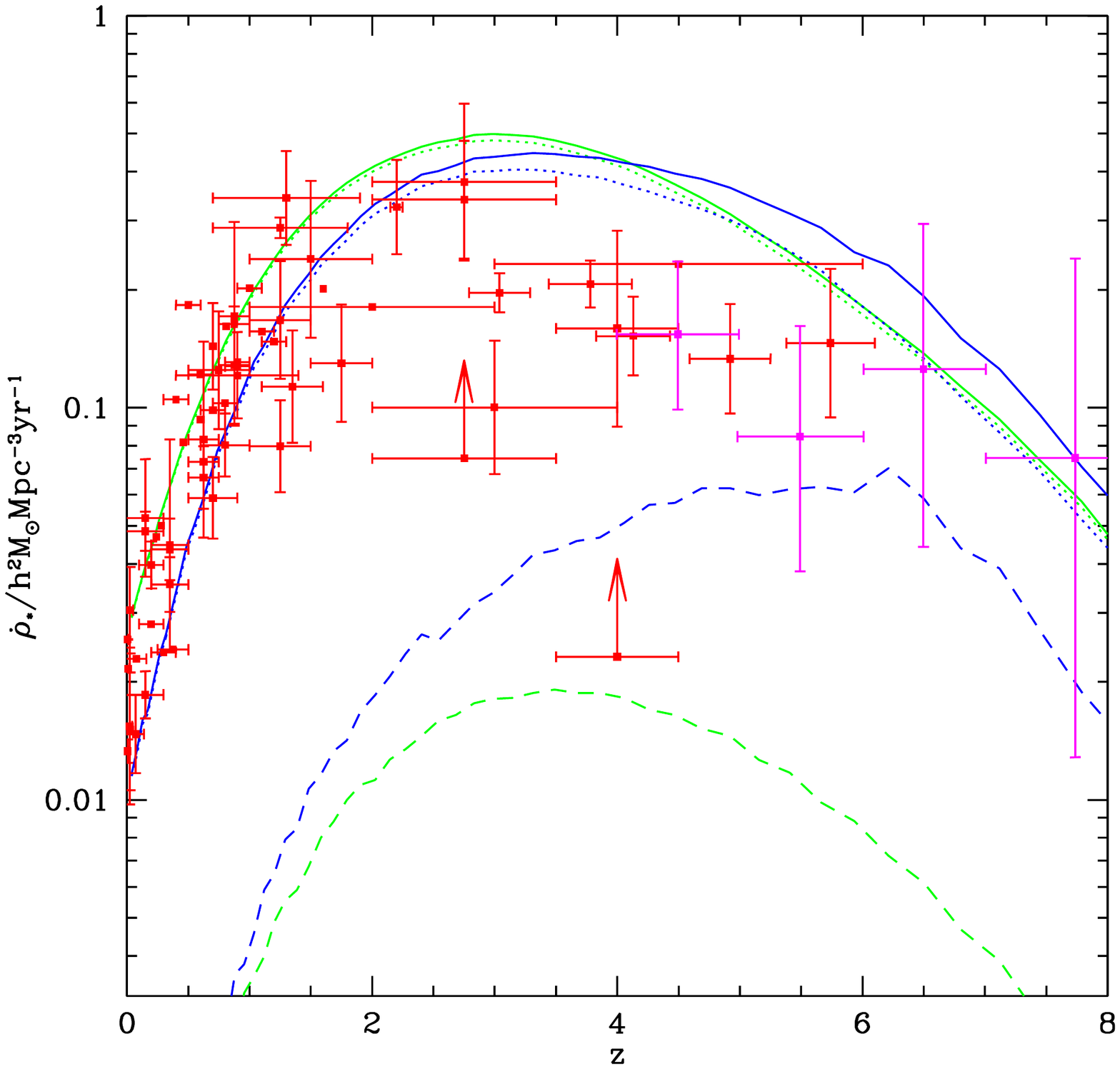}\\
 \vspace{-10mm}\hspace{65mm}c & \hspace{65mm}d\\
 \includegraphics[width=80mm,viewport=7mm 50mm 205mm 255mm,clip]{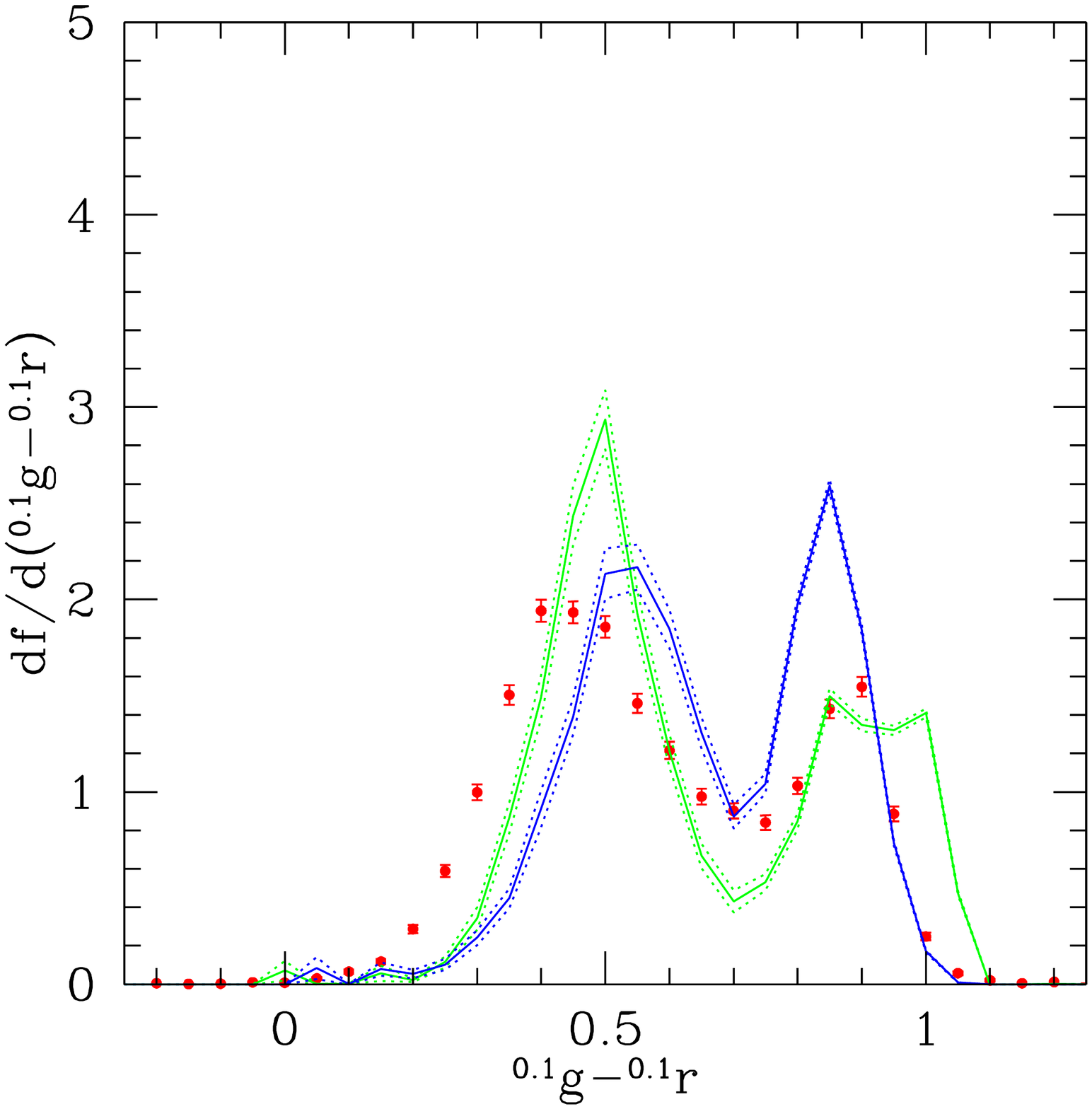} &
  \includegraphics[width=80mm,viewport=7mm 50mm 205mm 255mm,clip]{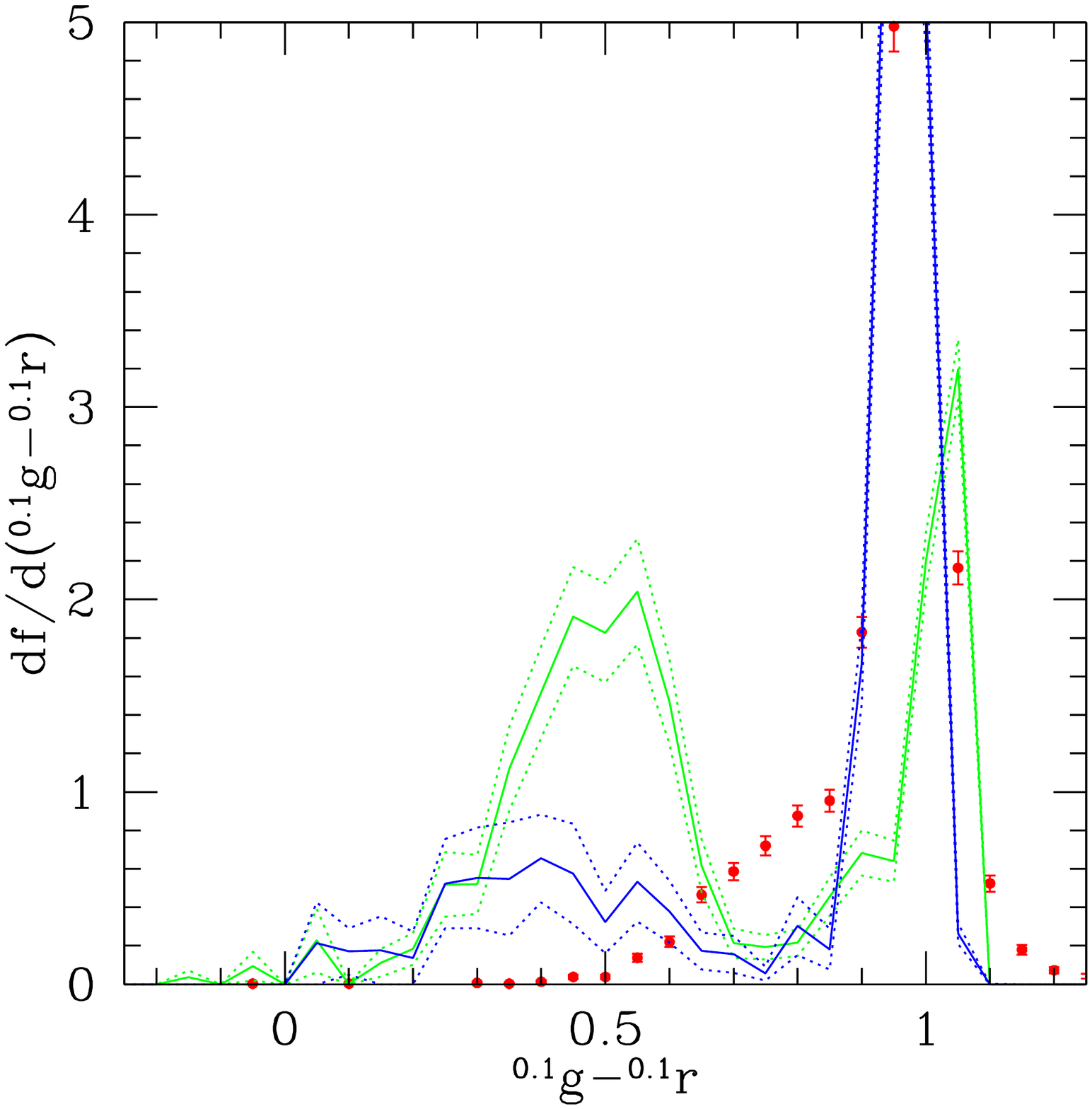}\\
 \end{tabular}
 \caption{Comparisons between our best-fit model (blue lines) and the same model using an instantaneous approximation for recycling, chemical enrichment and \protect\SNe\ feedback (green lines). \emph{Panel a:} The $z=0$ b$_{\rm J}$-band luminosity function as in Fig.~\protect\ref{fig:bJ_LF}. \emph{Panel b:} The star formation rate density as a function of redshift as in Fig.~\protect\ref{fig:SFH}. \emph{Panel c:} The $^{0.1}$g$-^{0.1}$r colour distribution for galaxies at $z=0.1$ with $-18<M_{^{0.1}\rm g}\le-17$ as in Fig.~\protect\ref{fig:SDSS_Colours}. \emph{Panel d:} The $^{0.1}$g$-^{0.1}$r colour distribution for galaxies at $z=0.1$ with $-22<M_{^{0.1}\rm g}\le-21$ as in Fig.~\protect\ref{fig:SDSS_Colours}.}
 \label{fig:NoNonInstant}
\end{figure*}

Figure~\ref{fig:NoTidalOrRam} compares results from the model with no tidal or ram pressure stripping (green lines) with our standard, best-fit model (blue lines). In panel ``a'' we show the b$_{\rm J}$-band luminosity function. At the faintest magnitudes, the model without stripping shows an excess of galaxies relative to the standard model. This is due to low mass galaxies in groups and clusters being stripped of a significant fraction of their stars in the standard model. Conversely, the model without stripping produces fewer of the brightest galaxies (or, more correctly, the bright galaxies that it produces are not quite as luminous as in the standard model). This is a consequence of the fact the ram pressure stripping is able to remove some gas from low mass galaxies, making it available for later accretion onto massive galaxies, allowing those massive galaxies to grow somewhat more luminous. In panel ``b'' we examine the colour distribution of faint galaxies. The model with no stripping produces a shift of galaxies to the blue cloud as expected---with stripping included these galaxies lose their gas supply and quickly turn red.

A further effect of stripping can be seen in panel ``c'' which shows the correlation function of faint galaxies. Without stripping, this is increased on small scales since a greater number of galaxies in massive halos now make it in to the luminosity range selected. Tidal stripping of stars (and, to some extent, ram pressure removal of gas) reduce the luminosities of cluster galaxies and thereby reduce the number of galaxy pairs on small scales in a given luminosity range, thereby helping to reduce small scale correlations. Finally, we show in panel ``d'' the gas to light ratio in a model without stripping. In low mass galaxies the resulting ratio is much higher than in our standard case, a direct result of this gas no longer being removed by ram pressure forces. In more massive galaxies there is, instead a reduction in the gas to light ratio relative to the standard model arising because much of the gas is now locked away in smaller systems and so not available for incorporation into larger galaxies.

Although not shown in Fig.~\ref{fig:NoTidalOrRam} stripping processes have an effect on Local Group galaxies---in the absence of stripping there is a modest increase (by around 50\%) in the number of galaxies brighter than $M_{\rm V}=-10$, but the total number of galaxies is mostly unchanged. Additionally, the sizes of Local Group satellites are larger when stripping processes are ignored as expected (many of the satellites lose their outer portions due to tidal stripping), while metallicities are mostly unaffected.

\subsection{Non-instantaneous Recycling, Enrichment and Supernovae Feedback}\label{sec:NonInstEffect}

\begin{figure*}
 \begin{tabular}{cc}
 \vspace{-10mm}\hspace{65mm}a & \hspace{65mm}b\\
 \vspace{10mm}\includegraphics[width=80mm,viewport=7mm 50mm 205mm 255mm,clip]{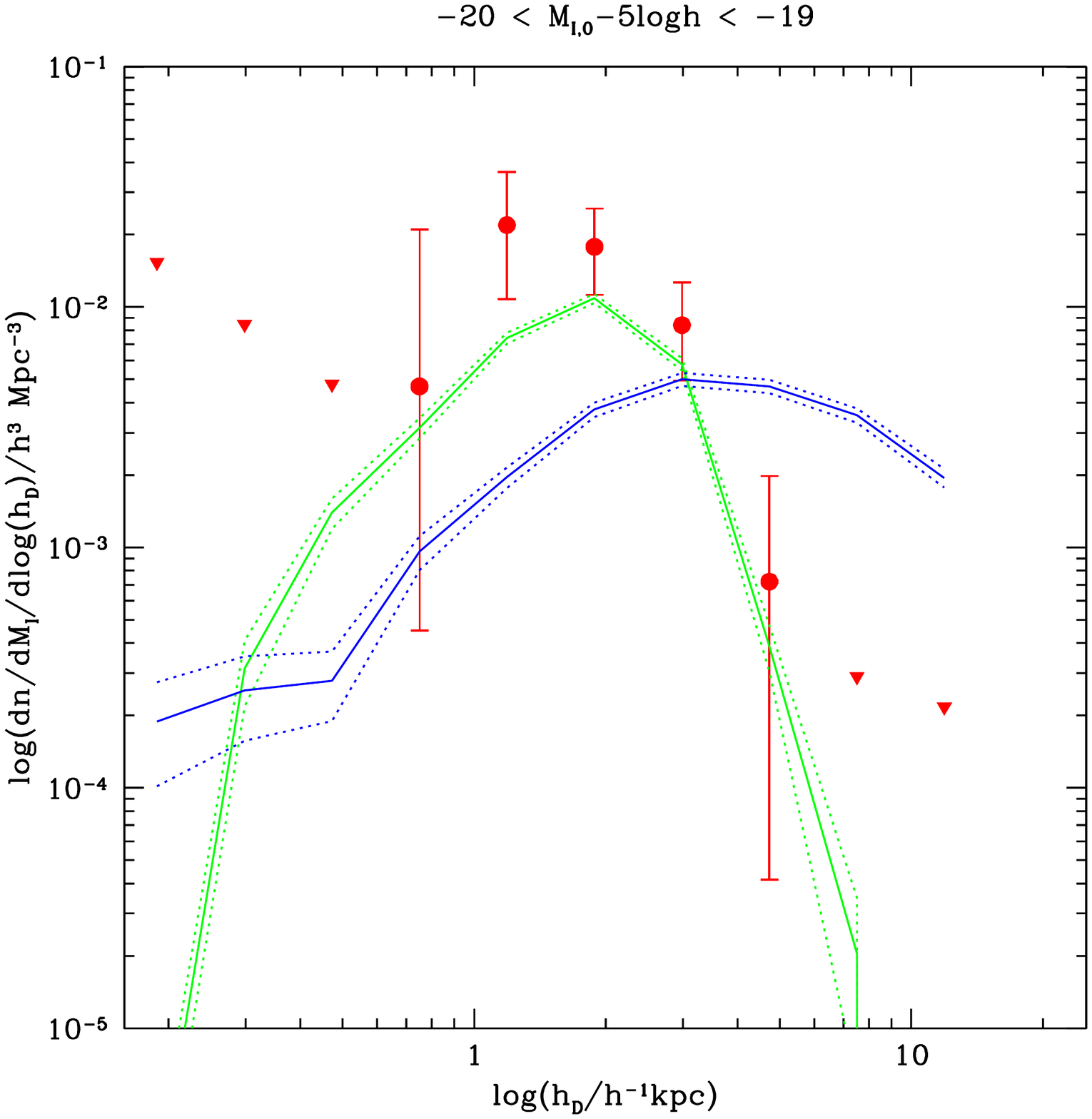} &
 \includegraphics[width=80mm,viewport=7mm 50mm 205mm 255mm,clip]{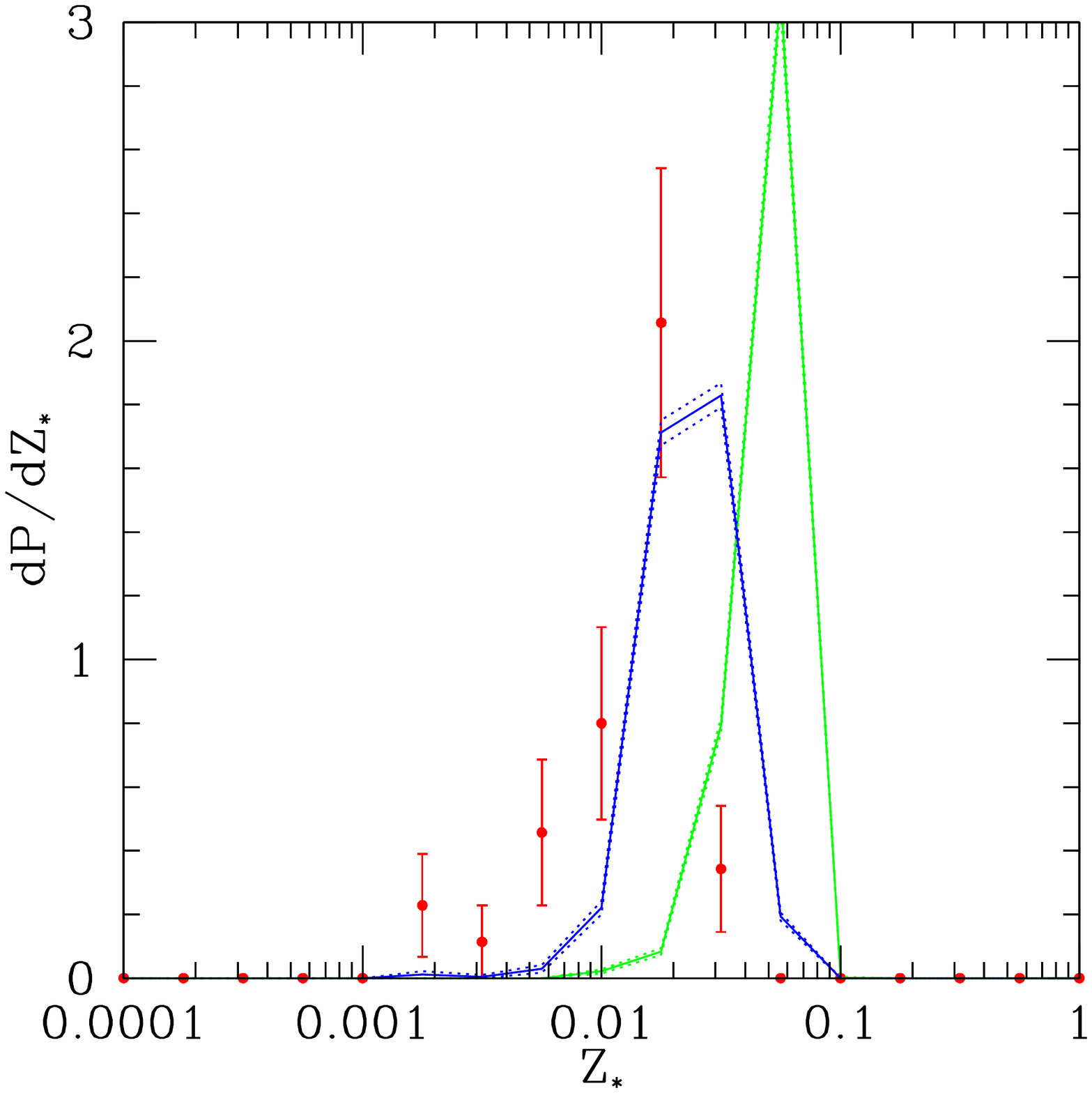} \\
 \vspace{-10mm}\hspace{65mm}\raisebox{0mm}{c} & \hspace{65mm}\begin{tabular}{c} d \\ \raisebox{-60mm}{e} \end{tabular}\vspace{-60mm}\\
 \includegraphics[width=80mm,viewport=7mm 50mm 205mm 255mm,clip]{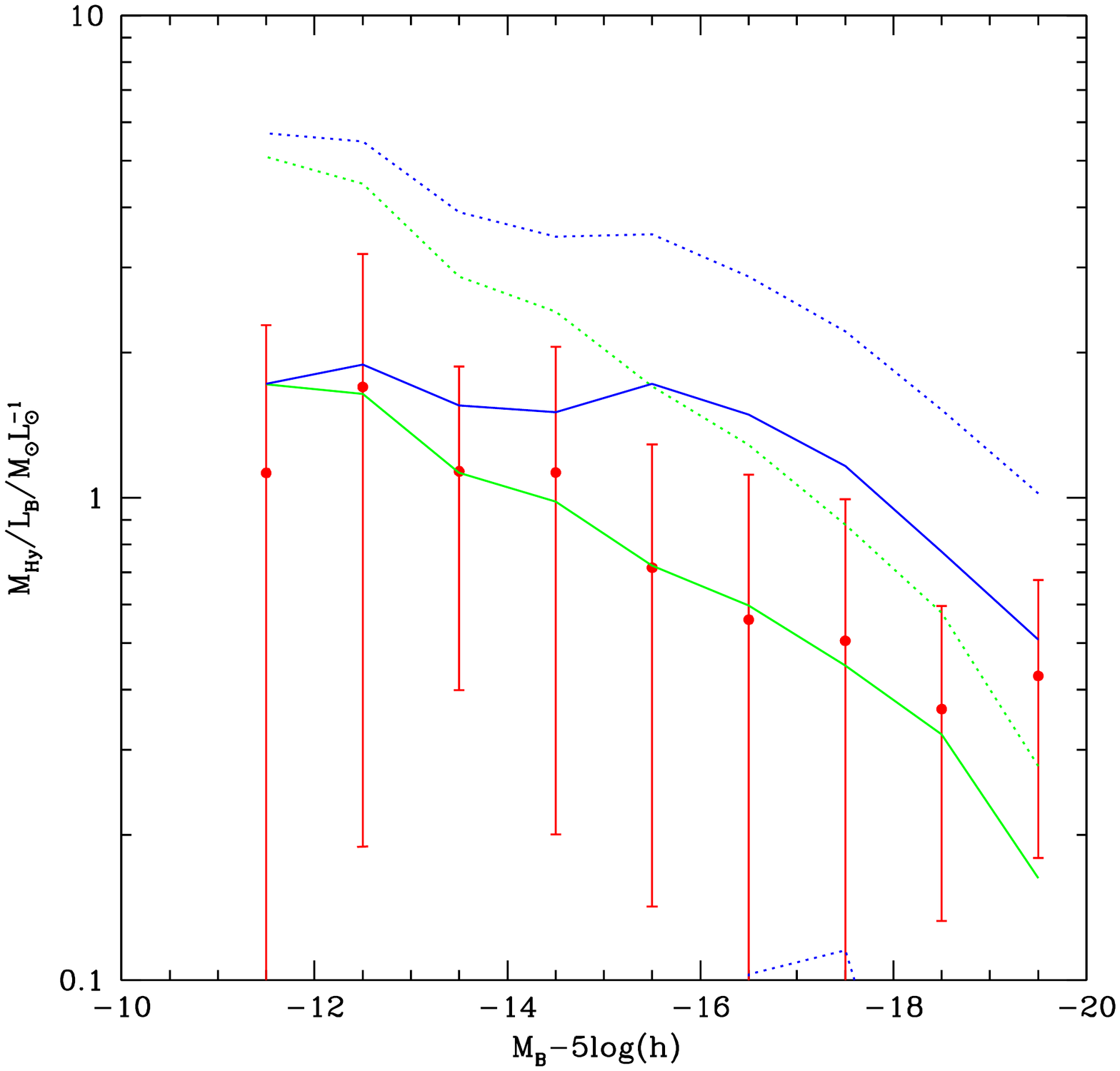} &
 \includegraphics[width=80mm,viewport=0mm 10mm 195mm 265mm,clip]{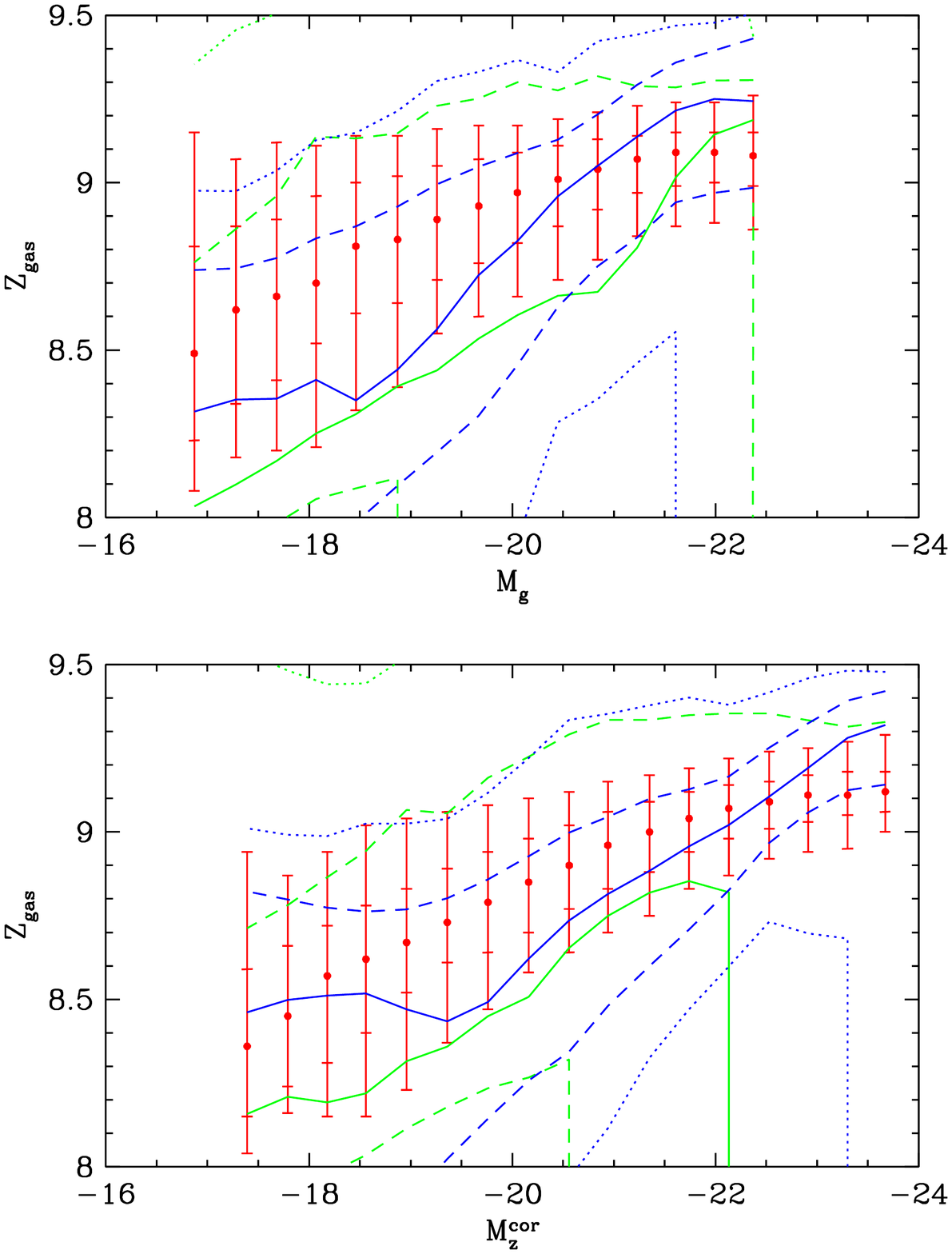}\\
 \end{tabular}
 \caption{Comparisons between our best-fit model (blue lines) and the same model with instantaneous recycling, chemical enrichment and \protect\SNe\ feedback (green lines). \emph{Panel a:} The distribution of disk sizes for galaxies in the range $-20<M_{I,0}-5\log_{10}h\le-19$ as in Fig.~\protect\ref{fig:Other_Sizes}. \emph{Panel b:} The distribution of stellar metallicities for galaxies in the range $-20<M_{\rm B}-5\log_{10}h\le-19$ as in Fig.~\protect\ref{fig:Zstar}. \emph{Panel c:} The ratio of hydrogen gas mass to B-band luminosity as in Fig.~\protect\ref{fig:Gas2Light}. \emph{Panels d \& e:} The gas phase metallicity as a function of absolute magnitude as in Fig.~\protect\ref{fig:SDSS_Zgas}.}
 \label{fig:NoNonInstant2}
\end{figure*}

Our standard model utilizes a fully non-instantaneous model of recycling and chemical enrichment from stellar populations and of feedback from \SNe. We compare this model with one in which the instantaneous recycling approximation is used and in which \SNe\ feedback occurs instantaneously after star formation. In this model, cooling rates are computed from the total metallicity (rather than accounting for the abundances of individual elements as described in \S\ref{sec:Cooling}) since we cannot track individual elements in this approximation. We adopt a yield of $p=0.04$ and a recycled fraction of $R=0.39$ for this instantaneous recycling model. (These values correspond approximately the values expected for a single stellar population with a Chabrier \IMF\ and an age of approximately 10~Gyr.)

Figures~\ref{fig:NoNonInstant} and \ref{fig:NoNonInstant2} compare the results of this model with our best-fit standard model. In Fig.~\ref{fig:NoNonInstant}, panel ``a'' shows that, at $z=0$ the bright-end of the b$_{\rm J}$-band luminosity function is shifted brightwards in the instantaneous model. This is a consequence of the increased metal enrichment in this model which increases cooling rates (which both increases the amount of gas that can cool and increases the mass scale at which \AGN\ feedback becomes effective). This trend is reversed at higher redshifts for the UV luminosity function that we consider. Here, the luminosity function is shifted fainter in the instantaneous model. This effect is due to increased dust extinction in the instantaneous model (which is able to build up metals more rapidly, particularly at high redshifts and so results in dustier galaxies).

Panel ``b'' shows the star formation rate density as a function of redshift. The instantaneous model shows a lower star formation rate at high redshift, and a higher rate at low redshift compared to our standard model. At high redshift this can be seen to be due almost entirely to a change in the rate of bursty star formation. The cause of this is rather subtle: in the non-instantaneous model gas is rapidly locked up into stars at high redshifts and is only slowly returned to the \ISM\ of galaxies. This, coupled with somewhat reduced feedback in the non-instantaneous model (since it takes some time for the \SNe\ to occur after star formation happens) makes disks more massive and therefore more prone to instabilities (see \S\ref{sec:MinorChanges}). The non-instantaneous model has more instability triggered bursts of star formation at high redshift and there is more gas availble to burst in those events. At low redshifts differences in metal enrichment in hot gas in the instantaneous model results in slightly less efficient \AGN\ feedback and, therefore, a higher star formation rate.

Instantaneous enrichment has a big effect on galaxy colours as indicated in panels ``c'' and ``d'' of Fig.~\ref{fig:NoNonInstant}. At faint magnitudes we find a somewhat better fit to the data in the instantaneous model (the blue and red peaks are more widely separated and the red peak is less populated). However, at bright magnitudes the instantaneous model produces too many blue galaxies and too few red ones, resulting in significant disagreement with the data.

Panel ``a'' of Fig.~\ref{fig:NoNonInstant2} shows the sizes of galaxy disks. Remarkably, the instantaneous models shows a much better match to the data than our standard model\footnote{It is worth noting that the \protect\cite{bower_breakinghierarchy_2006} model uses the instantaneous recycling approximation and also does better at matching galaxy sizes than our current best-fit model.}.  This can be traced to a corresponding difference in the distributions of specific angular momenta of disks in the two models, which, in turn, can be traced to the different rates of instability-triggered bursts at high redshifts in the two models. In the non-instantaneous model these happen at a high rate. As a result, the low angular momentum material of these disks is locked up into the spheroid components. Later accretion then results in the formation of disks from higher angular momentum material, resulting in disks that are too large. The stochasticity of this process likewise leads to a large dispersion in disk specific angular momenta and, therefore, sizes. In the instantaneous model the rate of instability-triggered bursts is greatly reduced, allowing disks to retain their early accreted, low angular momentum material, giving smaller disks with less variation in size.

Panel ``b'' shows an example of the distribution of stellar metallicities. Stars in the instantaneous model are enriched to higher metallicities as expected---in the non-instantaneous model it takes time for stars to evolve and produce metals, allowing less enrichment overall. Panels ``c'' and ``d'' show the effects on gas content and metallicity respectively. The gas content is reduced in the instantaneous model and is in excellent agreement with the data. This is a result of the late-time replenishment of the \ISM\ in the non-instantaneous model by material recycled from stars. The instantaneous model produces lower gas phase metallicities, again as a result of the lack of this late-time replenishment which consists of relatively low metallicity material.

\subsection{Adiabatic Contraction}\label{sec:ContractionEffect}

\begin{figure*}
 \begin{tabular}{cc}
 \vspace{-10mm}\hspace{65mm}a & \hspace{65mm}b\\
 \vspace{10mm}\includegraphics[width=80mm,viewport=7mm 50mm 205mm 255mm,clip]{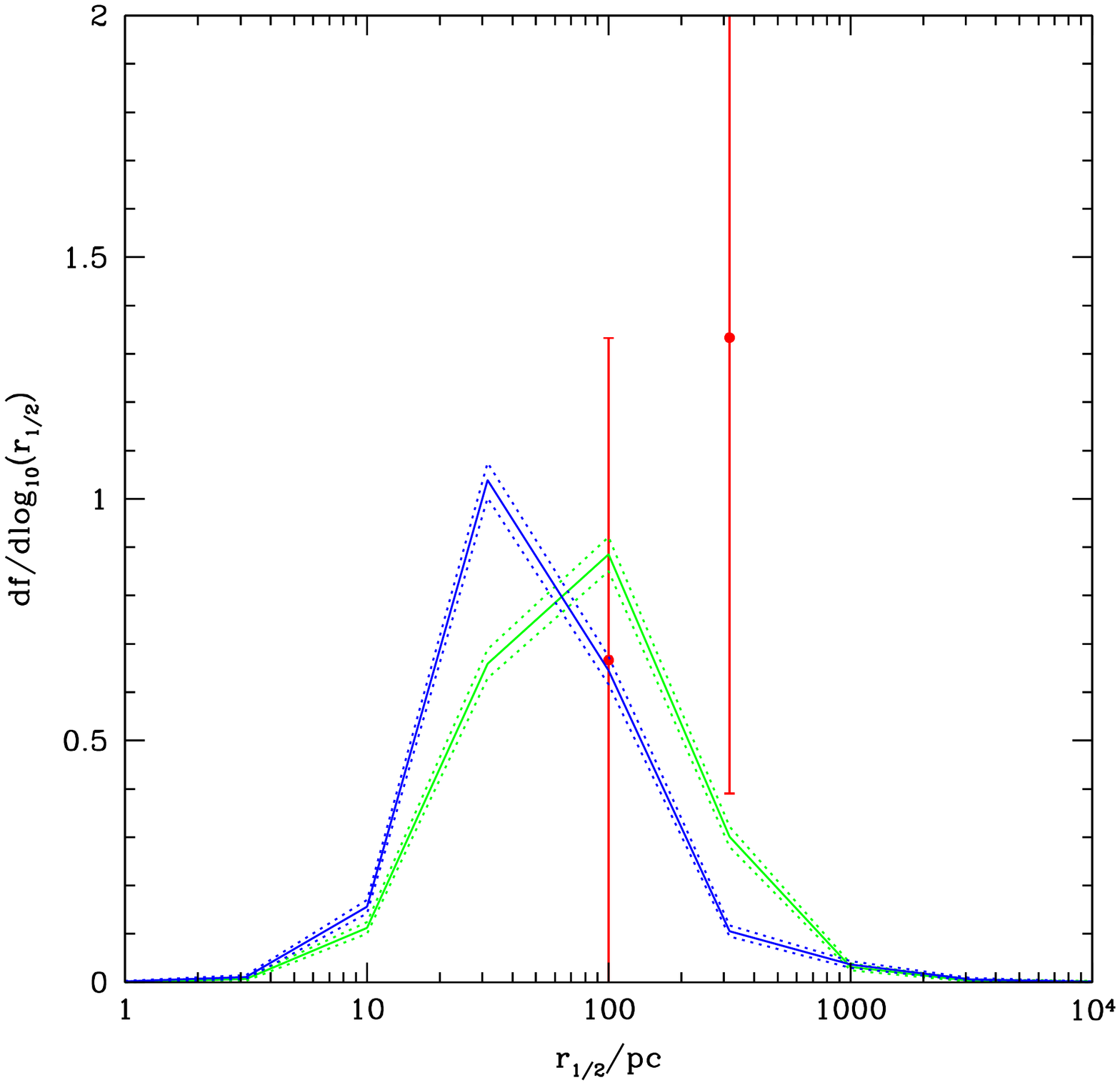} &
 \includegraphics[width=80mm,viewport=7mm 50mm 205mm 255mm,clip]{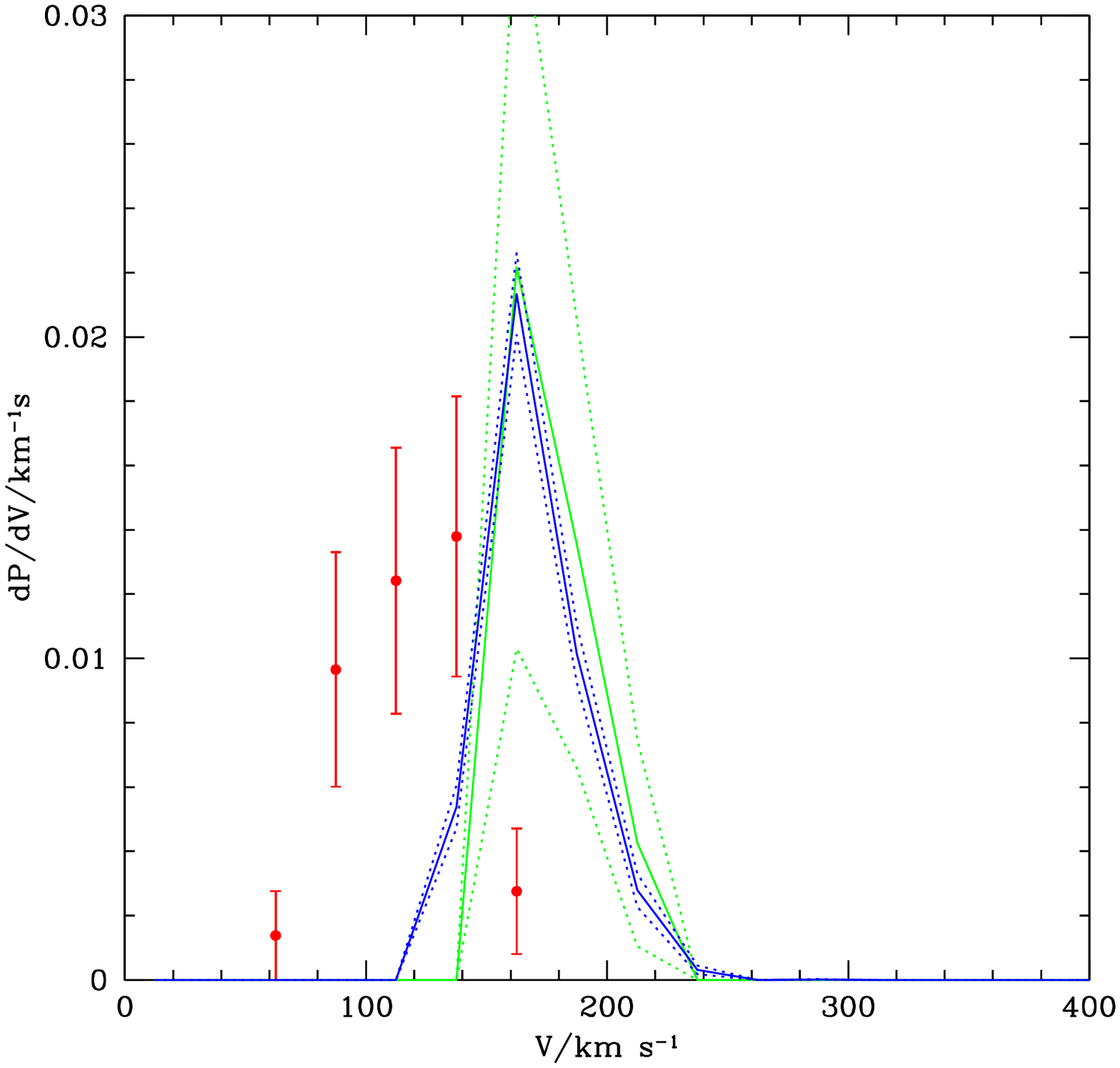}\\
 \end{tabular}
 \caption{Comparisons between our best-fit model (blue lines) and the same model without adiabatic contraction of dark matter halos (green lines). \emph{Panel a:} The distribution of half-light radii for Local Group satellites in the magnitude range $-15 < M_{\rm V} \le -10$ as in Fig.~\protect\ref{fig:LocalGroup_Sizes}. \emph{Panel b:} The Tully-Fisher relation for galaxies in the magnitude range $-21 < M_{i} \le -20$ as in Fig.~\protect\ref{fig:SDSS_TF}.}
 \label{fig:NoContraction}
\end{figure*}

Adiabatic contraction of dark matter halos in response to the condensation of baryons is included in our standard model as described in \S\ref{sec:Sizes}. In Fig.~\ref{fig:NoContraction} we compare our standard model with one in which this adiabatic contraction is switched off such that dark matter halos profiles are unchanged by the presence of baryons. Such a change may be expected to result in galaxies which are somewhat larger and more slowly rotating. Panel ``a'' shows the effects on Local Group satellite galaxy sizes. A slight increase in size is seen as expected.  For larger galaxies, we see a similar effect. Rotation speeds of galaxies are less affected though---panel ``b'' shows a slice through the Tully-Fisher and indicates that switching off adiabatic contraction has actually had little effect on this statistic.

\section{Discussion}\label{sec:Discussion}

We have described a substantially revised implementation of the \gf\ semi-analytic model of galaxy formation. This version incorporates the numerous developments in our understanding of galaxy formation since the last major review of the code \pcite{cole_hierarchical_2000}. Together with
changes to the code to implement black hole feedback \pcite{bower_breakinghierarchy_2006, bower_flip_2008},
ram-pressure stripping \pcite{font_colours_2008} and to track the formation of black holes \pcite{malbon_black_2007}, we have made
fundamental improvements to key physical processes (such as cooling, 
re-ionisation, galaxy merging and tidal stripping) and removed a number of limiting assumptions (in particular, instantaneous recycling and chemical enrichment are no longer assumed). 
In addition to computing the properties of galaxies, the model now self-consistently solves for the evolution of the intergalactic medium and its influence on later epochs of galaxy formation.

The goals of these changes have been three-fold. Firstly, a prime motivation
has been to remove the code's explicit dependence on discrete halo formation events. In the older code, the mass-doubling events were used to
reset halo properties and re-initalise the cooling and free fall
accretion calculations. In turn, this lead to abrupt changes in the
supply of cold gas to the central galaxy which was often not associated with any particular merging event in the haloes history.  The new method avoids such artificial dependencies and leads to smoothly varying gas accretion rates in haloes
with smooth accretion histories, and only leads to abrupt changes during
sufficiently important merging events.  The new scheme explicitly tracks
the energetics of material expelled from galaxies by feedback,
and also allows the angular momentum of the feedback and accreted material
to be self-consistently propagated through the code. 
Secondly, we have aimed to enhance the range of physical processes
treated in the code so that it incorporates the full range of 
effects that are likely to be key in determining galaxy properties. 
In particular, we now include careful treatments of galaxy-environment
interarctions (tidal and ram-pressure stripping), taking into
account the sub-halo hierarchy present within each halo; we take into
account the self-consistent re-ionisation of the IGM and the impact that
this has on gas supply to early galaxies; and we allow for material
to be ejected from haloes (both by star-formation and AGN), broadening the
range of plausible feedback schemes included in the model.  Finally,
the verison of the code described may be driven by accurate Monte-Carlo
realisations of halo merger trees. This allows the uncertainty in the
background cosmological parameters to be factored in to the model
parameter constraints.

We have also advanced the methodology by which we test the model's 
performance by simulataneously comparing the model to a wide range
of observational data. In addition to our conventional approach of 
primarily comparing to local optical and near-IR luminosity functions, we 
now include luminosity function data covering a much greater a range
of redshift and wavelength, the star formation history of the universe,
the distribution of galaxies in colour-space, their gas and metal content,
the Tully-Fisher relation and various observational measurements 
of the galaxy size distribution. In addition, to these galaxy properties
we also use the thermal evolution of the IGM as an additional
constraint.

The drawback of introducing additional physical processes is that this
introduces additional parameters into the model. However, we now
beleive that we have the tools to efficiently explore high dimensional
parameter spaces and thus identify strongly constrained parameter
combinations, and the additional model freedom is much less than the
sum of the observational constraints.
We performed an extensive search of the new model's parameter space utilizing the ``parameter pursuit'' methodology of \cite{bower_parameter_2010} to rapidly search the high-dimensional space. 

This allowed us to find a model which is an adequate description of many of the data sets which were used as constraints. In particular, the model is a good match to local luminosity functions and the overall rate of star formation in the Universe while simultaneously producing reasonable distributions of galaxy colours, metallicities, gas fractions and supermassive black hole masses all while predicting a plausible reionization history.
In many of the original data comparisons, the model gives comparable results to \cite{bower_breakinghierarchy_2006}. In other
comparisons (particularly, colours, metallicities and gas fractions) it greatly improves on the older model.  

Additionally, most of the model parameters
have shifted relatively little compared to the older model.
Where parameters have changed significantly, it is possible to identify
a direct cause. For example, the minimum timescale on which feedback 
material can be re-accreted by a galaxy (which is set  by $\alpha_{\rm reheat}$) is shorter for the new model. This makes good sense since a fraction
of feedback material is now expelled from the system through the new
expulsive feedback channel (see \S\ref{sec:Feedback}). Far from indicating a lack of progress, the comparability of the models is a tremendous success. We cannot emphasise enough how much many of the internal algorithms of the model have been revised: the near stability of the end results suggests a high
degree of convergence, and that adding additional detailing of many 
aspects of the model is not required.

Despite this encouraging success, significant discrepancies between the
model and the data remain in many areas. In particular, the sizes of galaxies are too large in our model (and there is too much dispersion in galaxy sizes). This may reflect a break down in certain model assumptions (e.g. the conservation of angular momentum of gas during the cooling and collapse phase), or that we are still lacking some key physics in this
part of the model model (e.g. dissipative effects during spheroid formation; \citealt{covington_predictingproperties_2008}). In addition to the sizes, our model continues to produce too many satellite galaxies in high mass halos, leading to an overprediction of the small scale clustering amplitude of faint galaxies; and predicts a Tully-Fisher relation offset from that which is observed, despite using the latest models of adiabatic contraction. (We note that \cite{dutton_revised_2007} have demonstrated the difficulty of obtaining a match to the Tully-Fisher relation quite clearly, and have advocated adiabatic expansion or transfer of angular momentum from gas to dark matter to alleviate this problem.) Additionally, at high redshifts the agreement with luminosity function data is relatively poor, but these results are highly sensitive to the very uncertain effects of dust on galaxy magnitudes.

The overall aim of this work was to construct a model that incorporates the majority of our current understanding of galaxy formation and explore the extent to which such a model can reproduce a large body of observational data spanning a range of physical properties, mass scales and redshifts. This is far from being the final word on the progress of this model. Numerous improvements remain to be made---such as the inclusion of a physics-based model of star formation. Nevertheless, the current version has been demonstrated to produce good agreement with a very wide range of observational data. Despite the large number of adjustable parameters current observational data is more than sufficient to constrain this model---the good agreement with that data should be seen as a confirmation of current galaxy formation theory.

We have not attempted, in this work, to explore in detail which physical processes are responsible for which observed phenomena. That, and an investigation of which data provided constraints on which parameters, will be the subject of a future work. The parameter space searching methodology described in this paper is quite efficient and successful, but is presently limited by two factors. The first is the available computing time and speed of model calculations which limits how fine-grained any parameter space search can be. Further optimization of our galaxy formation code coupled with more and faster computers will alleviate this problem, but it will remain a limitation for the near future. The second limitation is our ignorance about how best to combine constraints from different datasets. Some of the observational data that we would like to use is undoubtedly affected by poorly understood systematic errors. As a result it is unclear how a precidence 
should be assigned to each dataset. 
For example, given the robustness of the measurements, are we more
interested in the class of models that accurately match the $z=5$ luminosity, or those that perform better in clustering measurements?
Ideally the model would match both equally well, but underlying systematic
errors may make this impossible.
Furthermore, to utilize the obervational data in a statistically correct way we often require more information (e.g. the full covariance matrix rather than just errors on each data point) than is available. 

The most formidable challenge, however, is to better understand the uncertainty in each model prediction. This is a combination of the variance introduced by the limited number of dark matter halo merger trees that we are able to simulate and the accuracy of the approximations made in computing a given property in the model. The first of these is relatively straightforward to estimate (for example, via a bootstrap resampling approach), but the second is much more difficult. For example, we are quite sure that calculations of dust extinction in rapidly evolving high redshift galaxies are very uncertain, while calculations of galaxy stellar masses at $z=0$ are much more robust. The difficulty arises in assigning a numerical ``weight'' to the model predictions for these different constraints. Beyond simply making an educated guess, one might envisage comparing predictions of dust extinction from our model with a matched sample of simulated high redshift galaxies in which the complicated dynamics geometry and radiative transfer could be treated more accurately. The variance between the semi-analytic and numerical simulation results would then give a quantitative estimate of the model uncertainty. The problem with such an approach is that creating such a matched sample is extremely difficult and time consuming.

In addition to these uncertainties, we should really include uncertainties arising from non-galaxy formation aspects of the calculation. Good examples of these include the \IMF\ (which we are not explicitly trying to predict in our work, but which is uncertain and makes a significant difference to many of our results) and the spectra of stellar populations which have significant uncertainties in some regimes. Understanding these various model uncertainties is extremely challenging, but is crucial if serious parameter space searching in semi-analytic models is to take place.

However, even in the absense of a well synthesised approach, it is clear
from the data sets we have considered that
certain key problems remain to be tackled in order to produce a model of galaxy formation consistent with a broad range of observed data. 
Firstly, the sizes of model galaxies are too large, suggesting a lack of understanding of the physics of angular momentum in galaxies (see \S\ref{sec:Sizes}). It is known that the simple energy-conserving model for merger remnant sizes proposed by \cite{cole_hierarchical_2000} systematically overpredicts the sizes of spheroids and results in too much scatter in their sizes \pcite{covington_predictingproperties_2008}, but it remains unclear how much this will affect the sizes of disks\footnote{Disks feel the gravitational potential of any embedded spheroid, so their sizes will be somewhat reduced if the sizes of spheroids are systematically reduced.} and, furthermore, many spheroids in our model are formed through disk-instabilities rather than mergers---there is, as yet, no good systematic study of how to accurately determine the sizes of such instability-formed spheroids. The disk-instability process itself has signficant consequences for the angular momentum content of disks and, as such, a careful examination of this process is called for.
Secondly, despite the inclusion of tidal stripping and satellite-satellite merging, the number of satellite galaxies in high mass halos seems to remain too high, as evidenced by the clustering of galaxies (see \S\ref{sec:Clustering}). Thirdly, the clear tension between luminosity function constraints and those from the inferred star formation rate density must be reconciled. 

The model described in this work will provide the basis for further improvements to our modeling of galaxy formation. In the near future we intend to return to the following outstanding issues and examine their importance for the constraints and results presented here in greater detail:
\begin{itemize}
 \item when, exactly, do disk instabilities occur and precisely what effect do they have on the galaxies in which they happen;
 \item improved modeling of the sizes of galaxies and how different physical processes affect these sizes;
 \item the X-ray properties and hot gas fractions in halos and how these constrain the amount and type of feedback from galaxies;
 \item the effects of patchy reionization on Local Group galaxy properties and on the galaxy population as a whole;
 \item the importance of the cold mode of gas accretion and how this affects the build up of galaxies at high redshifts (c.f. \citealt{brooks_role_2009});
 \item improved modeling of \AGN\ feedback utilizing recent estimates of jet power, spin-up rates and the effects of mergers on black hole spin and mass \pcite{boyle_binary_2008,benson_maximum_2009};
 \item examination of physically motivated models of star formation and \SNe\ feedback utilizing the framework of \cite{stringer_formation_2007}.
\end{itemize}

\section{Conclusions}\label{sec:Conclusions}

In this paper we have presented recent developments of the galaxy formation model \gf. This extends the model presented in \cite{cole_hierarchical_2000} and
\cite{bower_breakinghierarchy_2006} adding many additional physical process
(such as environmental interactions and additional feedback channels), improving the treatment of other key processes (including cooling, 
re-ionisation and galaxy merging) and removing unnecessary limiting 
assumptions (such the instantaneous recycling approximation).

The new code is compared to wide range of observational constraints from
both the local and distant universe and across a wide range of wavelengths.
We navigate through the high dimesional parameter space using the 
``projection pursuit'' method suggested in \cite{bower_parameter_2010},
identifying a model that performs well in many of the observational
comparisons. 
We find it impossible to identify a model that matches
all the available datasets well and there are inherent tensions
between the datasets pointing to some remaining inadequacies in our
understanding and implementation. In particular, the model as it stands
fails to correctly account for the observed distribution of galaxy sizes
and the observed Tuly-Fisher relation.

Galaxy formation is an inherently complex and highly non-linear process. As such, it is clear that our understanding of it remains incomplete and our ability to model it imperfect. Nevertheless, huge progress has been made in both of these areas, and we expect that progress will continue at a rapid pace. The model described in this work provides an excellent match to many datasets and is in reasonable agreement with many others; it represents a solid foundation upon which to base further calculations of galaxy formation. In particular, with its parameters well constrained by current data it can be used to make predictions for as yet unprobed regimes of galaxy formation.

The present work is clearly not the last word on the subjects covered 
herein, however. In fact, we expect to constantly revise our model in response to new constraints and improved understanding of the physics\footnote{We intend to maintain a ``living document'' describing any such alterations at {\tt www.galform.org}, where we will also make available results from the model via an online database.}. This simply reflects the current state of galaxy formation theory---it is a rapidly developing field about which we are constantly gaining new insight.

\section*{Acknowledgements}

AJB acknowledges support from the Gordon and Betty Moore Foundation and would like to acknowledge the hospitality of the Kavli Institute for Theoretical Physics at the University of California, Santa Barbara where part of this work was completed. This research was supported in part by the National Science Foundation under Grant No. NSF PHY05-51164. We thank the \gf\ team (Carlton Baugh, Shaun Cole, Carlos Frenk, John Helly and Cedric Lacey) for allowing us to use the collaboratively developed \gf\ code in this work. This work has benefited from conversations with numerous people, including Juna Kollmeier, Aparna Venkatesan, Annika Peter, Alyson Brooks and Yu Lu. We thank Simon White and the anonymous referee for suggestions which helped improve the clarity of the original manuscript. We thank Shiyin Shen for providing data in electronic form. We are grateful to the authors of {\sc RecFast} and {\sc Cloudy} for making these valuable codes publicly available and to Charlie Conroy, Jim Gunn, Martin White and Jason Tumlinson for providing \SED s of single stellar populations. Lauren Porter and Tom Fox contributed code to compute galaxy clustering and \IGM\ evolution respectively. We gratefully acknowledge the Institute for Computational Cosmology at the University of Durham for supplying a large fraction of the computing time required by this project. This research was supported in part by the National Science Foundation through TeraGrid \pcite{catlett_teragrid:_2007} resources provided by the NCSA and by Amazon Elastic Compute Cloud resources provided by a generous grant from the Amazon in Education program.

\bibliographystyle{mn2e}
\bibliography{HGFv2.0}

\appendix

\end{document}

%% file: Data/data.tex
\def\NPCASearch {36,017}
\def\ChabrierIMFiZa {0.0001}
\def\ChabrierIMFiZb {0.0501}
\def\SpinModelNTree {51625}

\def\BestFitvcut {\multicolumn{2}{>{\columncolor[gray]{0.9}[1pt][2pt]}c}{N/A}}
\def\BestFitzcut {\multicolumn{2}{c}{N/A}}
\def\BestFitomega0 {0&284}
\def\BestFitlambda0 {0&716}
\def\BestFitomegab {0&04724}
\def\BestFith0 {0&691}
\def\BestFitnspec {0&933}
\def\BestFitsigma8 {0&807}
\def\BestFitcore {0&163}
\def\BestFitalphaCooledRemove {0&102}
\def\BestFitalphahot {3&36}
\def\BestFitvhotdisk {358&0}
\def\BestFitvhotburst {328&0}
\def\BestFitalphareheat {2&32}
\def\BestFitfellip {0&0214}
\def\BestFitfburst {0&335}
\def\BestFitbtburst {0&672}
\def\BestFitfgasburst {0&331}
\def\BestFitAGnedin {0&742}
\def\BestFitwGnedin {0&920}
\def\BestFitalphastar {-3&28}
\def\BestFitepsilonStar {0&0152}
\def\BestFitalphacool {0&550}
\def\BestFitepsilonSMBHEddington {0&0134}
\def\BestFitetaSMBH {0&0163}
\def\BestFitFSMBH {0&0125}
\def\BestFitstabledisk {0&743}
\def\BestFitlambdaExpelDisk {0&785}
\def\BestFitlambdaExpelBurst {7&36}
\def\BestFitRamPressureTransferFraction {0&335}

%% file: Data/PCA_Table_Ranges.tex
$h_0$ & 0&6750 & 0&7270 \\
$\Omega_{\rm b}$ & 0&04320 & 0&04920 \\
$\Lambda_0$ & 0&7142 & 0&7278 \\
$\sigma_8$ & 0&7650 & 0&8690 \\
$n_{\rm s}$ & 0&9320 & 0&9880 \\
$V_{\rm cut}$/km~s$^{-1}$ & 10&00 & 50&00 \\
$z_{\rm cut}$ & 5&000 & 13&00 \\
$\log_{10}$($\alpha_{\rm cool}$) & -1&523 & 0&4771 \\
$\log_{10}$($\alpha_{\rm remove}$) & -1&523 & 0&0000 \\
$\log_{10}$($a_{\rm core}$) & -2&000 & -0&5229 \\
$\log_{10}$($\epsilon_\star$) & -3&523 & -1&301 \\
$\alpha_\star$ & -4&000 & 1&000 \\
$V_{\rm hot,disk}$/km~s$^{-1}$ & 100&0 & 550&0 \\
$V_{\rm hot,burst}$/km~s$^{-1}$ & 100&0 & 550&0 \\
$\alpha_{\rm hot}$ & 1&000 & 3&700 \\
$\log_{10}$($\lambda_{\rm expel,disk}$) & -1&523 & 1&000 \\
$\log_{10}$($\lambda_{\rm expel,burst}$) & -1&523 & 1&000 \\
$\log_{10}$($\epsilon_\bullet$) & -2&398 & -1&000 \\
$\log_{10}$($\eta_\bullet$) & -3&000 & -1&000 \\
$\log_{10}$($F_\bullet$) & -3&000 & -1&523 \\
$\log_{10}$($\alpha_{\rm reheat}$) & -1&523 & 0&4771 \\
$\log_{10}$($f_{\rm ellip}$) & -2&000 & -0&3010 \\
$\log_{10}$($f_{\rm burst}$) & -2&000 & -0&3010 \\
$\log_{10}$($f_{\rm gas,burst}$) & -1&523 & -0&3010 \\
$B/T_{\rm burst}$ & 0&0000 & 1&000 \\
$A_{\rm ac}$ & 0&7000 & 1&000 \\
$w_{\rm ac}$ & 0&7000 & 1&000 \\
$\epsilon_{\rm d,gas}$ & 0&7000 & 1&150 \\
$\log_{10}$($\epsilon_{\rm strip}$) & -2&000 & 0&0000 \\

%% file: Data/Constraints_Table.tex
Star formation history & \S\ref{sec:SFH} & 1 & 00 \\ 
b$_{\rm J}$-band $z=0$ luminosity function & Fig. \ref{fig:bJ_LF} & 2 & 00 \\ 
K-band $z=0$ luminosity function & Fig. \ref{fig:K_LF} & 2 & 00 \\ 
Morphologically segregated $z=0$ luminosity function & Fig. \ref{fig:K_Morpho_LF} & 1 & 00 \\ 
$60\mu$m $z=0$ luminosity function & Fig. \ref{fig:60mu_z0_LF} & 1 & 00 \\ 
Evolving K-band luminosity function & Fig. \ref{fig:K20_Ks_LF} & 1 & 00 \\ 
$z=3$ UV luminosity function & Fig. \ref{fig:LyBreakLF} & 1 & 00 \\ 
$z=5$ UV luminosity functio & Fig. \ref{fig:z5_6_LF} & 0 & 75 \\ 
$z=6$ UV luminosity functio & Fig. \ref{fig:z5_6_LF} & 0 & 75 \\ 
Tully-Fisher relation & \S\ref{sec:TF} & 2 & 00 \\ 
Gas-phase metallicities & Fig. \ref{fig:SDSS_Zgas} & 1 & 00 \\ 
Colour distributions & \S\ref{sec:Colours} & 2 & 00 \\ 
Half-light radius distributions & Fig. \ref{fig:SDSS_Sizes} & 1 & 50 \\ 
Disk scale length distributions & Fig. \ref{fig:Other_Sizes} & 2 & 00 \\ 
Supermassive black hole mass distributions & \S\ref{sec:SMBH} & 1 & 00 \\ 
Stellar metallicities & Fig. \ref{fig:Zstar} & 1 & 00 \\ 
Gas-to-light ratios & Fig. \ref{fig:Gas2Light} & 1 & 00 \\ 
Clustering & \S\ref{sec:Clustering} & 1 & 50 \\ 
Local Group luminosity function & Fig. \ref{fig:LocalGroup_LF} & 1 & 00 \\ 
Local Group satellite galaxy sizes & Fig. \ref{fig:LocalGroup_Sizes} & 1 & 00 \\ 
Local Group satellite galaxy metallicities & Fig. \ref{fig:LocalGroup_Metallicities} & 1 & 00 \\ 

%% file: Data/PCA_Table_Part1.tex
{\bf PCA} & \multicolumn{2}{c}{\boldmath{$\sigma$}} & \multicolumn{2}{c}{\boldmath{$h_0$}} & \multicolumn{2}{c}{\boldmath{$\Omega_{\rm b}$}} & \multicolumn{2}{c}{\boldmath{$\Lambda_0$}} & \multicolumn{2}{c}{\boldmath{$\sigma_8$}} & \multicolumn{2}{c}{\boldmath{$n_{\rm s}$}} & \multicolumn{2}{c}{\boldmath{$V_{\rm cut}$}} & \multicolumn{2}{c}{\boldmath{$z_{\rm cut}$}} & \multicolumn{2}{c}{\boldmath{$\alpha_{\rm cool}$}} & \multicolumn{2}{c}{\boldmath{$\alpha_{\rm remove}$}} & \multicolumn{2}{c}{\boldmath{$a_{\rm core}$}} & \multicolumn{2}{c}{\boldmath{$\epsilon_\star$}} & \multicolumn{2}{c}{\boldmath{$\alpha_\star$}} & \multicolumn{2}{c}{\boldmath{$V_{\rm hot,disk}$}} & \multicolumn{2}{c}{\boldmath{$V_{\rm hot,burst}$}} & \multicolumn{2}{c}{\boldmath{$\alpha_{\rm hot}$}} \\
\hline
PCA 1 &  0&025 &  0&186 & -0&049 &  0&027 & -0&168 &  0&036 &  0&022 & {\bf -0}&{\bf 352} &  0&045 & -0&227 & -0&123 & -0&261 & {\bf  0}&{\bf 425} & -0&097 & -0&051 & {\bf -0}&{\bf 379} \\
PCA 2 &  0&043 &  0&115 & {\bf -0}&{\bf 627} & -0&283 & -0&040 & -0&283 & -0&032 &  0&182 &  0&152 &  0&050 &  0&095 & -0&111 &  0&011 & -0&143 &  0&097 &  0&008 \\
PCA 3 &  0&045 & -0&185 & -0&116 &  0&086 &  0&203 &  0&122 & {\bf  0}&{\bf 364} & -0&039 &  0&147 &  0&093 & -0&029 & -0&010 &  0&166 & {\bf  0}&{\bf 420} & -0&213 & -0&027 \\
PCA 4 &  0&051 & -0&138 &  0&004 & -0&032 & -0&032 &  0&008 & -0&055 & -0&056 & -0&011 & {\bf  0}&{\bf 410} &  0&047 & {\bf -0}&{\bf 424} & {\bf  0}&{\bf 392} &  0&030 & -0&023 & -0&284 \\
PCA 5 &  0&064 & -0&066 &  0&058 & -0&128 & -0&140 &  0&158 & -0&076 &  0&174 &  0&014 & -0&050 & -0&102 &  0&046 & -0&230 &  0&203 &  0&107 & -0&270 \\
PCA 6 &  0&083 &  0&045 & -0&117 &  0&031 &  0&135 &  0&141 &  0&217 & -0&081 &  0&112 & -0&159 & -0&260 & -0&129 & -0&092 &  0&216 &  0&053 & -0&060 \\
PCA 7 &  0&104 & -0&070 & -0&290 & {\bf -0}&{\bf 658} &  0&033 &  0&149 &  0&121 & -0&027 & -0&008 &  0&019 & -0&004 &  0&112 & -0&103 &  0&056 & -0&023 & -0&152 \\
PCA 8 &  0&106 & -0&137 & -0&109 & -0&059 &  0&041 &  0&009 &  0&145 & -0&225 & {\bf -0}&{\bf 616} &  0&028 & -0&149 & -0&205 &  0&040 &  0&108 &  0&271 & {\bf  0}&{\bf 471} \\
PCA 9 &  0&111 &  0&123 &  0&132 &  0&032 &  0&070 & -0&058 & {\bf  0}&{\bf 579} &  0&027 &  0&255 &  0&064 &  0&066 &  0&299 &  0&138 & -0&288 & {\bf  0}&{\bf 423} & -0&038 \\
PCA 10 &  0&114 & -0&054 &  0&098 & -0&050 & -0&047 & -0&053 &  0&089 & -0&152 & {\bf  0}&{\bf 411} &  0&222 & -0&032 & -0&236 & -0&116 & -0&296 & -0&212 & {\bf  0}&{\bf 407} \\
PCA 11 &  0&122 & -0&015 & -0&001 & -0&012 &  0&003 &  0&025 & -0&043 &  0&047 &  0&000 &  0&013 &  0&049 &  0&008 &  0&004 & -0&029 &  0&041 &  0&007 \\
PCA 12 &  0&128 &  0&092 &  0&043 & -0&037 & -0&072 & -0&263 &  0&037 & -0&006 & -0&010 &  0&018 & -0&122 &  0&069 &  0&000 &  0&034 & -0&012 & -0&001 \\
PCA 13 &  0&137 &  0&002 & -0&178 & -0&027 &  0&098 &  0&018 &  0&258 &  0&056 & -0&118 & -0&177 &  0&200 & -0&020 & -0&124 & -0&050 & {\bf -0}&{\bf 503} &  0&047 \\
PCA 14 &  0&142 &  0&128 & -0&031 &  0&077 &  0&063 &  0&035 & -0&022 & -0&017 & -0&032 & {\bf  0}&{\bf 467} & -0&158 &  0&195 & -0&015 & -0&022 & {\bf -0}&{\bf 482} &  0&032 \\
PCA 15 &  0&147 &  0&112 & -0&036 & -0&059 &  0&038 & -0&074 & -0&307 & {\bf -0}&{\bf 680} &  0&101 & -0&105 &  0&256 &  0&151 & -0&224 &  0&015 & -0&007 & -0&033 \\
PCA 16 &  0&155 &  0&223 &  0&021 &  0&037 &  0&016 & -0&091 &  0&139 &  0&140 & -0&162 & {\bf -0}&{\bf 505} &  0&194 & -0&079 &  0&231 & -0&135 & -0&287 &  0&101 \\
PCA 17 &  0&157 & -0&089 &  0&080 &  0&040 &  0&014 &  0&075 & -0&110 &  0&011 & -0&003 &  0&014 &  0&039 &  0&096 &  0&011 & -0&076 & -0&030 & -0&017 \\
PCA 18 &  0&168 &  0&030 & -0&122 & -0&123 & -0&013 &  0&068 & -0&216 & -0&025 &  0&302 & -0&074 & -0&131 &  0&061 & {\bf  0}&{\bf 384} &  0&147 &  0&080 & {\bf  0}&{\bf 474} \\
PCA 19 &  0&174 & {\bf  0}&{\bf 727} & -0&017 &  0&029 &  0&123 &  0&160 &  0&000 &  0&093 & -0&191 &  0&313 &  0&204 & -0&075 & -0&074 &  0&143 &  0&106 &  0&007 \\
PCA 20 &  0&182 & -0&099 & {\bf -0}&{\bf 458} & {\bf  0}&{\bf 551} & {\bf  0}&{\bf 334} & -0&087 & -0&014 & -0&077 &  0&094 & -0&033 & -0&075 & -0&117 & -0&237 & -0&063 &  0&135 & -0&109 \\
PCA 21 &  0&189 & -0&131 &  0&204 & -0&150 & -0&162 & -0&023 &  0&290 & -0&146 & -0&050 &  0&032 &  0&032 & {\bf -0}&{\bf 375} & {\bf -0}&{\bf 360} & -0&178 & -0&011 & -0&064 \\
PCA 22 &  0&196 & -0&017 & -0&322 &  0&291 & {\bf -0}&{\bf 773} &  0&318 &  0&130 & -0&047 & -0&016 &  0&062 &  0&150 &  0&134 & -0&040 &  0&009 &  0&002 &  0&085 \\
PCA 23 &  0&204 & -0&152 &  0&001 &  0&058 & -0&110 & -0&015 & -0&195 & {\bf  0}&{\bf 339} &  0&005 & -0&107 & -0&246 & -0&008 & -0&044 & -0&068 & -0&066 & -0&028 \\
PCA 24 &  0&231 &  0&099 &  0&151 &  0&011 & -0&043 &  0&053 & -0&033 &  0&157 & {\bf  0}&{\bf 346} & -0&148 &  0&292 & {\bf -0}&{\bf 374} & -0&176 & {\bf  0}&{\bf 431} &  0&074 &  0&132 \\
PCA 25 &  0&242 & -0&122 & -0&032 & -0&067 &  0&232 & {\bf  0}&{\bf 359} & -0&154 & -0&005 & -0&022 & -0&068 &  0&091 &  0&041 &  0&022 & -0&266 &  0&025 & -0&027 \\
PCA 26 &  0&279 & -0&049 &  0&084 &  0&018 &  0&111 &  0&314 & -0&011 &  0&007 & -0&015 & -0&018 &  0&154 &  0&051 &  0&086 & -0&077 & -0&005 &  0&042 \\
PCA 27 &  0&343 &  0&131 & -0&042 & -0&041 &  0&120 & {\bf  0}&{\bf 596} &  0&005 & -0&042 &  0&111 & -0&097 & -0&233 & -0&082 & -0&051 & -0&183 &  0&009 &  0&054 \\
PCA 28 &  0&366 &  0&111 & -0&046 &  0&031 &  0&011 &  0&071 & -0&151 &  0&227 & -0&039 &  0&040 & -0&047 & -0&326 & -0&063 & {\bf -0}&{\bf 333} &  0&069 &  0&023 \\
PCA 29 &  0&436 & {\bf -0}&{\bf 345} & -0&051 &  0&034 &  0&125 &  0&098 & -0&041 &  0&081 & -0&047 &  0&109 & {\bf  0}&{\bf 598} & -0&009 &  0&155 & -0&038 &  0&092 & -0&004 \\
\hline

%% file: Data/PCA_Table_Part2.tex
{\bf PCA} & \multicolumn{2}{c}{\boldmath{$\lambda_{\rm expel,disk}$}} & \multicolumn{2}{c}{\boldmath{$\lambda_{\rm expel,burst}$}} & \multicolumn{2}{c}{\boldmath{$\epsilon_\bullet$}} & \multicolumn{2}{c}{\boldmath{$\eta_\bullet$}} & \multicolumn{2}{c}{\boldmath{$F_\bullet$}} & \multicolumn{2}{c}{\boldmath{$\alpha_{\rm reheat}$}} & \multicolumn{2}{c}{\boldmath{$f_{\rm ellip}$}} & \multicolumn{2}{c}{\boldmath{$f_{\rm burst}$}} & \multicolumn{2}{c}{\boldmath{$f_{\rm gas,burst}$}} & \multicolumn{2}{c}{\boldmath{$B/T_{\rm burst}$}} & \multicolumn{2}{c}{\boldmath{$A_{\rm ac}$}} & \multicolumn{2}{c}{\boldmath{$w_{\rm ac}$}} & \multicolumn{2}{c}{\boldmath{$\epsilon_{\rm d,gas}$}} & \multicolumn{2}{c}{\boldmath{$\epsilon_{\rm strip}$}} \\
\hline
PCA 1 &  0&093 &  0&047 & -0&162 &  0&033 & -0&168 &  0&297 & -0&288 &  0&187 & -0&024 &  0&006 &  0&009 & -0&016 &  0&249 & -0&003 \\
PCA 2 &  0&093 & {\bf -0}&{\bf 486} & -0&053 & -0&059 &  0&073 &  0&109 &  0&145 &  0&048 &  0&025 & -0&105 & -0&052 &  0&052 &  0&092 & -0&054 \\
PCA 3 &  0&199 &  0&006 & {\bf  0}&{\bf 395} &  0&049 &  0&285 &  0&138 &  0&208 &  0&297 &  0&054 &  0&037 & -0&033 & -0&020 &  0&194 &  0&007 \\
PCA 4 & -0&296 & -0&021 &  0&006 & -0&073 &  0&230 & {\bf -0}&{\bf 405} &  0&060 & -0&233 &  0&019 & -0&008 & -0&027 &  0&010 & -0&117 & -0&057 \\
PCA 5 & -0&234 & -0&234 & -0&128 & {\bf  0}&{\bf 639} &  0&147 & -0&173 & -0&158 &  0&258 & -0&045 & -0&008 &  0&033 &  0&012 &  0&135 &  0&046 \\
PCA 6 &  0&050 & -0&241 &  0&231 &  0&209 & -0&265 &  0&072 & -0&142 & {\bf -0}&{\bf 629} & -0&018 & -0&011 & -0&065 & -0&025 & -0&200 & -0&117 \\
PCA 7 &  0&102 & {\bf  0}&{\bf 575} & -0&116 & -0&001 &  0&057 & -0&024 & -0&088 & -0&102 & -0&049 & -0&007 & -0&006 &  0&013 & -0&071 & -0&074 \\
PCA 8 & -0&053 & -0&044 & -0&157 &  0&066 &  0&115 &  0&029 & -0&068 & -0&043 & -0&001 & -0&001 &  0&008 &  0&002 &  0&289 & -0&046 \\
PCA 9 & -0&080 &  0&015 & -0&030 &  0&000 &  0&049 & -0&219 & -0&070 & -0&109 &  0&009 & -0&026 &  0&063 & -0&022 &  0&311 & -0&016 \\
PCA 10 & -0&154 &  0&074 & -0&051 &  0&267 &  0&287 &  0&319 & -0&237 & -0&018 &  0&007 & -0&021 & -0&006 & -0&024 & -0&131 &  0&008 \\
PCA 11 & -0&007 & -0&019 & -0&006 & -0&022 &  0&010 &  0&001 & -0&002 &  0&033 & -0&308 &  0&108 & {\bf -0}&{\bf 453} & {\bf -0}&{\bf 817} &  0&040 & -0&085 \\
PCA 12 & -0&066 &  0&010 &  0&064 & -0&005 &  0&043 & -0&012 & -0&038 &  0&025 & -0&252 & {\bf  0}&{\bf 678} & {\bf -0}&{\bf 411} & {\bf  0}&{\bf 409} &  0&003 & -0&139 \\
PCA 13 & {\bf -0}&{\bf 560} & -0&061 &  0&107 & -0&171 & -0&252 & -0&140 & -0&219 &  0&092 & -0&009 & -0&001 &  0&036 & -0&016 &  0&166 & -0&045 \\
PCA 14 &  0&327 & -0&087 & {\bf -0}&{\bf 335} &  0&153 & -0&192 & -0&179 &  0&019 & -0&178 & -0&037 &  0&027 &  0&003 & -0&011 &  0&317 &  0&043 \\
PCA 15 & -0&012 & -0&062 &  0&255 &  0&034 &  0&205 & -0&194 &  0&067 & -0&149 & -0&112 & -0&039 &  0&047 &  0&003 &  0&252 &  0&025 \\
PCA 16 &  0&237 &  0&043 & -0&109 &  0&303 & {\bf  0}&{\bf 352} & -0&250 &  0&124 & -0&140 & -0&022 &  0&025 &  0&022 & -0&029 & -0&157 & -0&025 \\
PCA 17 &  0&047 & -0&044 & -0&004 &  0&024 &  0&032 &  0&042 & -0&011 &  0&058 &  0&112 & -0&024 &  0&134 & -0&001 &  0&049 & {\bf -0}&{\bf 950} \\
PCA 18 &  0&055 &  0&030 &  0&134 &  0&068 & -0&279 & {\bf -0}&{\bf 435} & -0&202 &  0&236 & -0&011 &  0&007 &  0&038 & -0&007 & -0&085 & -0&018 \\
PCA 19 &  0&051 &  0&023 &  0&189 & -0&023 &  0&124 &  0&049 & -0&322 &  0&124 &  0&024 &  0&007 &  0&050 & -0&019 & -0&143 & -0&042 \\
PCA 20 &  0&041 &  0&269 & -0&219 &  0&046 &  0&057 & -0&203 & -0&140 &  0&147 & -0&049 &  0&005 & -0&047 &  0&034 & -0&118 & -0&047 \\
PCA 21 & {\bf  0}&{\bf 430} & -0&168 &  0&116 & -0&116 & -0&161 & {\bf -0}&{\bf 335} & -0&089 &  0&269 & -0&010 &  0&001 & -0&022 &  0&005 & -0&127 & -0&029 \\
PCA 22 &  0&010 &  0&082 &  0&058 &  0&001 &  0&034 & -0&031 &  0&038 & -0&093 &  0&085 &  0&058 & -0&072 &  0&013 & -0&048 & -0&012 \\
PCA 23 &  0&183 &  0&006 &  0&245 & -0&293 & {\bf  0}&{\bf 357} & -0&062 & {\bf -0}&{\bf 484} & -0&190 & -0&174 & -0&071 &  0&109 &  0&020 &  0&306 &  0&051 \\
PCA 24 &  0&032 &  0&104 & {\bf -0}&{\bf 348} & -0&174 & -0&035 & -0&025 &  0&055 & -0&146 &  0&105 &  0&077 & -0&072 &  0&042 & {\bf  0}&{\bf 338} & -0&057 \\
PCA 25 &  0&074 & -0&126 &  0&020 & -0&014 &  0&073 & -0&025 & -0&144 & -0&036 & {\bf  0}&{\bf 657} &  0&322 & -0&280 & -0&020 &  0&074 &  0&127 \\
PCA 26 &  0&030 & -0&066 & -0&052 & -0&015 &  0&009 &  0&021 & -0&041 &  0&034 & -0&270 & {\bf -0}&{\bf 521} & {\bf -0}&{\bf 577} & {\bf  0}&{\bf 384} &  0&002 & -0&034 \\
PCA 27 & -0&083 & -0&179 & -0&158 & -0&232 &  0&139 & -0&017 &  0&281 &  0&070 & {\bf -0}&{\bf 359} &  0&236 &  0&277 &  0&031 & -0&044 &  0&016 \\
PCA 28 & -0&025 & {\bf  0}&{\bf 335} & {\bf  0}&{\bf 414} &  0&301 & -0&249 &  0&003 &  0&329 & -0&031 & -0&042 & -0&006 & -0&025 &  0&053 & {\bf  0}&{\bf 350} & -0&013 \\
PCA 29 &  0&173 & -0&080 &  0&017 &  0&176 & -0&176 &  0&158 & -0&200 & -0&085 & {\bf -0}&{\bf 344} &  0&257 &  0&270 &  0&055 & -0&020 &  0&077 \\
\hline

%% file: Data/BestModelsTable0.tex
{\bf Parameter} &  \multicolumn{2}{c}{\begin{sideways}{\bf Overall}\end{sideways}} & \multicolumn{2}{c}{\begin{sideways}{\bf Star Formation Rate}\end{sideways}} & \multicolumn{2}{c}{\begin{sideways}{\bf b{\boldmath $_{\rm J}$} \& K LFs}\end{sideways}} & \multicolumn{2}{c}{\begin{sideways}{\bf 60$\mu$m LF}\end{sideways}} & \multicolumn{2}{c}{\begin{sideways}{\bf K20 LFs}\end{sideways}} & \multicolumn{2}{c}{\begin{sideways}{\bf Morphological LFs}\end{sideways}} & \multicolumn{2}{c}{\begin{sideways}{\bf {\boldmath $z=3$} UV LF}\end{sideways}} & \multicolumn{2}{c}{\begin{sideways}{\bf {\boldmath $z=5$} \& 6 UV LFs}\end{sideways}} & \multicolumn{2}{c}{\begin{sideways}{\bf Colours}\end{sideways}} & \multicolumn{2}{c}{\begin{sideways}{\bf Tully-Fisher}\end{sideways}} \\
\hline
\rowcolor[gray]{1.0}[1pt][2pt]
$\Lambda_0$ & \BestFitlambda0 & 0&723 & 0&723 & 0&717 & 0&721 & 0&720 & 0&717 & 0&721 & 0&723 & 0&722 \\
\rowcolor[gray]{0.9}[1pt][2pt]
$\Omega_{\rm b}$ & \BestFitomegab & 0&0445 & 0&0465 & 0&0479 & 0&0471 & 0&0491 & 0&0452 & 0&0477 & 0&0482 & 0&0441 \\
\rowcolor[gray]{1.0}[1pt][2pt]
$h_0$ & \BestFith0 & 0&677 & 0&711 & 0&714 & 0&707 & 0&726 & 0&700 & 0&703 & 0&689 & 0&724 \\
\rowcolor[gray]{0.9}[1pt][2pt]
$\sigma_8$ & \BestFitsigma8 & 0&799 & 0&786 & 0&805 & 0&785 & 0&788 & 0&779 & 0&765 & 0&808 & 0&783 \\
\rowcolor[gray]{1.0}[1pt][2pt]
$n_{\rm s}$ & \BestFitnspec & 0&955 & 0&957 & 0&939 & 0&952 & 0&960 & 0&947 & 0&946 & 0&951 & 0&959 \\
\rowcolor[gray]{0.9}[1pt][2pt]
$\alpha_{\rm reheat}$ & \BestFitalphareheat & 2&68 & 1&98 & 1&47 & 1&76 & 2&34 & 2&38 & 2&16 & 2&26 & 1&91 \\
\rowcolor[gray]{1.0}[1pt][2pt]
$\alpha_{\rm cool}$ & \BestFitalphacool & {\bf 0}&{\bf 571} & {\bf 2}&{\bf 31} & 1&12 & 1&50 & 2&67 & 2&10 & 2&81 & 0&588 & 1&06 \\
\rowcolor[gray]{0.9}[1pt][2pt]
$\alpha_{\rm remove}$ & \BestFitalphaCooledRemove & 0&842 & 0&692 & 0&133 & 0&0986 & 0&547 & 0&0607 & 0&228 & 0&162 & 0&125 \\
\rowcolor[gray]{1.0}[1pt][2pt]
$a_{\rm core}$ & \BestFitcore & 0&128 & 0&168 & 0&155 & 0&0695 & 0&187 & 0&0515 & 0&142 & 0&109 & 0&216 \\
\rowcolor[gray]{0.9}[1pt][2pt]
$A_{\rm ac}$ & \BestFitAGnedin & 0&920 & 0&819 & 0&764 & 0&780 & 0&770 & 0&804 & 0&746 & 0&880 & 0&860 \\
\rowcolor[gray]{1.0}[1pt][2pt]
$w_{\rm ac}$ & \BestFitwGnedin & 0&792 & 0&954 & 0&941 & 0&957 & 0&989 & 0&968 & 0&972 & 0&868 & 0&817 \\
\rowcolor[gray]{0.9}[1pt][2pt]
$\epsilon_\star$ & \BestFitepsilonStar & {\bf 0}&{\bf 0695} & 0&0453 & 0&0375 & 0&00520 & 0&0281 & 0&0223 & {\bf 0}&{\bf 300} & 0&0153 & 0&0427 \\
\rowcolor[gray]{1.0}[1pt][2pt]
$\alpha_\star$ & \BestFitalphastar & -2&11 & -2&73 & -2&65 & {\bf -2}&{\bf 05} & -3&43 & -2&95 & {\bf -3}&{\bf 68} & {\bf -0}&{\bf 392} & -1&69 \\
\rowcolor[gray]{0.9}[1pt][2pt]
$\epsilon_{\rm d,gas}$ & \BestFitstabledisk & 0&734 & 0&743 & 0&726 & 0&829 & 0&774 & 0&804 & 0&957 & 0&808 & 0&812 \\
\rowcolor[gray]{1.0}[1pt][2pt]
$V_{\rm hot,disk}$ & \BestFitvhotdisk & 425&0 & 532&0 & 421&0 & 491&0 & 411&0 & 506&0 & {\bf 546}&{\bf 0} & 459&0 & 389&0 \\
\rowcolor[gray]{0.9}[1pt][2pt]
$V_{\rm hot,burst}$ & \BestFitvhotburst & 130&0 & 470&0 & 413&0 & 539&0 & 498&0 & 544&0 & 488&0 & 242&0 & 370&0 \\
\rowcolor[gray]{1.0}[1pt][2pt]
$\alpha_{\rm hot}$ & \BestFitalphahot & 2&61 & 2&81 & 2&73 & {\bf 3}&{\bf 57} & 3&50 & 3&53 & 3&15 & 2&95 & 2&58 \\
\rowcolor[gray]{0.9}[1pt][2pt]
$\lambda_{\rm expel,disk}$ & \BestFitlambdaExpelDisk & 0&738 & 0&252 & 0&920 & 0&273 & 0&477 & 0&571 & 0&266 & 0&607 & 0&551 \\
\rowcolor[gray]{1.0}[1pt][2pt]
$\lambda_{\rm expel,burst}$ & \BestFitlambdaExpelBurst & 6&49 & 7&90 & 5&39 & 5&23 & 7&46 & 6&62 & 9&23 & 6&55 & {\bf 2}&{\bf 13} \\
\rowcolor[gray]{0.9}[1pt][2pt]
$\epsilon_{\rm strip}$ & \BestFitRamPressureTransferFraction & 0&951 & 0&696 & 0&248 & 0&207 & 0&0997 & 0&739 & 0&355 & 0&145 & 0&101 \\
\rowcolor[gray]{1.0}[1pt][2pt]
$f_{\rm ellip}$ & \BestFitfellip & 0&184 & 0&0946 & 0&0658 & 0&0250 & {\bf 0}&{\bf 327} & 0&0246 & 0&118 & 0&0250 & 0&308 \\
\rowcolor[gray]{0.9}[1pt][2pt]
$f_{\rm burst}$ & \BestFitfburst & 0&260 & 0&310 & 0&477 & 0&297 & 0&286 & 0&281 & 0&451 & 0&183 & 0&263 \\
\rowcolor[gray]{1.0}[1pt][2pt]
$f_{\rm gas,burst}$ & \BestFitfgasburst & 0&209 & 0&0817 & {\bf 0}&{\bf 0553} & 0&164 & 0&349 & 0&236 & 0&225 & 0&452 & 0&160 \\
\rowcolor[gray]{0.9}[1pt][2pt]
${\rm B/T}_{\rm burst}$ & \BestFitbtburst & 0&538 & 0&889 & 0&890 & 0&517 & 0&367 & 1&00 & 0&215 & 0&928 & 0&409 \\
\rowcolor[gray]{1.0}[1pt][2pt]
$\epsilon_\bullet$ & \BestFitepsilonSMBHEddington & 0&0437 & 0&00596 & 0&0363 & 0&00877 & 0&0542 & 0&00423 & 0&0407 & 0&0130 & 0&0857 \\
\rowcolor[gray]{0.9}[1pt][2pt]
$\eta_\bullet$ & \BestFitetaSMBH & 0&0596 & 0&00476 & 0&00711 & 0&00188 & 0&0137 & 0&00728 & 0&0307 & 0&00788 & 0&0893 \\
\rowcolor[gray]{1.0}[1pt][2pt]
$F_\bullet$ & \BestFitFSMBH & 0&00818 & 0&00289 & 0&00628 & 0&0206 & 0&0233 & 0&00190 & 0&0256 & 0&00271 & 0&0164 \\
\rowcolor[gray]{0.9}[1pt][2pt]
$V_{\rm cut}$ & \BestFitvcut & 17&0 & 36&5 & 28&4 & 38&7 & 43&9 & 32&9 & 27&7 & 34&5 & 12&7 \\
\rowcolor[gray]{1.0}[1pt][2pt]
$z_{\rm cut}$ & \BestFitzcut & 10&1 & 11&7 & 10&9 & 12&7 & 12&4 & 10&8 & 11&9 & 12&8 & 10&2 \\
\hline

%% file: Data/BestModelsTable1.tex
{\bf Parameter} &  \multicolumn{2}{c}{\begin{sideways}{\bf Overall}\end{sideways}} & \multicolumn{2}{c}{\begin{sideways}{\bf Tully-Fisher}\end{sideways}} & \multicolumn{2}{c}{\begin{sideways}{\bf Sizes}\end{sideways}} & \multicolumn{2}{c}{\begin{sideways}{\bf Metallicity}\end{sideways}} & \multicolumn{2}{c}{\begin{sideways}{\boldmath $M_{\rm Hy}/L_{\rm B}$}\end{sideways}} & \multicolumn{2}{c}{\begin{sideways}{\bf Clustering}\end{sideways}} & \multicolumn{2}{c}{\begin{sideways}{\bf SMBHs}\end{sideways}} & \multicolumn{2}{c}{\begin{sideways}{\bf Local Group LF}\end{sideways}} & \multicolumn{2}{c}{\begin{sideways}{\bf Local Group Sizes}\end{sideways}} & \multicolumn{2}{c}{\begin{sideways}{\bf Local Group {\boldmath $Z$}'s}\end{sideways}} \\
\hline
\rowcolor[gray]{1.0}[1pt][2pt]
$\Lambda_0$ & \BestFitlambda0 & 0&722 & 0&720 & 0&724 & 0&715 & 0&714 & 0&716 & 0&722 & 0&718 & 0&722 \\
\rowcolor[gray]{0.9}[1pt][2pt]
$\Omega_{\rm b}$ & \BestFitomegab & 0&0441 & 0&0447 & 0&0453 & 0&0458 & 0&0470 & 0&0437 & 0&0475 & 0&0492 & 0&0467 \\
\rowcolor[gray]{1.0}[1pt][2pt]
$h_0$ & \BestFith0 & 0&724 & 0&698 & 0&720 & 0&711 & 0&688 & 0&699 & 0&710 & 0&682 & 0&685 \\
\rowcolor[gray]{0.9}[1pt][2pt]
$\sigma_8$ & \BestFitsigma8 & 0&783 & 0&809 & 0&775 & 0&788 & 0&795 & 0&778 & 0&771 & 0&769 & 0&773 \\
\rowcolor[gray]{1.0}[1pt][2pt]
$n_{\rm s}$ & \BestFitnspec & 0&959 & 0&961 & 0&948 & 0&935 & 0&957 & 0&945 & 0&938 & 0&933 & 0&942 \\
\rowcolor[gray]{0.9}[1pt][2pt]
$\alpha_{\rm reheat}$ & \BestFitalphareheat & 1&91 & 2&33 & 2&37 & 2&52 & 0&922 & 2&96 & 2&43 & 2&62 & 1&81 \\
\rowcolor[gray]{1.0}[1pt][2pt]
$\alpha_{\rm cool}$ & \BestFitalphacool & 1&06 & {\bf 0}&{\bf 0955} & 2&11 & 1&48 & 0&855 & 1&28 & 2&30 & 2&25 & 1&49 \\
\rowcolor[gray]{0.9}[1pt][2pt]
$\alpha_{\rm remove}$ & \BestFitalphaCooledRemove & 0&125 & 0&917 & 0&334 & 0&146 & 0&466 & 0&848 & 0&0814 & 0&0825 & 0&0508 \\
\rowcolor[gray]{1.0}[1pt][2pt]
$a_{\rm core}$ & \BestFitcore & 0&216 & 0&0905 & 0&0772 & 0&0281 & 0&105 & 0&222 & 0&127 & 0&0940 & 0&0210 \\
\rowcolor[gray]{0.9}[1pt][2pt]
$A_{\rm ac}$ & \BestFitAGnedin & 0&860 & {\bf 0}&{\bf 964} & 0&765 & 0&766 & 0&795 & 0&876 & 0&741 & 0&736 & 0&737 \\
\rowcolor[gray]{1.0}[1pt][2pt]
$w_{\rm ac}$ & \BestFitwGnedin & 0&817 & 0&809 & 0&945 & 0&989 & 0&871 & 0&919 & 0&908 & 0&928 & 0&985 \\
\rowcolor[gray]{0.9}[1pt][2pt]
$\epsilon_\star$ & \BestFitepsilonStar & 0&0427 & 0&00735 & {\bf 0}&{\bf 00272} & {\bf 0}&{\bf 00329} & 0&0295 & 0&0420 & 0&00751 & 0&0322 & 0&0175 \\
\rowcolor[gray]{1.0}[1pt][2pt]
$\alpha_\star$ & \BestFitalphastar & -1&69 & -2&83 & {\bf -3}&{\bf 60} & -3&07 & -2&65 & -2&51 & -3&32 & -2&65 & {\bf -1}&{\bf 52} \\
\rowcolor[gray]{0.9}[1pt][2pt]
$\epsilon_{\rm d,gas}$ & \BestFitstabledisk & 0&812 & {\bf 0}&{\bf 716} & 0&774 & 0&773 & 0&736 & 0&743 & 0&957 & 0&784 & 0&800 \\
\rowcolor[gray]{1.0}[1pt][2pt]
$V_{\rm hot,disk}$ & \BestFitvhotdisk & 389&0 & 341&0 & 497&0 & 449&0 & 353&0 & 393&0 & 374&0 & 452&0 & 543&0 \\
\rowcolor[gray]{0.9}[1pt][2pt]
$V_{\rm hot,burst}$ & \BestFitvhotburst & 370&0 & {\bf 125}&{\bf 0} & 498&0 & 496&0 & 341&0 & 271&0 & 507&0 & 533&0 & 467&0 \\
\rowcolor[gray]{1.0}[1pt][2pt]
$\alpha_{\rm hot}$ & \BestFitalphahot & 2&58 & 3&12 & 3&32 & {\bf 3}&{\bf 53} & 2&37 & 3&18 & 3&25 & 3&14 & {\bf 2}&{\bf 48} \\
\rowcolor[gray]{0.9}[1pt][2pt]
$\lambda_{\rm expel,disk}$ & \BestFitlambdaExpelDisk & 0&551 & 0&412 & 0&283 & 0&380 & 1&06 & 0&646 & 0&438 & 0&659 & 0&622 \\
\rowcolor[gray]{1.0}[1pt][2pt]
$\lambda_{\rm expel,burst}$ & \BestFitlambdaExpelBurst & {\bf 2}&{\bf 13} & 5&62 & 8&97 & 7&87 & 7&24 & 9&86 & 9&60 & 8&16 & 6&38 \\
\rowcolor[gray]{0.9}[1pt][2pt]
$\epsilon_{\rm strip}$ & \BestFitRamPressureTransferFraction & 0&101 & 0&607 & 0&0184 & 0&200 & 0&288 & 0&359 & 0&0787 & 0&975 & 0&595 \\
\rowcolor[gray]{1.0}[1pt][2pt]
$f_{\rm ellip}$ & \BestFitfellip & 0&308 & 0&360 & 0&0925 & 0&0204 & 0&107 & 0&203 & 0&454 & 0&0672 & 0&0212 \\
\rowcolor[gray]{0.9}[1pt][2pt]
$f_{\rm burst}$ & \BestFitfburst & 0&263 & 0&242 & 0&348 & 0&483 & 0&239 & 0&435 & 0&379 & 0&388 & 0&436 \\
\rowcolor[gray]{1.0}[1pt][2pt]
$f_{\rm gas,burst}$ & \BestFitfgasburst & 0&160 & 0&0937 & 0&171 & 0&264 & 0&361 & 0&120 & 0&410 & 0&225 & 0&450 \\
\rowcolor[gray]{0.9}[1pt][2pt]
${\rm B/T}_{\rm burst}$ & \BestFitbtburst & 0&409 & 0&681 & 0&734 & 0&825 & 0&500 & 0&695 & 0&545 & 0&251 & 0&718 \\
\rowcolor[gray]{1.0}[1pt][2pt]
$\epsilon_\bullet$ & \BestFitepsilonSMBHEddington & 0&0857 & 0&0232 & 0&0266 & 0&0914 & 0&0201 & 0&0560 & 0&0419 & 0&00481 & 0&00823 \\
\rowcolor[gray]{0.9}[1pt][2pt]
$\eta_\bullet$ & \BestFitetaSMBH & 0&0893 & 0&0588 & 0&00928 & 0&0912 & 0&0216 & 0&0248 & 0&0139 & 0&0119 & 0&00538 \\
\rowcolor[gray]{1.0}[1pt][2pt]
$F_\bullet$ & \BestFitFSMBH & 0&0164 & 0&00970 & 0&00807 & 0&0293 & 0&00352 & 0&0287 & 0&0133 & 0&0279 & 0&00585 \\
\rowcolor[gray]{0.9}[1pt][2pt]
$V_{\rm cut}$ & \BestFitvcut & 12&7 & 27&2 & 26&5 & 43&0 & 42&9 & 28&8 & {\bf 45}&{\bf 5} & {\bf 47}&{\bf 5} & 35&5 \\
\rowcolor[gray]{1.0}[1pt][2pt]
$z_{\rm cut}$ & \BestFitzcut & 10&2 & 9&31 & 11&0 & 12&5 & 12&7 & 11&0 & {\bf 12}&{\bf 8} & {\bf 12}&{\bf 9} & {\bf 12}&{\bf 7} \\
\hline